\documentclass[%
 aip,
 jmp,%
 amsmath,amssymb,
reprint,%
]{revtex4-2}
\usepackage[margin=1in]{geometry}
\usepackage{comment,ulem}
\usepackage{graphicx}
\usepackage{subfigure}
\usepackage{float}
\usepackage{amsmath,amssymb,amsfonts}
\usepackage{mathtools}
\usepackage{enumitem}
\usepackage{dcolumn}
\usepackage{bm}
\usepackage{xcolor}
\usepackage{environ}
\usepackage{microtype}
\usepackage{booktabs}
\usepackage{multirow}
\usepackage{soul}
\usepackage{color}
\newcommand{\hlll}[1]{{#1}}

\newcommand{\cmnt}[1]{\ignorespaces}

\begin{document}

\preprint{AIP/123-QED}

\title[{Wavy Wall Turbulence}]{Slope Dependent Turbulence over Two-dimensional Wavy Surfaces}



\author{Balaji Jayaraman}
 \email{jayaramanbalaji@gmail.com}
\author{Saadbin Khan}%
\affiliation{ 
School of Mechanical and Aerospace Engineering, Oklahoma State University, Stillwater, OK 74078. 
}%

\date{\today}

\begin{abstract}
Knowledge of turbulent flows over non-flat surfaces is of major practical interest in diverse applications. Significant work continues to be reported in the roughness regime at high Reynolds numbers where the cumulative effect of surface undulations on the averaged and integrated turbulence quantities is well documented. Even for such cases, the surface topology plays an important role for transitional roughness Reynolds numbers that is hard to characterize. In this work, we attempt to develop a bottom up understanding of the mechanisms underlying turbulence generation and transport, particularly within the region of the turbulent boundary layer (TBL) affected by the surface. We relate surface characteristics with turbulence generation mechanisms, Reynolds stress transport and the resulting drag increase. To this end, we perform a suite of direct numerical simulations of fully developed turbulent flow between two infinitely wide, two-dimensional sinusoidally wavy surfaces at a friction Reynolds number, $Re_{\tau}=180$, with different mean surface slopes, $\zeta$ (and fixed inner-scaled undulation height, $a^+=13$) corresponding to the `waviness' regime. The increase in wave slope enhances near surface turbulent mixing resulting in increased total drag, higher fraction of form drag, faster approach to isotropy and thereby, modulation of the buffer layer. The primary near-surface streamwise and vertical turbulence generation occurs in the leeward and windward side of the wave. In contrast, the spanwise variance is produced through pressure-rate-of-strain mechanism, primarily, along the windward side of the wave for flows with very little flow separation. We also observe significant dispersion effects in the production structure.
\end{abstract}

\maketitle




\section{Introduction \label{sec:intro}}
Understanding turbulent boundary layers over complex surface undulations has immense practical value in  both environmental and engineering applications. Common examples of engineering applications include internal flows in pipes and turbomachinery, external flows over fouled ship hulls~\cite{schultz2007effects}, wind turbine blades and other aerodynamic surfaces. In the environment, the earth's surface offers a wide variety of surface heterogenieities  including water waves, sand dunes, and shrubs on the smaller end; tree canopies and buildings at the medium scale; and mountains and hills at the larger scale. Being ubiquitous, understanding this flow dynamics over such wide range of heterogeneity becomes a necessity. Given the need for accurate drag estimation, a significant amount of research has been devoted to understanding turbulent flows over pipe roughness, for example, the work of Darcy~\cite{darcy1857recherches} nearly two hundred years ago, in the early half of twentieth century by Nikuradse~\cite{nikuradse1950laws}, Colebrook~\cite{colebrook1939correspondence} and Moody~\cite{moody1944friction} and more recently by various research groups~\cite{shockling2006roughness,hultmark2013logarithmic,chan2015systematic}. 
By roughness, one typically refers to undulations with characteristic scales that are much smaller (say, $k/\delta \lesssim O(0.01)$) than the outer layer scales and much larger (say, $k/L_{v}=k^+\gtrsim O(100)$) than the inner layer (i.e. viscous layer) scales. Here $k$ is roughness length scale, $L_v$ is the viscous length scale and $\delta$ represents outer layer scale or boundary layer height. Recent fundamental investigation of turbulent flows over uniform roughness embedded in flat surface has been undertaken through a series of experimental studies~\cite{coleman2007spatially,flack2005experimental,schultz2005outer,schultz2007rough,schultz2009turbulent,flack2014roughness,flack2007examination,flack2010review} as reviewed in ~\cite{jimenez2004turbulent,flack2014roughness}. 
In addition, there exists an extensive body of experimental~\cite{coleman2007spatially} and simulation-based research of turbulent boundary layers (TBLs) over systematically designed roughness using direct numerical simulation (DNS) ~\cite{de1997direct, cherukat1998direct,bhaganagar2004effect,chau2012understanding,napoli2008effect,chan2015systematic,leonardi2007properties,leonardi2010channel} and large eddy simulation (LES)~\cite{de2012effects}. These studies invariably focused on relatively large-scale surface undulations, i.e. $k/\delta \lesssim O(0.01-0.1)$, $k/L_{v}=k^+\gtrsim O(10)$ and low to moderate Reynolds numbers, $Re_{\tau}\sim 100-700$ owing to computational considerations. 
From a roughness characterization perspective, there has been interesting recent efforts to mimic Nikuradse-type sand grain roughness using DNS at moderately high Reynolds numbers~\cite{thakkar2018direct,busse2017reynolds} ($Re_{\tau}\sim 400-700$). While these studies more realistic, they are limited in their ability to elucidate low-level eddying structure, coherence characteristics and provide insight into physical mechanisms. Overcoming this gap in knowledge requires investigation of the surface (or roughness) layer physics to generate a bottom-up understanding of the relationship between surface texture, roughness layer dynamics and in turn, averaged statistics including drag. This is being accomplished through many studies that (i) characterize the roughness sublayer scale and its relationship to other flow scales; (ii) characterize the roughness layer dynamics for a given parameterized shape; (iii)  characterize the interaction of sublayer dynamics  with inertial layer turbulence and (iv) develop strategies for physics-aware predictive models. Knowledge of the surface layer dynamics \cmnt{in the presence of surface undulations} from such fundamental studies also serve the additional purpose of understanding resolvable surface heterogeneity-driven lower atmosphere dynamics such as those pursued in our group~\cite{jayaraman2014transition,jayaraman2018transition,khan2019statistical} and elsewhere~\cite{coceal2006mean}. Therefore, in this study we primarily focus on in-depth investigation of turbulence statistical structure over parameterized two-dimensional wavy surfaces and their implications to roughness characterization.

\paragraph{Roughness Characterization:} Early efforts~\cite{nikuradse1950laws,colebrook1939correspondence,moody1944friction} related drag with roughness classified as hydraulically smooth, transitional or fully rough regimes based on the roughness scales. These regimes correspond to flows with purely viscous, combined viscous and form drag and purely form drag respectively. In the fully rough limit with minimal viscous effects, the drag depends only on the roughness scale, $k$ (and not on the Reynolds number) unlike the transitional regime. The roughness function~\cite{hama1954boundary}, $\Delta \langle u \rangle^+$ quantifies the effect of roughness on outer layer flow as the downward displacement in the mean velocity profile (when plotted in a semi-log scale) due to increased drag from surface inhomogeneities. 
The classical view is that roughness tends to impact the turbulence structure only up to a few roughness lengths from the surface beyond which the outer layer flow is unaffected except for appropriate scales, i.e. scale similarity~\cite{townsend1980structure}.
A consequence of such `wall similarity'~\cite{raupach1991rough} is that the shape of the mean velocity in the overlap and outer layers remains unaffected (relative to a smooth wall) by the surface texture. 
 The criteria for the universal existence of wall similarity (likely for $k/\delta \ll 1$ and $k^+ \gg 1$) is actively being explored~\cite{flack2014roughness,jimenez2004turbulent}. Further, wall similarity is often demonstrated using first order statistics and not for higher order or eddying turbulent structure~\cite{bhaganagar2004effect}. In fact, a single critical roughness height does not exist as the influence of surface texture on outer layer statistics decays gradually and quantification dependent~\cite{flack2010review,flack2014roughness,bhaganagar2004effect}.
The challenge is roughness modeling, say, $\Delta \langle u \rangle^+=f(k^+,\dots)$ is two fold: (i) identifying the different independent variables and (ii) characterizing the nature of the function dependencies, i.e. $f$. The common models such as the correlations of Nikuradse~\cite{nikuradse1950laws} ($\Delta \langle u \rangle^+ = (1/{\kappa})\log(k^+)+B-8.5$) and Colebrook~\cite{colebrook1939correspondence} ($\Delta \langle u \rangle^+ = (1/{\kappa})\log(1+0.3k^+)$) use a single roughness parameter, $k^+$ that represents the entire complexity of the surface topology.    
Topology dependence shows up in many ways, for example, two-dimensional wavy roughness are known to generate stronger vertical disturbances (and reduced outer layer similarity) due to the absence of significant spanwise motions in the roughness sublayer as compared to three-dimensional wavy surfaces~\cite{bhaganagar2004effect,napoli2008effect}. Volino et al.~\cite{volino2011turbulence,volino2009turbulence} corroborate these trends for sharp-edged roughness such as square bars or cubes. It turns out that two-dimensional surface undulations require a larger scale separation (higher $\delta/k$ and $Re_{\tau}$) to match the characteristics of thee-dimensional surface undulations~\cite{krogstad2012turbulence}. Therefore, all of this highlights the need for improved understanding of the roughness layer dynamics for better characterization of outer layer responses to the detailed surface topology.

\paragraph{Turbulence Structure Over Regular Surface Undulations:}
Understanding the influence of topology on averaged turbulence statistics requires deep insight into the underlying turbulence generation mechanisms and the Reynolds stress evolution that determines the overall drag.
Turbulent flows over smooth sinusoidal surface undulations have been explored experimentally~\cite{thorsness1977turbulent,zilker1977influence,zilker1979influence,hudson1996turbulence} since the seventies. Zilker and co-authors~\cite{zilker1977influence,zilker1979influence} study small amplitude effects of wavy surface abutting a turbulent flow and note that the dynamics is well approximated by linear theory for cases with very little to no separation. Larger amplitude wavy surface turbulence\cite{buckles1984turbulent} show significant flow separation with strong nonlienar dynamics, especially for wave slopes, $\zeta=2a/\lambda \gtrsim 0.05$ where $a$ is the wave amplitude and $\lambda$, the wavelength. For flows with incipient separation or when fully attached ($\zeta=2a/\lambda \lesssim 0.03$), the pressure variations show linear dependence with wave height. With advances in computing, DNS of turbulent flow over two-dimensional sinusoidal waves~\cite{de1997direct} (for $\zeta=0.05,0.1$ and $Re_{\tau}\sim 85$) was employed to analyze the detailed turbulence structure. This study explores the asymmetric pressure field along the wave and represents one of the first analysis of turbulence energy budgets and production mechanisms. The near-surface turbulence mechanism is summarized as follows. The wavy surface undulations generate alternating and asymmetric bands of favorable (upslope) and adverse (downslope) pressure gradients which in turn cause regions of alternating shear (contributes to dissipation) and Reynolds stresses (contributes to production) that complement each other. In the downslope, the higher momentum fluid goes away from the surface (ejection-like events) while in the upslope it moves toward the wall (sweep-like events). This results in shape-induced turbulent mixing, increased Reynolds stresses near the surface and pressure-strain generation of spanwise turbulent motions (through splat events). 
DNS~\cite{bhaganagar2004effect} was also used to characterize rough-wall TBL structure over three-dimensional egg-carton-shaped sinusoidal undulations ($\zeta=2a/\lambda \sim O(0.1)$ and $Re_{\tau}\sim 180,400$) where it was observed that the effect of surface undulations on the outer layer is primarily felt by large-scale motions\cmnt{ while the effect on small scale (vorticity) motions is negligible}. 
Further, the study also reports numerical experiments to assess the role of surface-induced production over wavy undulations. These conclusions align with outcomes from the current study where we directly isolate the contributions to turbulence production from the different mechanisms\cmnt{, i.e. surface-induced and shear instability}. 
Surface undulations impact the near-wall coherent structures in a manner consistent with the horizontal scale of the undulations as observed using two-point correlation measures~\cite{de1997direct,bhaganagar2004effect} that show appropriate streamwise coherence reduction over non-flat surfaces.    
This can be related to traditional classifications~\cite{perry1987experimental} of ``d-type'' or ``k-type'' roughness for sharp shaped surface undulations such as pyramids~\cite{schultz2009turbulent}, cubes or square prisms~\cite{leonardi2003direct,leonardi2007properties,leonardi2010channel} depending on which length scale influences the flow, i.e. the roughness scale $k$ or the boundary layer scale, $\delta$.
Similarly, when dealing with wavy undulations, there exist multiple length scales~\cite{nakato1985resistance} ($\lambda_{x,z}$ in the horizontal and $a$ in the vertical). It turns out that strong separated flow increases the dependence of the turbulence structure on roughness height, $k$ (i.e. form drag dominates) to mimic sand-type roughness. However, the topology determines the extent of flow separation which for smooth wavy-type undulations is characterized by an effective slope ($ES$) metric~\cite{napoli2008effect} which for two-dimensional waves is given by $ES=2\zeta$. 

In this study we explore the near-wall turbulence structure over two-dimensional wavy surfaces using direct numerical simulation of wavy channel flow at a friction Reynolds number, $Re_{\tau}=\delta^+\approx 180$ with higher-order spectral like accuracy~\cite{laizet2009high} and surface representation using an immersed boundary method~\cite{peskin1972flow,parnaudeau2004combination}.
The focus of our current analysis is to better understand the mechanisms underlying drag increase and turbulence generation at low enough slopes where viscous and form drag are both important. For the sinusoidal two-dimensional surfaces considered in this study, the effective slope, ES is directly related to the non-dimensional ratio of the amplitude ($a$) and wavelength ($\lambda$) of the sinusoid, i.e. ES $=4a/\lambda=2\zeta$ where $\zeta$ is the steepness factor. 
Here $\zeta$ is varied from $0$ to $0.044$ (ES$~ \sim 0-0.088$) while keeping the roughness Reynolds number, $a^+$ constant.  This $\zeta$ range represents the transition from attached flow to incipient to weakly separated flow. 
The roughness height or wave amplitude, $a$,  is chosen to generate moderate scale separation, i.e. $\delta/a \approx 15$ ($a^+ \approx 12-13$) that \cmnt{is representative of geophysical flows and} falls in the range of values capable of generating outer layer similarity as per Flack et al.~\cite{flack2007examination,flack2014roughness}.
%
%
The rest of the article is organized as follows. In section~\ref{sec:num_method}, we describe the numerical methods, simulation design, quantification of statistical convergence and validation efforts. In section~\ref{sec:results-firstorder}, we present the streamwise averaged first order turbulence structure. In section~\ref{sec:results-secondorder} we characterize the second-order turbulence structure, namely, the components of the Reynolds stress tensor and their generation mechanisms as modulated by the wavy surface undulations. In section~\ref{sec:conclusion} we summarize the major findings from this study.

\section{Numerical Methods \label{sec:num_method}}
In this study, we adopt a customized version of the Incompact3D~\cite{laizet2009high} code framework to perform our DNS study. The dynamical system is the incompressible Navier-Stokes equation for Newtonian flow described in a Cartesian coordinate system with $x$,$y$,$z$ corresponding to streamwise, vertical and spanwise directions respectively. The skew-symmetric vector form of the equations are given by
\begin{align}
\frac{\partial \mathbf{u}}{\partial t}=-\mathbf{\nabla} p-\frac{1}{2}\big[\mathbf{\nabla} (\mathbf{u}\oplus \mathbf{u})+(\mathbf{u}\mathbf{\nabla})\mathbf{u})\big]+\nu \mathbf{\nabla}^2\mathbf{u}+\mathbf{f} \textrm{ and }
\end{align}
\begin{align}
\mathbf{\nabla}.\mathbf{u}=0, 
\end{align}
where $\mathbf{f}$ and $p$ represents the body force and pressure field respectively. As we consider the fluid to be incomrpesisble, the fluid density ($\rho$) is taken to as one without loss of generality \cmnt{ fluid as we solve the non-dimensional system of equations}. Naturally, the above system of equations can be rewritten to generate a separate Poisson equation for pressure. 

The system of equations are advanced in time using a 3$^{rd}$ order Adam-Bashforth (AB3) time integration with pressure-velocity coupling using a fractional step method \cite{kim1985application}. For the channel flow, the body force term, $\mathbf{f}$ is dropped. The velocity is staggered by half a cell to the pressure variable for exact mass conservation. A $6^{th}$ Order Central Compact Scheme (6OCCS) with quasi-spectral accuracy is used to calculate the first and second derivative terms in the transport equation. The pressure Poisson equation (PPE) is efficiently solved using a spectral approach. 
The right hand side of the PPE is computed using a quasi-spectral accuracy using 6OCCS and then transformed to Fourier space. To account for the discrepancy between the spectrally accurate derivative for the pressure gradient and a quasi-spectral accuracy for the divergence term, the algorithm uses a modified wavenumber in the pressure solver.

A major downside to the use of higher order schemes as above is the representation of complex geometry. In particular, the boundary conditions for higher order methods are hard to implement without loss of accuracy near the surface. In this work, we adopt an immersed boundary method (IBM)  framework where the solid object is represented through a force field in the governing equations to be solved  on a Cartesian grid. 
In this study, we leverage the higher order, direct forcing IBM implementation in Incompact3D requiring reconstruction of the velocity field inside the solid object so as to enforce zero velocity at the interface
Therefore, this velocity reconstruction step is the key to success of this approach. The numerous different IBM implementations~\cite{parnaudeau2004combination} differ in the details of this reconstruction. In the current study, we adopt the one-dimensional higher order polynomial reconstruction of~\citet{gautier2014dns} and refer to~\citet{khan2019statistical} for a more detailed presentation of the method. The reconstructed velocity field is directly used to estimate the derivatives in the advection and diffusion terms of the transport equation which is shown to be reasonably accurate as per Section~\ref{subsec:assess_accuracy}.  

\subsection{Simulation Design \label{subsec:sim_design}}

\begin{figure}[ht!]
\centering
\mbox{
\subfigure[\label{schematic}]{\includegraphics[width=0.4\textwidth]{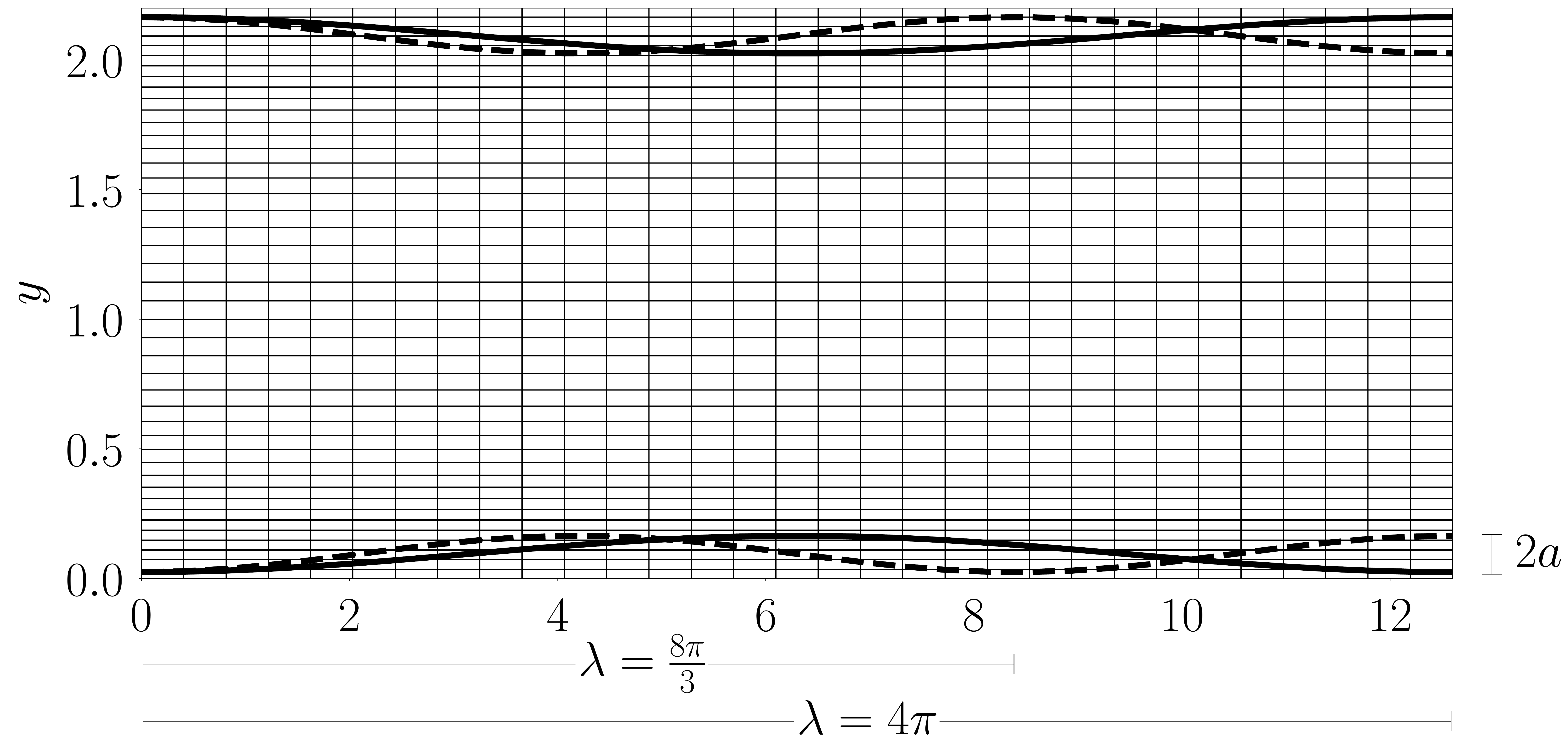}}
\subfigure[\label{schematic_zoomed}]{\includegraphics[width=0.4\textwidth]{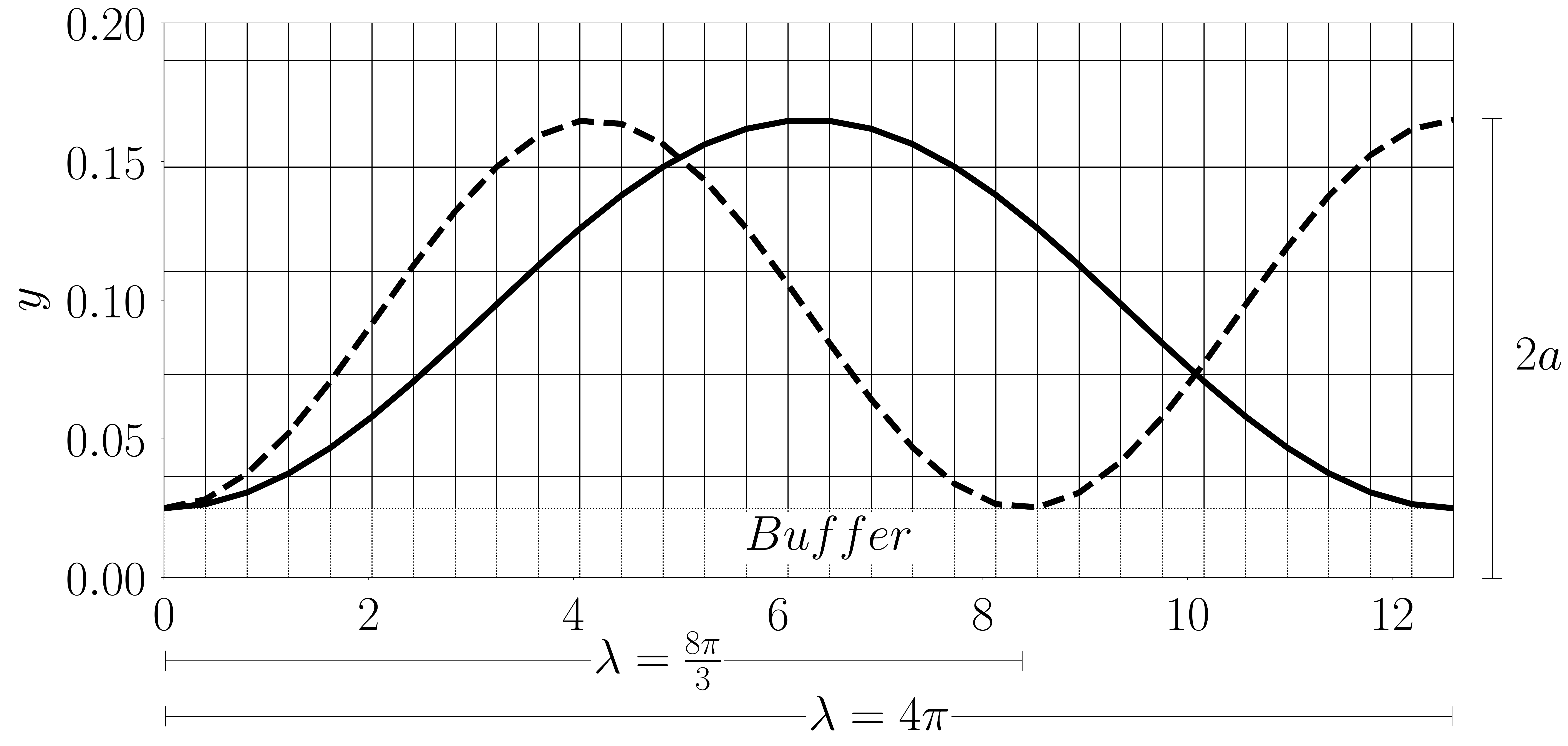}}
}
\caption{Schematic illustration of the Cartesian grid with the immersed boundaries of different shapes in (a) and a close-up of the buffer region in (b). The solid thick curve represents the wave for $\lambda=4\pi$ and the dashed line for $\lambda=\frac{8\pi}{3}$. A  similar setup is used for other surface shapes as well.  
}
\end{figure}

We carry out six different simulations of turbulent channel flows with flat and wavy surfaces of different steepness levels ($\zeta$), but constant wave amplitude ($a$) as shown in figure~\ref{schematic}. Here $\zeta$ is the average steepness for this sinusoidal shape given by $\zeta=2a/\lambda$, where $\lambda$ is the wavelength.  In this study $\zeta$ is varied from $0-0.044$ with $0$ corresponding to a flat surface and $\zeta=0.044$ corresponding to four sinusoidal waves over the streamwise length of the simulation domain.  In all these cases, care was taken to ensure that the realized friction Reynolds number is sufficiently close to the targeted value of $\sim 180$ by modifying the corresponding bulk Reynolds number. The bulk Reynolds number, $Re_b=\frac{u_b\delta}{\nu}$,  for the flat channel is chosen as $\sim 2800$. 
For the different wavy channel flows with the same effective flow volume and mean channel heights, using the same flow rate (or bulk Reynolds number) increases the friction Reynolds number, $Re_{\tau}=\frac{u_{\tau}\delta}{\nu}$, due to increase in $u_{\tau}$ with wave steepness, $\zeta$. Therefore, to achieve a constant value of the friction Reynolds number, the bulk Reynolds number was appropriately reduced through an iterative process so that the increment in $u_\tau$ does not significantly affect the friction Reynolds number, $Re_{\tau}=\frac{u_{\tau}\delta}{\nu}$. The simulation parameters for the different cases are summarized in Table \ref{table_cases}. Although one could perform these studies at much higher Reynolds numbers, the choice of $Re_{\tau}=180$ was chosen to balance computational cost, storage requirements and yet, maximize resolution in the roughness sublayer. 


\begin{table}[ht!]
  \begin{center}
    \begin{tabular}{c|c|c|c|c|c|c|c|c|c|c|c|c|c|c|r} 
      \textbf{Case} & $\boldsymbol{\lambda}$ &$\boldsymbol{{\lambda}^+}$& $\mathbf{a^+}$ & $\boldsymbol{\zeta}$ & $\mathbf{\Delta x^+}$ & $\mathbf{\Delta y^+_{w}}$ & $\mathbf{\Delta y^+_{cl}}$ & $\mathbf{\Delta z^+}$ & $\boldsymbol{Re_{cl}}$ & $\boldsymbol{Re_{b}}$ & $\boldsymbol{Re_{\tau}}$ & $\boldsymbol{u_{\tau}\times 10^{3}}$ \\
      \hline
      A & $\infty$ & $\infty$ & 0 & 0 & 8.94 & 1.05 & 2.00 & 4.55 & 3263 & 2800 & 180.9 & 43.07\\
      B & $4\pi$ & 2281 & 12.67 & 0.011 & 8.94 & 1.12 & 2.18 & 4.56 & 3148 & 2700 & 181.0 & 44.70\\
      C & $\frac{8}{3}\pi$ & 1516 & 12.64 & 0.017 & 9.08 & 1.12 & 2.17 & 4.54 & 3070 & 2620 & 180.5 & 45.94\\
      D & $2\pi$ & 1143 & 12.70 & 0.022 & 8.97 & 1.12 & 2.18 & 4.57 & 3002 & 2540 & 181.5 & 47.64\\
      E & $\frac{4}{3}\pi$ & 773 & 12.88 & 0.033 & 9.09 & 1.14 & 2.21 & 4.63 & 2833 & 2387 & 183.9 & 51.38\\
      F & $\pi$ & 578 & 12.84 & 0.044 & 9.07 & 1.13 & 2.21 & 4.62 & 2689 & 2240 & 183.5 & 54.61\\
    \end{tabular}
    \caption{Tabulation of different design parameters for the simulations such as: wavelength ($\lambda$), amplitude ($a$) and steepness ($\zeta=\frac{2a}{\lambda}$) of the wavy surface, friction velocity ($u_\tau$), Reynolds numbers ($Re$) based on boundary layer height ($\delta$) and different velocities expressed as the subscripts ('cl'=centerline velocity, 'b'=bulk velocity, '$\tau$'=friction velocity) and the grid spacing in different directions ('$\Delta x$'=streamwise, '$\Delta z$'=spanwise, '$\Delta y_w$'=wall normal near the wall, '$\Delta y_{cl}$'=wall normal near the flow centerline). Superscript '+' refers to inner scaled quantity (scaled with respect to dynamic viscosity ($\nu$) and friction velocity ($u_\tau$)).
    }\label{table_cases}
  \end{center}
\end{table}
The simulation domain is chosen as $4\pi\delta \times 2.2\delta \times 4\pi\delta/3$ (including the buffer zone for the IBM) where $\delta$ is the boundary layer height set to unity for these runs. This volume is discretized using a resolution of $256\times 257\times 168$ grid points which is more than adequate for the purposes of this study. In the streamwise and spanwise directions, periodic boundary conditions are enforced while a uniform grid distribution is adopted. In wall normal direction, no slip condition representing the presence of the solid wall causes inhomogeneity. To capture the viscous layers accurately, a stretched grid is used. The grid stretching in the inhomogeneous direction is carefully chosen using a mapping function that operates well with the spectral solver for the pressure Poisson equation. The different inner scaled grid spacing are also included in Table \ref{table_cases}. 

\subsection{Convergence of Turbulence Statistics \label{subsec:convergence}}
\begin{figure}[ht!]
\centering
\mbox{
\includegraphics[width=0.5\textwidth]{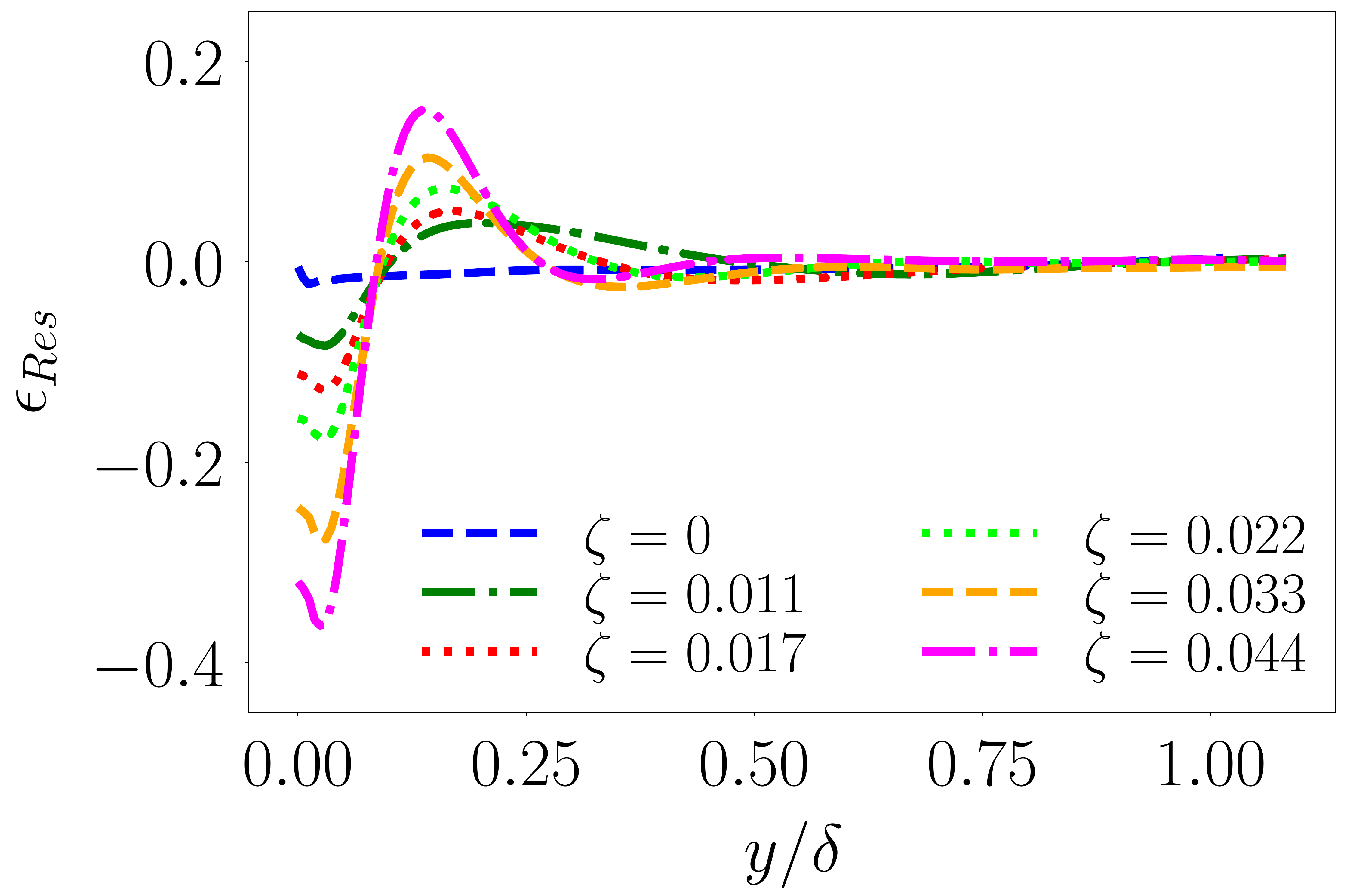}
}
\caption{\label{converror} Quantification of statistical stationarity for the different DNS datasets using the residual of mean horizontal stress from 2500 samples collected over $\sim 12\frac{\delta}{u_{\tau}}$.}
\end{figure}

To quantify the statistical stationarity of the turbulence simulation data, we consider the streamwise component of the inner scaled mean (spatial and temporally averaged) horizontal stress (including the viscous and Reynolds stress), $\tau_{H,x}=\langle\frac{\partial u}{\partial y}\rangle^+_{x,z,t}-\langle u^{\prime}v^{\prime}\rangle^+_{x,z,t}$.  Here $\langle \rangle_{x,z,t}$ represents the averaging operation with subscripts denoting averaging directions. In the limit of statistically stationary and horizontally homogeneous turbulence, $\tau_{H,x}(y)$ converges to a linear profile, $1-\frac{y}{\delta}$ as derived from mean momentum conservation. Using this, we estimate a residual convergence error, $\epsilon_{Res}=\langle\frac{\partial u}{\partial y}\rangle^+_{x,z,t}-\langle u^{\prime}v^{\prime}\rangle^+_{x,z,t}-(1-\frac{y}{\delta})$
%
whose variation with $y/\delta$ is shown in figure~\ref{converror}.
As expected, this error is sufficiently small for the flat channel ($\zeta=0$) with magnitudes approaching $O(0.001-0.01)$ through the turbulent boundary layer (TBL). The plot also shows similar quantification for wavy channel turbulence data with large residual errors near the surface indicative of deviations from equilibrium due to local homogeneity. Such quantifications also provide insight into the height of the roughness sublayer beyond which the mean horizontal stress approaches equilibrium values.


\subsection{Assessment of Simulation Accuracy \label{subsec:assess_accuracy}}

\begin{figure}[ht!]
\centering
\mbox{
\subfigure[Normalized and averaged streamwise velocity distribution in wall coordinates\label{valid_up}]{\includegraphics[width=0.48\textwidth]{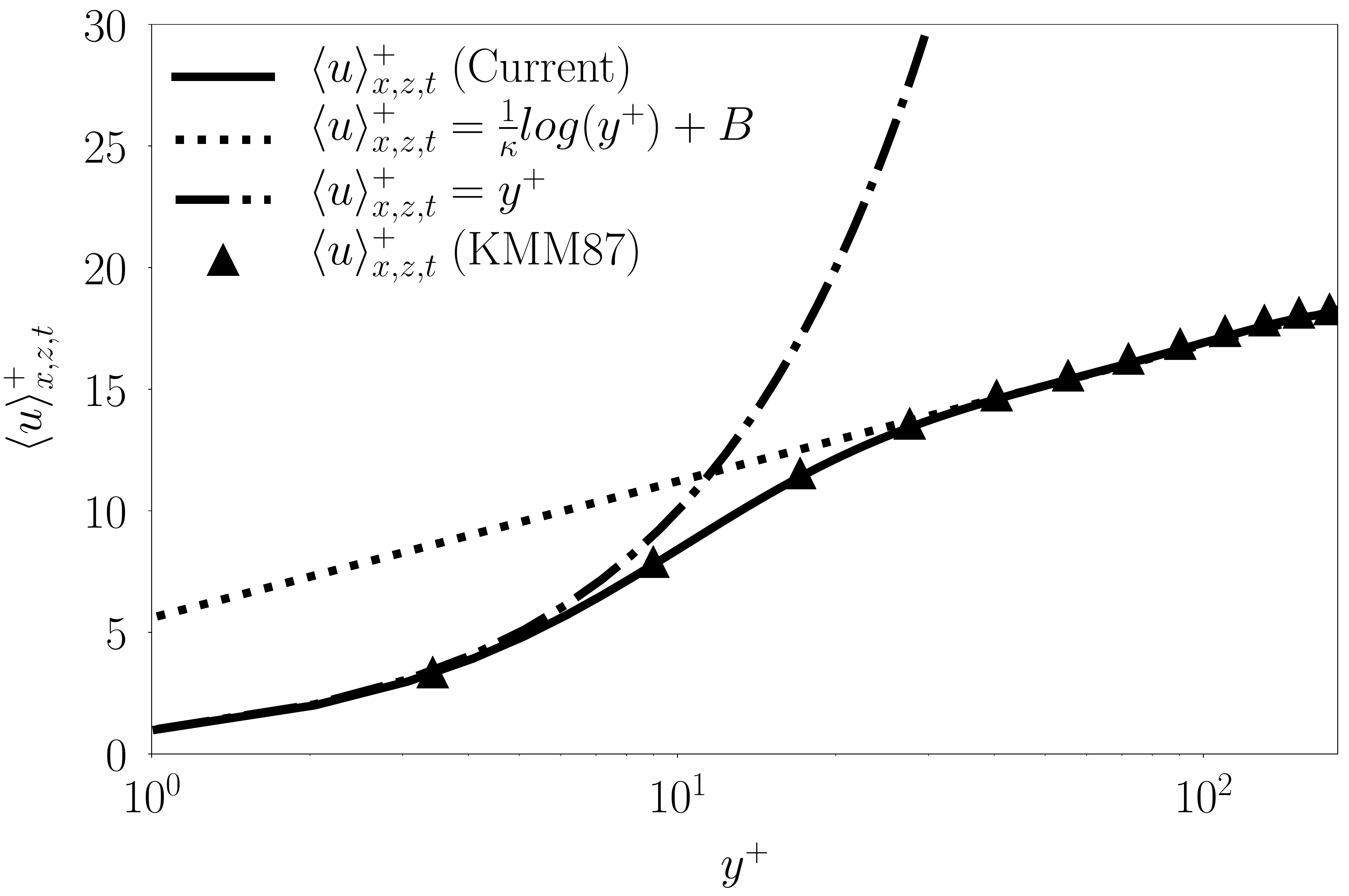}}
\subfigure[RMS normalized velocity fluctuation profiles in wall coordinates\label{valid_rms}]{\includegraphics[width=0.48\textwidth]{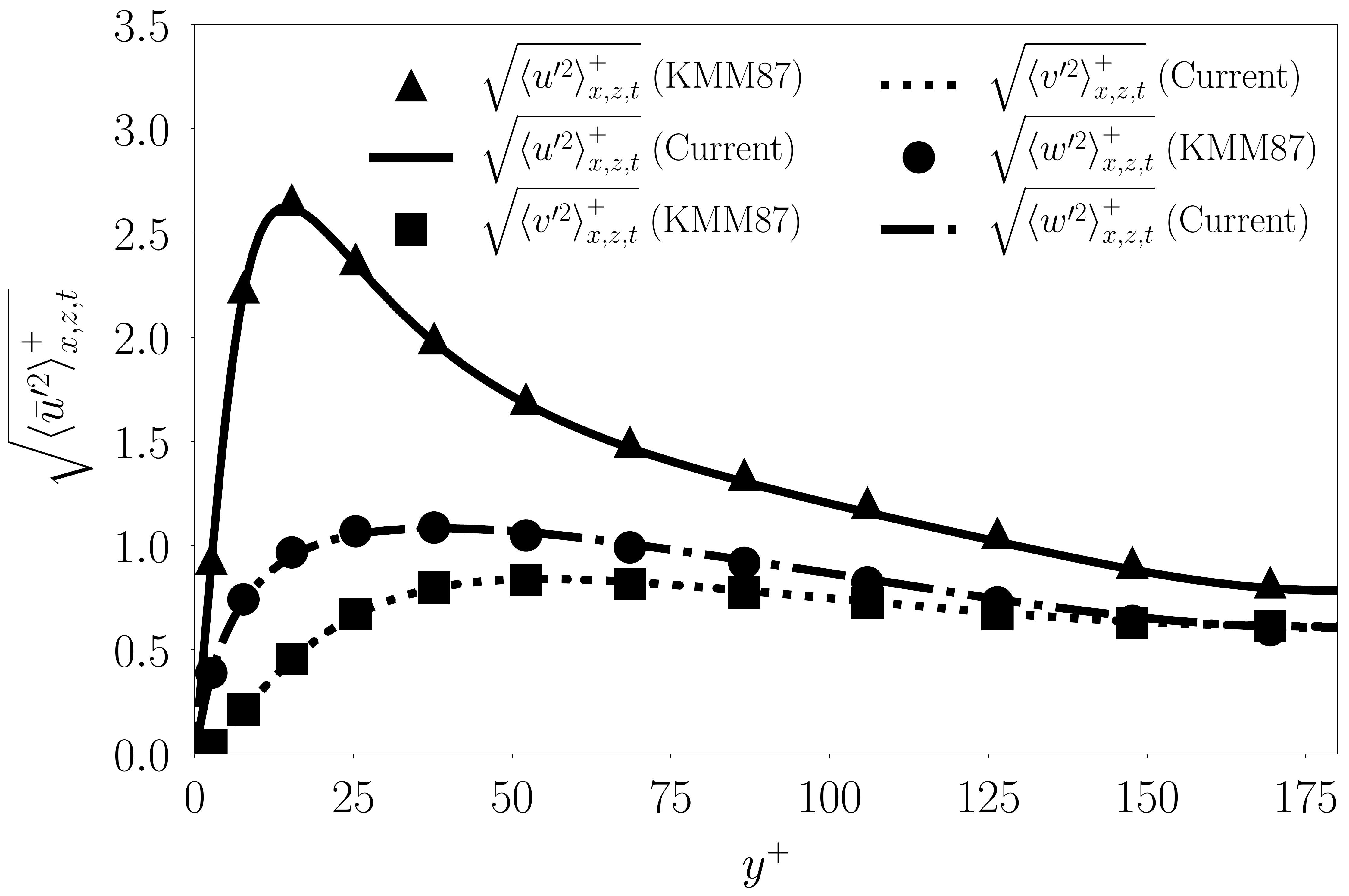}}
}
\caption{\label{fig:validation} Comparison of mean velocity and RMS velocity fluctuation between DNS of flat channel turbulent flow with IBM and the \citet{kim1987turbulence} DNS without IBM.} 
\end{figure}

We perform a baseline assessment of the computational accuracy for the turbulent channel flow using an immersed flat channel surface before adopting it for more complex surface shapes. 
We compare statistics from the current DNS with the well known work of \citet{kim1987turbulence} (KMM87 here onwards) which corresponds to a bulk Reynolds number, $Re_b\approx2800$, mean centerline velocity Reynolds number, $Re_{cl}\approx 3300$ and a friction Reynolds number, $Re_{\tau}\approx 180$. KMM87 used nearly $4\times10^6 \  (128 \times 129 \times 128)$ grid points and solved the flow equations by advancing modified variables, namely, wall-normal vorticity and Laplacian of the wall-normal velocity without explicitly considering pressure.  They adopt a Chebychev-tau scheme in the wall-normal direction, Fourier representation in the horizontal and Crank-Nicholson scheme for the time integration. In our work, we adopt spectrally accurate $6^{th}$ order compact scheme in space and a third order Adam-Bashforth time integration~\cite{laizet2009high,khan2019statistical}.
Figure~\ref{fig:validation} shows that the inner-scaled mean (figure~\ref{valid_up}) and  root mean square of the fluctuations (figure~\ref{valid_rms}) from the current DNS match that of KMM87. 
 
\section{Mean First-order Turbulence Structure \label{sec:results-firstorder}}

\begin{figure}[ht!]
\centering
\mbox{
\subfigure[$\zeta=0$\label{sep-0-wave}]{\includegraphics[width=0.33\textwidth]{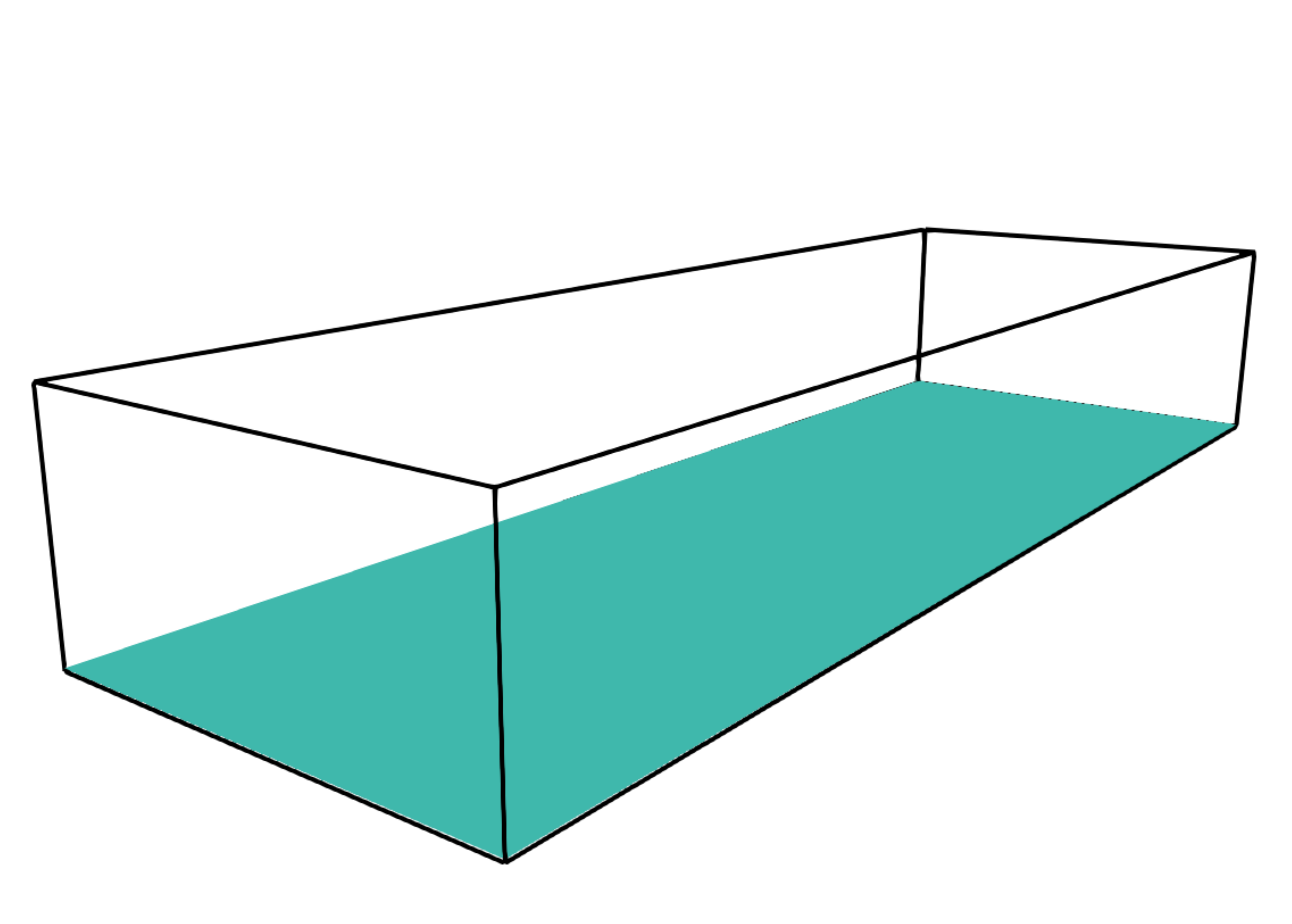}}
\subfigure[$\zeta=0.011$\label{sep-1-wave}]{\includegraphics[width=0.33\textwidth]{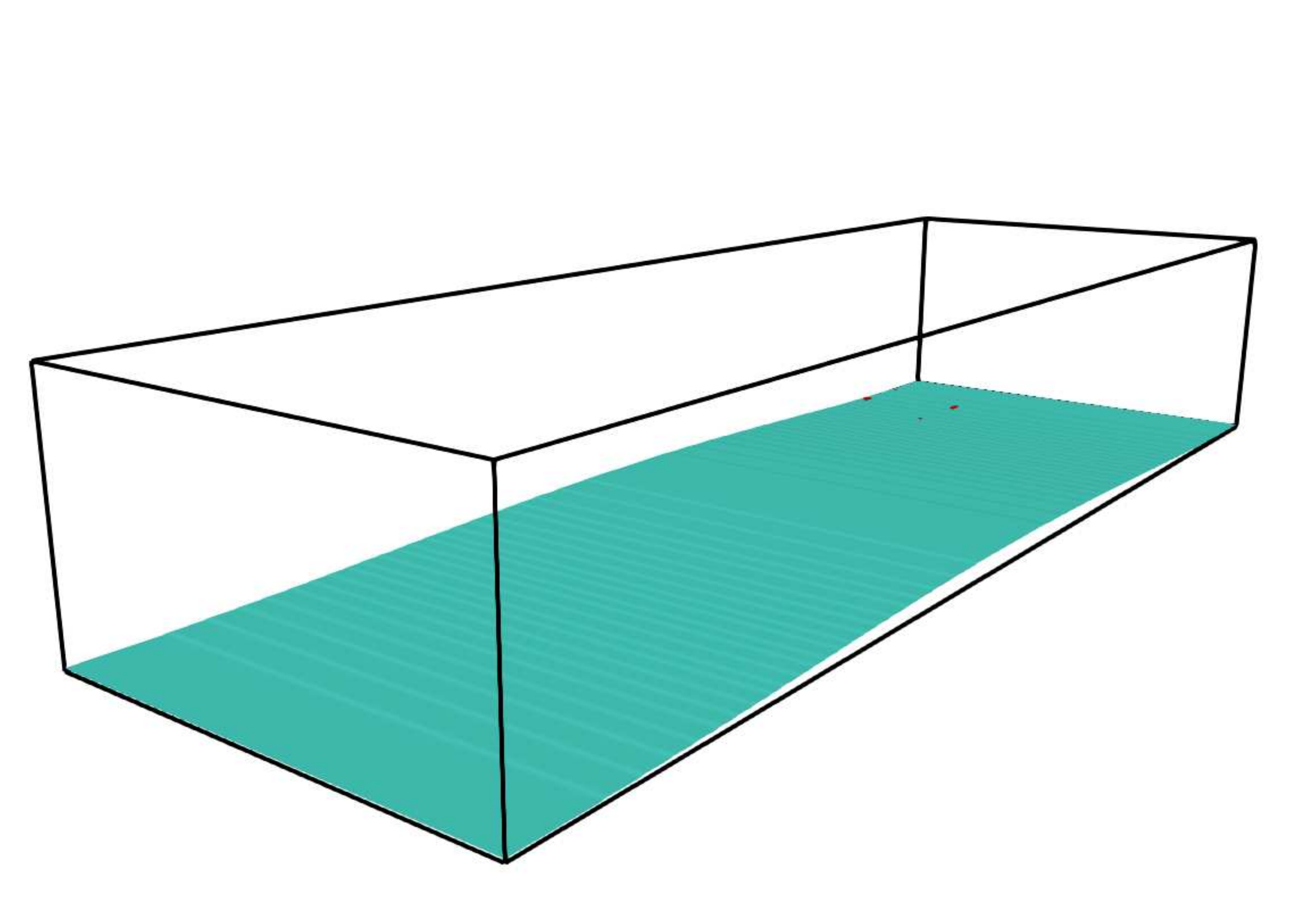}}
}
\mbox{
\subfigure[$\zeta=0.017$\label{sep-15-wave}]{\includegraphics[width=0.33\textwidth]{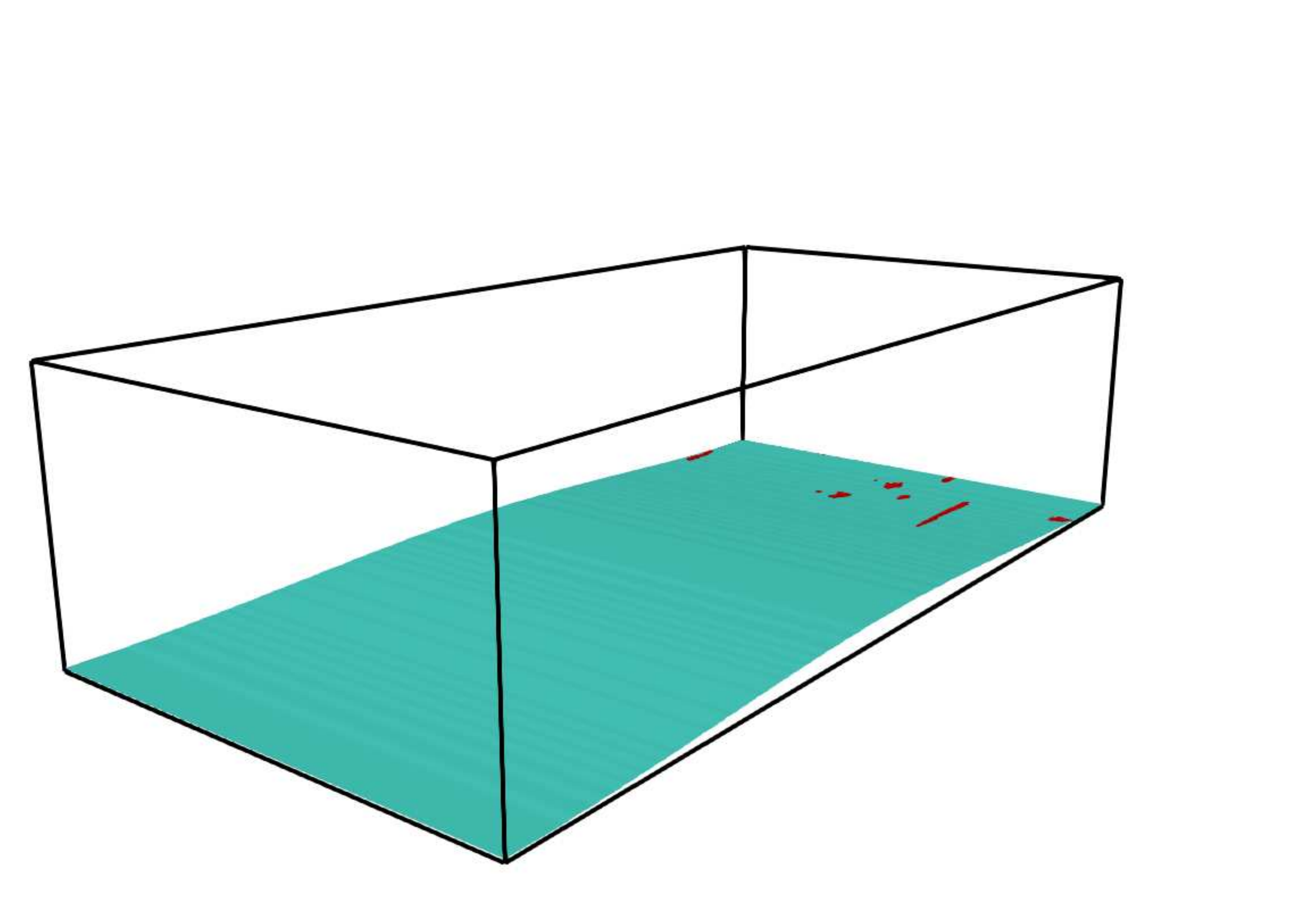}}
\subfigure[$\zeta=0.022$\label{sep-2-wave}]{\includegraphics[width=0.33\textwidth]{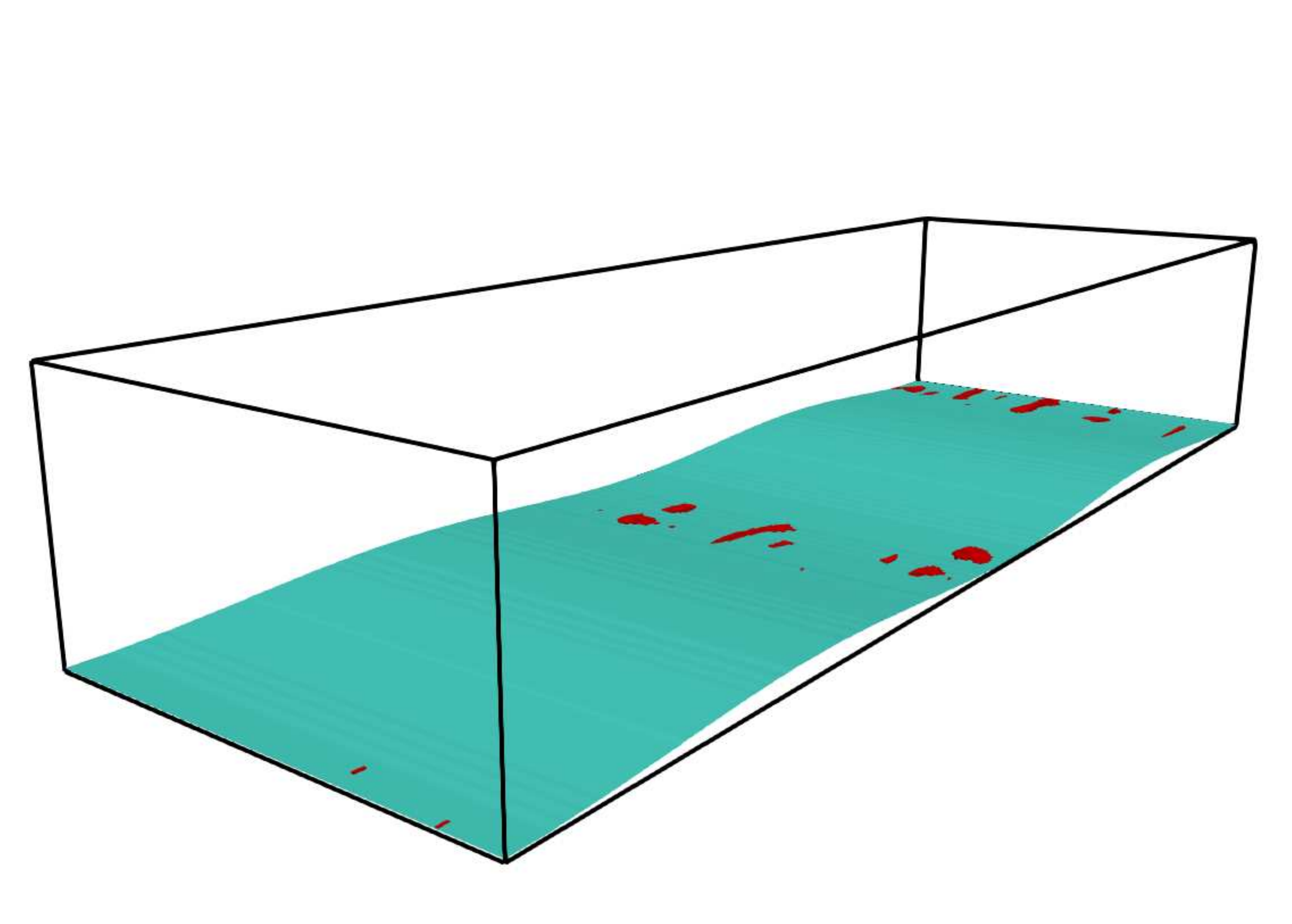}}
}
\mbox{
\subfigure[$\zeta=0.033$\label{sep-3-wave}]{\includegraphics[width=0.33\textwidth]{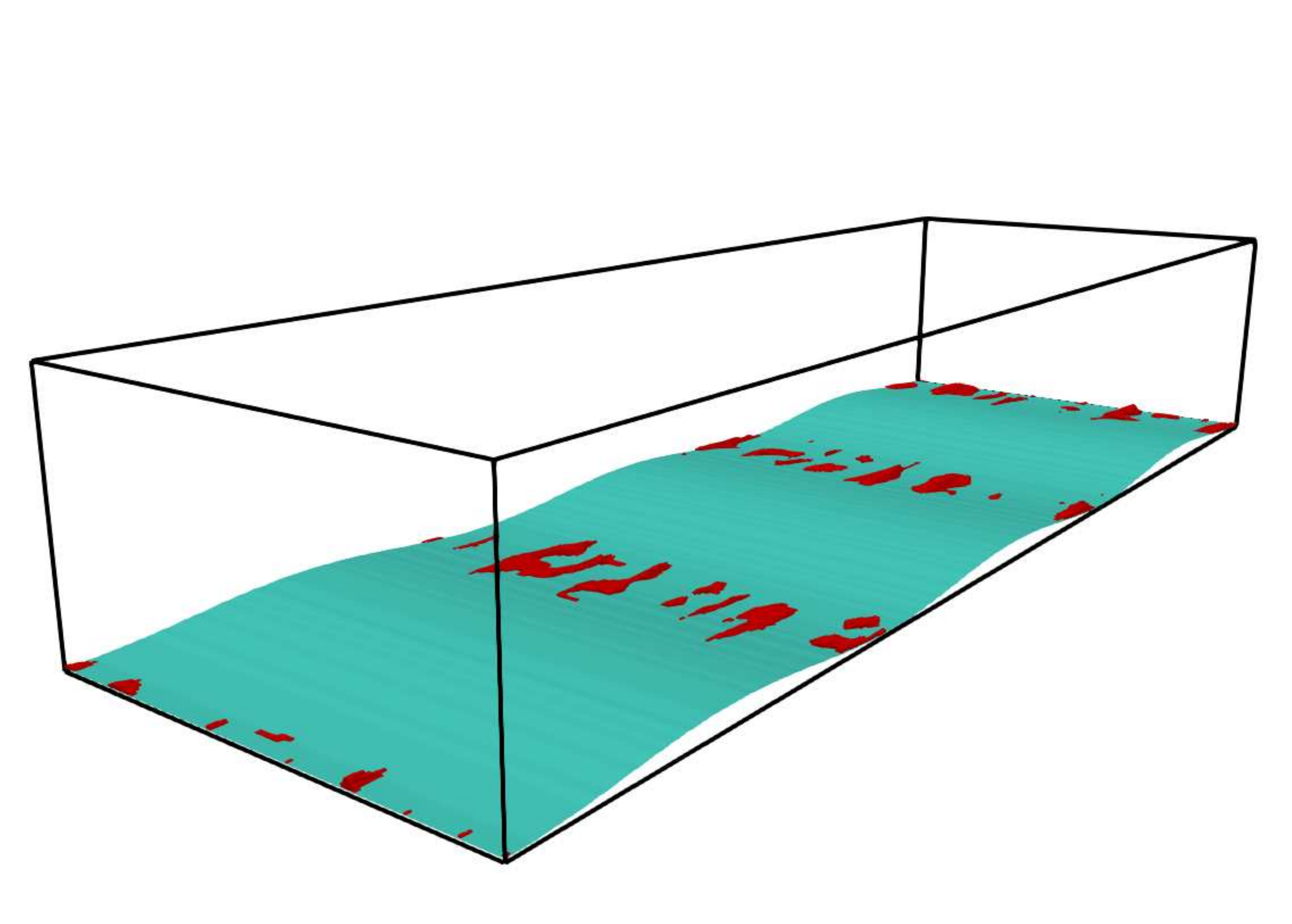}}
\subfigure[$\zeta=0.044$\label{sep-4-wave}]{\includegraphics[width=0.33\textwidth]{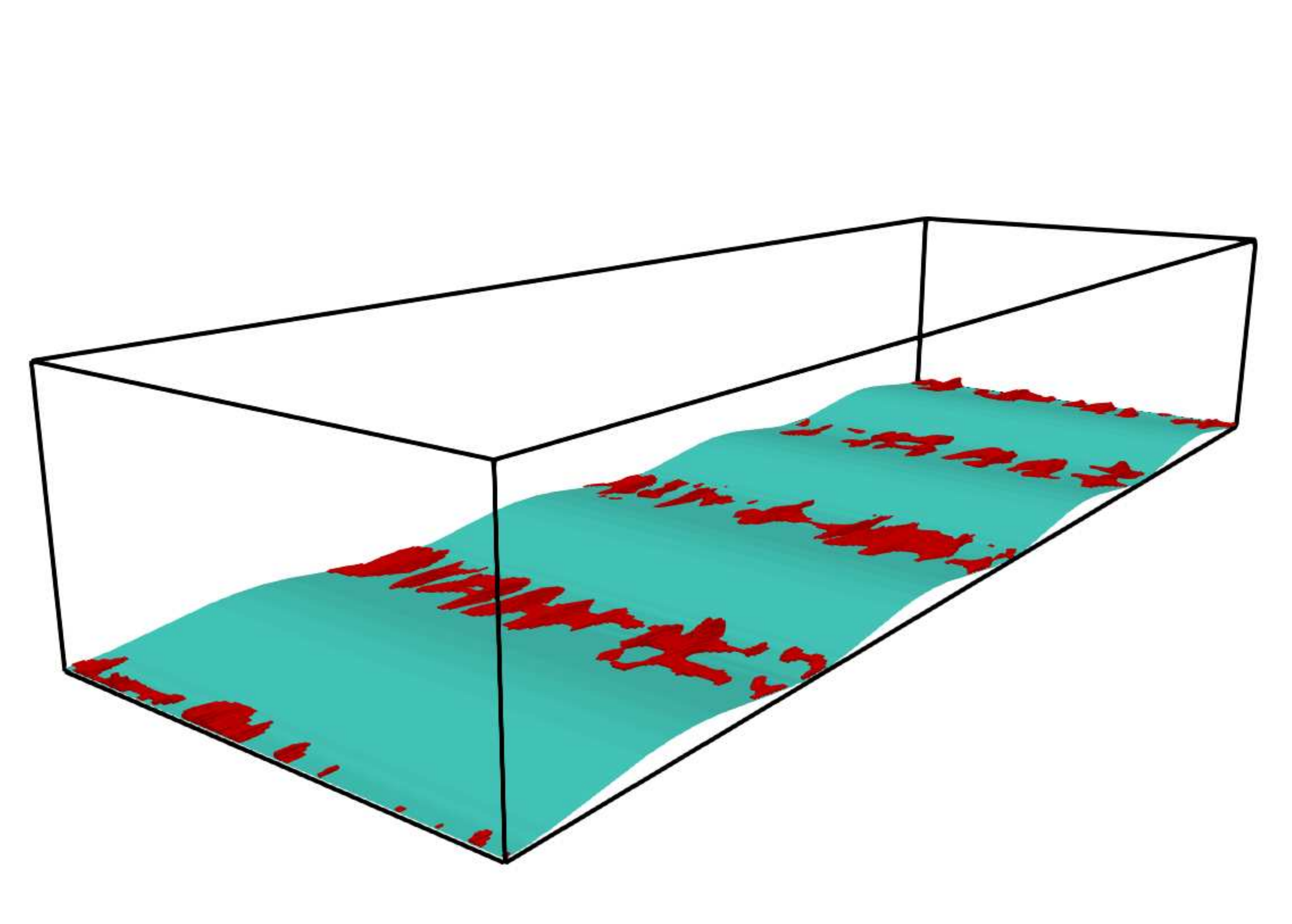}}
}
\caption{\label{fig:Separation} Comparison of instantaneous flow separation for the different wave steepness, $\zeta$. The wavy surface is denoted in cyan and the separation is denoted in red. 
}
\end{figure}

The primary goal of this study is to explore the non-equilibrium, near-surface turbulence structure over systematically varied sinusoidal undulations with particular emphasis on delineating the shape dependent turbulence generation mechanisms. Naturally, this involves characterization of deviations in (streamwise-averaged) first order turbulence structure from equilibrium phenomenology as evidenced in flat channel turbulence, assess the extent of outer layer similarity and try to relate these observation to roughness induced turbulence effects as appropriate. 
For $\zeta \ll 1$ considered here, separation is minimal as shown in figure~\ref{fig:Separation} using isosurfaces of instantaneous negative velocity. We note that separation is inconsistent, but becomes prominent at higher $\zeta$. 
The streamwise-averaged or `double-averaged' turbulence structure denoted by $\langle \rangle_{x,z,t}$ is a function of solely the wall normal distance and refers to averaging along both homogeneous ($z,~t$) and inhomogeneous ($x$) directions.
Temporal ($t$) averaging is performed using $2500$ three-dimensional snapshots over $20$ flow through times for the chosen friction Reynolds number. 

\subsection{Streamwise Averaging of Turbulence Statistics \label{subsec:str_avg_turb_stat}}
While spatial averaging along the homogeneous $z$ direction is straightforward,  averaging along the streamwise wavy surface can be done using multiple approaches. In this study, we define a local vertical coordinate, $d$ at each streamwise location with $d=0$ at the wall. Its maximum possible value corresponds to the mid channel location and changes with streamwise coordinate. We then perform streamwise averaging along constant values of $d$, to generate mean statistical profiles. This approach works well for $\frac{a}{\delta}<<1$ as it tries to approximate the terrain as nearly flat with a large local radius of curvature and therefore, nearly homogeneous. To justify this approach, we note that in our study ${a}=0.07{\delta}$ ($a^+ \approx 12$,$\delta=1$) which is an order of magnitude larger than the typical viscous length scale, $L_v={\nu}/{u_\tau}=1/Re_{\tau} \approx 0.0055$, but smaller than the start of the log layer ($y^+ \approx 50$).

\begin{figure}[ht!]
\centering
\mbox{
\subfigure[\label{fig:prof_umean_avg2}]{\includegraphics[width=0.29\textwidth]{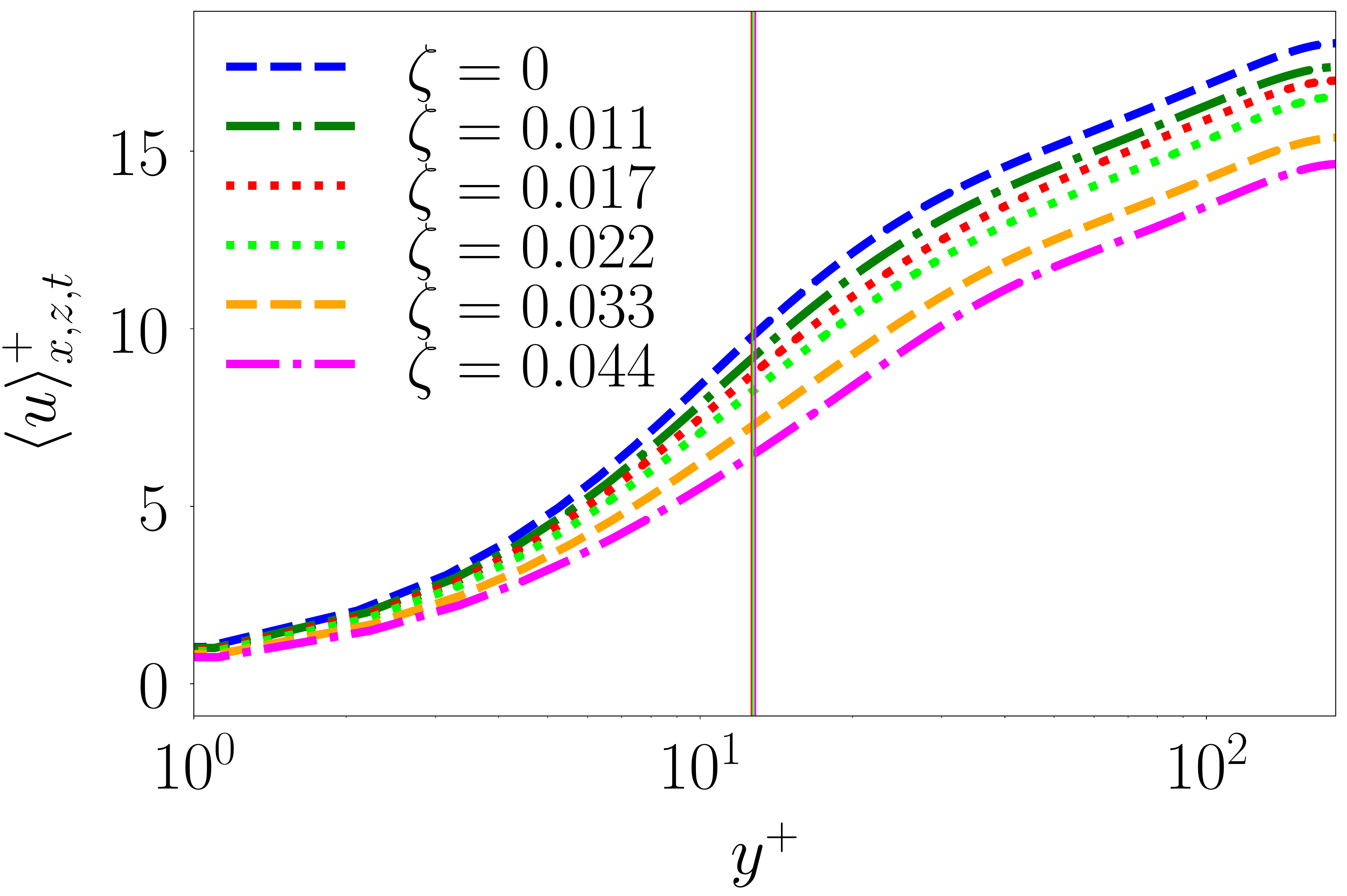}}\hspace{0.2em}
\subfigure[\label{fig:prof_vmean_avg2}]{\includegraphics[width=0.295\textwidth]{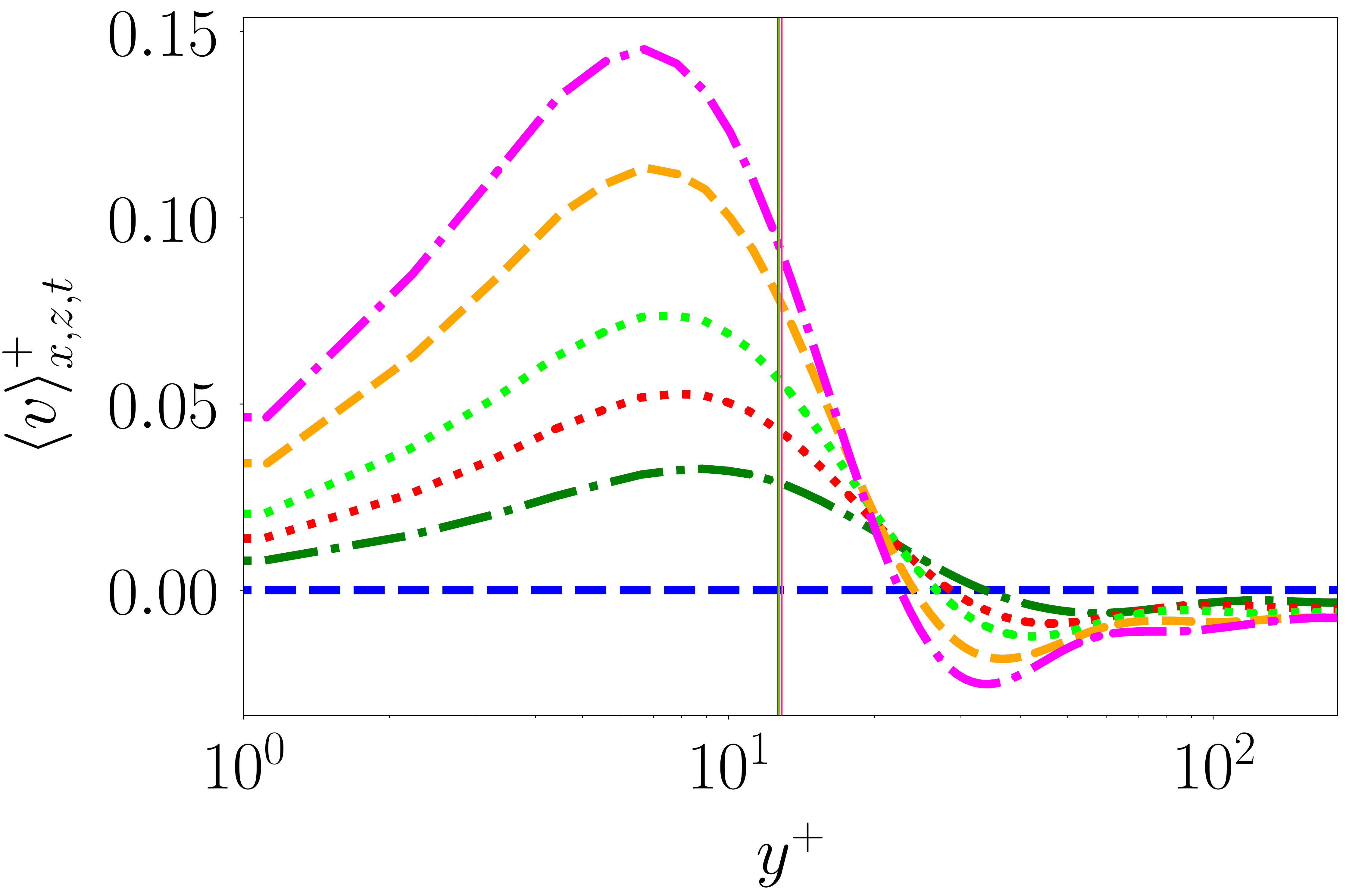}}\hspace{0.2em}
\subfigure[\label{fig:prof_deffectu_avg2}]{\includegraphics[width=0.29\textwidth]{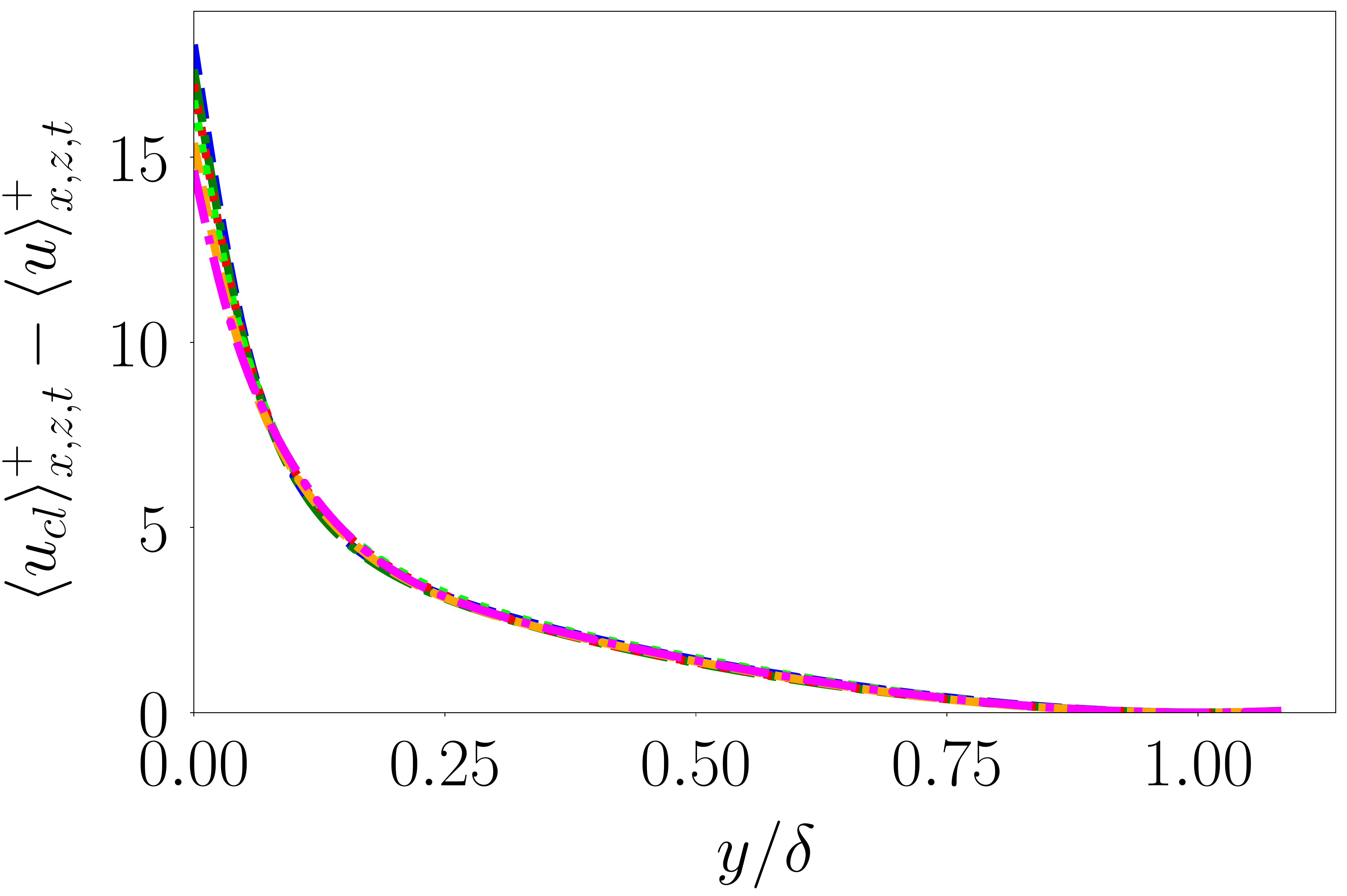}}
\subfigure[\label{fig:prof_deffectu_zoomed_avg2}]{\includegraphics[width=0.0638\textwidth]{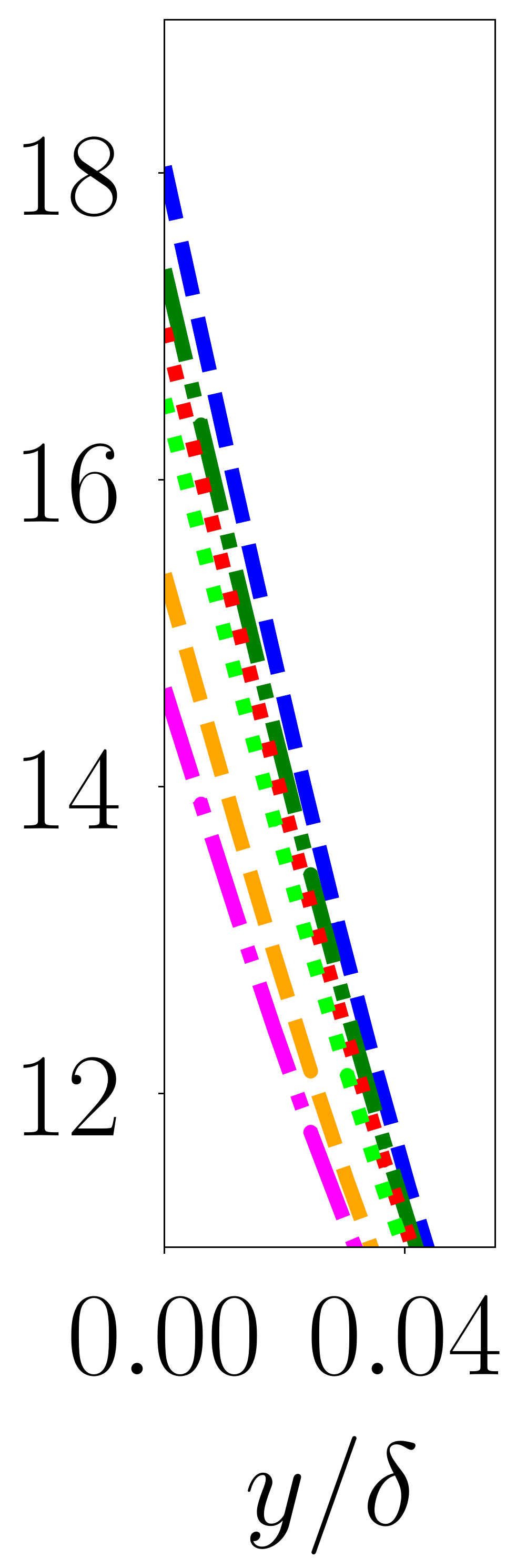}}
}
\caption{Inner scaled mean (a) streamwise velocity (b) vertical velocity, and (c) defect velocity computed using local coordinate-based average. A magnified version of (c) focusing on the near surface region is presented in (d). 
	Three vertical straight lines correspond to the different $a^+$ for $\zeta > 0$ (see Table \ref{table_cases}). 
\label{fig:profiles_avg2_meanvelocity}}
\end{figure}


\subsection{Outer Layer Similarity and Mean Velocity Profiles \label{subsec:outlay_simi_mean_velo}}
As the mean channel height, $\delta$ (for wavy geometry) is kept constant across all the different steepness, $\zeta$, the observed  changes in the mean statistics are only due to surface slope effects\cmnt{ and not the outer layer dynamics}.  In figure~\ref{fig:profiles_avg2_meanvelocity} we show the inner-scaled, double averaged streamwise and vertical velocity for the different surfaces. 
The different colors in the plot, namely, blue, green, red, lime, orange, and magenta correspond to different wave steepness, $\zeta=0$, $\zeta=0.011$, $\zeta=0.017$, $\zeta=0.022$, $\zeta=0.033$, and $\zeta=0.044$ respectively.
The double-averaged, inner-scaled streamwise velocity shows the well known downward shift of the log-region in the $u^+-y^+$ plot \cmnt{of the logarithmic region} for increasing wave steepness, $\zeta$ (figure~\ref{fig:prof_umean_avg2})and is indicative of the flow slowing down near the surface from increased drag due to steeper undulations. The double-averaged, inner-scaled vertical velocity structure (figure~\ref{fig:prof_vmean_avg2})  show increasingly stronger net upward flow close to the surface with increase in $\zeta$.  Away from the surface in the logarithmic region, $\langle v \rangle^+_{x,z,t}$ shows downward flow so that there is no net vertical transport. For the horizontally homogeneous flat channel ($\zeta=0$) the mean vertical velocity is zero. Therefore, these well established vertical motions in the mean over wavy surfaces (although small, i.e. $\langle v \rangle^+ = O(0.1)$), represent the obvious form of surface-induced deviations from equilibrium. 
In spite of these near surface deviations, the dynamics outside the roughness sublayer tend to be similar when normalized and shifted appropriately as illustrated through the defect velocity profiles in figures~\ref{fig:prof_deffectu_avg2} and \ref{fig:prof_deffectu_zoomed_avg2} that indicate little to no deviation between $\zeta=0$ and $\zeta=0.044$. 

\subsection{Quantification of Mean Velocity Gradients and Inertial Sublayer \label{subsec:quant_mean_velograd_inertsub}}
\begin{figure}[ht!]
\centering
\mbox{
\subfigure[\label{fig:profile_gamma_2}]{\includegraphics[width=0.42\textwidth]{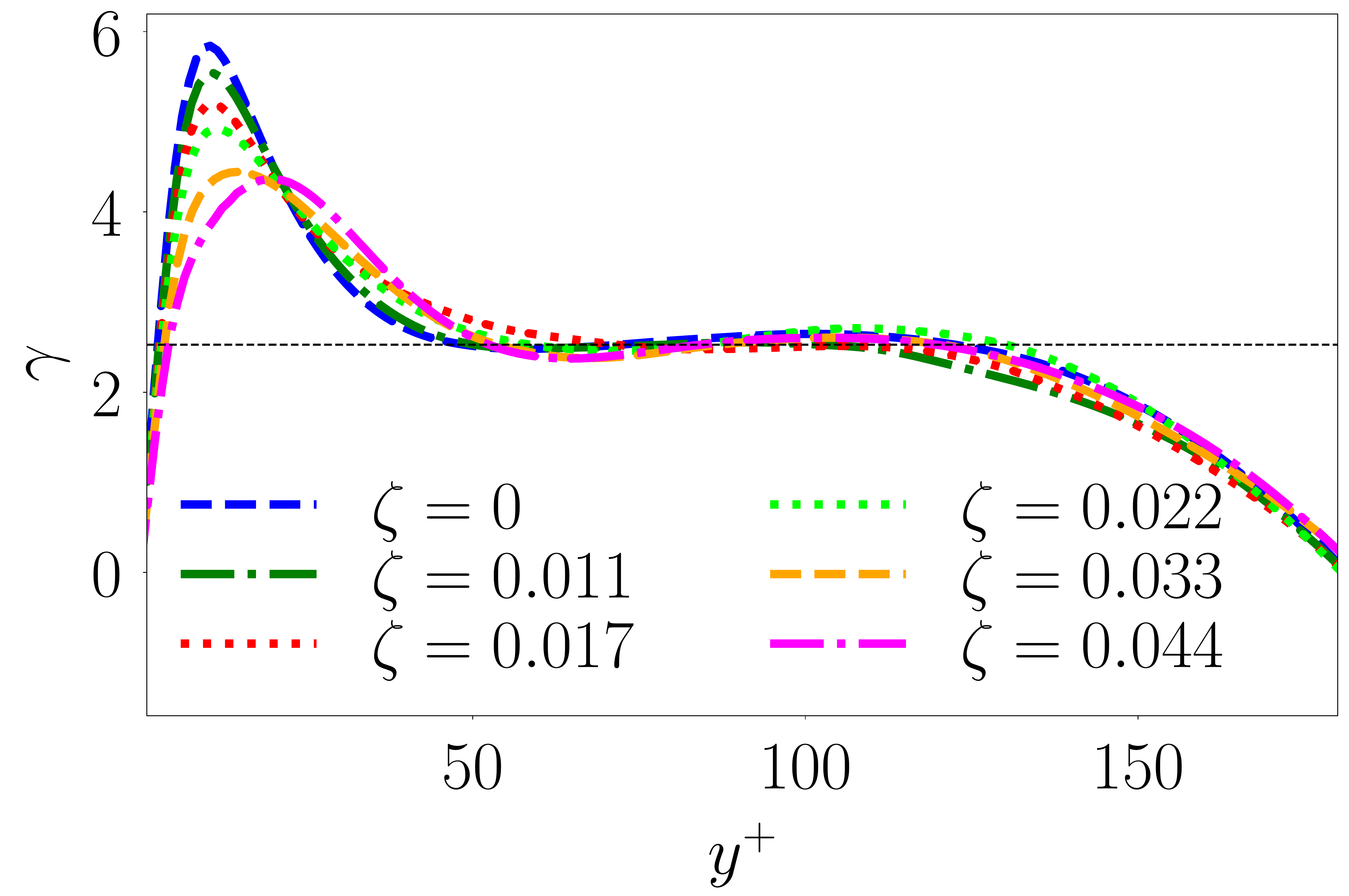}}
\subfigure[\label{fig:profile_phi_2}]{\includegraphics[width=0.42\textwidth]{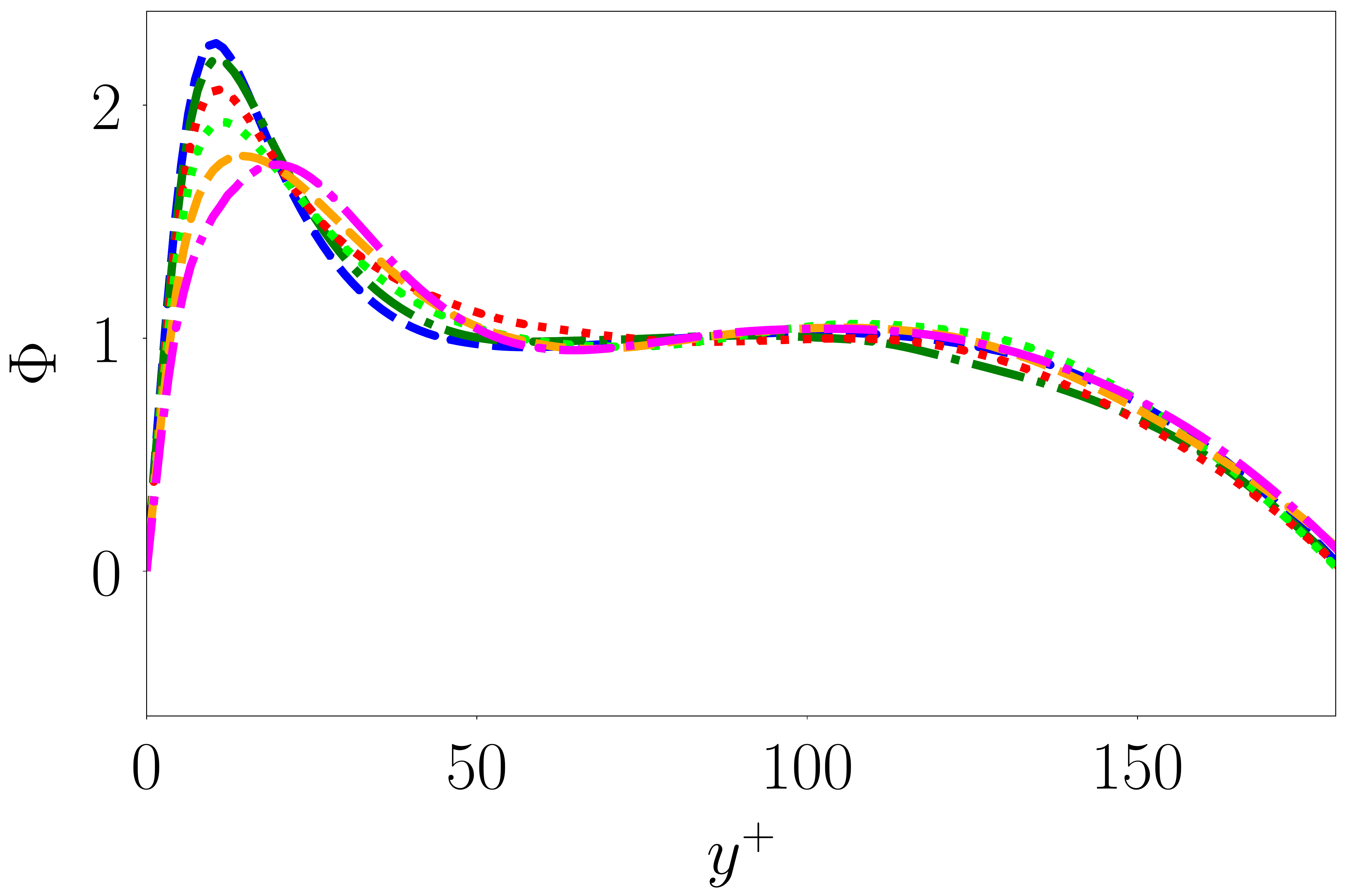}}
}
\caption{Variation of non-dimensional mean streamwise velocity gradients, (a) $\gamma = y^+ \frac{d{\langle u \rangle^+_{x,z,t}}}{dy^+}$ and (b) $\Phi=\frac{\kappa y}{u_{\tau}}\frac{d{\langle u \rangle_{x,z,t}}}{dy}$. The thin dashed black line in (a) corresponds to the mean $\gamma$ valued 2.5315 computed based on $y^+=60-110$. 
} 
\label{fig:MeanVelocityDeviationQuantification}
\end{figure}

The normalized mean streamwise velocity gradient helps characterize the different regions of a TBL, especially the inertial (or logarithmic) region\cmnt{ and the von K\'arm\'an constant}. In this study, we estimate the normalized premultiplied inner-scaled mean gradient, $\gamma = y^+ \frac{d{\langle u \rangle^+_{x,z,t}}}{dy^+}$ (shown in figure~\ref{fig:profile_gamma_2}) which achieves a near constant value of $1/\kappa$ (where $\kappa$ is the von K\'arm\'an constant) in the inertial sublayer due to normalization of the mean gradient by characteristic law of the wall variables, i.e., surface layer velocity ($u_{\tau}$) and distance from the wall ($y$). In this study, for the chosen friction Reynolds numbers $Re_{\tau}$ we observe that the inertial layer exists over $y^+\sim 60-110$ for $\zeta=0$ and shifts to $y^+\sim 70-120$ for $\zeta=0.044$, i.e. an upward (rightward) shift in the log layer with increase in wave steepness, $\zeta$.
\cmnt{The proof for the above statement is that this shift is observed in the phi_m profiles too where the the friction velocity does not enter the y-coordinate. }
The estimated von K\'arm\'an constants are tabulated in Table~\ref{table_kappa} and show a range of $0.39-0.40$ for the different runs with no discernible trend. In this study, we use the appropriate value of $\kappa$ to compute the different metrics. 
This is in contrast to the flow dependent $\kappa$ values reported in \citet{leonardi2010channel} for cube roughness, i.e. $\kappa$ is reported to decrease from $0.41$ in smooth channels to $\sim 0.35$ in rough-wall TBL~\cite{nagib2008variations}. 

\begin{table}[ht!]
  \begin{center}
    \begin{tabular}{l|c|c|c|c|c|c} 
      $\boldsymbol{\zeta}$ & 0.000 & 0.011 & 0.017 & 0.022 & 0.033 & 0.044 \\
      \hline
      $\boldsymbol{\kappa}$ & 0.3878 & 0.3965 & 0.3975 & 0.3917 & 0.4010 & 0.3996 \\
    \end{tabular}
    \caption{Tabulation of estimated von K\'arm\'an constants ($\kappa$) for different steepness factor ($\zeta$).
    }\label{table_kappa}
  \end{center}
\end{table}

\begin{figure}[ht!]
\centering
\mbox{
\subfigure[\label{fig:tke_Pii_notnorm_new}]{\includegraphics[width=0.21\textwidth]{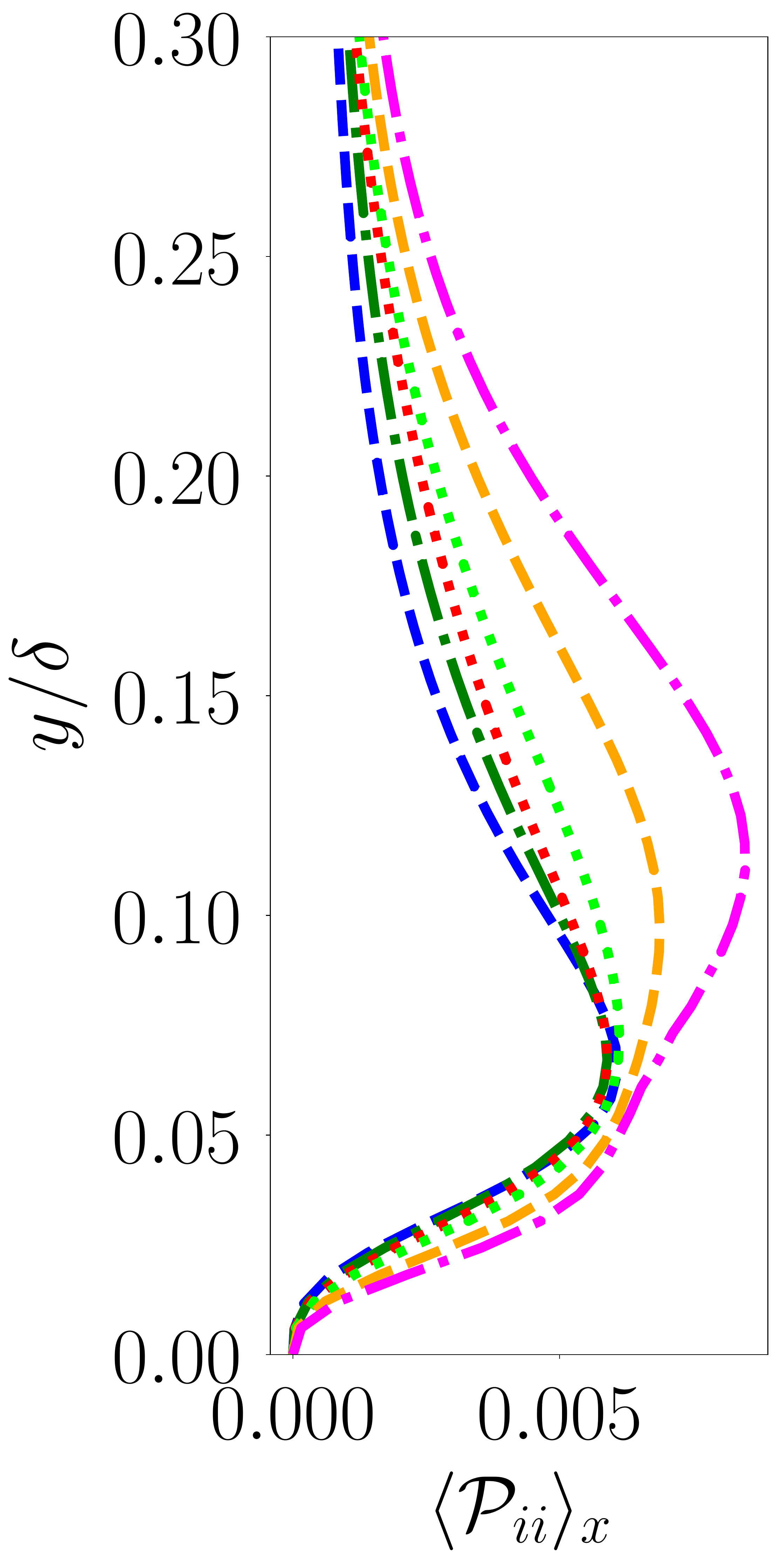}}\hspace{2em}
\subfigure[\label{fig:tke_Pii_new}]{\includegraphics[width=0.21\textwidth]{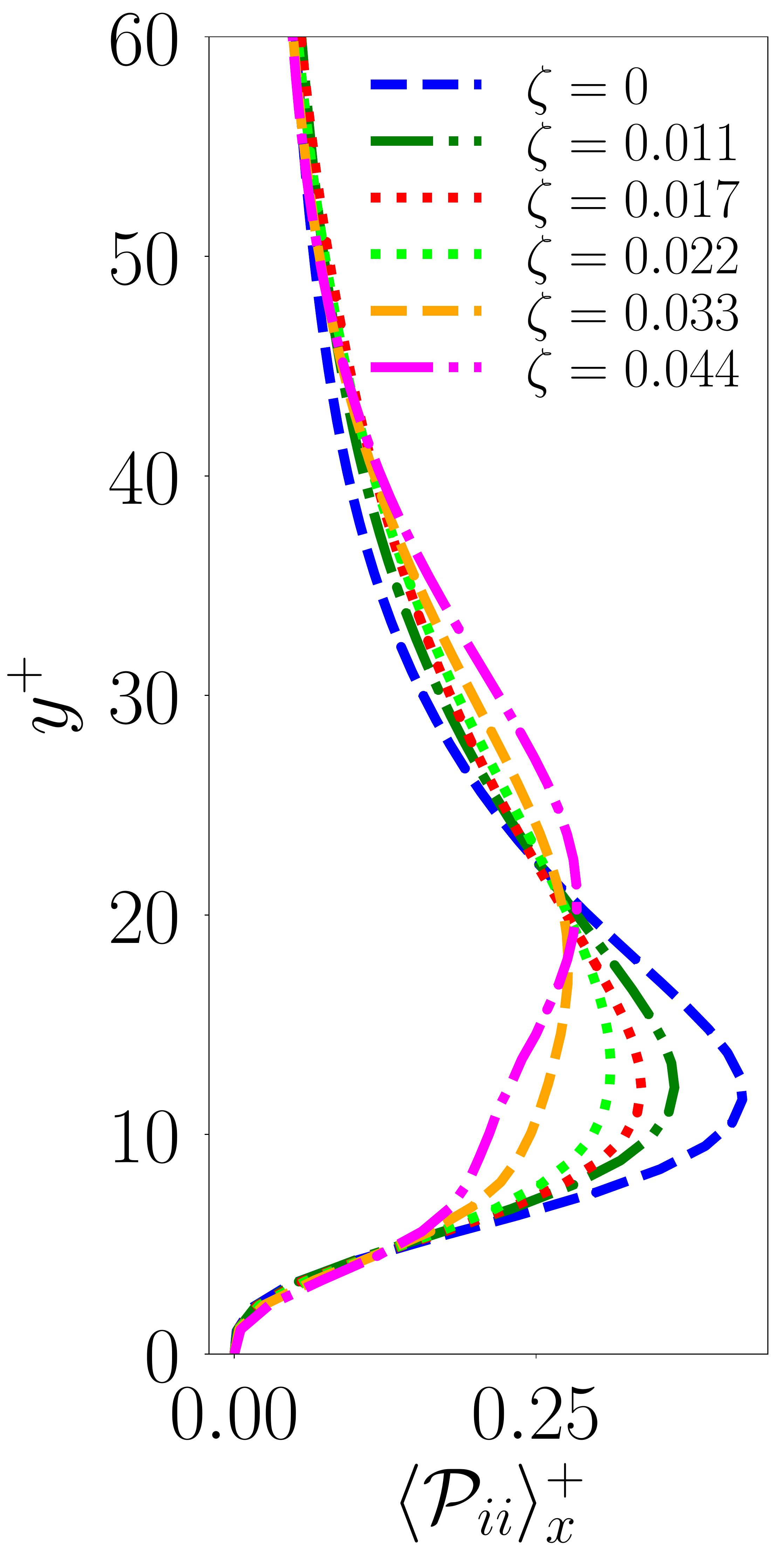}}\hspace{2em}
\subfigure[\label{fig:tke_PiibyEii_new}]{\includegraphics[width=0.21\textwidth]{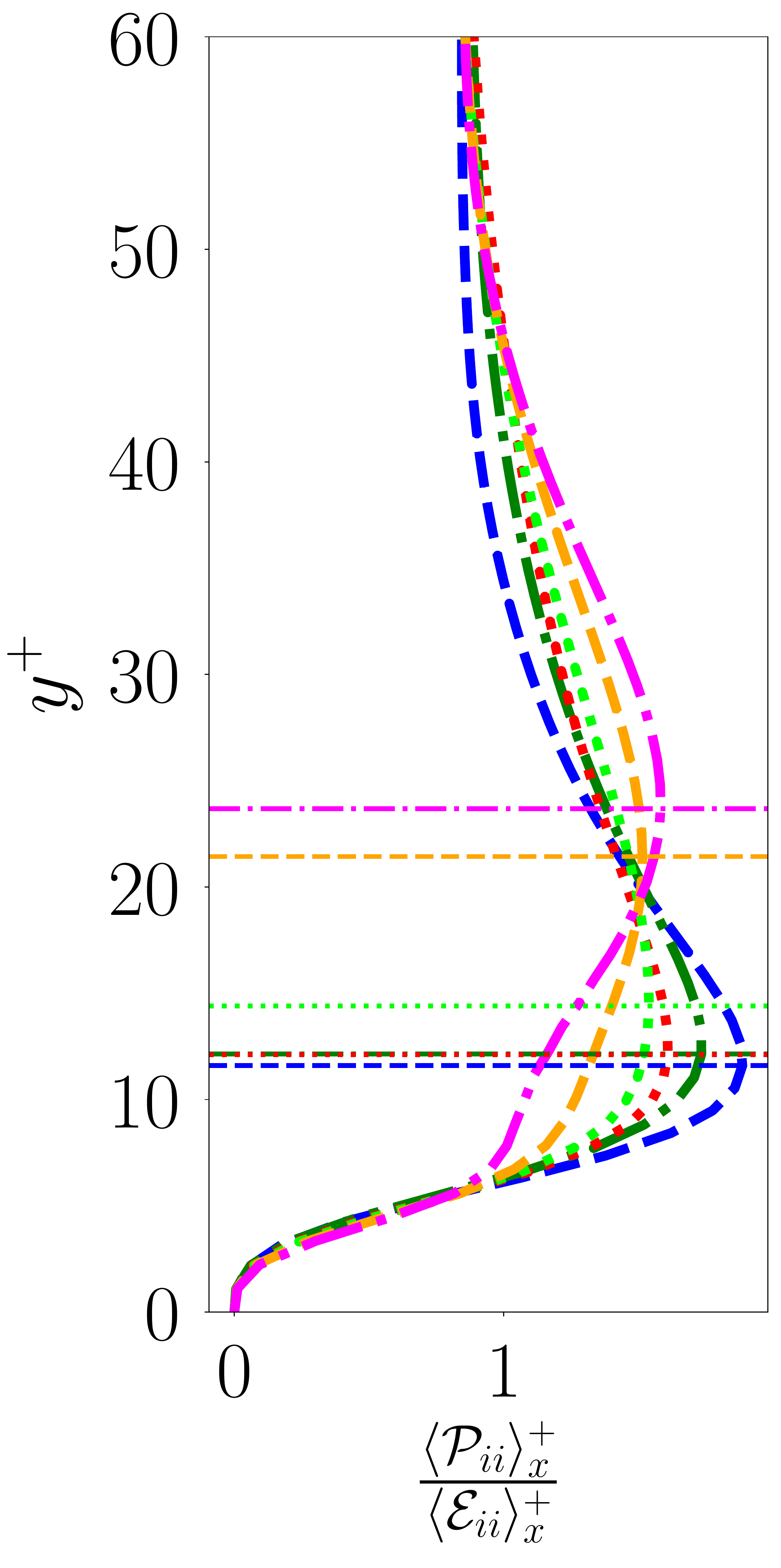}}
}
\caption{\label{fig:tke_production_new} Schematic illustrating the influence of surface undulations on near-surface turbulence structure, namely, wall-normal variation of streamwise averaged production of turbulent kinetic energy in (a) inner variable in dimensional ($m^2/s^3$) and (b) non-dimensional forms. In (c) we show the ratio of double-averaged production to dissipation. 
}
\end{figure}

For completeness, we also show the non-dimensional mean streamwise velocity gradient, $\Phi=\frac{\kappa y}{u_{\tau}}\frac{d{\langle u \rangle_{x,z,t}}}{dy}=\gamma \kappa$ in figure~\ref{fig:profile_phi_2}. The $\Phi$ profiles for different $\zeta$ mimic the characteristic equilibrium structure starting from zero at the wall followed by a peak at the edge of viscous layer and subsequently, a gradual decrease in the buffer layer to a value of one in the inertial sublayer indicative of overall shape similarity in $\Phi$. 
The shape of the above curves, namely the upward shift in the log region (figure~\ref{fig:profile_gamma_2}) and the smaller peak in $\Phi$ with increase in $\zeta$, indicate that the buffer layer becomes increasingly thicker for steeper waves. The `buffer layer' is a region of high turbulence production~\cite{pope2001turbulent} where both the viscous and Reynolds stresses are significant. Therefore, the expansion of the buffer layer with $\zeta$ is tied to the expansion of the turbulence production zone due to the wavy surface as evidenced from figure~\ref{fig:tke_production_new} where the turbulence kinetic energy (TKE) production grows and decays slower for higher $\zeta$ in the buffer region ($y^+ \approx 10-50$) in both inner-scaled (figure~\ref{fig:tke_Pii_new}) and dimensional (figure~\ref{fig:tke_Pii_notnorm_new}) forms. In fact, this is explicitly seen from the production-dissipation ratio in figure~\ref{fig:tke_PiibyEii_new} where the horizontal lines clearly show the upward shift in $\langle\mathcal{P}_{11}\rangle^+_x/\langle\mathcal{E}_{11}\rangle^+_x$ with $\zeta$. 

\section{Mean Second-order Turbulence Structure \label{sec:results-secondorder}}

\subsection{Overview of Reynolds Stress Transport \label{subsec:ReynoldsTSressTransport-Overview}}
\begin{figure}[ht!]
\centering
\mbox{
\subfigure[\label{fig:prof_uvar_avg2_new}]{\includegraphics[width=0.21\textwidth]{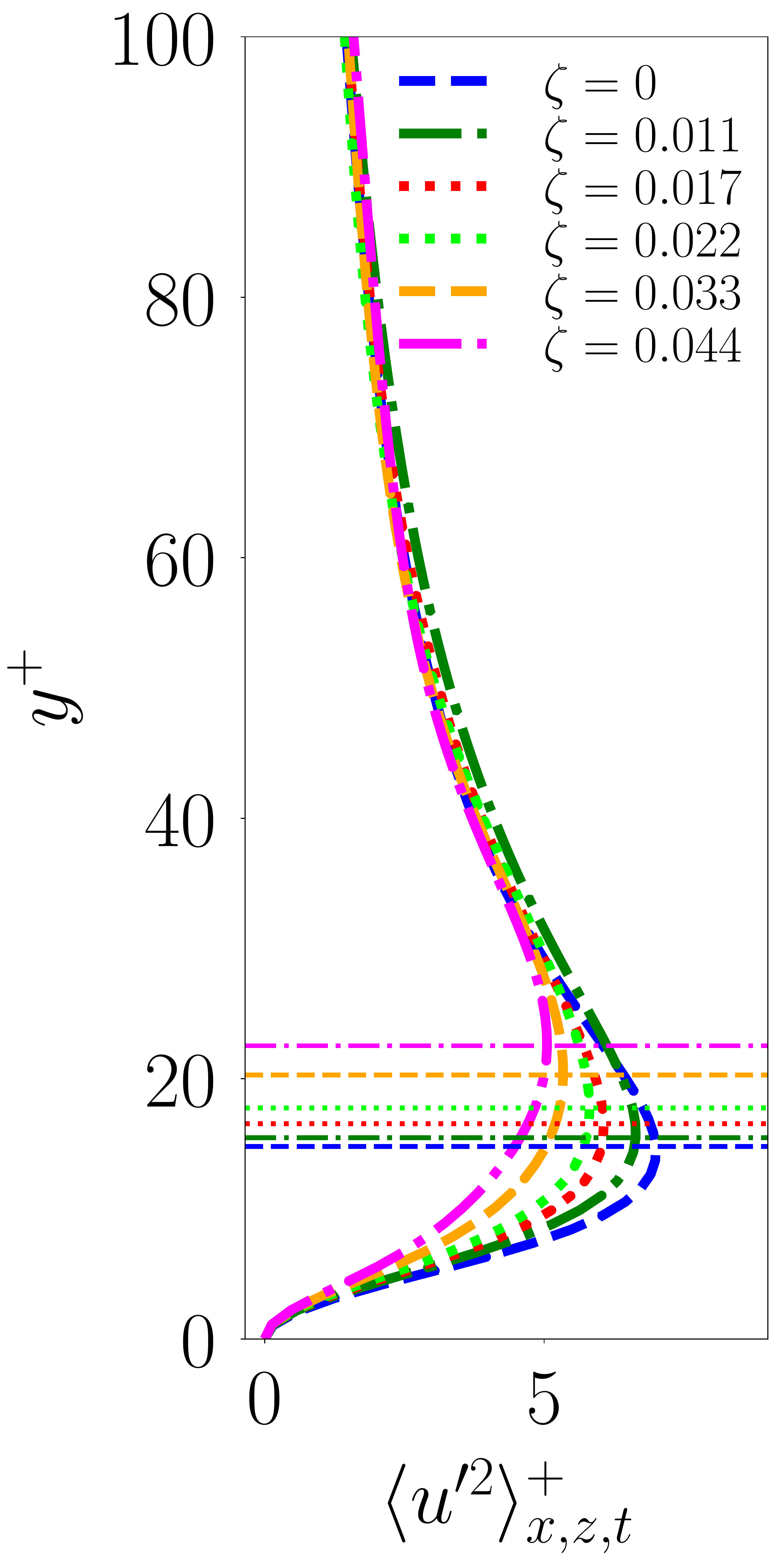}}
\subfigure[\label{fig:prof_vvar_avg2_new}]{\includegraphics[width=0.21\textwidth]{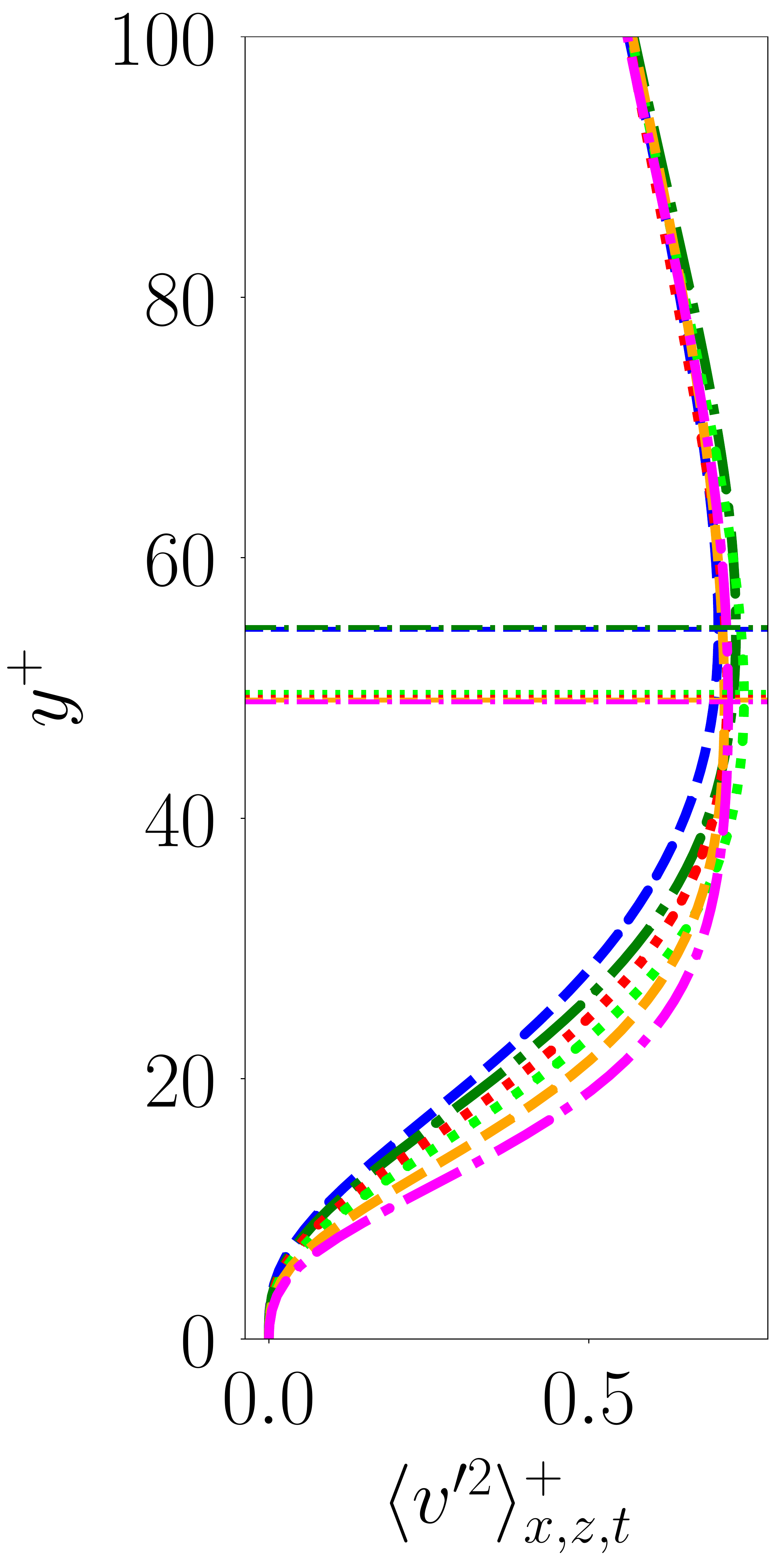}}
\subfigure[\label{fig:prof_wvar_avg2_new}]{\includegraphics[width=0.21\textwidth]{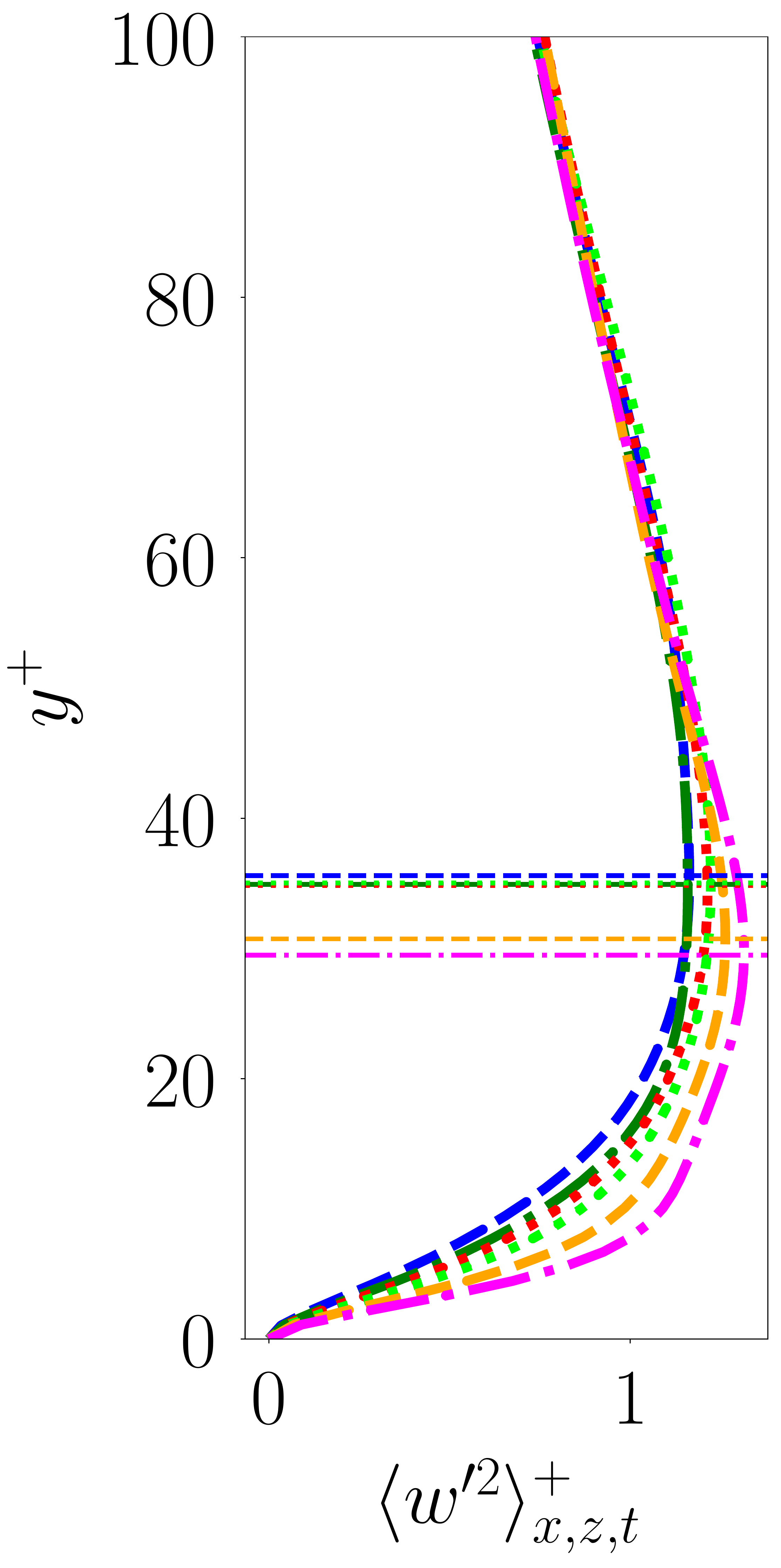}}
\subfigure[\label{fig:prof_tke_avg2_new}]{\includegraphics[width=0.21\textwidth]{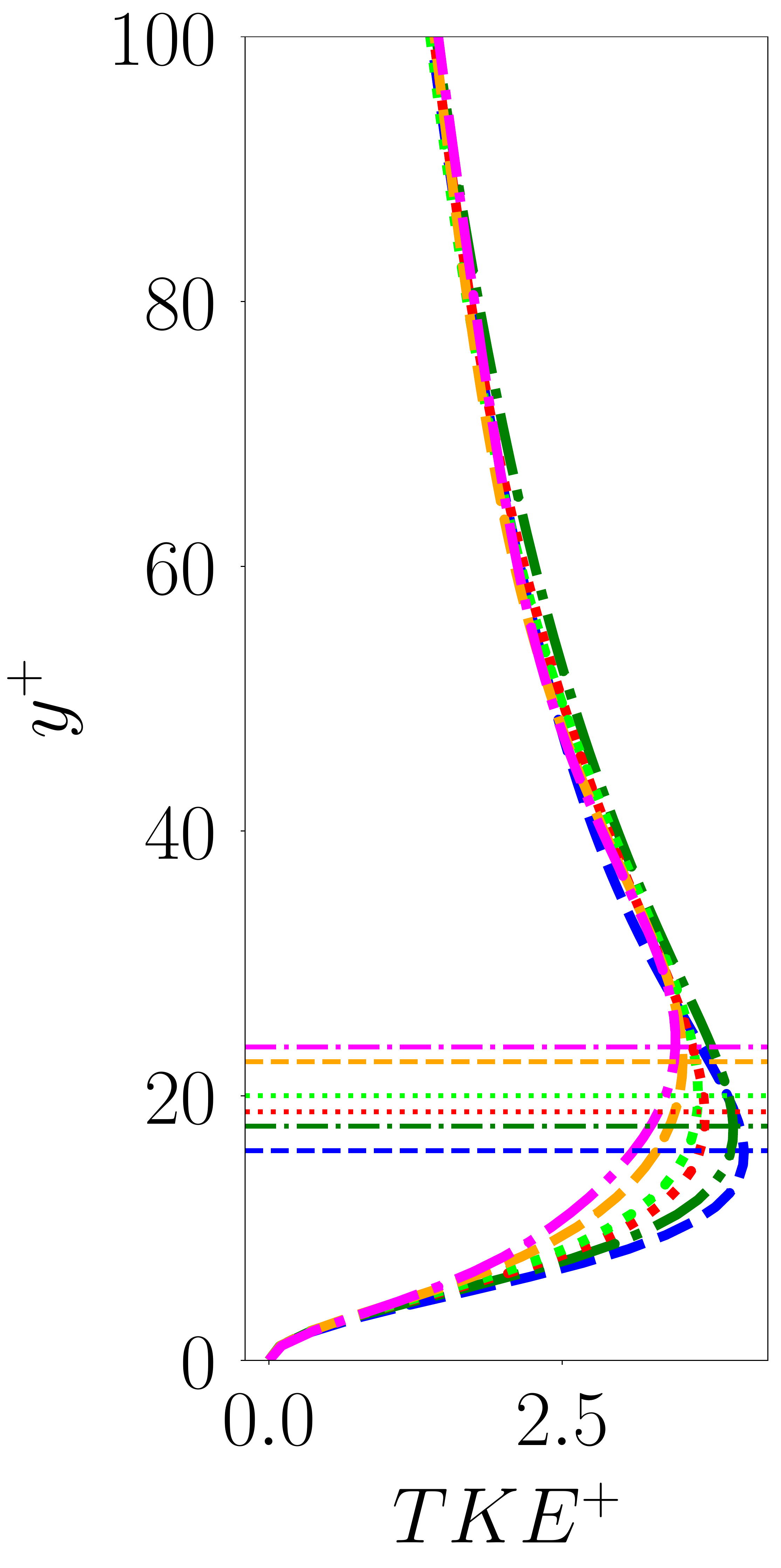}}
}
\caption{\label{fig:prof_variances_avg2_new} Inner scaled mean (a) streamwise variance, (b) vertical variance, (c) spanwise variance and (d) turbulent kinetic energy (TKE). The horizontal lines correspond to height with maximum value of the statistics along the profile.}
 
\label{fig:profiles_avg2_new_set1_RSelements}
\end{figure}

In order to interpret the structure of the components of the Reynolds stress tensor, we also need to study its evolution mechanisms. Below, we provide an overview of Reynolds stress transport and the nomenclature adopted over the rest of this manuscript.
The  Reynolds stress transport equation is shown below in equation~\eqref{eq:trans0}. Here $\mathcal{L}_{ij}$ is the local rate of change, $\mathcal{C}_{ij}$ and $\mathcal{D}_{ij}$  represent advective and diffusive transport respectively. The local terms in the evolution equation are $\mathcal{P}_{ij}$ representing production, $\mathcal{E}_{ij}$ representing dissipation and $\mathcal{R}_{ij}$ is the pressure-rate-of-strain correlation contributing to the redistribution of Reynolds stress. All the above terms are estimated using averaged quantities along the only homogeneous direction ($z$) and over a stationary window ($t$) and given by the notation, $<>_{z,t}$. In this study, the stationary window of time is sampled using 2500 temporal snapshots collected over 20 flow through times. Therefore, each of the terms in the equation below vary over the $(x,y)$ space. 
\begin{align}
    \underbrace{\frac{\partial}{\partial t}\langle{u_i^\prime u_j^\prime}\rangle_{z,t}}_{\mathcal{L}_{ij}} 
    + \underbrace{\langle{u_k}\rangle_{z,t}\frac{\partial\langle{u_i^\prime u_j^\prime}\rangle_{z,t}}{\partial x_k}}_{\mathcal{C}_{ij}} 
    = \underbrace{-\langle{u_k^\prime u_j^\prime}\rangle_{z,t}\frac{\partial\langle{u_i}\rangle_{z,t}}{\partial x_k} - \langle{u_k^\prime u_i^\prime}\rangle_{z,t}\frac{\partial\langle{u_j}\rangle_{z,t}}{\partial x_k}}_{\mathcal{P}_{ij}} 
    - \underbrace{2\nu\Big\langle{\frac{\partial u_j^\prime}{\partial x_k}\frac{\partial u_i^\prime}{\partial x_k}}\Big\rangle_{z,t}}_{\mathcal{E}_{ij}}  \nonumber \\
    + \underbrace{\Big\langle{\frac{p^\prime}{\rho}\Big(\frac{\partial u_i^\prime}{\partial x_j}+\frac{\partial u_j^\prime}{\partial x_i}\Big)}\Big\rangle_{z,t}}_{\mathcal{R}_{ij}}
    + \underbrace{\frac{\partial}{\partial x_k}\bigg[\langle{-u_i^\prime u_j^\prime u_k^\prime}\rangle_{z,t} + \nu \frac{\partial}{\partial x_k}\langle{u_i^\prime u_j^\prime}\rangle_{z,t} - \Big\langle{\frac{p^\prime}{\rho}(\delta_{ki}u_j^\prime+\delta_{kj}u_i^\prime)}\Big\rangle_{z,t} \bigg]}_{\mathcal{D}_{ij}}
    \label{eq:trans0}
\end{align}
In the above, the indices $i,j=1,2,3$ correspond to the streamwise ($x$), vertical ($y$) and spanwise ($z$) directions respectively. Also, $u_1=u$,$u_2=v$ and $u_3=w$.
On account of statistical stationarity, $\mathcal{L}_{ij}=0$ which allows us to rewrite equation~\eqref{eq:trans0} as
\begin{align}
    \mathcal{C}_{ij}=\mathcal{P}_{ij}-\mathcal{E}_{ij}+\mathcal{R}_{ij}+\mathcal{D}_{ij}.
    \label{eq:trans1}
\end{align}
We further average these individual terms along the inhomogeneous streamwise ($x$) direction, i.e. double averaging. The cumulative effect of the locally acting terms in the transport equation, namely the sum of production, dissipation and pressure-rate-of-strain is denoted by $\langle \Lambda_{ij} \rangle_x$ expressed as
\begin{align}
    \langle\Lambda_{ij}\rangle_x = \langle\mathcal{P}_{ij}\rangle_x-\langle\mathcal{E}_{ij}\rangle_x+\langle\mathcal{R}_{ij}\rangle_x.
    \label{eq:trans2}
\end{align}
Using the above, we can indirectly compute the streamwise-averaged diffusive transport term $\langle\mathcal{D}\rangle_x$ as
\begin{align}
    \langle\mathcal{D}_{ij}\rangle_x=\langle\mathcal{C}_{ij}\rangle_x-\langle\Lambda_{ij}\rangle_x.
    \label{eq:trans3}
\end{align}
\cmnt{For the benefit of readers, we note that the streamwise averaging can not be interchanged with the spanwise and temporal averaging because of the streamwise inhomogeneity (i.e. $\langle\mathcal{P}_{ij}\rangle_x \neq \langle{u_k^\prime u_j^\prime}\rangle_{x,z,t}\frac{\partial\langle{u_i}\rangle_{x,z,t}}{\partial x_k}$).} 
%
In the following subsections, we explore the structure of the diagonal elements of the Reynolds stress tensor, $\langle {u_i}'{u_j}' \rangle_{x,z,t}, i=j$ with particular focus on the streamwise (double) averaged second order turbulence structure (figure~\ref{fig:profiles_avg2_new_set1_RSelements}) and their underlying generation (both double- and single-averaged) in response to the surface undulations. 


\begin{figure}[ht!]
\centering
\mbox{
\subfigure[\label{fig:prof_uvar_avg2_new60}]{\includegraphics[width=0.21\textwidth]{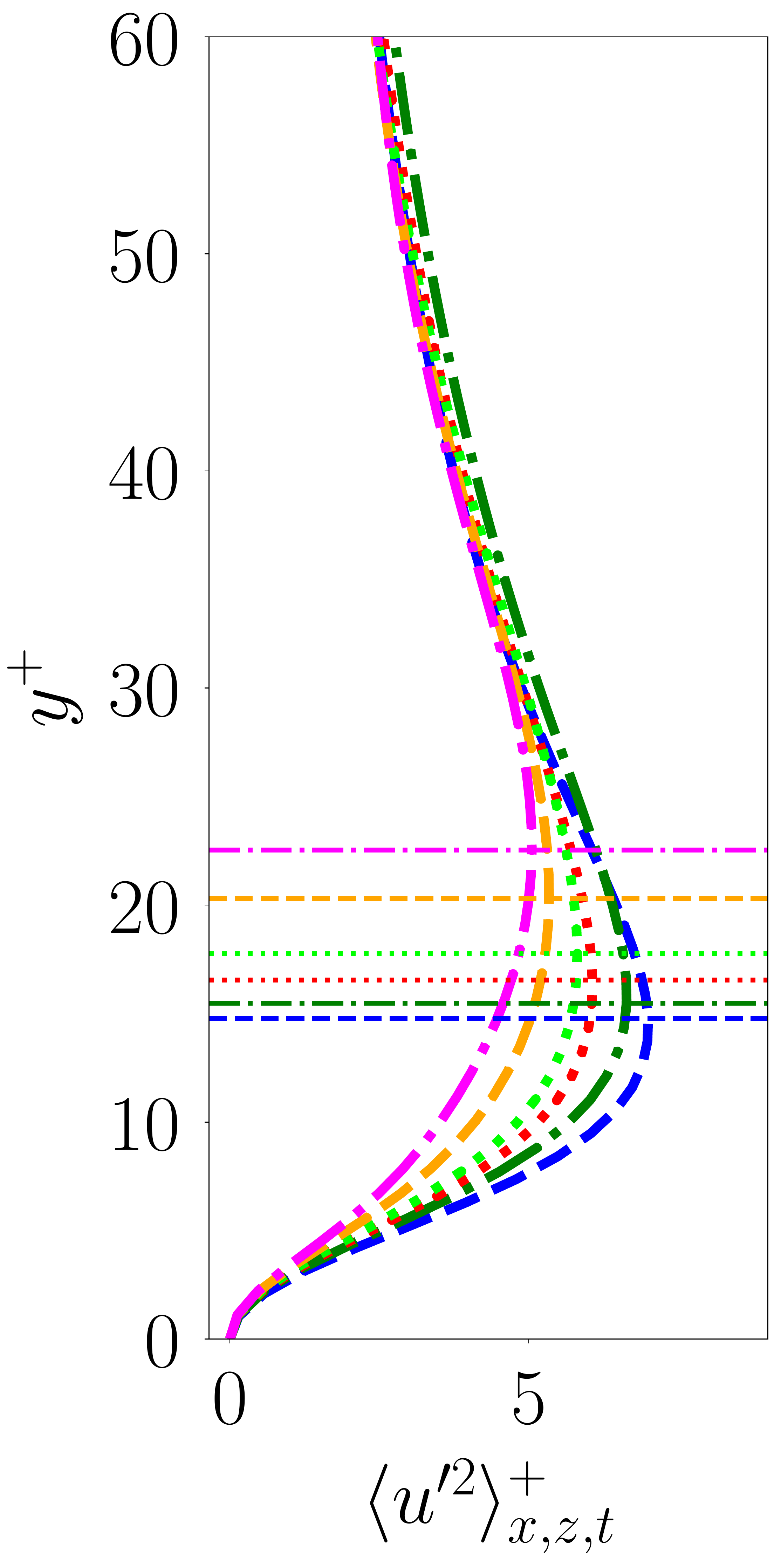}}
\subfigure[\label{fig:tke_P11_new}]{\includegraphics[width=0.21\textwidth]{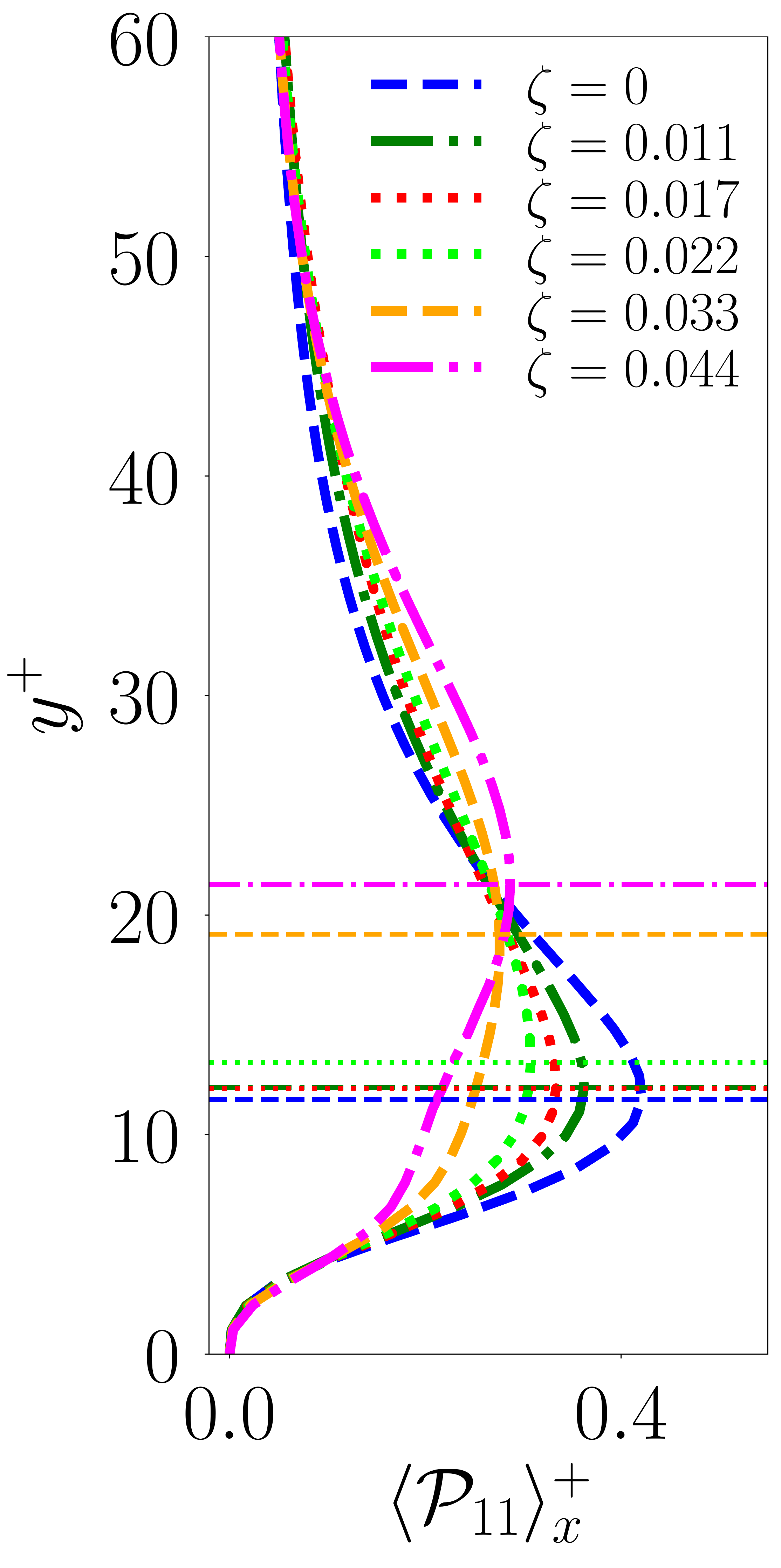}}
\subfigure[\label{fig:tke_P11_dudx_new}]{\includegraphics[width=0.21\textwidth]{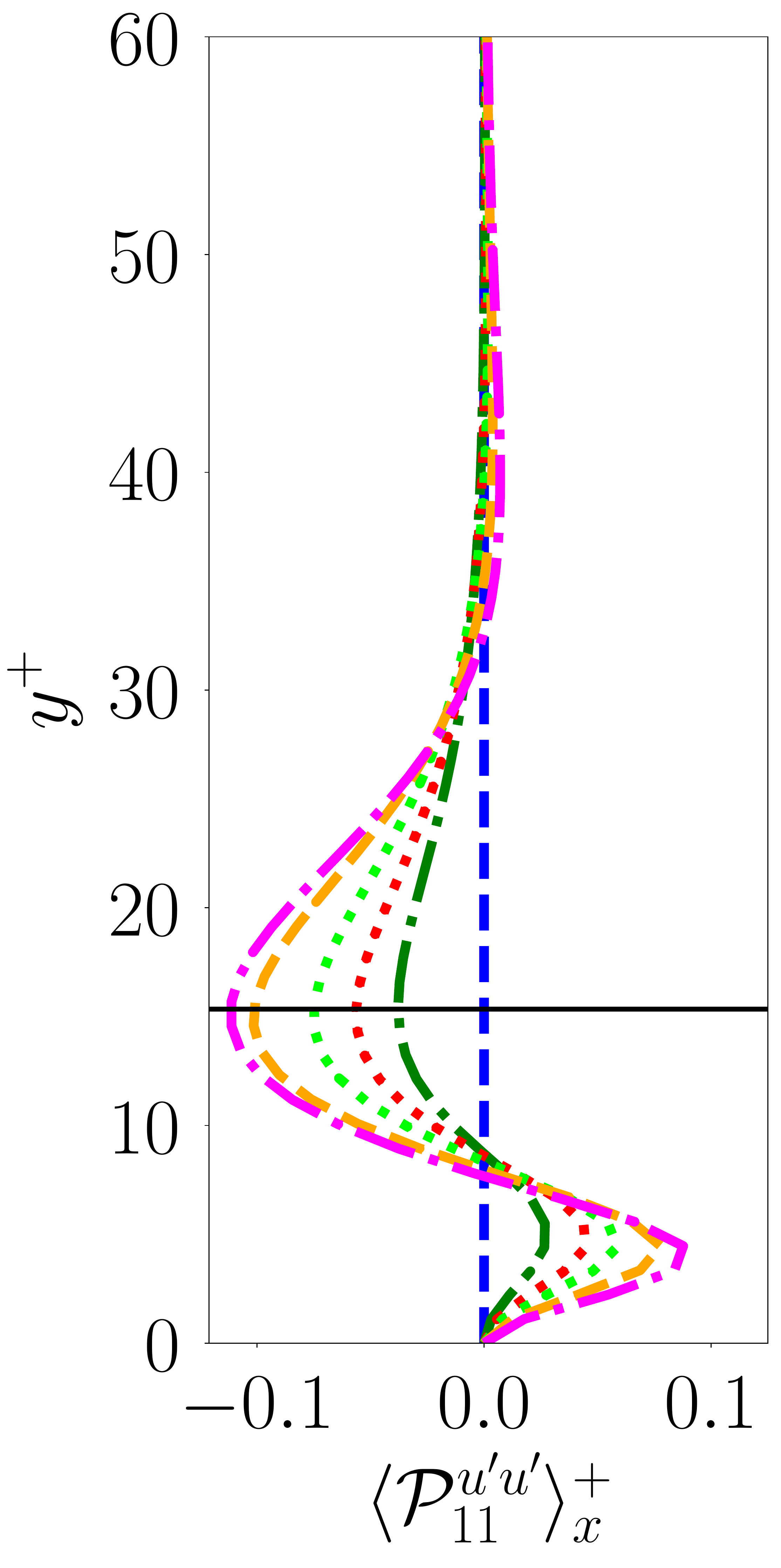}}
\subfigure[\label{fig:tke_P11_dudy_new}]{\includegraphics[width=0.21\textwidth]{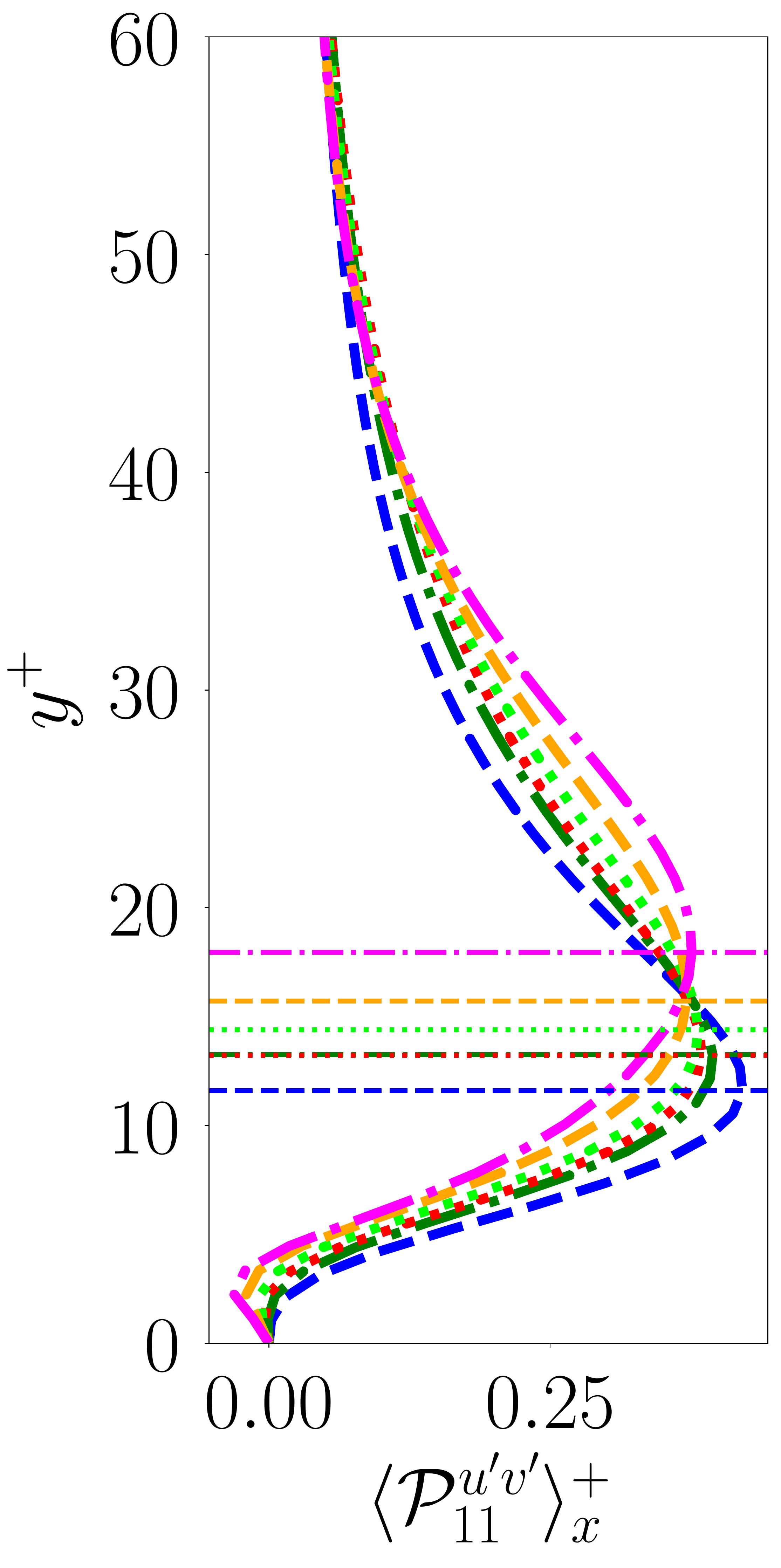}}
}
\mbox{
\subfigure[\label{fig:tke_E11_new}]{\includegraphics[width=0.21\textwidth]{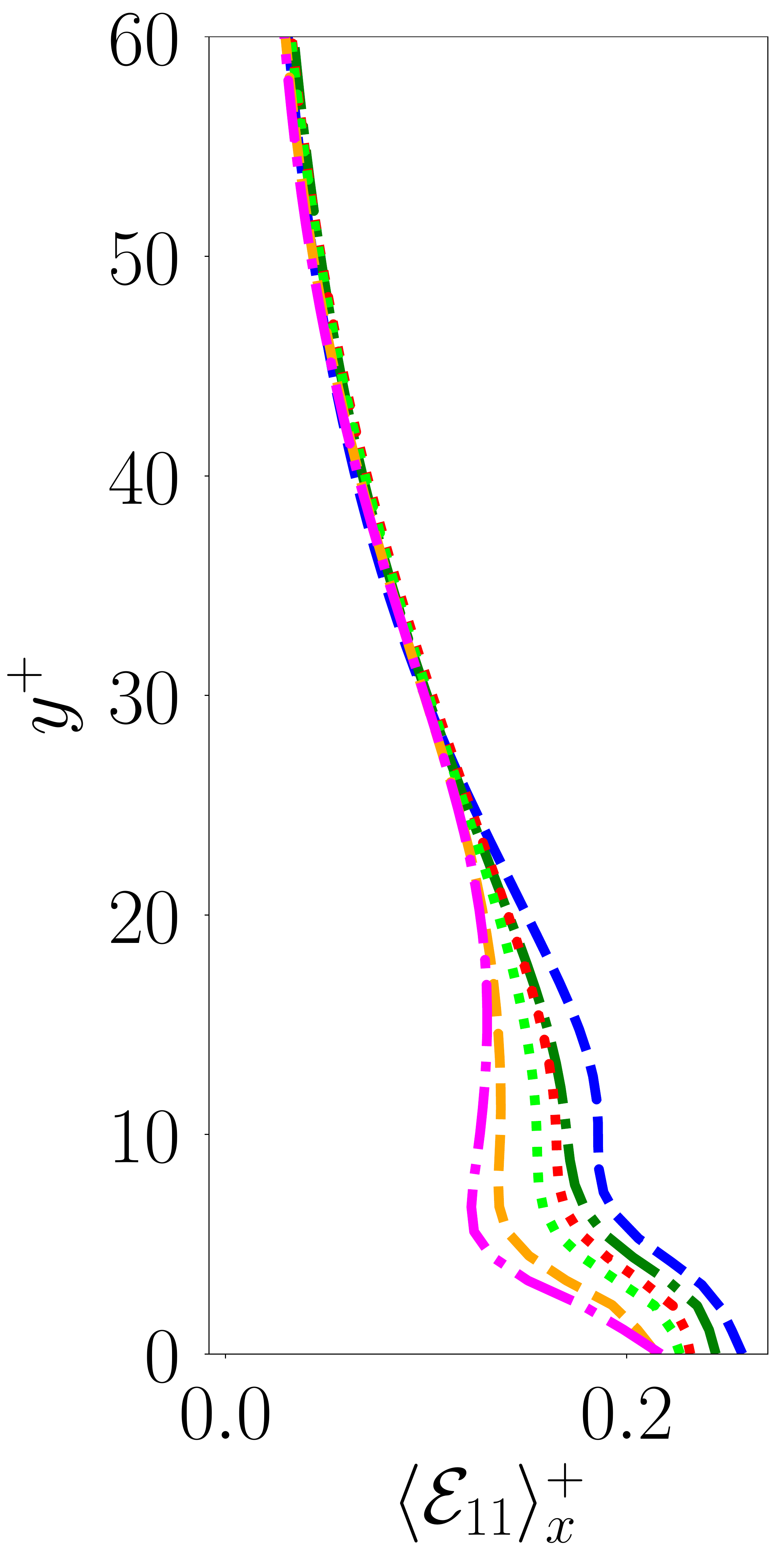}}
\subfigure[\label{fig:tke_R11_new}]{\includegraphics[width=0.21\textwidth]{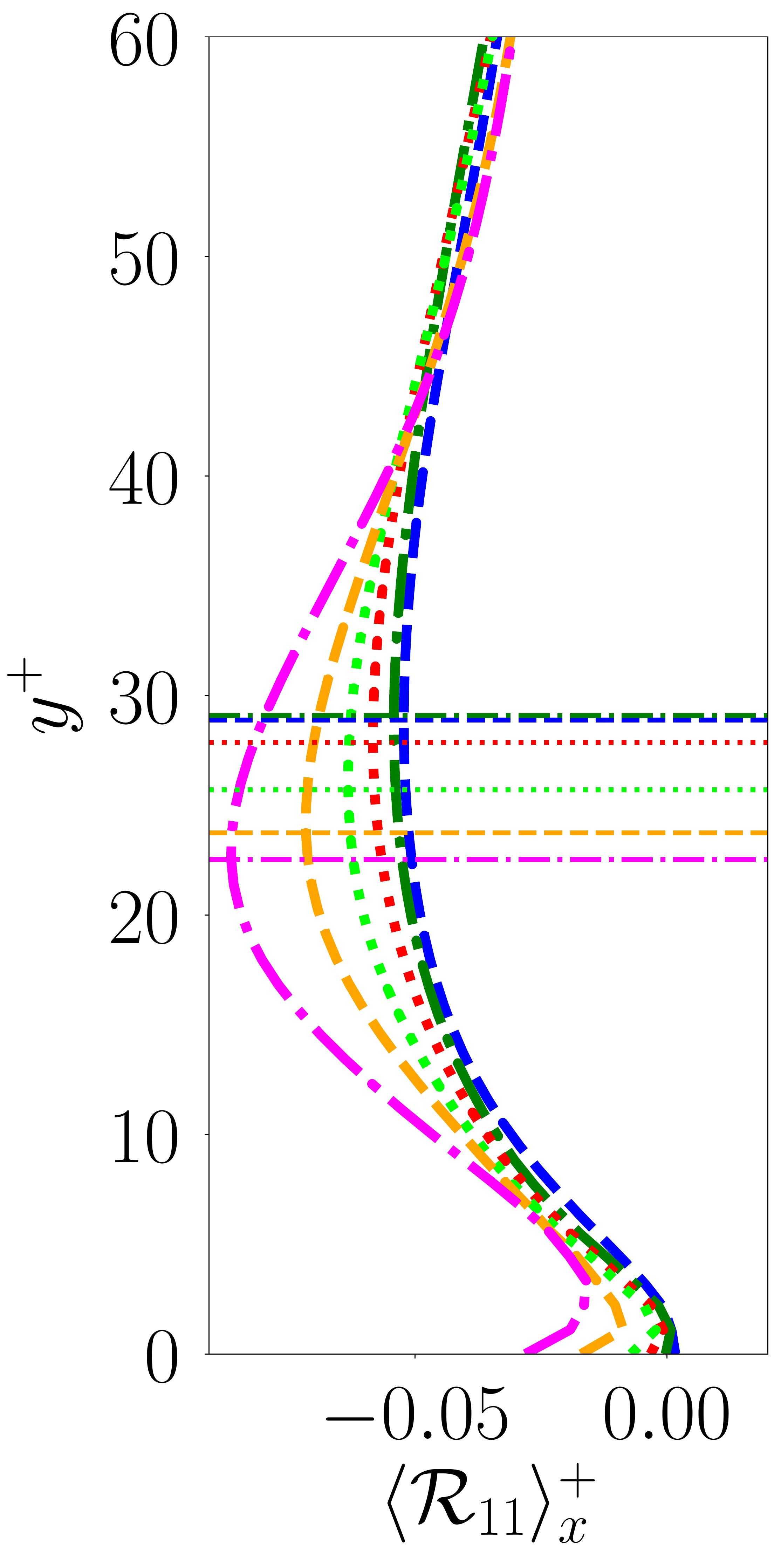}}
\subfigure[\label{fig:tke_L11_new}]{\includegraphics[width=0.21\textwidth]{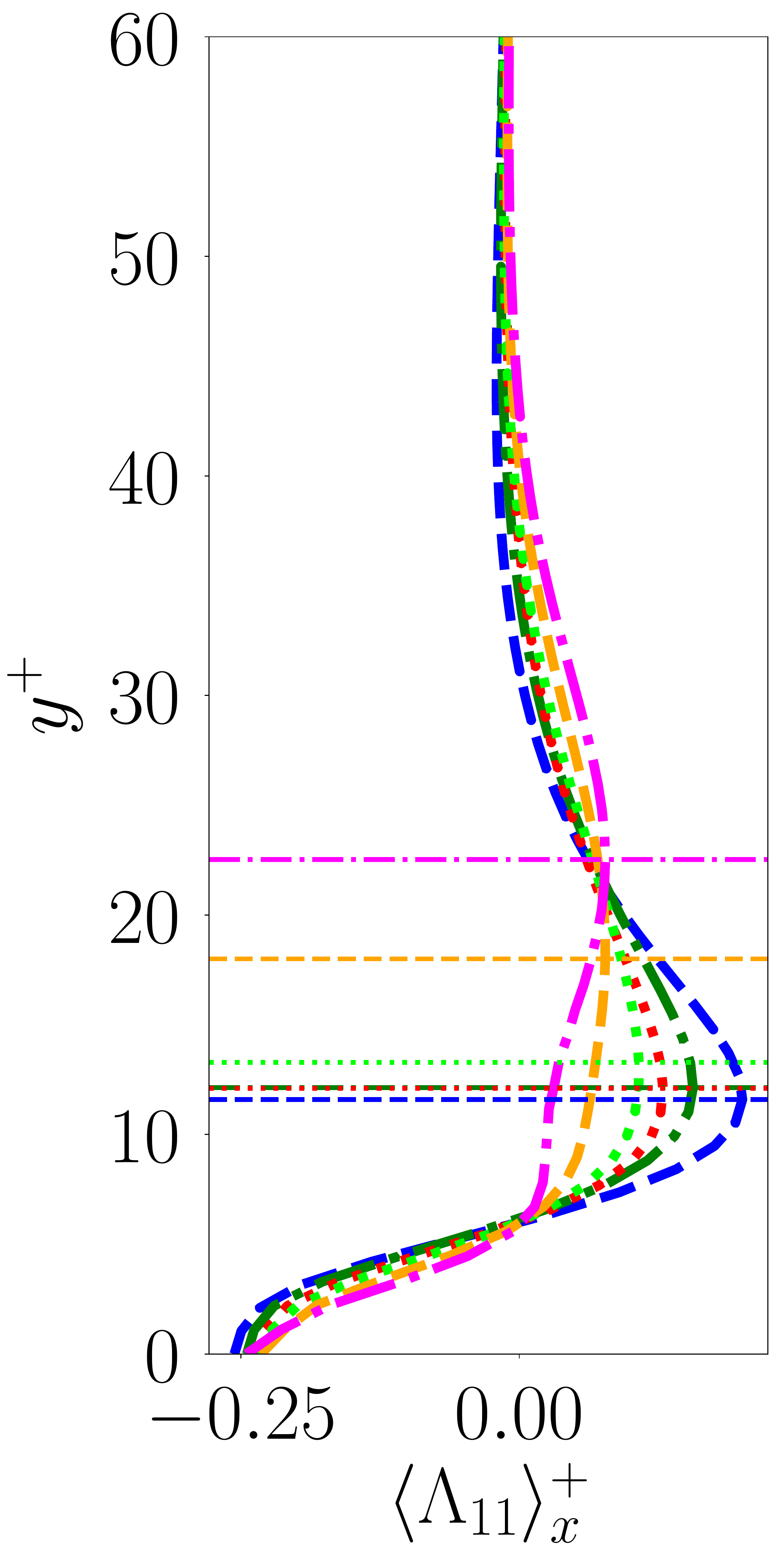}}
\subfigure[\label{fig:tke_D11_new}]{\includegraphics[width=0.21\textwidth]{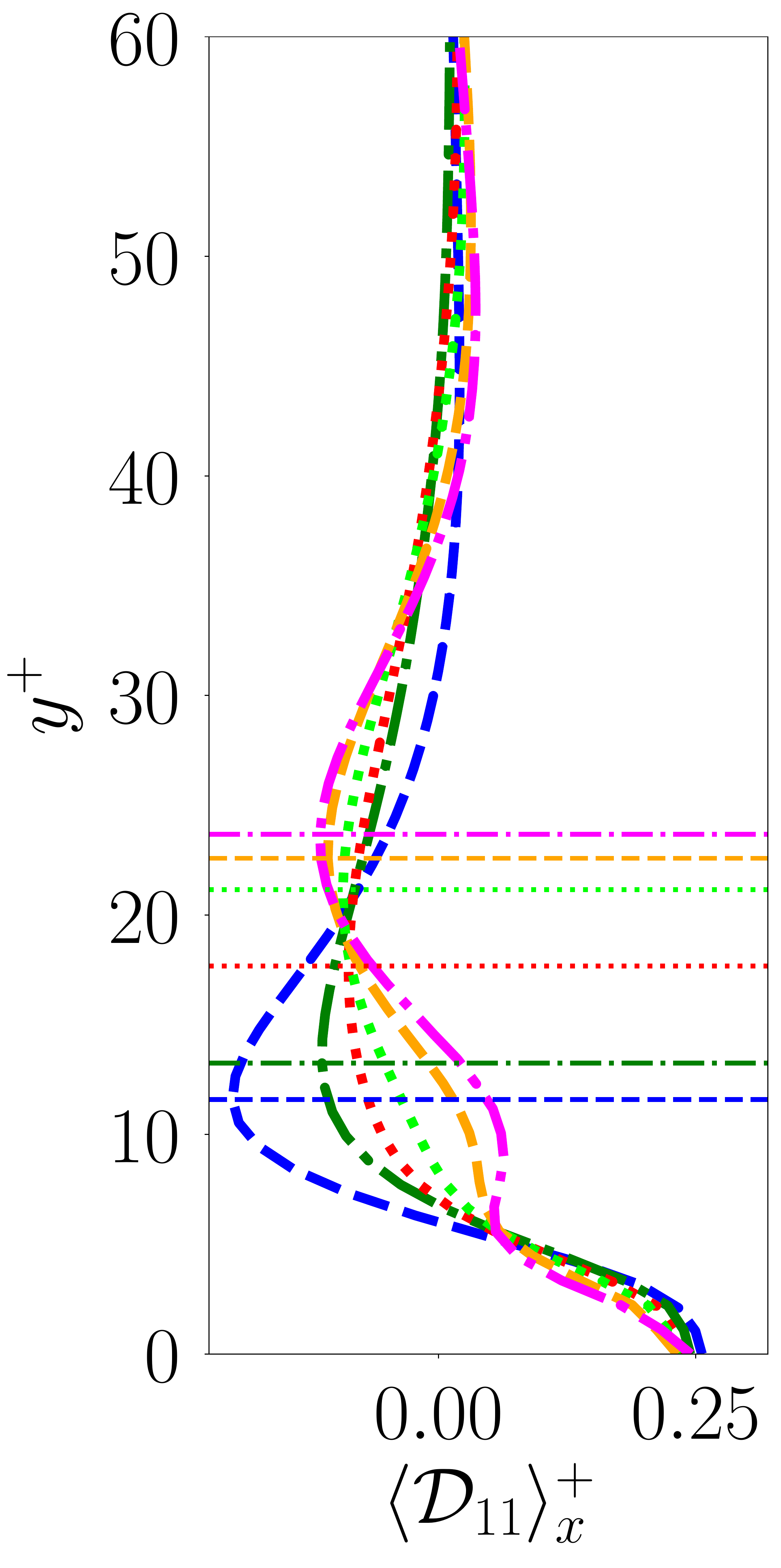}}
}
\caption{\label{fig:uvar_production_new} 
Schematic showing wall-normal variation of inner-scaled, double-averaged vertical variance (a) along with averaged production (b), dissipation (e), pressure-rate-of-strain (f), cumulative local generation $\langle \Lambda_{11} \rangle^+_{x}=\langle\mathcal{P}_{11}\rangle^+_x-\langle\mathcal{E}_{11}\rangle^+_x+\langle\mathcal{R}_{11}\rangle^+_x$, and diffusion (g). We further split the production term $\langle P_{11} \rangle^+_{x}$ into $\langle P^{u'u'}_{11} \rangle^+_{x}$ (c) and $\langle P^{u'v'}_{11} \rangle^+_{x}$ (d). The horizontal lines correspond to the vertical location of maximum/minimum value for a chosen statistic. If the peak locations are different, we color match the horizontal lines with the corresponding curves.
}
\end{figure}

\subsection{Streamwise Variance Structure, $\langle u'^2 \rangle^+_{x,z,t}$  \label{subsec:streamwise_var}}
Noticeable deviations from equilibrium in wavy wall turbulence occur in the second order statistics. 
We observe in figure~\ref{fig:prof_uvar_avg2_new} the \hlll{inner-scaled streamwise variance that peaks in the buffer layer and this peak value decreases systematically with increase in $\zeta$. This $\zeta$ dependence is localized to the roughness sublayer as the inner scaled profiles collapse in the outer region for all the different cases. The location of peak streamwise variance shifts systematically upward with increase in $\zeta$ which is enhanced further due to consistent flow separation at larger wave slopes}. These observations are consistent with \cite{khan2019statistical} which shows that the primary influence of effective wave slope, $\zeta$ on the near surface turbulence structure is to accelerate the transition to isotropy of the Reynolds stress tensor as one moves away from the surface. 
Most literature~\cite{ganju2019direct} suggest the existence of such an upward shift only in response to increases in roughness scale ($a^+$) but not so with the effective slope (i.e. $\lambda^+$ for fixed $a^+$). This aligns with the classical view of fully rough wall turbulence~\cite{pope2001turbulent,jimenez2004turbulent} where the peak variances are expected to occur at nearly the roughness height, $a$\cmnt{ in the limit of a very thin viscous region}. In fact, wall stress boundary conditions for large eddy simulation over rough surfaces are designed to approximate this behavior, especially at high Reynolds numbers with large scale separation ($\delta/a$).
 We remind the reader, we note that in our study, the wave amplitude, $a$ is fixed while the wavelength, $\lambda$ is decreased to increase $\zeta=2a/\lambda$. For a fixed friction Reynolds number, the decrease in $\lambda$ (or increase in $\zeta$) increases the net drag and in turn the friction velocity, ${u_\tau}$. The resulting viscous length scale, $L_v=\frac{\nu}{u_\tau}$ changes very little (because for this fully developed channel flow, $\delta$ and $\delta^+=\delta/L_v$ are constant by design) as does the inner-scaled wave height, $a^+=a/L_v$ (which varies from $12.6-12.9$ for $\zeta$ changing from $0.011 \textrm{ to } 0.044$ as shown in Table~\ref{table_cases}). 
\hlll{Therefore, it is evident that this systematic upward shift in the location of peak streamwise variance arises solely from the changes in $\lambda^+$ and $\zeta$}. In the following, we explore this in great detail by studying the different terms of the transport equation in Figure~\ref{fig:uvar_production_new}.
\subsubsection{Dynamics of Streamwise Variance Transport\label{subsubsec:streamwise_var_transport}}
To dissect the above trends in inner-scaled double-averaged streamwise variance, we quantify the inner-scaled double-averaged terms in the transport equation~\eqref{eq:trans0}-\eqref{eq:trans1} as shown in Figure~\ref{fig:uvar_production_new}. For ease of readability, we include the profile of the streamwise variance (Figure~\ref{fig:prof_uvar_avg2_new60}) along with profile plots of double-averaged transport equation terms (shown in Figures~\ref{fig:tke_P11_new}-\ref{fig:tke_D11_new}). Looking at the magnitudes, the dynamics is controlled by the locally operating terms such as production ($\langle \mathcal{P}_{11} \rangle_x^+$), dissipation ($\langle \mathcal{E}_{11} \rangle_x^+$) and pressure-rate-of-strain ($\langle \mathcal{R}_{11} \rangle_x^+$) which are collectively denoted by $\langle {\Lambda}_{11} \rangle_x^+$. Of these, we note that production and dissipation are dominant while the pressure-rate-of-strain term is relatively less important for this streamwise variance transport. Further, the double-averaged cumulative local variance generation rate, $\langle {\Lambda}_{11} \rangle_x^+$ matches the double-averaged diffusive transport, $\langle \mathcal{D}_{11} \rangle_x^+$. Combining this with equations~\ref{eq:trans1},\ref{eq:trans2} and \ref{eq:trans3}, \hlll{we infer that the (double-averaged) advective transport, $\langle \mathcal{C}_{11} \rangle_x^+$ has little impact on the evolution of (double-averaged) streamwise variance} (see figure~\ref{fig:prof_ConvectiveTransport_C11} in Appendix). 
\hlll{Further, the positive variance production term, $\langle \mathcal{P}_{11} \rangle_x^+$ shows the same trend as $\langle u'^2 \rangle_x^+$ with its peak location trending upwards with increase in $\zeta$ while the peak magnitude decreases. Conversely, the small negative growth rate term, $\langle \mathcal{R}_{11} \rangle_x^+$ displays a downward shift in the location of its peak magnitude with $\zeta$ so that the effective local turbulence generation, $\langle {\Lambda}_{11} \rangle_x^+$ displays an exaggerated upward shift (of its peak). 
Mechanistically, $\langle \mathcal{R}_{11} \rangle_x^+$ represents conversion of streamwise variance to other component variances and is therefore, negative with peak magnitudes occurring away from the surface due to wall-damping of the vertical turbulent motions. The inner-scaled dissipation rate, $\langle \mathcal{E}_{11} \rangle_x^+$ is always positive with magnitudes peaking at the surface and decreasing with $\zeta$ through the boundary layer.} Taking this order of importance into account among the different terms, we focus primarily on the production structure in the following discussions.

\subsubsection{Mechanisms of Streamwise Variance Generation\label{subsubsec:streamwise_var_production}}
We dissect the inner scaled streamwise variance production (figure~\ref{fig:tke_P11_new}) term $\langle \mathcal{P}_{11} \rangle^+_{x}$, in the $\langle {u^{\prime}}^2 \rangle^+_{x,z,t}$ transport equation. 
In the viscous layer ($y^+ \lesssim 7$) $\langle \mathcal{P}_{11} \rangle^+_{x}$ is nearly zero and independent of $\zeta$. Beyond this viscous force dominated region, the production increases monotonically to peak in the buffer layer followed by the subsequent decrease in the log-layer where the different curves collapse (once again independent of $\zeta$). As expected from knowledge of $\langle {u^{\prime}}^2 \rangle^+_{x,z,t}$ profiles, the magnitude and location of the peak $\langle \mathcal{P}_{11} \rangle^+_{x}$ depends on $\zeta$. 
\hlll{We split this component variance production into its dominant contributions, $\langle \mathcal{P}^{u'u'}_{11} \rangle^+_{x}=\langle \langle u'u'\rangle^+_{z,t} d\langle u\rangle^+_{z,t}/dx^+  \rangle_{x}$ that can be attributed to surface undulations and $\langle \mathcal{P}^{u'v'}_{11} \rangle^+_{x}=\langle \langle u'v'\rangle^+_{z,t} d\langle u\rangle^+_{z,t}/dy^+  \rangle_{x}$ representing \cmnt{$\langle {u^{\prime}}^2 \rangle^+_{x,z,t}$} production from flow shear} as shown in figures~\ref{fig:tke_P11_dudx_new} and \ref{fig:tke_P11_dudy_new} respectively.

\begin{figure}[ht!]
	\centering
	\mbox{
		\subfigure[\label{fig:prof_upvp_avg2_new}]{\includegraphics[width=0.21\textwidth]{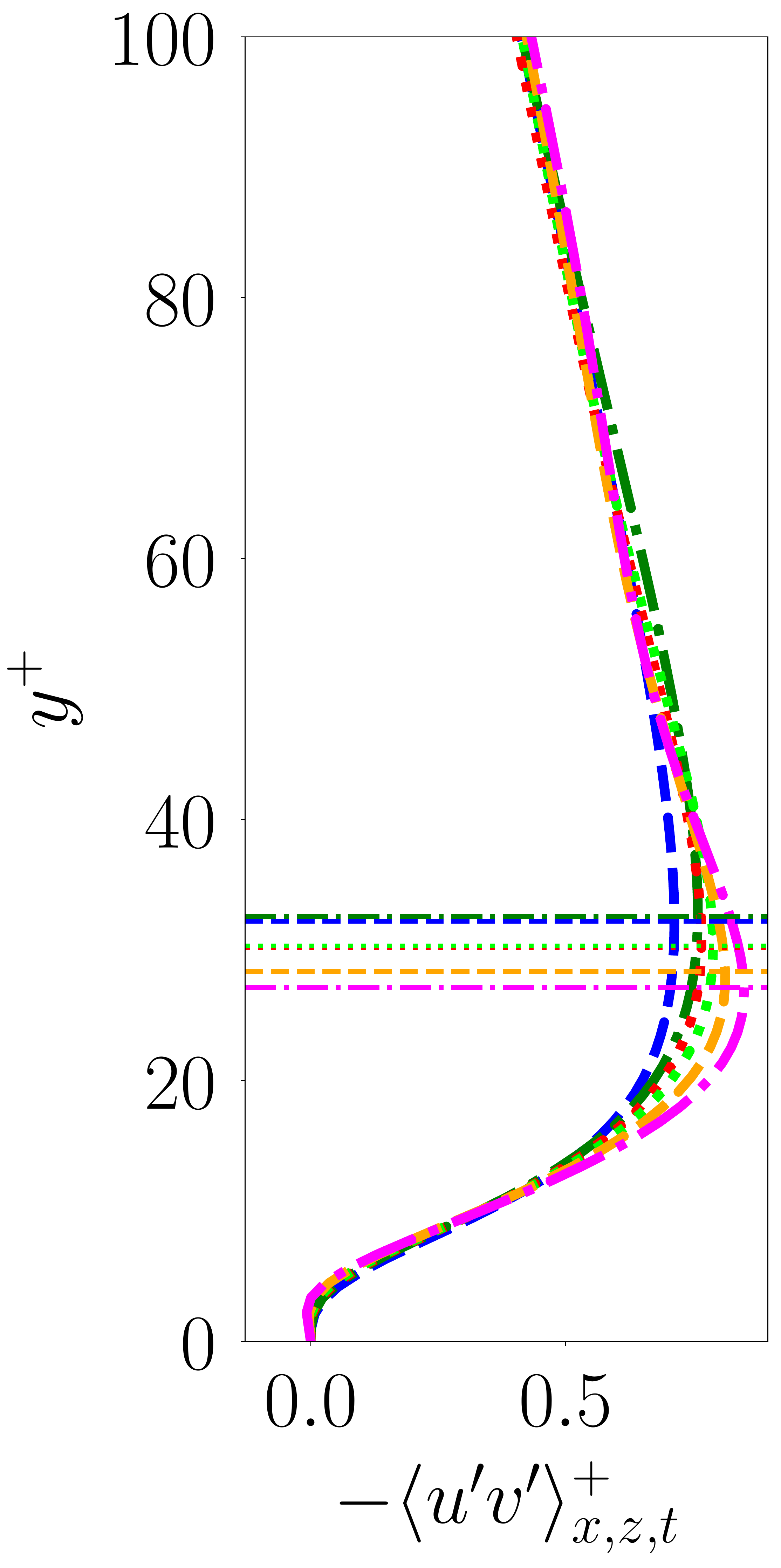}}
\subfigure[\label{fig:prof_upvpzoom_avg2_new}]{\includegraphics[width=0.21\textwidth]{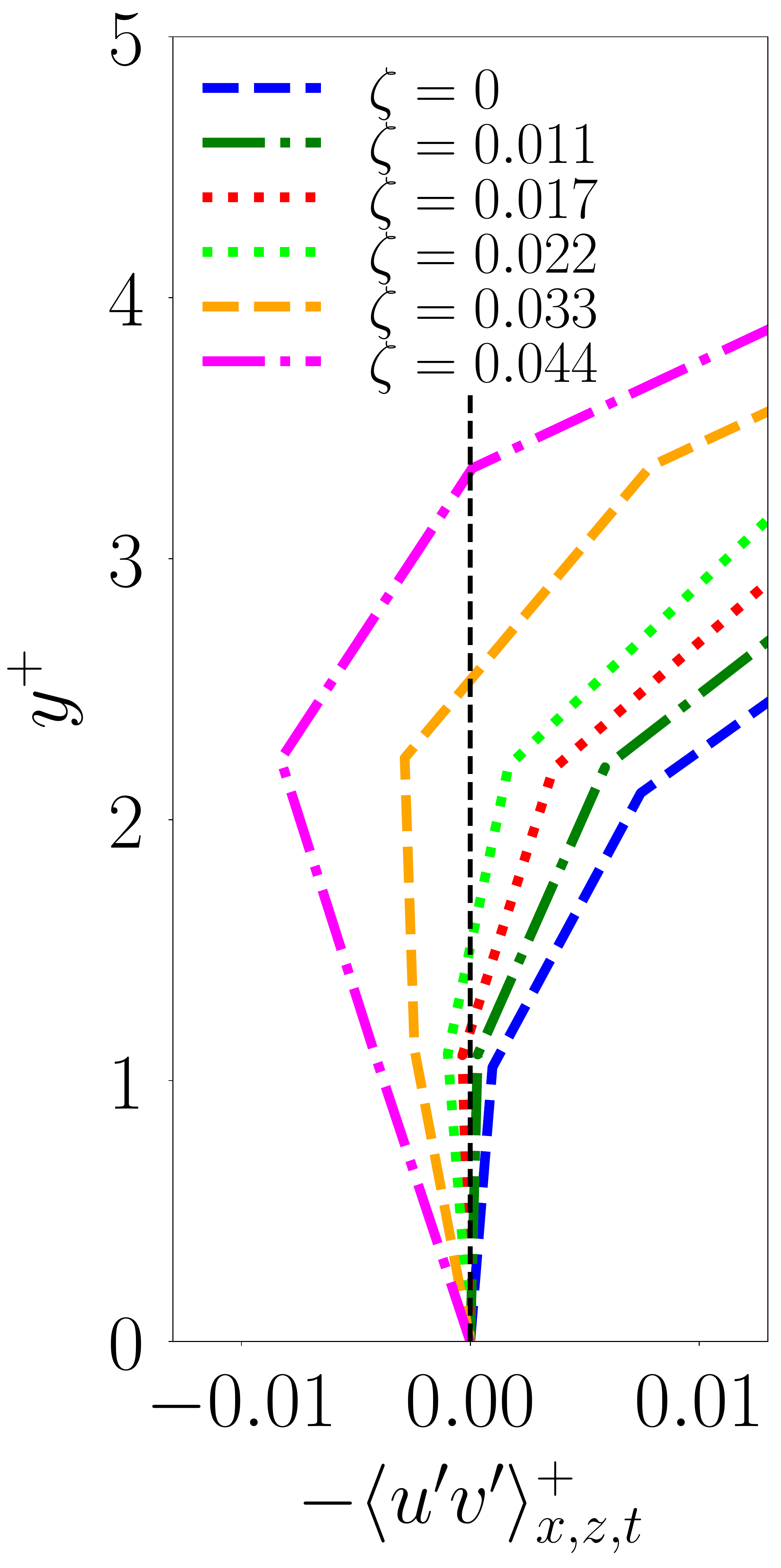}}
	\subfigure[\label{fig:profile_upvp_with_dudy_new}]{\includegraphics[width=0.21\textwidth]{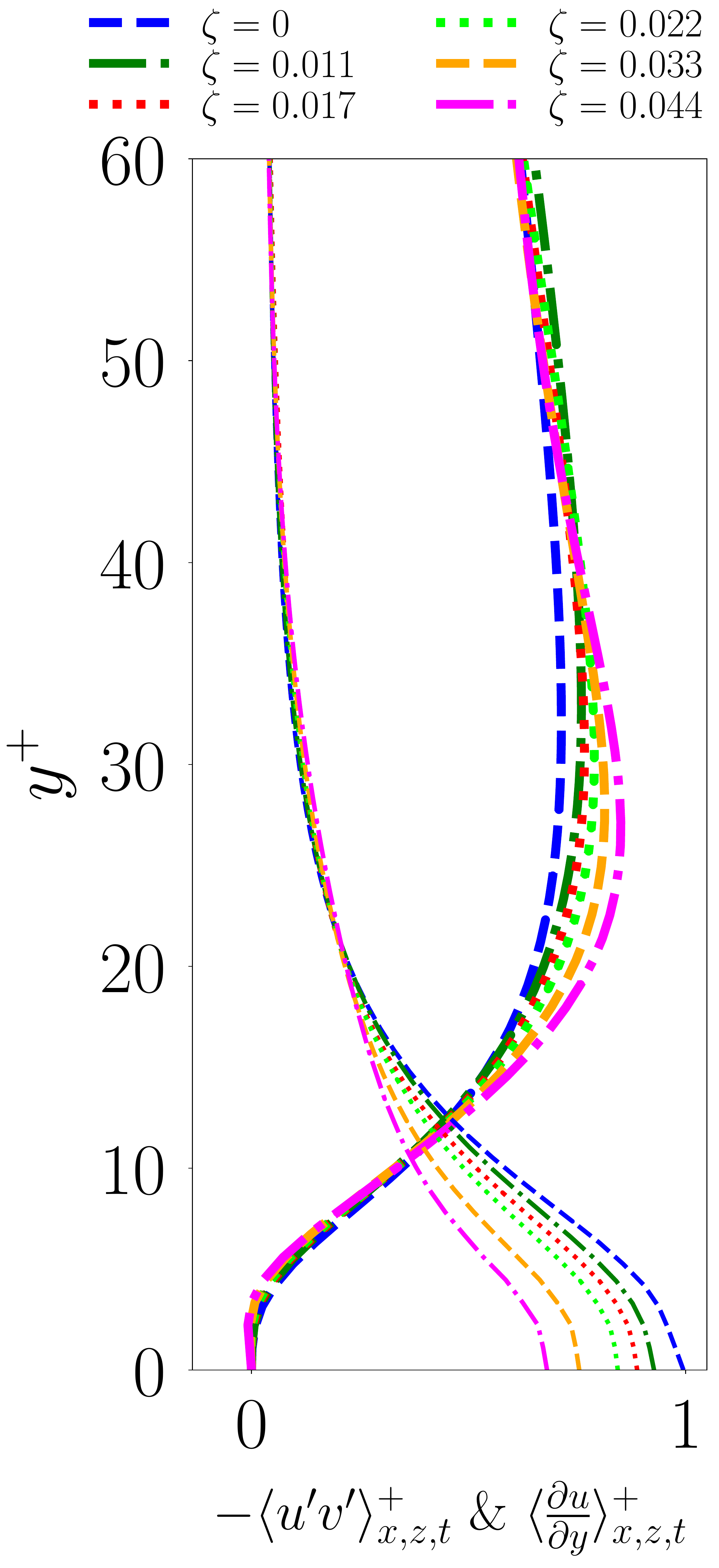}}
	\subfigure[\label{fig:profile_upvp_with_dudy_new_zoomed}]{\includegraphics[width=0.21\textwidth]{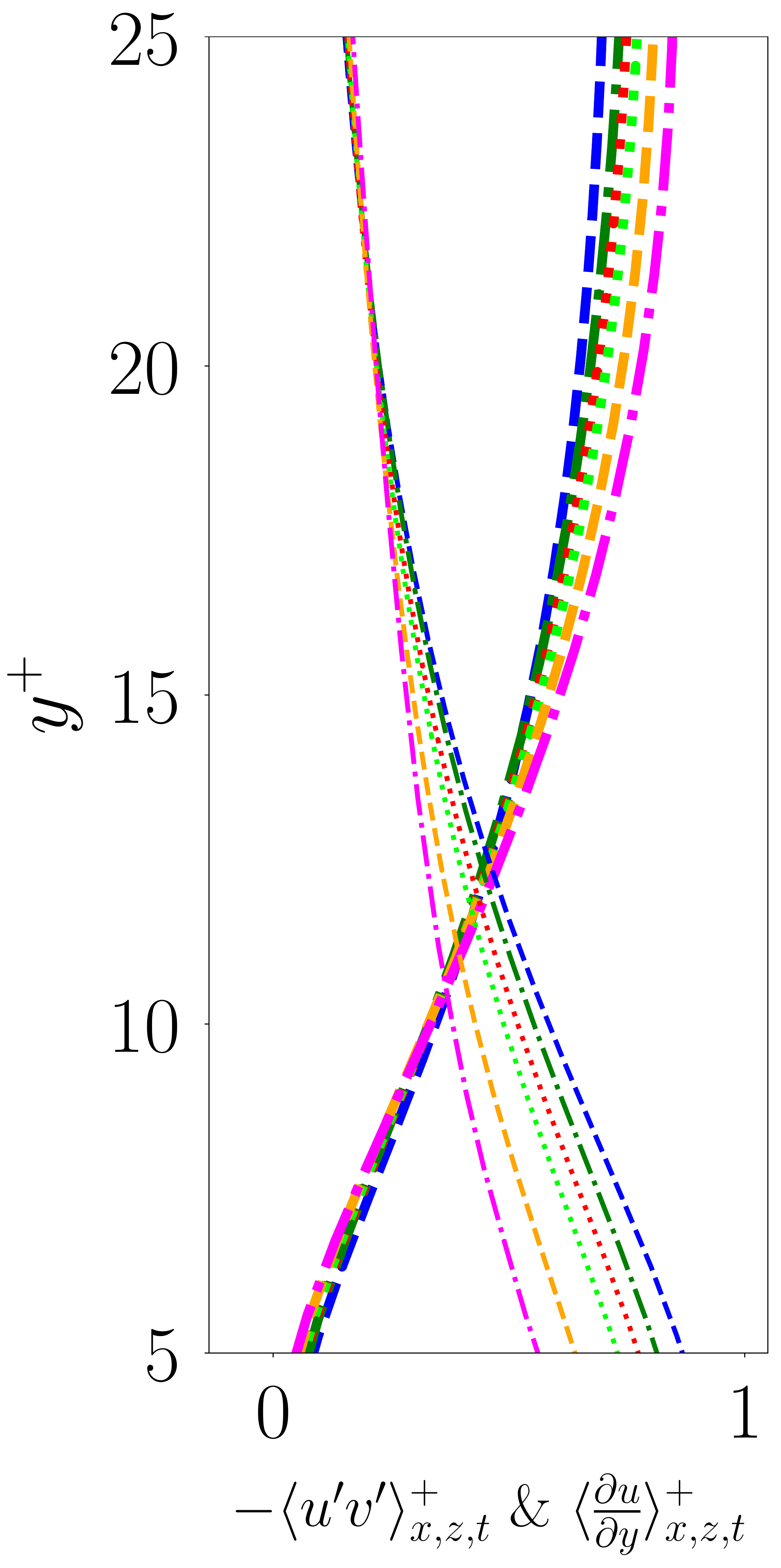}}
}
	\caption{The schematic shows the inner scaled (a) covariance $\langle u'v' \rangle^+_{x,z,t}$,  (b) covariance $\langle u'v' \rangle^+_{x,z,t}$ zoomed near the surface to highlight the positive values under the influence of separation, (c) covariance and mean strain rate to highlight crossover viscous and Reynolds stresses and (d) covariance and mean strain rate zoomed near the surface. 
		The horizontal lines in (a) correspond to the vertical location of maximum magnitude of $\langle u'v' \rangle^+_{x,z,t}$ for different $\zeta$ color  matched with the corresponding curve. The vertical dashed line in (b) corresponds to zero covariance.}
	\label{fig:profiles_avg2_new_set2_RSelements}
\end{figure}

\begin{figure}[ht!]
\centering
\mbox{
 \subfigure[\label{fig:prof_dudx_avg2_new}]{\includegraphics[width=0.21\textwidth]{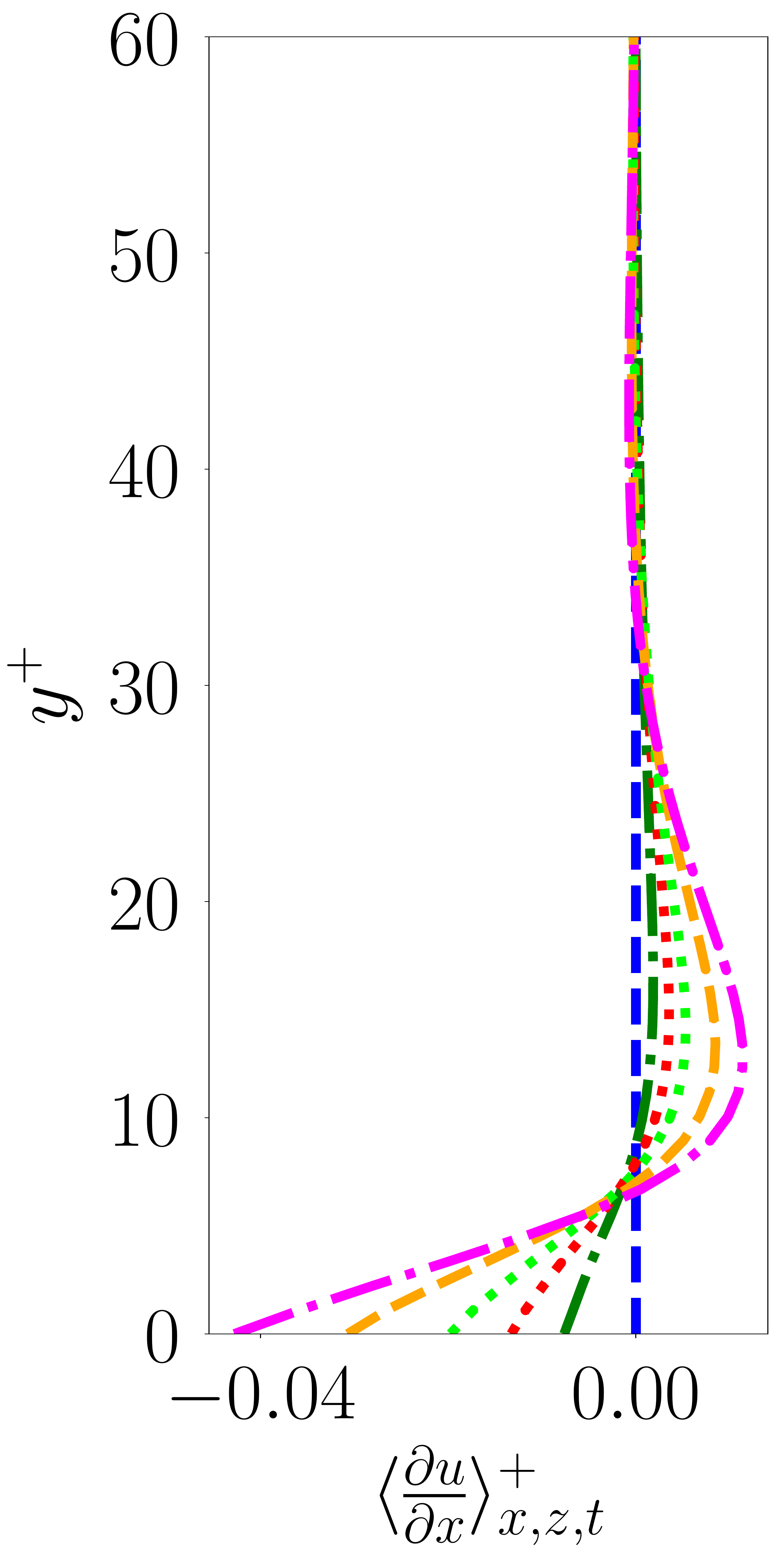}}
 \subfigure[\label{fig:prof_dudy_avg2_new}]{\includegraphics[width=0.21\textwidth]{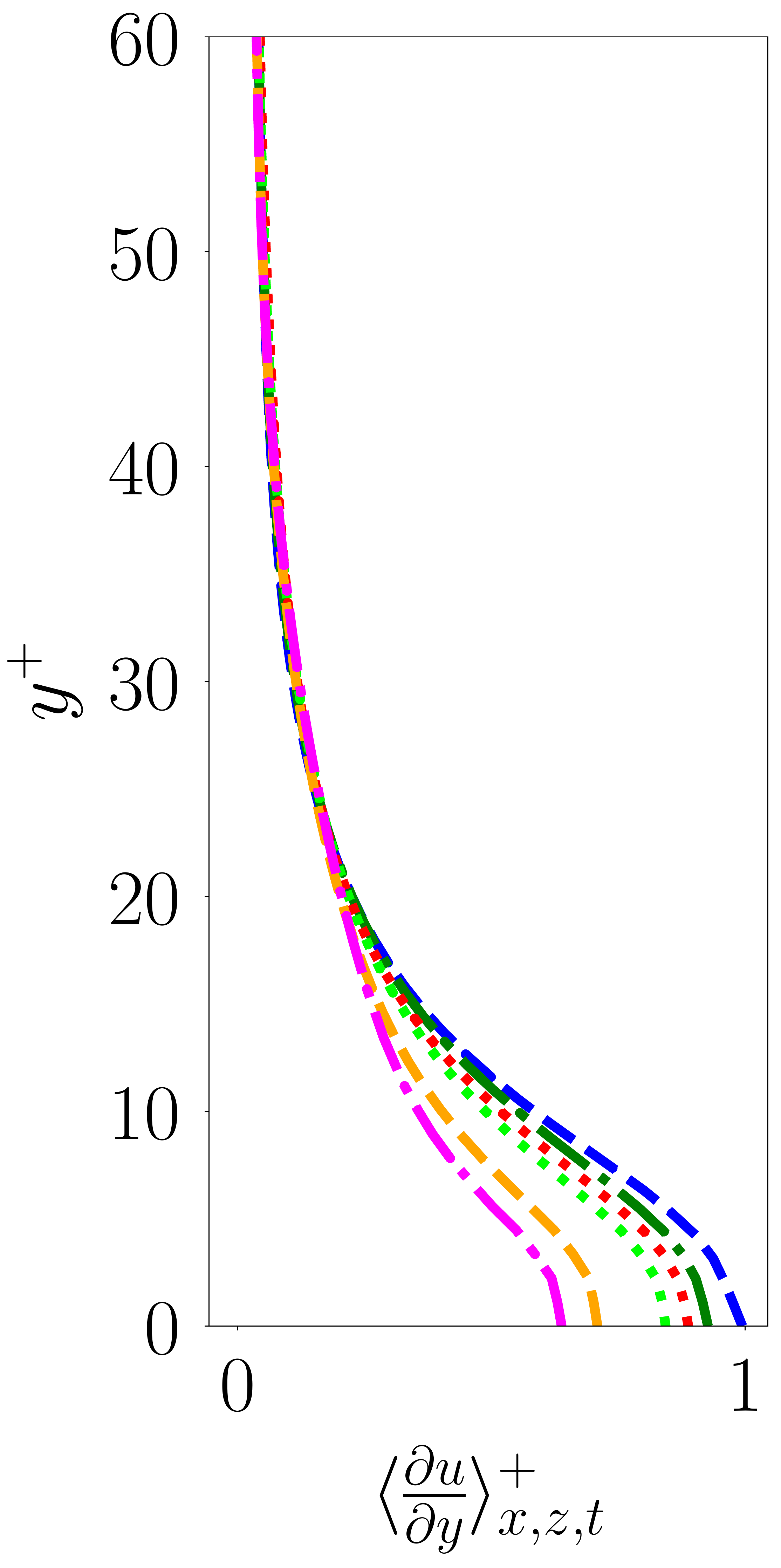}}
 \subfigure[\label{fig:prof_dvdx_avg2_new}]{\includegraphics[width=0.21\textwidth]{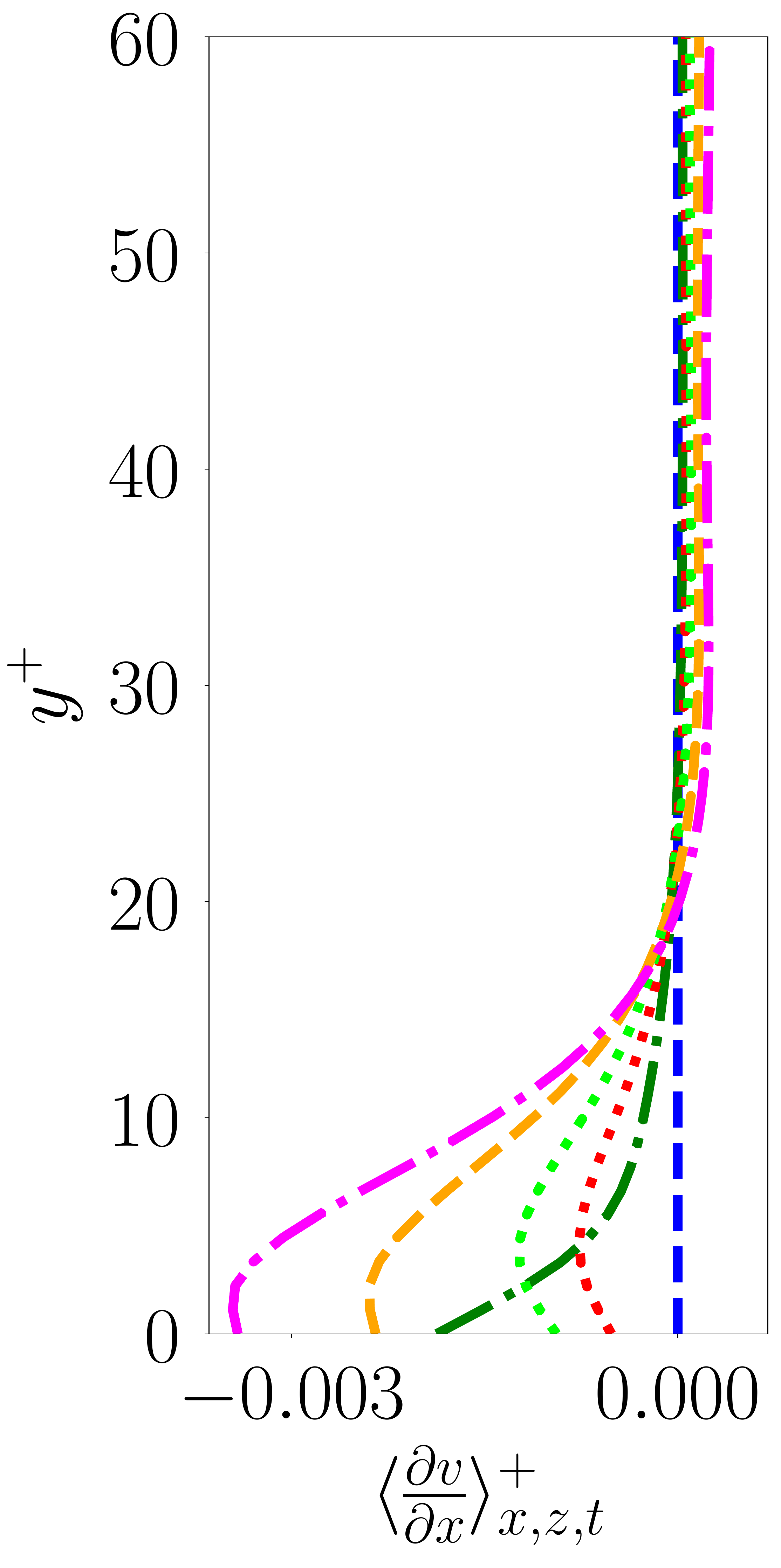}}
 \subfigure[\label{fig:prof_dvdy_avg2_new}]{\includegraphics[width=0.21\textwidth]{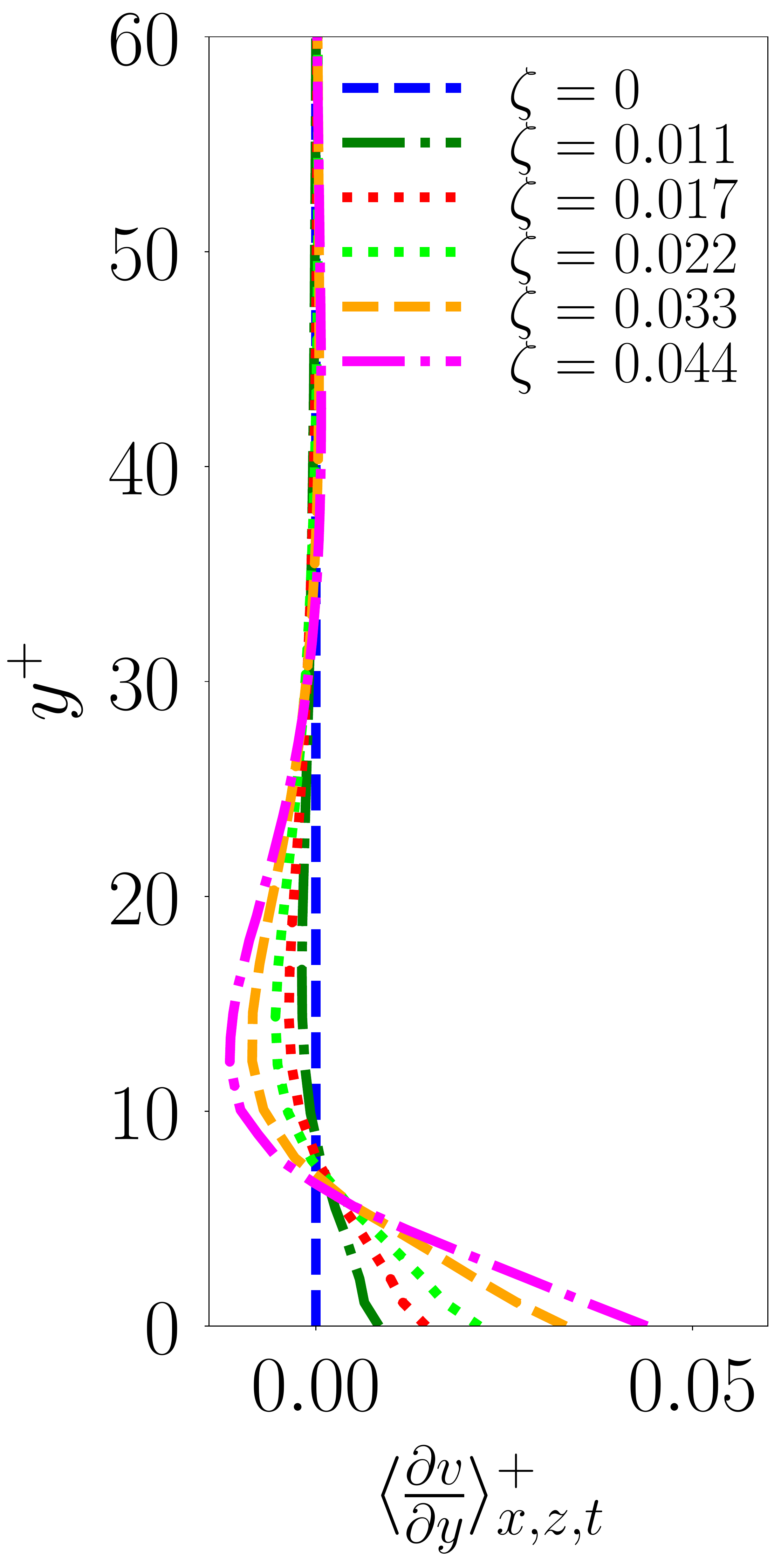}}
}
\caption{Inner scaled mean (a) streamwise gradient of streamwise velocity, $d\langle u^+ \rangle_{x,z,t}/{dx^+}$, (b) vertical gradient of streamwise velocity, $d\langle u^+ \rangle_{x,z,t}/{dy^+}$ (c) streamwise gradient of vertical velocity, $d\langle v^+ \rangle_{x,z,t}/{dx^+}$ and (d) vertical gradient of vertical velocity, $d\langle u^+ \rangle_{x,z,t}/{dy^+}$.
}
\label{fig:profiles_avg2_new_set2_velgradients}
\end{figure}

\begin{figure}[ht!]
\centering
\mbox{
\subfigure[\label{fig:SurfaceDispersion_in_ABL-turbulence_P11_a}]{\includegraphics[width=0.21\textwidth]{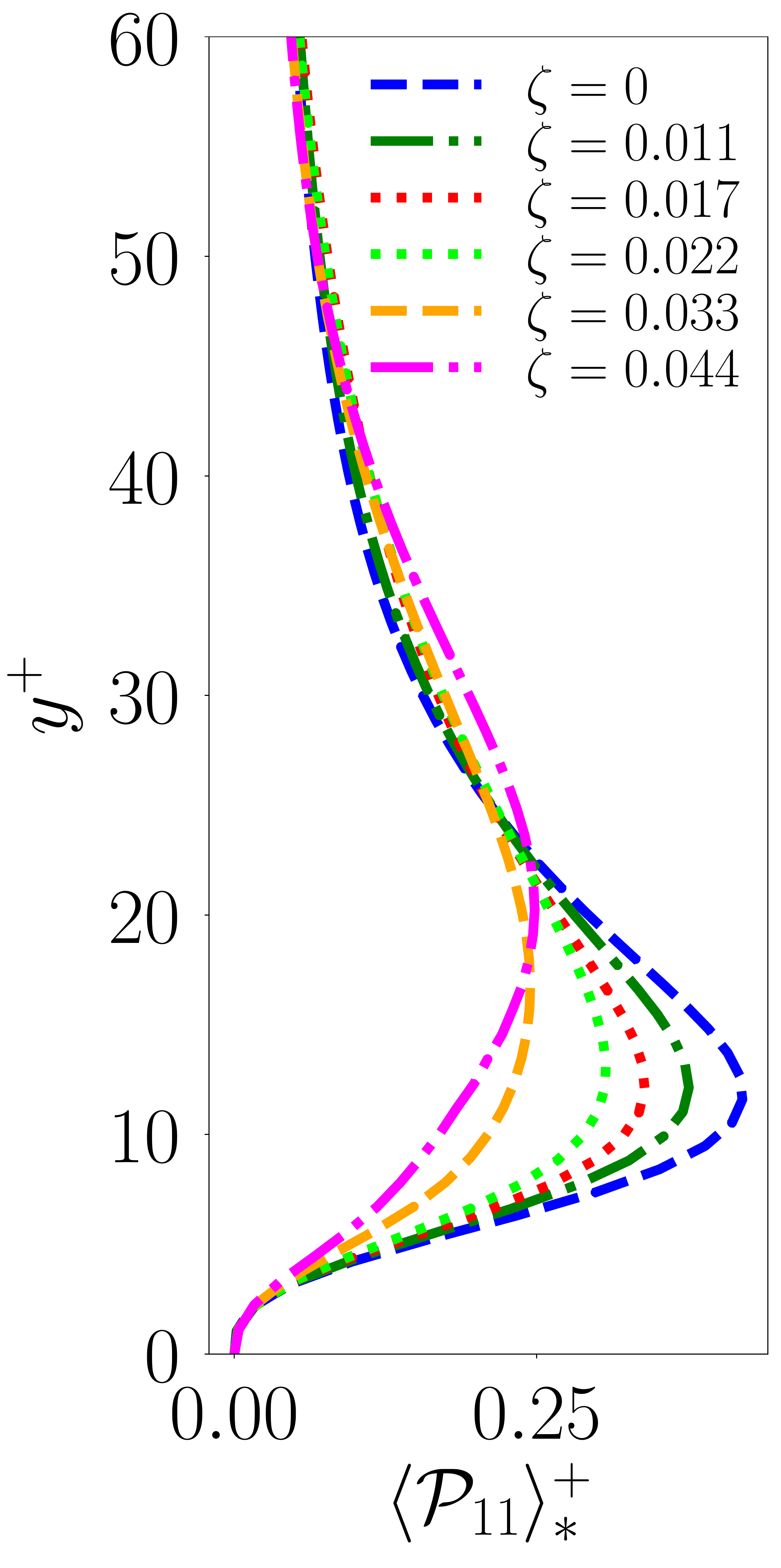}}
\subfigure[\label{fig:SurfaceDispersion_in_ABL-turbulence_P11_b}]{\includegraphics[width=0.21\textwidth]{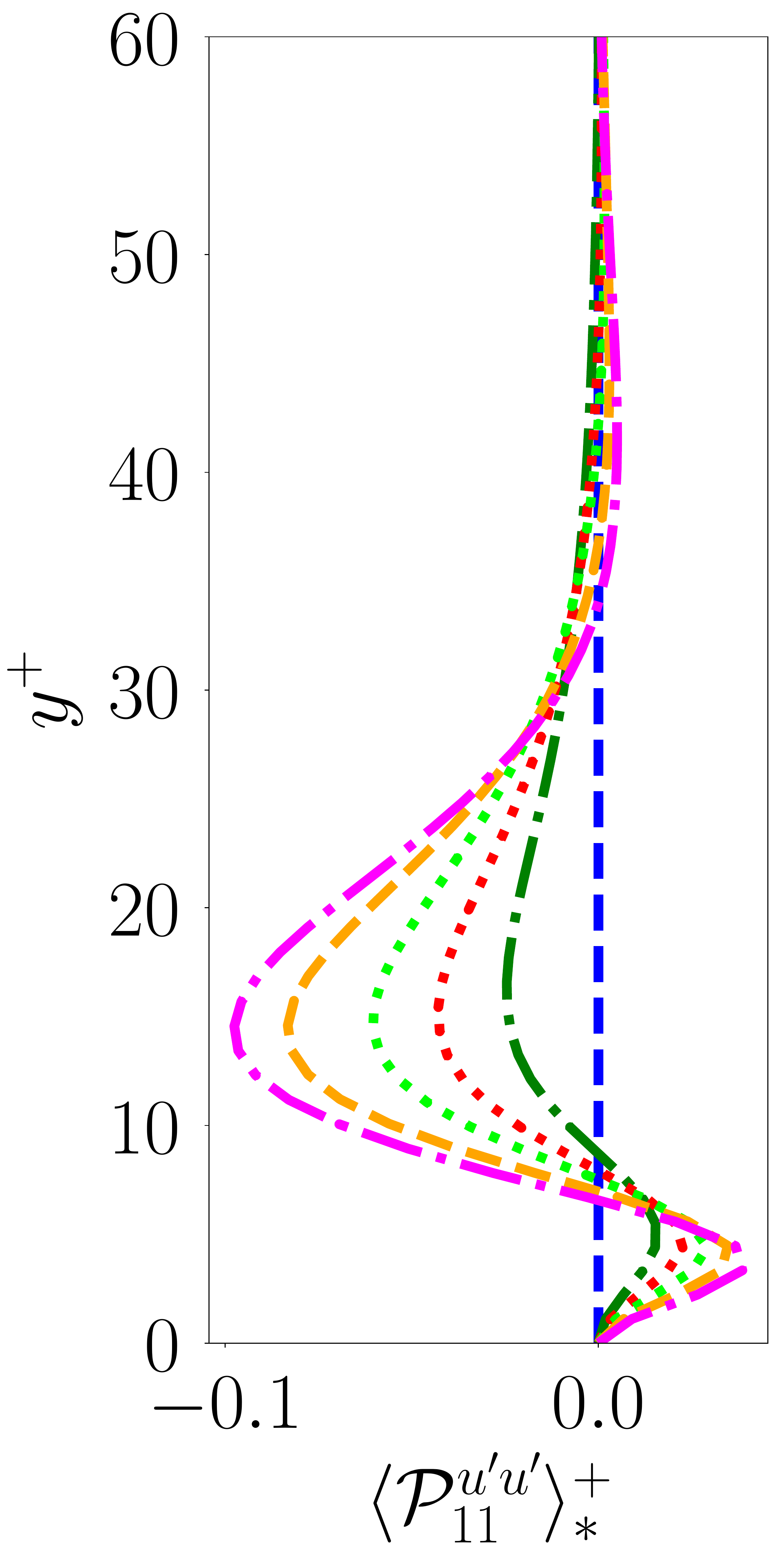}}
\subfigure[\label{fig:SurfaceDispersion_in_ABL-turbulence_P11_c}]{\includegraphics[width=0.21\textwidth]{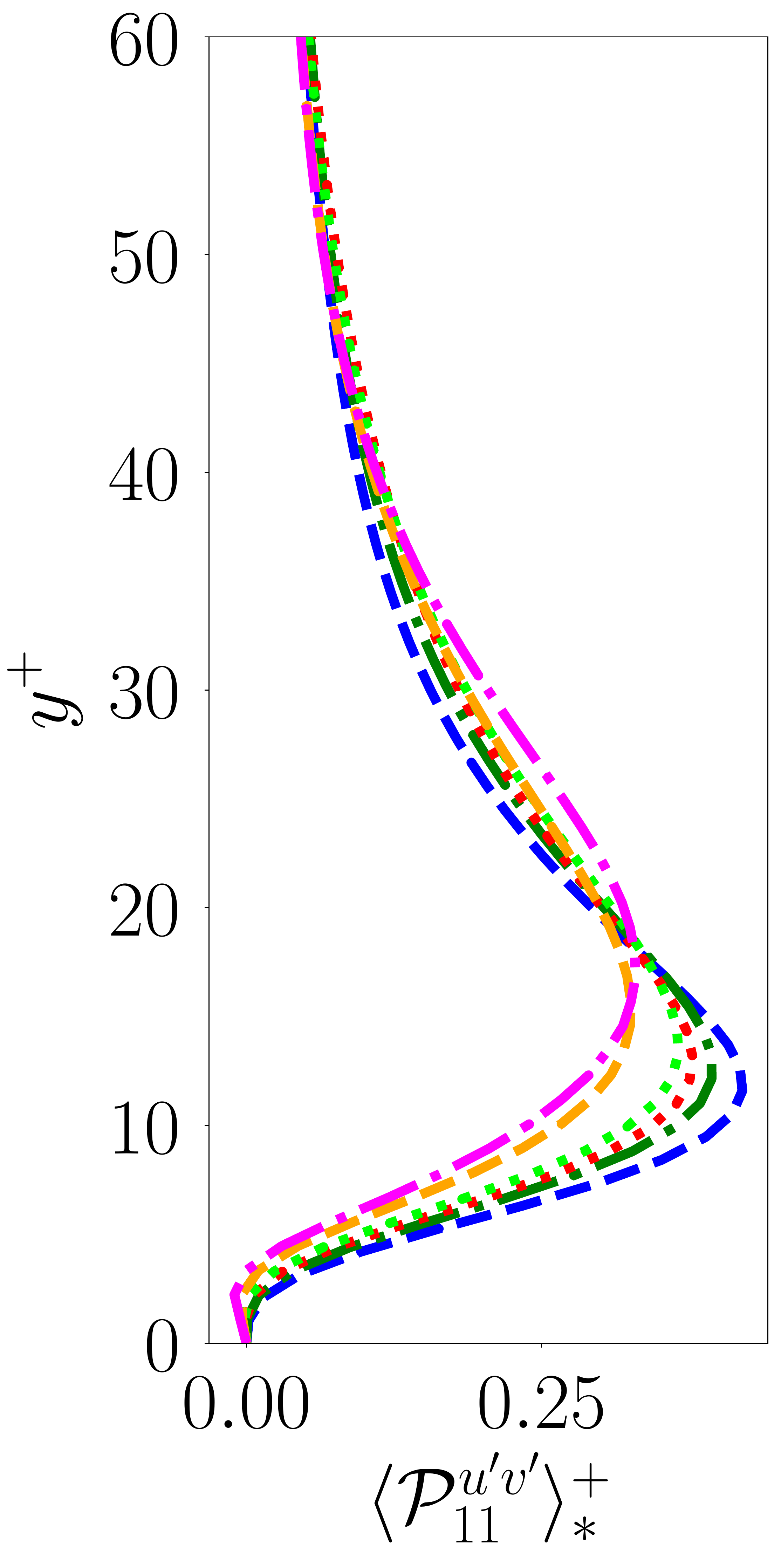}}
}
\mbox{
\subfigure[\label{fig:SurfaceDispersion_in_ABL-turbulence_P11_d}]{\includegraphics[width=0.21\textwidth]{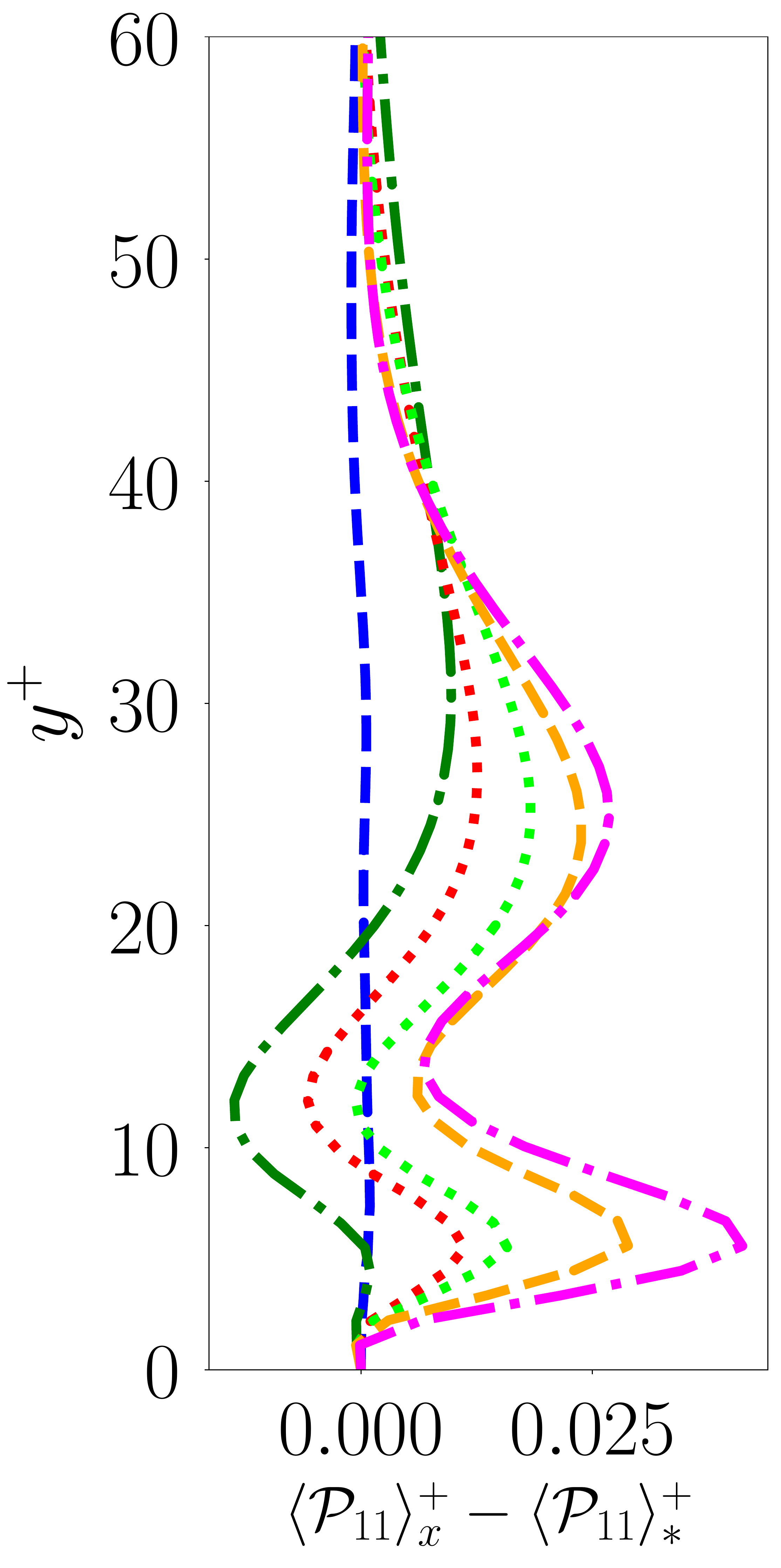}}
\subfigure[\label{fig:SurfaceDispersion_in_ABL-turbulence_P11_e}]{\includegraphics[width=0.21\textwidth]{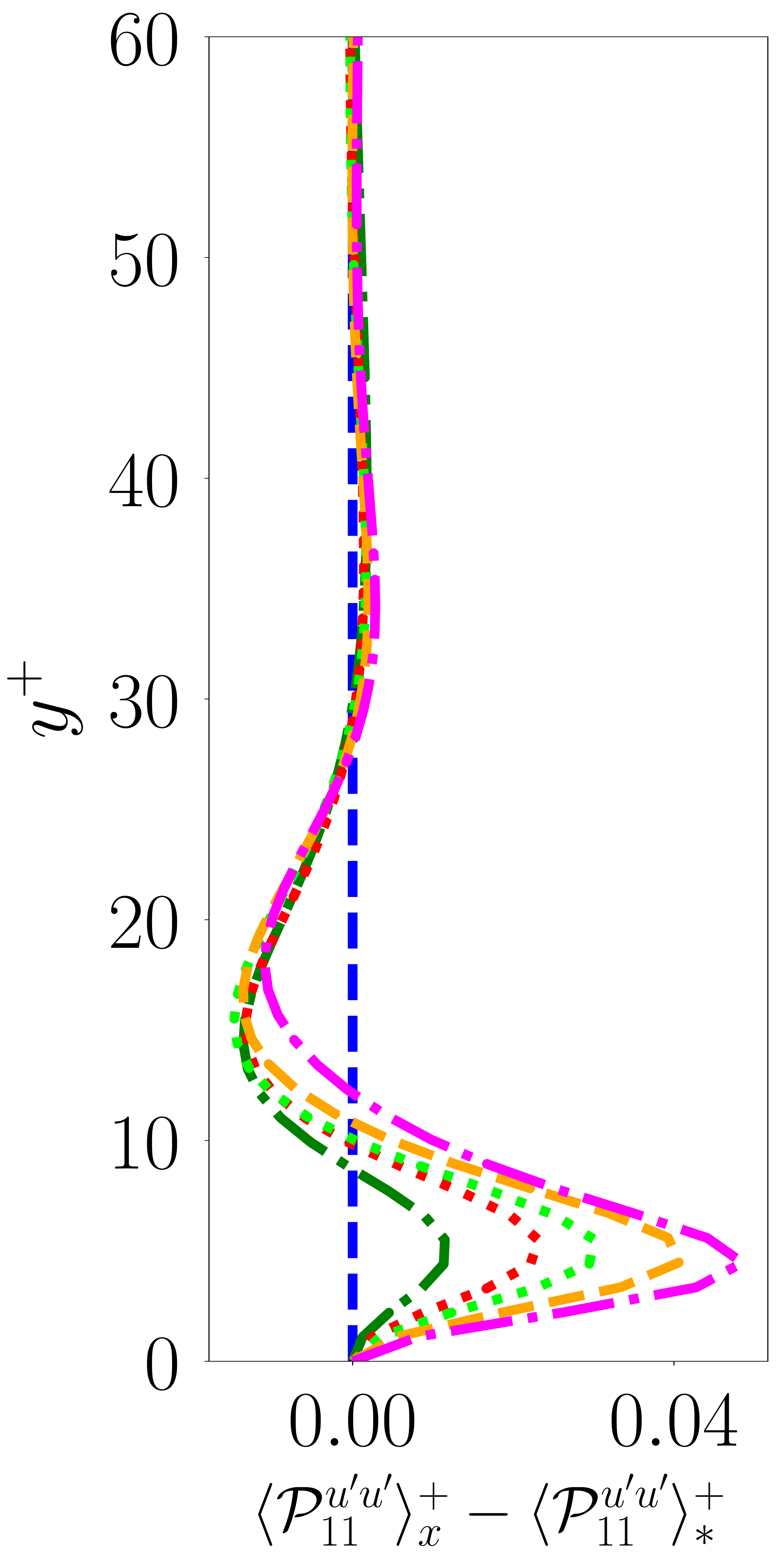}}
\subfigure[\label{fig:SurfaceDispersion_in_ABL-turbulence_P11_f}]{\includegraphics[width=0.21\textwidth]{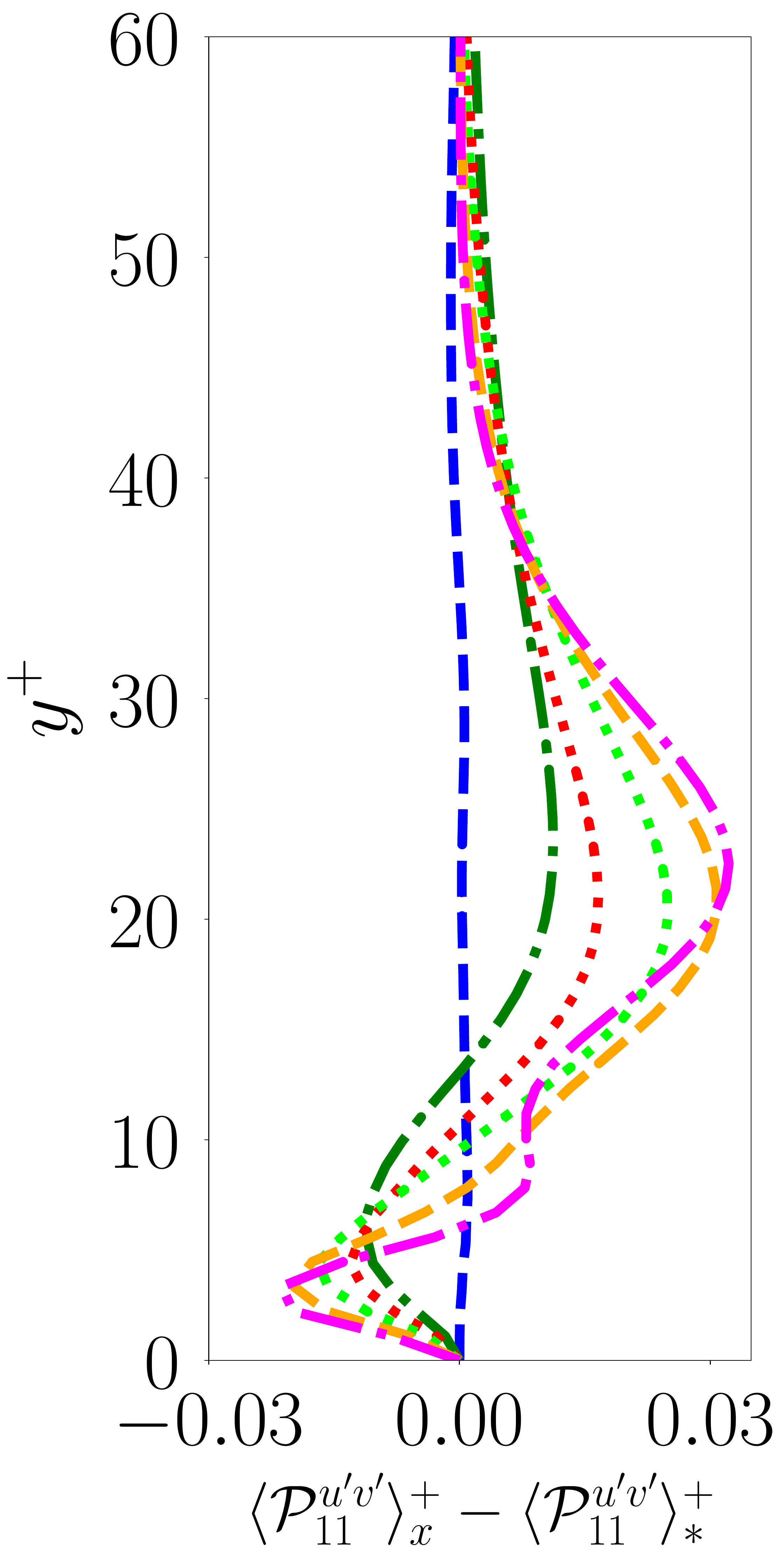}}
}
\caption{\label{fig:SurfaceDispersion_in_ABL-turbulence_P11}
{Schematic showing pseudo-production estimates of streamwise variance (top row) using product of double-averaged Reynolds stress and mean strain rate (denoted by a `$*$' subscript) and their deviations from true double-averaged production (bottom row). Top row: (a) total pseudo-production, $\langle \mathcal{P}_{11} \rangle_*^+$  (b) component, $\langle \mathcal{P}^{u'u'}_{11} \rangle_*^+$ and (c) component, $\langle \mathcal{P}^{u'v'}_{11} \rangle_*^+$. Bottom row: surface disperison-induced deviations (d) $\langle \mathcal{P}_{11} \rangle_x^+$-$\langle \mathcal{P}_{11} \rangle_*^+$, (e) $\langle \mathcal{P}^{u'u'}_{11} \rangle_x^+$-$\langle \mathcal{P}^{u'u'}_{11} \rangle_*^+$ and (f) $\langle \mathcal{P}^{u'v'}_{11} \rangle_x^+$-$\langle \mathcal{P}^{u'v'}_{11} \rangle_*^+$ .}
}
\end{figure}

\paragraph{Averaged Production from Flow Shear:} 
The $\langle {u^{\prime}}^2 \rangle^+_{x,z,t}$ production arising from the interaction of the inner scaled mean shear, $d\langle u\rangle^+_{z,t}/dy^+$ with the inner scaled vertical momentum flux ($\langle u'v'\rangle^+_{z,t}$) denoted by $\langle P^{u'v'}_{11} \rangle^+_{x}$ peaks in the buffer layer ($y^+ \approx 12-18$ as seen in figure~\ref{fig:tke_P11_dudy_new}) and this peak shifts systematically upward (see horizontal lines) as $\zeta$ varies over $0-0.044$. This trend can be interpreted approximately through figures~\ref{fig:prof_upvp_avg2_new} and \ref{fig:prof_dudy_avg2_new} representing the double averaged profiles of normalized covariance, $\langle u'v'\rangle^+_{x,z,t}$ and mean shear, $d\langle u\rangle^+_{x,z,t}/dy^+$ respectively. This interpretation is inexact for non-flat wavy surfaces as the streamwise production from the interaction of mean shear with mean vertical turbulent flux given by $\langle \mathcal{P}^{u'v'}_{11} \rangle^+_{x}=\langle \langle u'v'\rangle^+_{z,t} d\langle u\rangle^+_{z,t}/dy^+  \rangle_{x}$ is not the same as $\langle u'v'\rangle^+_{x,z,t}d\langle u\rangle^+_{x,z,t}/dy^+$. 
We denote the latter expression as surrogate or pseudo-production, $\langle \mathcal{P}^{u'v'}_{11} \rangle^+_{*}$ that is accurate only in the homogeneity limit.\cmnt{ These two expressions are separated by a dispersion or commutation error given by $\langle \mathcal{P}^{u'v'}_{11} \rangle^+_{x}-\langle \mathcal{P}^{u'v'}_{11} \rangle^+_{*}=\langle \langle u'v'\rangle^+_{z,t} d\langle u\rangle^+_{z,t}/dy^+  \rangle_{x}-\langle u'v'\rangle^+_{x,z,t}d\langle u\rangle^+_{x,z,t}/dy^+$ (quantified in figure~\ref{fig:SurfaceDispersion_in_ABL-turbulence_P11}) which is zero only for flat surfaces ($\zeta=0$). The significance of this dispersion in production depends on the role it plays on the qualitative and quantitative structure of the turbulence.} 

The double-averaged covariance $\langle u'v'\rangle^+_{x,z,t}$ peaks at the edge of the buffer layer at $y^+ \approx 28-34$ with the peak height decreasing with $\zeta$ (see horizontal lines in figure~\ref{fig:prof_upvp_avg2_new}) whereas the normalized mean shear (figure~\ref{fig:prof_dudy_avg2_new}), achieves its maximum value near the surface. In addition, the peak magnitude of $\langle u'v'\rangle^+_{x,z,t}$ increases with $\zeta$ while the maximum for $d\langle u\rangle^+_{x,z,t}/dy^+$ decreases.
\hlll{This is consistent with the notion of turbulent stresses forming a higher fraction of the total drag at higher wave slopes, i.e. form drag becomes increasingly important relative to viscous drag, especially for roughness Reynolds numbers, $a^+$ being an order of magnitude smaller than the fully rough regime and kept constant\cmnt{ through this study}} (Table~\ref{table_cases}). This aligns with observations for the transitional roughness regime~\cite{napoli2008effect}, where the surface increasingly moves from `waviness' to `roughness' regime with increase in effective slope ($2\zeta$) resulting in higher form drag.
\hlll{The combined influence of the surface-induced trends in $\langle u'v'\rangle^+_{x,z,t}$ and $d\langle u\rangle^+_{z,t}/dy^+$} (as summarized in figures~\ref{fig:prof_upvpzoom_avg2_new}-\ref{fig:profile_upvp_with_dudy_new_zoomed} and \ref{fig:prof_dudy_avg2_new}) show that 
\hlll{(i) the Reynolds stress dominates the viscous stresses increasingly closer to the surface at higher $\zeta$ and 
(ii) $\langle \mathcal{P}^{u'v'}_{11} \rangle^+_{x}$ \cmnt{or $\langle \mathcal{P}^{u'v'}_{11} \rangle^+_{*}$ (given by the product of the two mean profiles and} }
(shown in figure~\ref{fig:tke_P11_dudy_new}) \hlll{shows peak production occurring farther from the surface at higher $\zeta$ while decreasing in magnitude. This represents an interesting effect of surface heterogeneity where the peak production does not coincide with the crossover location of viscous and Reynolds stresses as the growth of the latter is steeper compared to decrease in the former with height} (see figure~\ref{fig:profile_upvp_with_dudy_new}).

To decipher the production mechanisms very close to the surface in the viscous layer we look at figures~\ref{fig:prof_upvpzoom_avg2_new}, \ref{fig:prof_dudy_avg2_new} and \ref{fig:tke_P11_dudy_new}. We see that for $\zeta$ sufficiently larger than $0$ the vertical turbulent momentum flux, $\langle u'v'\rangle^+_{x,z,t} > 0$ (figure~\ref{fig:prof_upvpzoom_avg2_new}) for $y^+ \lesssim 7$ resulting in $\langle \mathcal{P}^{u'v'}_{11} \rangle^+_{x} \lesssim 0$, i.e. small variance destruction close to the wall. While $\langle u'v'\rangle^+_{x,z,t} \approx 0$ close to the surface over flat surfaces (from wall damping), \hlll{the presence of surface undulations causes larger and positively correlated $u'$ and $v'$ resulting in $\zeta$ dependent destruction of $\langle {u^{\prime}}^2 \rangle^+_{x,z,t}$}. However, the net $\langle {u^{\prime}}^2 \rangle^+_{x,z,t}$ production, $\langle \mathcal{P}_{11} \rangle^+_{x}$ (in figure~\ref{fig:tke_P11_new}) is nearly zero for $y^+ \lesssim 7$ with little dependence on $\zeta$ due to $\langle \mathcal{P}^{u'v'}_{11} \rangle^+_{x}$ being balanced by surface induced production,  $\langle \mathcal{P}^{u'u'}_{11} \rangle^+_{x}=\langle \langle u'^2\rangle^+_{z,t} d\langle u\rangle^+_{z,t}/dx^+  \rangle_{x}$, i.e. $\langle {\mathcal{P}^{u'u'}_{11}}\rangle^+ > 0$ and $\langle {\mathcal{P}^{u'v'}_{11}}\rangle^+ < 0$ .

\paragraph{Averaged Production from Surface Undulations:}
We know that $\langle \mathcal{P}^{u'u'}_{11} \rangle^+_{x}=\langle \langle u'^2\rangle^+_{z,t} d\langle u\rangle^+_{z,t}/dx^+  \rangle_{x}=0$ (shown in figure~\ref{fig:tke_P11_dudx_new}) in a flat channel ($\zeta =0$) whereas for non-flat surfaces ($\zeta > 0$), the streamwise gradient of the mean streamwise velocity (${d\langle u \rangle^+_{x,z,t}}/{dx}$) is non-zero resulting in production close to the surface (viscous layer) and its destruction above it in the buffer layer before approaching zero in the logarithmic region. 
 Consequently, there is no significant net turbulence generation over the entire TBL from $\langle \mathcal{P}^{u'u'}_{11} \rangle^+_{x}$ except for pockets of local production and destruction that helps control the shape of overall production $\langle \mathcal{P}_{11} \rangle^+_{x}$. 
 \hlll{
Away from the surface, $\langle \mathcal{P}^{u'u'}_{11} \rangle^+_{x}$ and $\langle \mathcal{P}^{u'v'}_{11} \rangle^+_{x}$ are still complementary but the different terms do no cancel out as $\langle \mathcal{P}^{u'v'}_{11} \rangle^+_{x} > \langle \mathcal{P}^{u'u'}_{11} \rangle^+_{x}$. Later, we explore whether this is simply a consequence of $d\langle u\rangle^+_{z,t}/dy^+ \gg d\langle u\rangle^+_{z,t}/dx^+$ (due to $\zeta \ll 1$) or is more complicated. 
The different components contribute to the overall production trends as follows.
The peak location of $\langle \mathcal{P}^{u'v'}_{11} \rangle^+_{x}$ shows systematic upward shifts with $\zeta$ while its magnitude displays very little sensitivity to wave slope. In contrast, the profiles of $\langle \mathcal{P}^{u'u'}_{11} \rangle^+_{x}$ show very strong $\zeta$-dependence of the peak destruction magnitudes in the buffer layer, but its location does not. 
Therefore, $\langle \mathcal{P}_{11} \rangle^+_{x}$ dependence on $\zeta$ in both magnitude and shape arise from $\langle \mathcal{P}^{u'u'}_{11} \rangle^+_{x}$ and $\langle \mathcal{P}^{u'v'}_{11} \rangle^+_{x}$ respectively.}
%
One can use $\langle \mathcal{P}^{u'u'}_{11} \rangle^+_{x}$ (figure~\ref{fig:tke_P11_dudx_new}) to characterize the roughness sublayer height which in this case is $\sim 3a^+$ and nearly independent of $\zeta$.

\begin{figure}[ht!]
\centering
\mbox{
\subfigure[$\mathcal{P}_{11}^+$\label{fig:P11-contours-a}]{\includegraphics[width=0.27\textwidth]{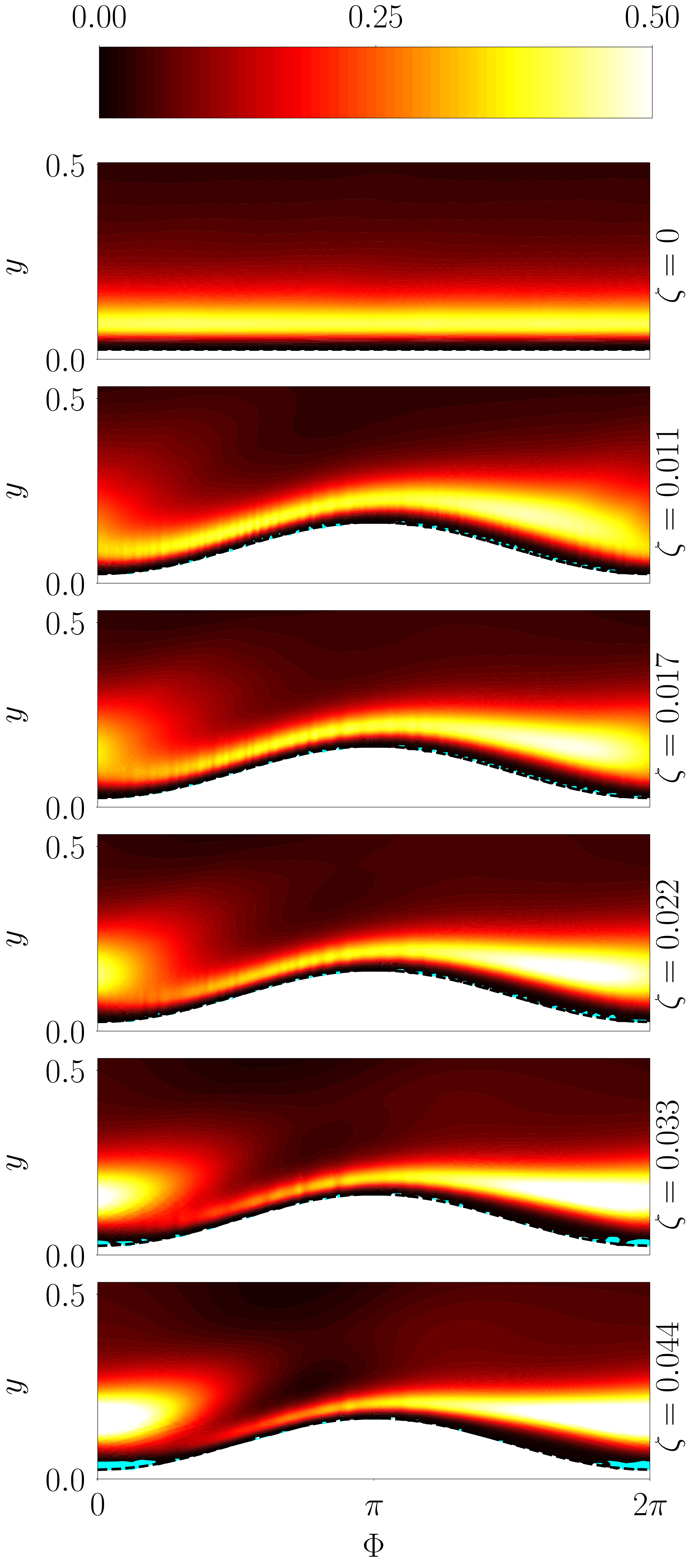}}
\subfigure[${\mathcal{P}_{11}^{u'u'}}^+$\label{fig:P11-contours-b}]{\includegraphics[width=0.27\textwidth]{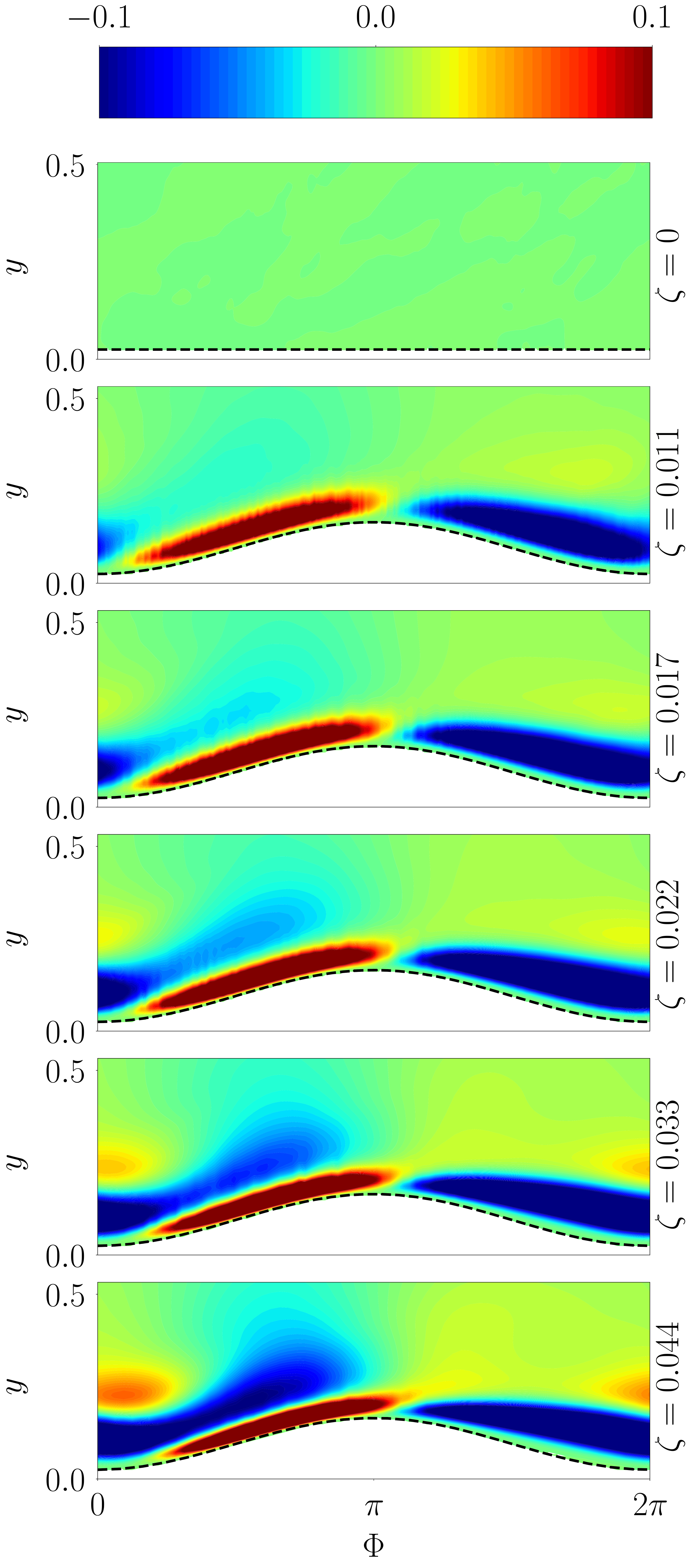}}
\subfigure[${\mathcal{P}_{11}^{u'v'}}^+$\label{fig:P11-contours-c}]{\includegraphics[width=0.27\textwidth]{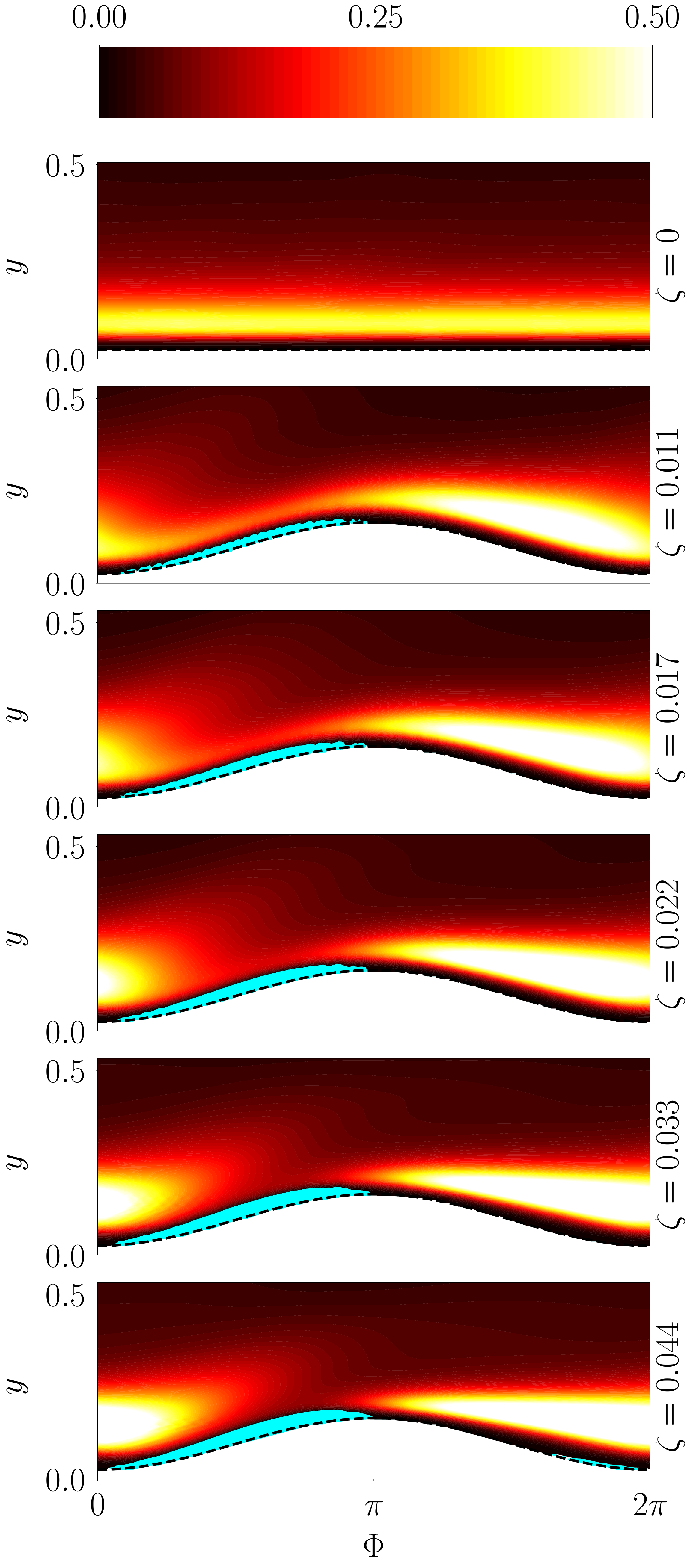}}
}
\caption{Contours of inner--scaled (single) averaged streamwise variance production, $\mathcal{P}_{11}^+$ (a) and its components, ${\mathcal{P}_{11}^{u'u'}}^+$ (b) and ${\mathcal{P}_{11}^{u'v'}}^+$ (c) as a function of $\Phi$ and $y$, where, $\Phi=2\pi x/\lambda$. In the above plots, the cyan colored region represents negative production of streamwise variance. \label{fig:P11-contours}}
\end{figure}

\paragraph{Two-dimensional Structure of $\langle u'^2\rangle^+_{z,t}$ Production:}
$\mathcal{P}^+_{11}(x,y)$ represents inner-scaled variance production based on averaging along homogeneous direction ($z$) and over a stationary window ($t$) of the turbulent flow.  It is this $\mathcal{P}^+_{11} (x,y)$ shown in figure~\ref{fig:P11-contours} that is averaged along the inhomogeneous streamwise ($x$) direction to estimate the averaged production, $\langle \mathcal{P}_{11} \rangle^+_{x}$ shown in figure~\ref{fig:tke_P11_new}. Interpretation of $\langle \mathcal{P}_{11} \rangle^+_{x}$ requires understanding the structure of $\mathcal{P}^+_{11} (x,y)$ and its components. 
Figure~\ref{fig:P11-contours} shows inner-scaled production over the $y-\phi$ space with $\phi$ being the non-dimensional streamwise phase coordinate ($\phi = 2\pi x/\lambda$) and $y=y/\delta$ for unit half channel height ($\delta=1$). 
We clearly see that the structure of averaged  production contours are qualitatively invariant in this $y-\phi$ space for different $\zeta=2a/\lambda > 0$ while, the magnitude depends on the wave slope with higher $\zeta$ producing stronger peaks and troughs. To identify the different layers of the TBL, we define $d$ as the local vertical coordinate relative to the wall at each streamwise location. 

\begin{figure}[ht!]
\centering
\hspace{-2em}
\mbox{
\subfigure[$\langle u^{\prime 2}\rangle_{z,t}^+$\label{fig:cont_uvar}]{\includegraphics[width=0.25\textwidth]{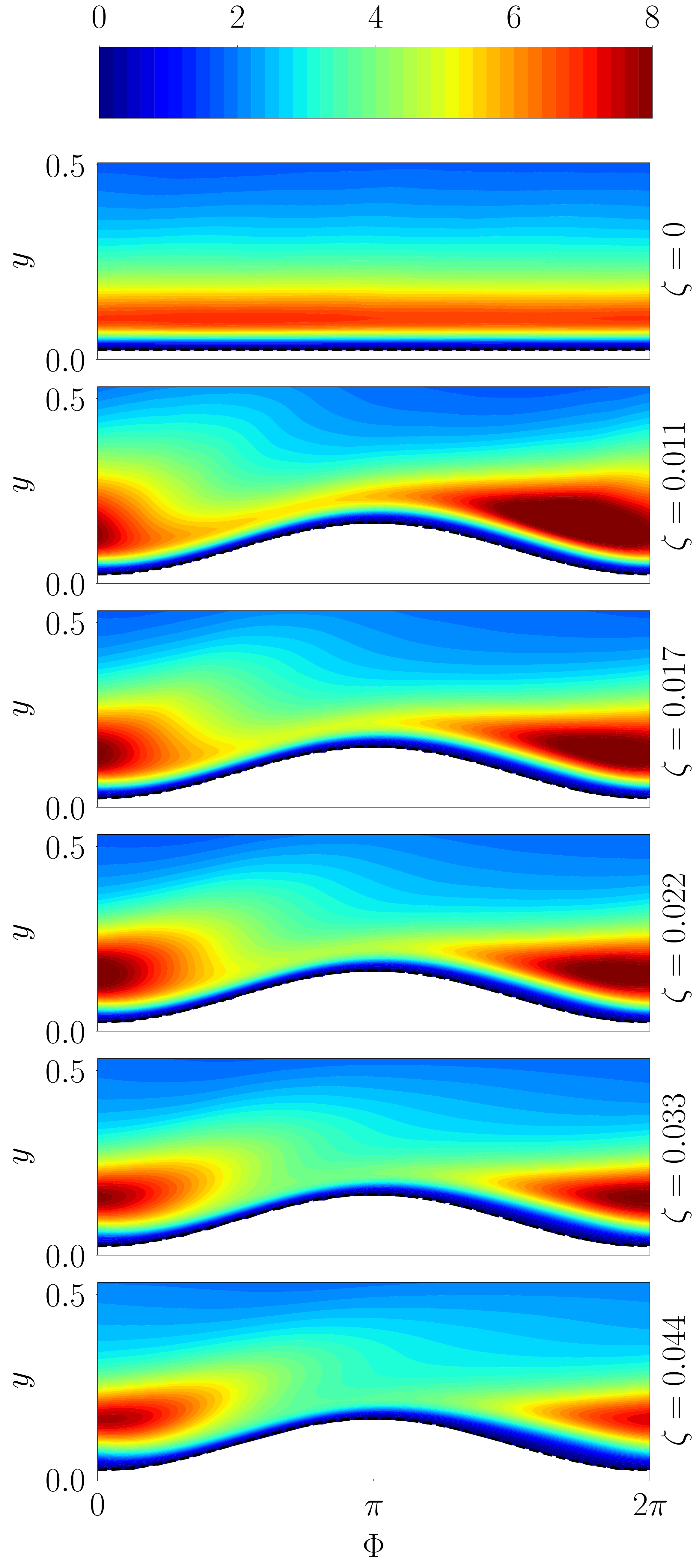}}
\subfigure[$\langle du/dx \rangle_{z,t}^+$\label{fig:cont_dudx}]{\includegraphics[width=0.25\textwidth]{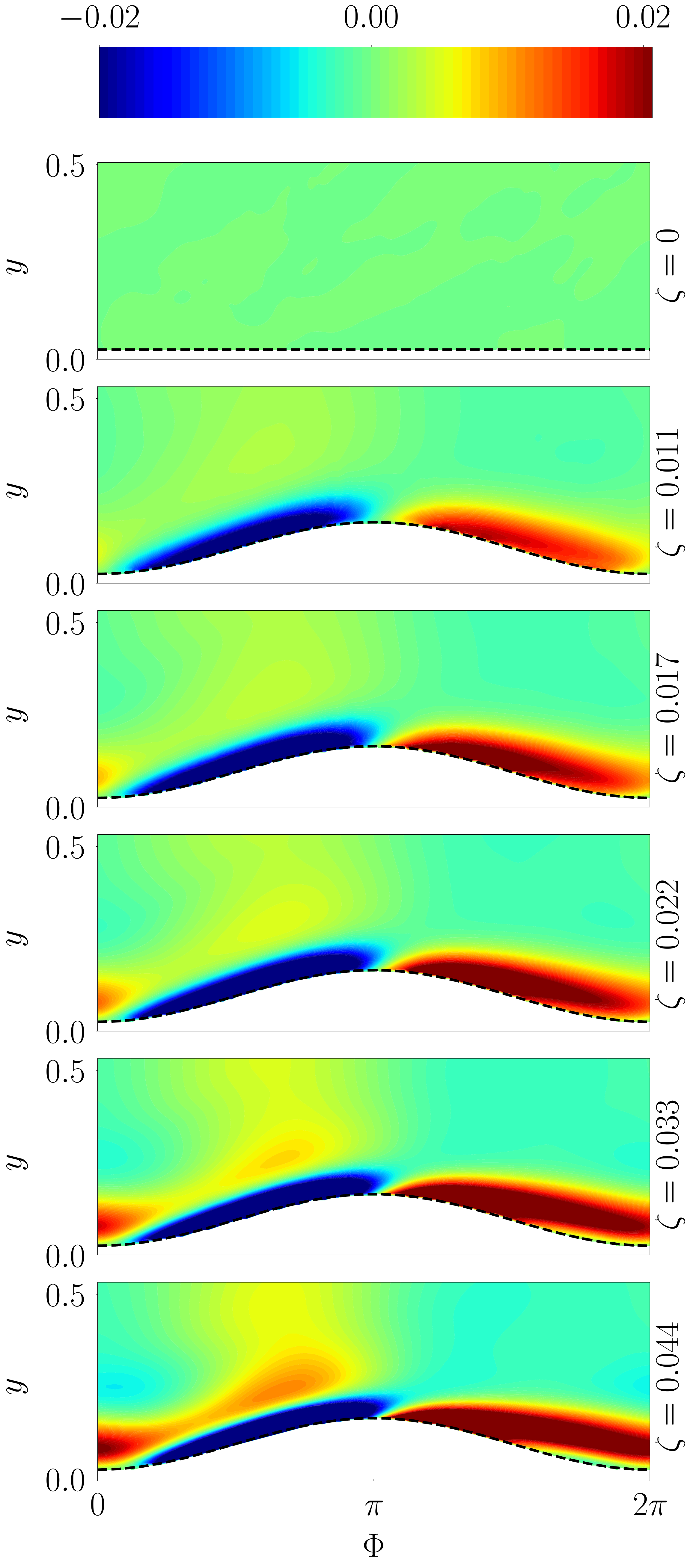}}
\subfigure[$\langle u'v'\rangle_{z,t}^+$\label{fig:cont_uvcovar}]{\includegraphics[width=0.25\textwidth]{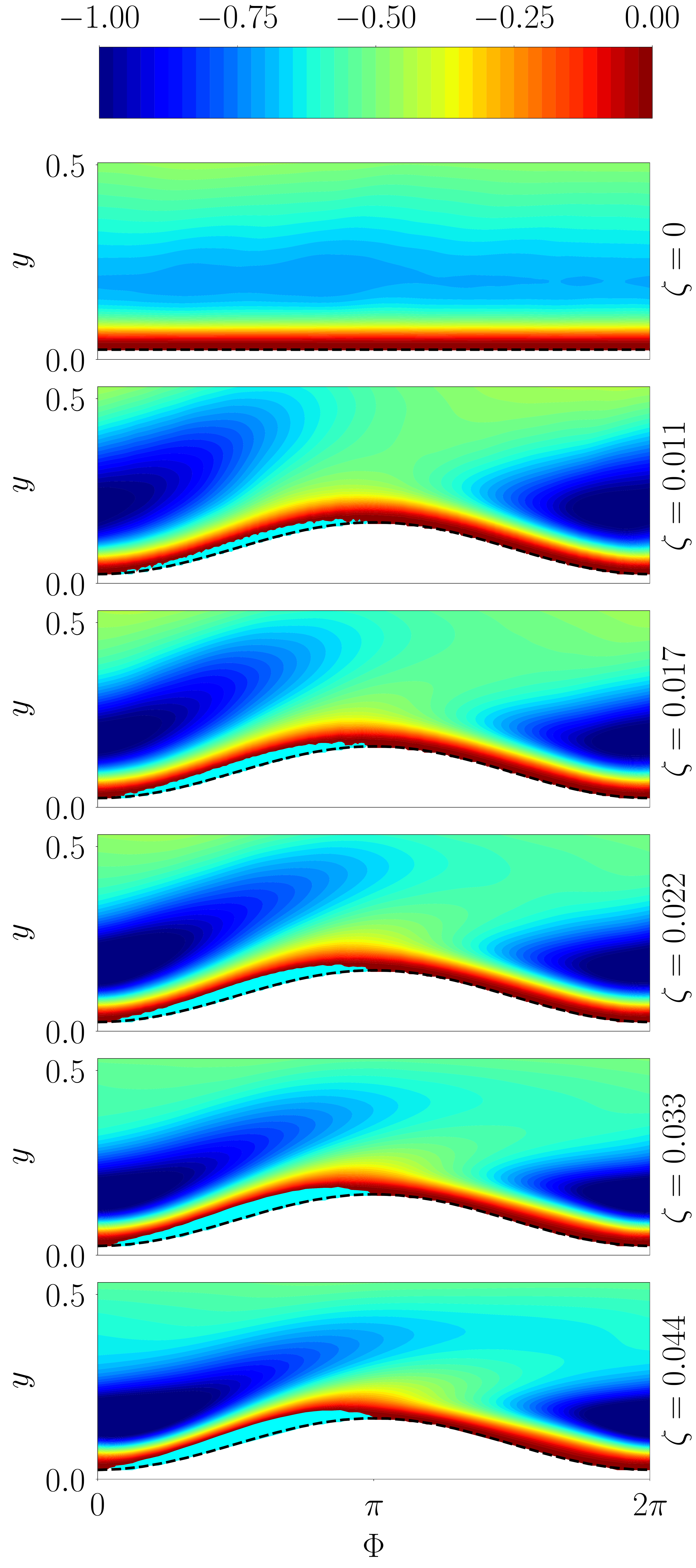}}
\subfigure[$\langle du/dy \rangle_{z,t}^+$\label{fig:cont_dudy}]{\includegraphics[width=0.25\textwidth]{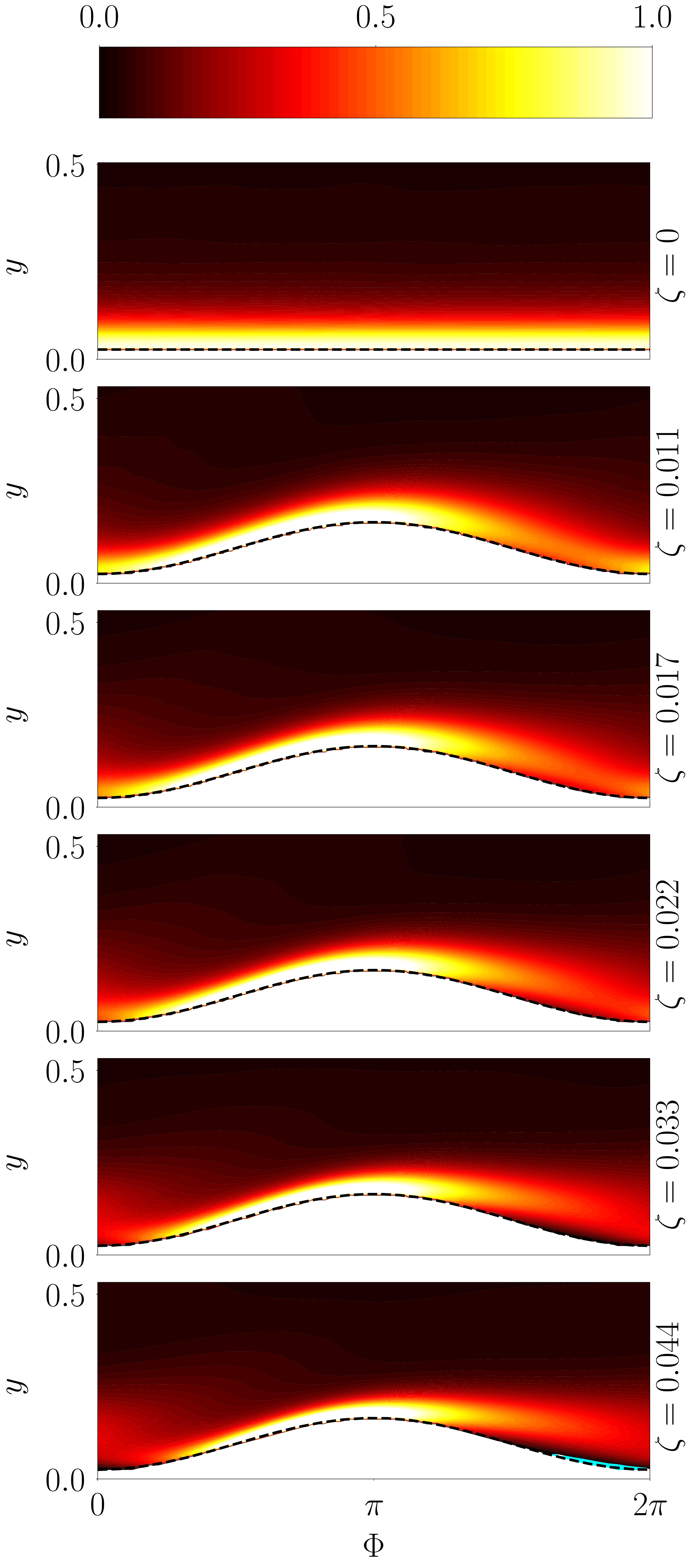}}
}
\caption{\label{fig:cont_ReynoldsStressStrainRate_uvariance} Contours of inner-scaled spanwise and temporally averaged (a) streamwise variance, (b) streamwise gradient of $\langle u \rangle^+_{z,t}$, (c) covariance, $\langle u'v' \rangle^+_{z,t}$ and (d) vertical gradient of $\langle u \rangle^+_{z,t}$. The cyan region closer to the wall in (c) identifies regions of $\langle u'v' \rangle_{z,t} > 0$ while that in (d) represents flow separation at higher $\zeta$.
}
\end{figure}

Looking at the isocontours in figures~\ref{fig:P11-contours-b} and \ref{fig:P11-contours-c} \hlll{we note that both $ {\mathcal{P}^{u'v'}_{11}}^+ (x,y)$ and  ${\mathcal{P}^{u'u'}_{11}}^+ (x,y)$ indeed play complementary roles of production and destruction in different regions of TBL, especially, close to the surface ($d \lesssim 0.1$)}. Specifically, $ {\mathcal{P}^{u'v'}_{11}}^+ (x,y) <0$ (destruction) along the windward side of the wavy surface (i.e. $\phi = 0-\pi$ and $d \lesssim 0.1$) as indicated by the cyan region in figure~\ref{fig:P11-contours-c} 
while along the leeward side (i.e. $\phi = \pi-2\pi$ and $d \lesssim 0.1$) there exists very little turbulence production (shown in black). The exception to this being a small negative production (cyan) zone in the trough due to flow separation at larger $\zeta$. 
This is complemented by ${\mathcal{P}^{u'u'}_{11}}^+ (x,y) > 0$ (in figure~\ref{fig:P11-contours-b}) along the windward (i.e. $\phi \approx 0-\pi$ and $d \lesssim 0.1$) and near-zero \cmnt{or negative} values along the leeward side (i.e. $\phi \approx \pi-2\pi$ and $d \lesssim 0.1$). \hlll{This complementary \cmnt{production and destruction} structure of ${\mathcal{P}^{u'u'}_{11}}^+$ and ${\mathcal{P}^{u'v'}_{11}}^+$ when averaged along the streamwise direction yield net production,   $\langle \mathcal{P}^{u'u'}_{11} \rangle^+_{x} > 0$} (figure~\ref{fig:tke_P11_dudx_new}) and destruction, $\langle \mathcal{P}^{u'v'}_{11} \rangle^+_{x} < 0$ (figure~\ref{fig:tke_P11_dudy_new}) respectively. 
Just as $\langle \mathcal{P}^{u'u'}_{11} \rangle^+_{x}$ and $\langle \mathcal{P}^{u'v'}_{11} \rangle^+_{x}$ cancel each other in the inner layer, $ {\mathcal{P}^{u'v'}_{11}}^+ (x,y)$ and $ {\mathcal{P}^{u'u'}_{11}}^+ (x,y)$ also show overlapping regions of production and destruction so that there is no net variance generation in the viscous dominated region over the surface for all $\zeta$ (see ${\mathcal{P}_{11}}^+ (x,y)$ in figure~\ref{fig:P11-contours-a}). Away from the surface in the outer layer, ${\mathcal{P}^{u'v'}_{11}}^+ (x,y)$ (figure~\ref{fig:P11-contours-c}) dominates and controls the large-scale structure of ${\mathcal{P}_{11}}^+ (x,y)$.
%
To interpret this structure, we make the following observations regarding the relevant strain rate and Reynolds stress tensor terms.

\begin{enumerate}[label=\roman*.,noitemsep,leftmargin=1cm]
  \item On average, $\| d\langle u\rangle^+_{z,t}/dx^+ \| \ll \| d\langle u\rangle^+_{z,t}/dy^+ \|$ by a factor of $\sim O(20)$ (as seen in figures~\ref{fig:cont_dudx} and \ref{fig:cont_dudy}) which is comparable to $O(1/\zeta)$.
  \item $\| \langle u'^2 \rangle^+_{z,t} \| > \| \langle u'v' \rangle^+_{z,t} \|$ by a factor of $\sim O(5)$ (in figures~\ref{fig:cont_uvar}-\ref{fig:cont_uvcovar}) and achieve their maximum magnitudes along the surface in the buffer layer.
  \item It is well known that as streamlines wrap around the wave crest, the flow accelerates creating a local low pressure region over the hump. This shape effect on the streamwise velocity field is felt away from the surface.
  \item $d\langle u\rangle^+_{z,t}/dx^+$ has an asymmetric structure (figure~\ref{fig:cont_dudx}) that is different away and close to the surface. Away from the surface, the accelerating and decelerating flow before and after the crest results in $d\langle u\rangle^+_{z,t}/dx^+ > 0$ and $d\langle u\rangle^+_{z,t}/dx^+ <0$ respectively. In the viscous region, this trend is reversed due to the surface slope-induced blockage causing $d\langle u\rangle^+_{z,t}/dx^+<0$ and $d\langle u\rangle^+_{z,t}/dx^+>0$ in the windward and leeward sides.
  \item $d\langle u\rangle^+_{z,t}/dy^+$ originates primarily from flow shear and therefore, is positive all along the surface while decreasing rapidly with height (figure~\ref{fig:cont_dudy}). The exception being flow separation at large enough $\zeta$ that cause $d\langle u\rangle^+_{z,t}/dy^+<0$ near the trough.  
  
  \item While the magnitudes of $d\langle u\rangle^+_{z,t}/dx^+$ and $d\langle u\rangle^+_{z,t}/dy^+$ are large over a thin layer along the windward side, we see a more diffused layer (thickness ($O(a^+)$)) along the leeward side due to wake mixing behind the crest.
\end{enumerate}

\hlll{Obviously, variance production/destruction is large in regions of strong correlation between the appropriate component of Reynolds stress and strain rate tensors. 
Given that $d\langle u\rangle^+_{z,t}/dx^+$, $d\langle u\rangle^+_{z,t}/dy^+$ (figures~\ref{fig:cont_dudx}-\ref{fig:cont_dudy}) are strong near the surface while $\langle u'^2 \rangle^+_{z,t}$, $\langle u'v' \rangle^+_{z,t}$ achieve larger magnitudes away from the wall (figures~\ref{fig:cont_uvar}-\ref{fig:cont_uvcovar}), the strong destruction/production zone in both ${\mathcal{P}^{u'u'}_{11}}^+$ and ${\mathcal{P}^{u'v'}_{11}}^+$ exists (figures~\ref{fig:P11-contours-a}-\ref{fig:P11-contours-b}) in the lower buffer region. 
However, as $d\langle u\rangle^+_{z,t}/dx^+$ and $d\langle u\rangle^+_{z,t}/dy^+$ assume a more diffused structure behind the wave crest due to wake mixing, the resulting production/destruction zone is also thicker with larger magnitudes in this leeward region and grows with slope, $\zeta$.
For such wavy surfaces, the leeward side production (as well for the windward side away from the surface) is negative for ${\mathcal{P}^{u'u'}_{11}}^+ $ and positive for ${\mathcal{P}^{u'v'}_{11}}^+$. 
Given that $\| {\mathcal{P}^{u'u'}_{11}}^+ \| < \| {\mathcal{P}^{u'v'}_{11}}^+ \|$ over most of the TBL, we see that the structure of net production, ${\mathcal{P}_{11}}^+$ is dominated by ${\mathcal{P}^{u'v'}_{11}}^+$ that depends on $d\langle u\rangle^+_{z,t}/dy^+$ (figure~\ref{fig:cont_dudy}) and $\langle {u'v'} \rangle^+_{z,t}$ (figure~\ref{fig:cont_uvcovar}). Specifically, the primary generation of $\langle u'^2 \rangle^+_{z,t}$ (see ${\mathcal{P}_{11}}^+$ in figure~\ref{fig:P11-contours-a}) occurs in the thick wake-induced buffer region along the leeward slope through shear production, ${\mathcal{P}^{u'v'}_{11}}^+$. However, the windward slope is responsible for surface induced near surface variance generation through ${\mathcal{P}^{u'u'}_{11}}^+$ (i.e. $d\langle u\rangle^+_{z,t}/dx^+$ (figure~\ref{fig:cont_dudx}) to overcome the \cmnt{surface-induced} destruction contained in ${\mathcal{P}^{u'v'}_{11}}^+$ due to positive covariance, $\langle {u'v'} \rangle^+_{z,t} > 0$ (cyan region in figure~\ref{fig:cont_uvcovar}). 
In addition to increasing strain rates at higher $\zeta$, the flow also involves unsteady separation dynamics with  $d\langle u\rangle^+_{z,t}/dy^+ <0$ (see cyan regions near the wave trough shown in figure~\ref{fig:cont_dudy}) and turbulence destruction zones (see cyan region in figures~\ref{fig:P11-contours-a} and \ref{fig:P11-contours-c} for $\zeta=0.044$) that grow thicker with $\zeta$. 
These results suggest that the influence of $\zeta$ shows up in at least three ways: (i) larger mixing and mean shear in the leeward side of the wave resulting in enhanced production; (ii) surface induced near-wall variance generation along the windward side and (iii) separation-induced destruction in the trough.
}


\hlll{
} 

\paragraph{Dispersion Effects in Production:}
\hlll{The two-dimensional surface undulations generate a complex production structure in $\mathcal{P}^+_{11}(x,y)$ (figure~\ref{fig:P11-contours}) for $\zeta>0$ that is submerged within its one-dimensional surrogate, $\langle \mathcal{P}_{11} \rangle^+_{x}$ (figure~\ref{fig:tke_P11_new}). There exist multiple ways to characterize the surface dispersion effects on turbulent statistics~\cite{florens2013defining} in order to estimate the roughness sublayer height. Here, we characterize the surface dispersion effects using the surrogate or pseudo-production estimate which computes the Reynolds stress-strain rate interaction as if homogeneity exists, i.e.  as product of the double averaged Reynolds stress tensor (i.e. $\langle u'u'\rangle^+_{x,z,t}, \langle u'v'\rangle^+_{z,t}$) and the mean strain rate tensor ($d\langle u\rangle^+_{x,z,t}/dy^+,d\langle u\rangle^+_{x,z,t}/dx^+, \dots$). The formal error contained in this production estimate is given by $\langle \mathcal{P}^{u'v'}_{11} \rangle^+_{x}-\langle \mathcal{P}^{u'v'}_{11} \rangle^+_{*}=\langle \langle u'v'\rangle^+_{z,t} d\langle u\rangle^+_{z,t}/dy^+  \rangle_{x}-\langle u'v'\rangle^+_{x,z,t}d\langle u\rangle^+_{x,z,t}/dy^+$ represents surface dispersion effects contained in the computation of turbulence production. This is quantified in figure~\ref{fig:SurfaceDispersion_in_ABL-turbulence_P11} with top row representing the pseudo-estimates and the bottom row representing the deviations. Obviously, in the homogeneous limit ($\zeta=0$), the dispersion in production is zero while it grows systematically with wave slope, $\zeta$. The dispersion errors are concentrated closer to the surface (i.e. $y^+ \lesssim 50 \sim 4a^+$) and decrease gradually to zero in the outer region. The surface influence on the TBL ($\sim 4a^+$) extends further than that observed for the surface-induced production, $\langle \mathcal{P}^{u'u'}_{11} \rangle^+_{x}$ ($\sim 3a^+$). While the different pseudo-production estimates under- and over-predict depending on the region of the TBL, it shares qualitative similarity with the true estimates. }

\subsection{Vertical Variance, $\langle v'^2 \rangle^+_{x,z,t}$ \label{subsec:vertical_var}}

The effect of surface undulations with increasing slope, $\zeta$ on vertical variance profiles (figure~\ref{fig:prof_vvar_avg2_new60}) is to enhance its growth near the surface, especially in the buffer layer  i.e. $y^+ \approx 7-40$. The steeper surface undulations increase the fraction of the vertical variance contributing to the turbulent kinetic energy (TKE) as was observed in \cite{khan2019statistical}.  It is well known that the vertical variance near the surface is smaller compared to the streamwise variance due to the wall damping effect which in turn causes the variance to peak further away from the surface. For these cases, the peak vertical variance occurs in the buffer-log transition region ($y^+ \approx 50-55 \gtrsim 4a^+$) which is outside the roughness sublayer (i.e. $\approx 3a^+$). Therefore, it is not surprising that the peak vertical variance magnitude and location is relatively insensitive to $\zeta$ as farther away from the surface (beyond the roughness sublayer) the wall effects have died down. In this region, the turbulence structure is insensitive to $\zeta$ with all the different curves collapsing on top of each other and thereby supporting outer layer similarity. 

\subsubsection{Dynamics of Vertical Variance Transport\label{subsubsec:vertical_var_transport}}
\begin{figure}[ht!]
\centering
\mbox{
\subfigure[\label{fig:prof_vvar_avg2_new60}]{\includegraphics[width=0.21\textwidth]{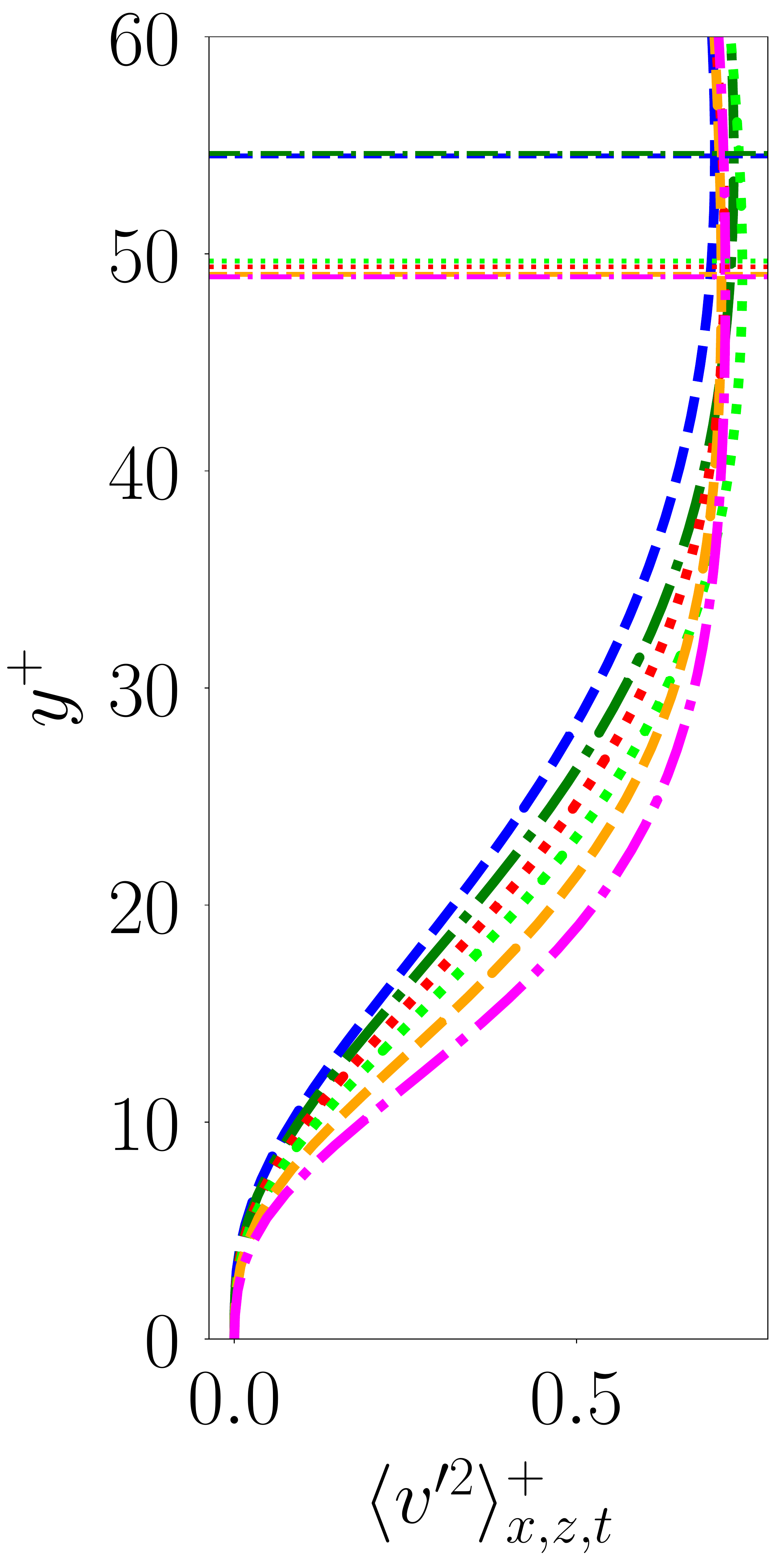}}
\subfigure[\label{fig:tke_P22_new}]{\includegraphics[width=0.21\textwidth]{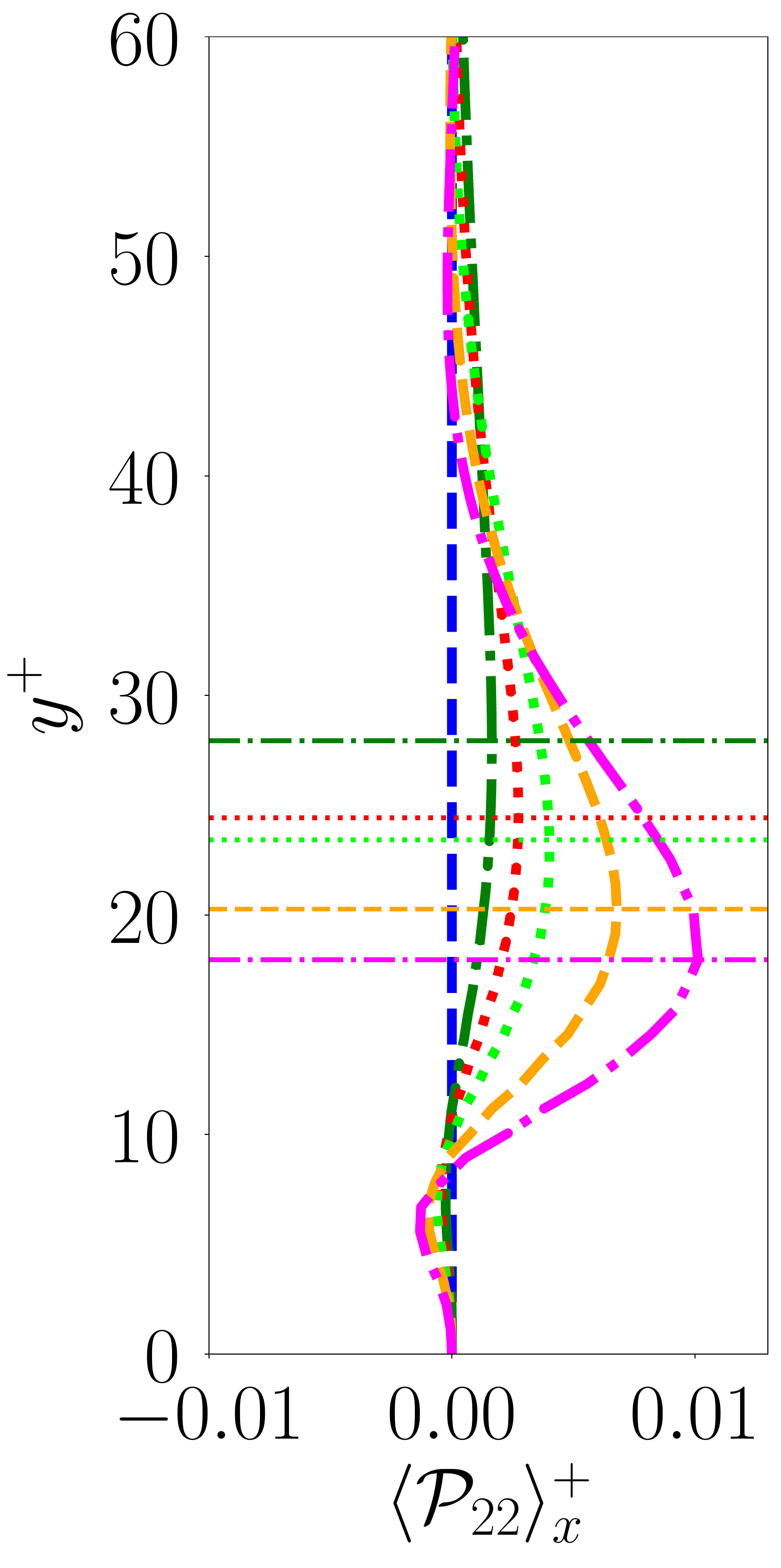}}
\subfigure[\label{fig:tke_P22_dvdx_new}]{\includegraphics[width=0.21\textwidth]{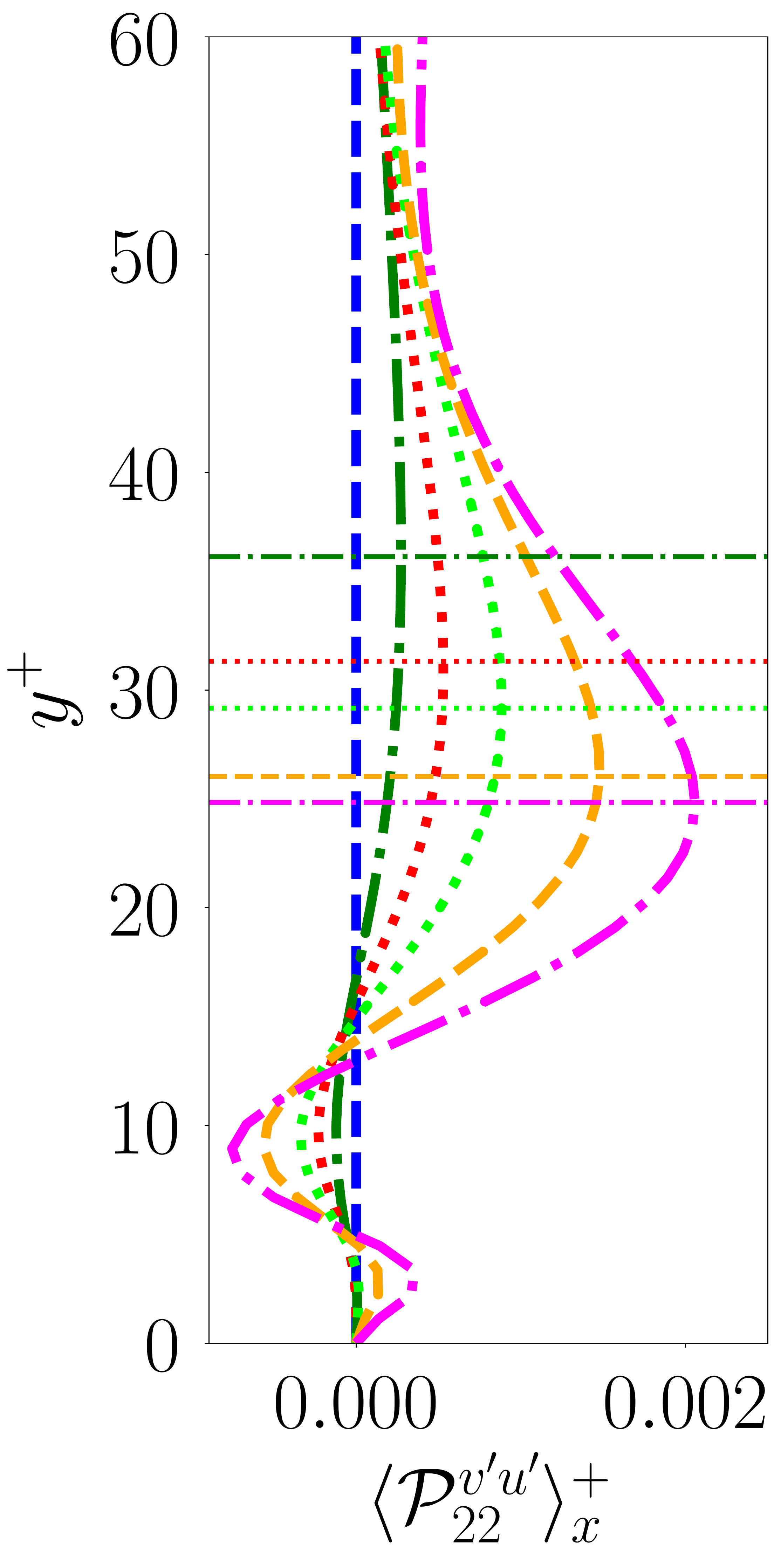}}
\subfigure[\label{fig:tke_P22_dvdy_new}]{\includegraphics[width=0.21\textwidth]{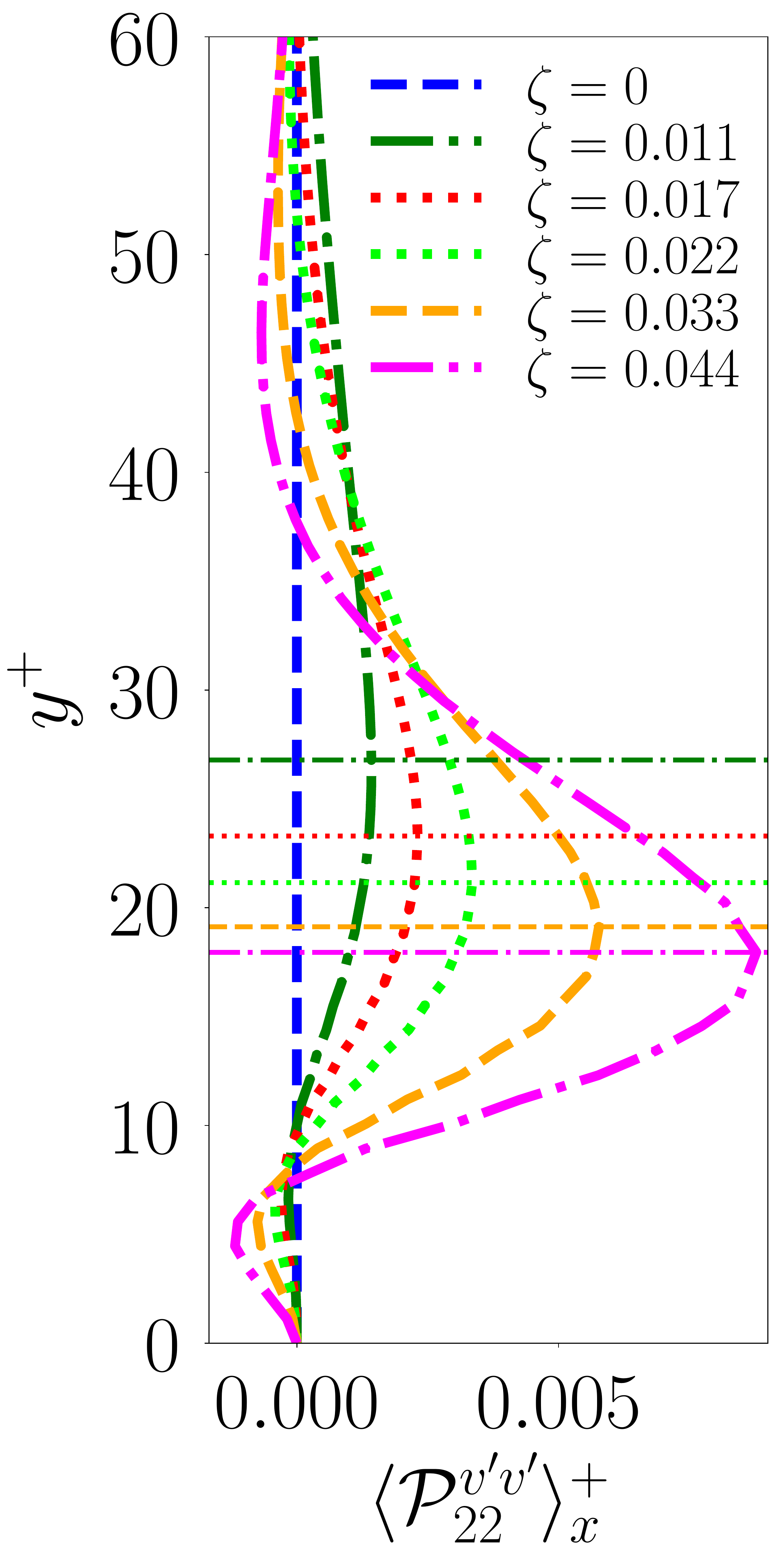}}
}
\mbox{
\subfigure[\label{fig:tke_E22_new}]{\includegraphics[width=0.21\textwidth]{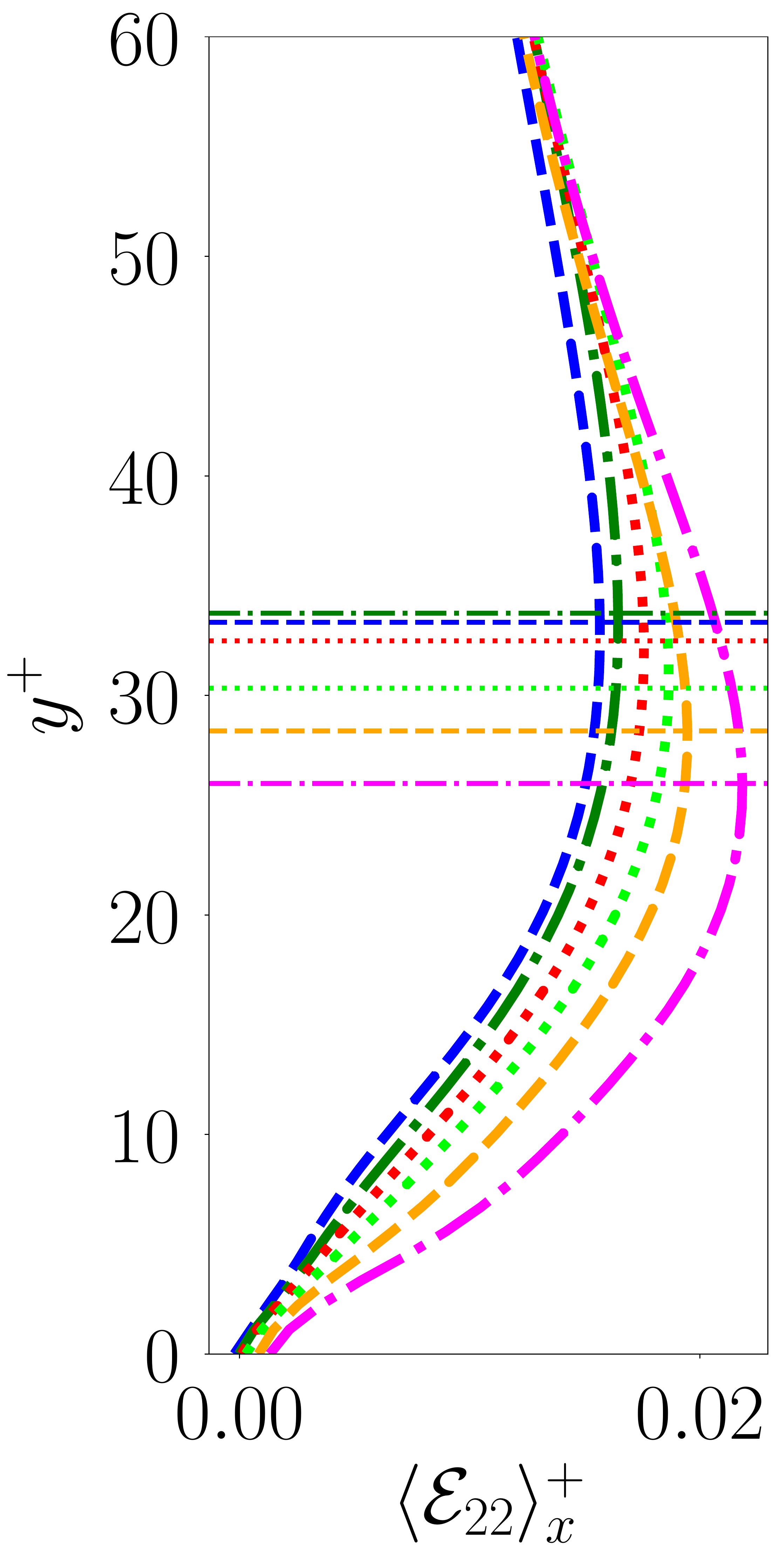}}\hspace{0em}
\subfigure[\label{fig:tke_R22_new}]{\includegraphics[width=0.21\textwidth]{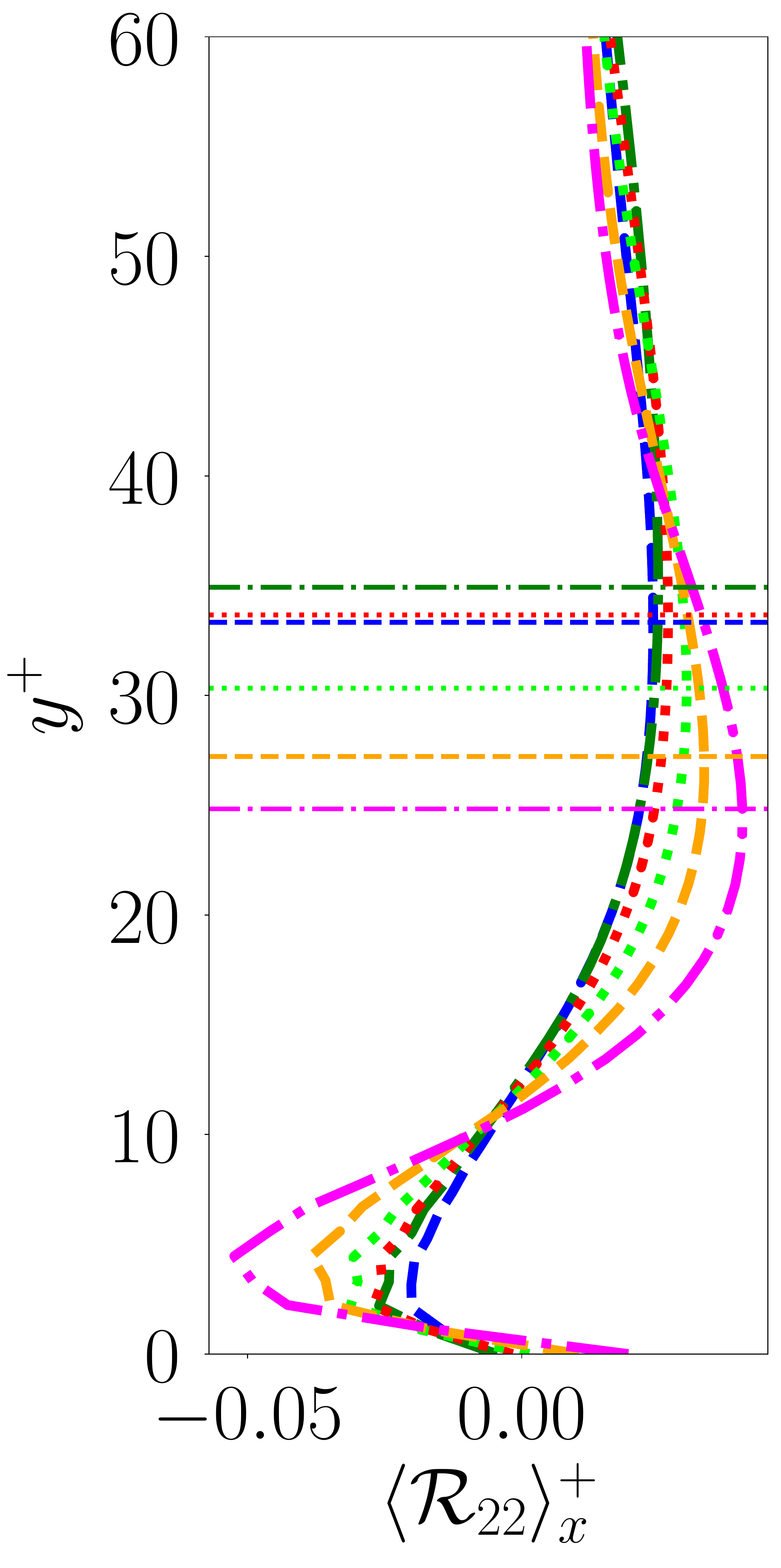}}\hspace{0em}
\subfigure[\label{fig:tke_L22_new}]{\includegraphics[width=0.21\textwidth]{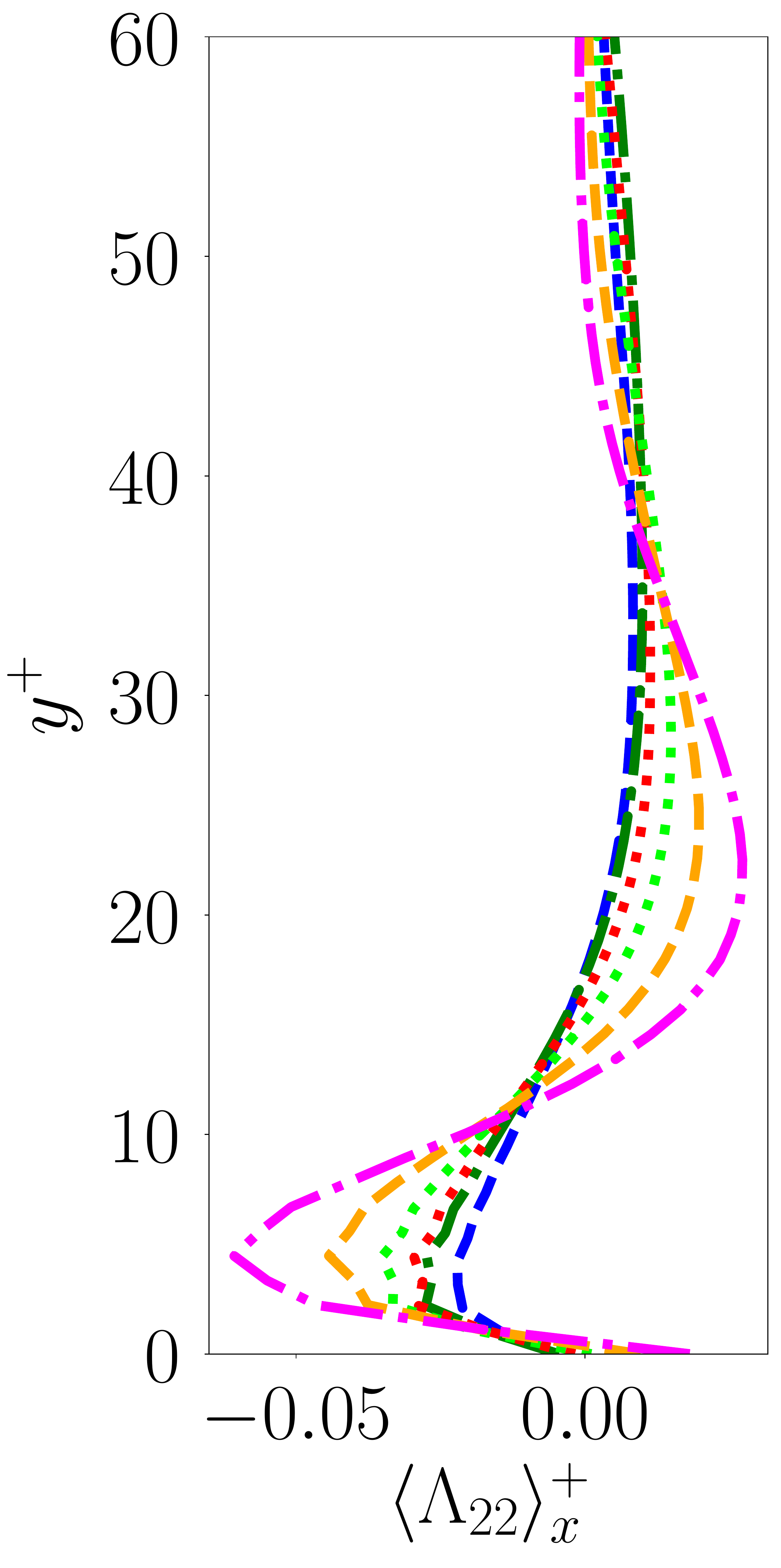}}\hspace{0em}
\subfigure[\label{fig:tke_D22_new}]{\includegraphics[width=0.21\textwidth]{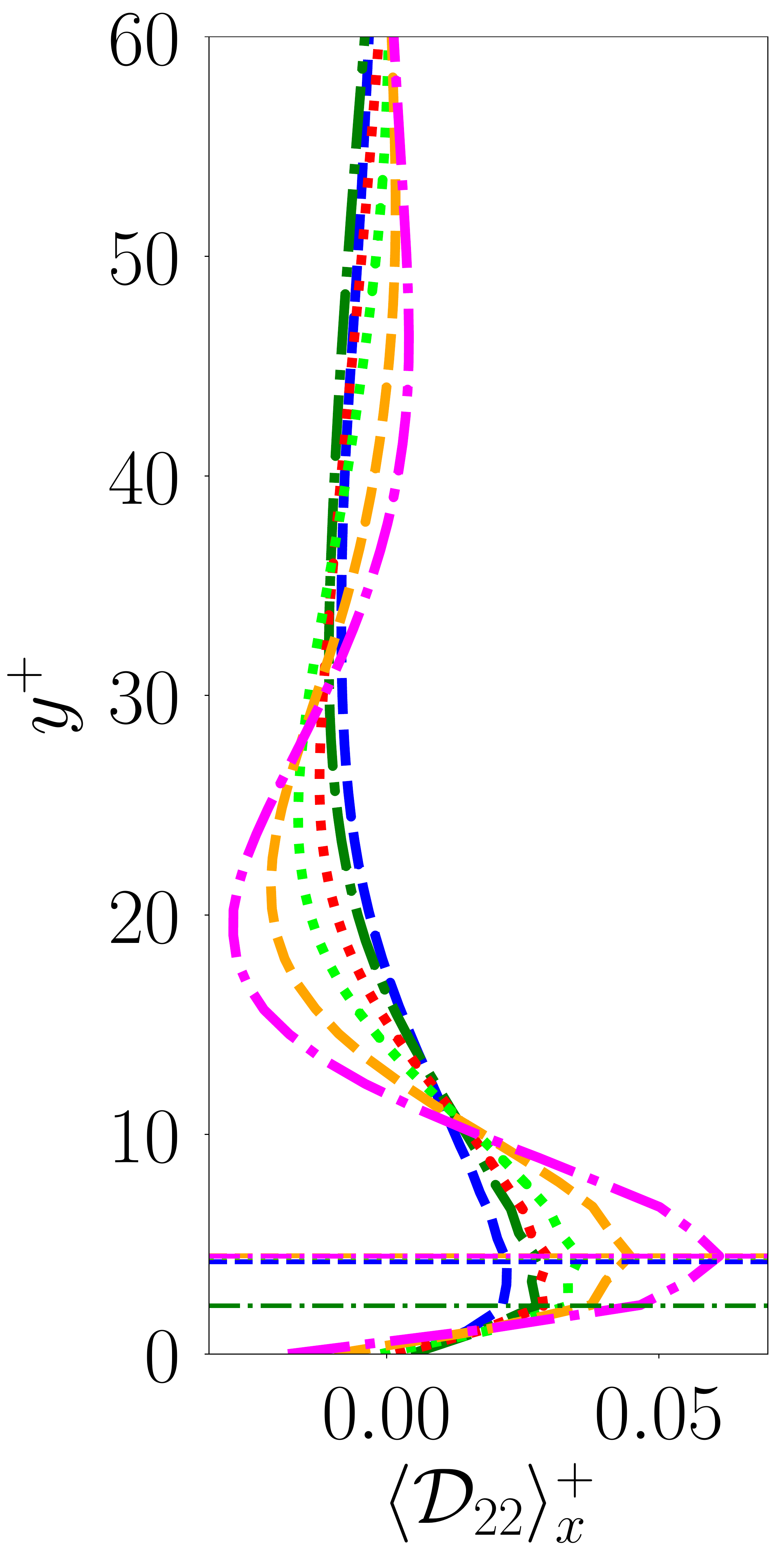}}
}
\caption{\label{fig:vvar_production_new} 
{Schematic showing wall-normal variation of inner-scaled, double-averaged vertical variance (a) along with averaged production (b), dissipation (e), pressure-rate-of-strain (f), cumulative local generation $\langle \Lambda_{22} \rangle^+_{x}=\langle\mathcal{P}_{22}\rangle^+_x-\langle\mathcal{E}_{22}\rangle^+_x+\langle\mathcal{R}_{22}\rangle^+_x$, and diffusion (g). We further split the production term $\langle P_{22} \rangle^+_{x}$ into $\langle P^{v'u'}_{22} \rangle^+_{x}$ (c) and $\langle P^{v'v'}_{22} \rangle^+_{x}$ (d). The horizontal lines correspond to the vertical location of maximum/minimum value for a chosen statistic. If the peak locations are different, we color match the horizontal lines with the corresponding curves.}
}
\end{figure}

In a TBL over flat surface, the  production of vertical variance from the interaction of the mean strain rate with the Reynolds stresses, $\langle \mathcal{P}_{22} \rangle_x^+$ (blue curve in figures~\ref{fig:tke_P22_new},\ref{fig:tke_P22_dvdx_new} and \ref{fig:tke_P22_dvdy_new}) is zero. Instead, $\langle v'^2 \rangle^+_{x,z,t}$ is generated through a conversion process of streamwise turbulent fluctuations through the pressure-rate-of-strain term, $\langle \mathcal{R}_{22} \rangle_x^+$ (figure~\ref{fig:tke_R22_new}) and modulated further by the diffusive transport, $\langle \mathcal{D}_{22} \rangle_x^+$  (figure~\ref{fig:tke_D22_new}) along with turbulent dissipation, $\langle \mathcal{E}_{22} \rangle_x^+$  (figure~\ref{fig:tke_E22_new}). However, for non-flat surfaces with $\zeta>0$, the inner-scaled averaged production, $\langle \mathcal{P}_{22} \rangle_x^+$ assumes non-zero values in the buffer layer (non-blue curves in figures~\ref{fig:tke_P22_new},\ref{fig:tke_P22_dvdx_new} and \ref{fig:tke_P22_dvdy_new}) due to surface inhomogeneities. 
\hlll{ Therefore, the dynamics of $\langle v'^2 \rangle^+_{x,z,t}$ is controlled by surface induced production along with mechanisms such as return-to-isotropy, dissipation and diffusion. The relative magnitudes of the different terms in the variance transport equation show that production ($\langle \mathcal{P}_{22} \rangle_x^+$), dissipation ($\langle \mathcal{E}_{22} \rangle_x^+$) and pressure-rate-of-strain ($\langle \mathcal{R}_{22} \rangle_x^+$) as cumulatively depicted by $\langle {\Lambda}_{22} \rangle_x^+$ dominate variance generation with production being the smallest.  This local generation of vertical variance is balanced only by diffusive transport of the double-averaged, inner-scaled variance, $\langle \mathcal{D}_{22} \rangle_x^+$ as the averaged advective transport is insignificant} (see figure~\ref{fig:prof_ConvectiveTransport_C22} in Appendix) for this statistically stationary system. The larger vertical variance in the buffer layer for higher $\zeta$ (figure~\ref{fig:prof_vvar_avg2_new60}) is also observed in production, $\langle \mathcal{P}_{22} \rangle_x^+$ (figure~\ref{fig:tke_P22_new}), dissipation, $\langle \mathcal{E}_{22} \rangle_x^+$ (figure~\ref{fig:tke_E22_new}), pressure-rate-of-strain, $\langle \mathcal{R}_{22} \rangle_x^+$ (figure~\ref{fig:tke_R22_new}) and consequently, $\langle {\Lambda}_{22} \rangle_x^+$ (figure~\ref{fig:tke_L22_new}) and $\langle \mathcal{D}_{22} \rangle_x^+$. 
\hlll{In the following, we delve into the mechanisms underlying vertical variance generation including the relatively small, but, dynamically important surface-induced production.
}

\subsubsection{Mechanisms of Vertical Variance Generation\label{subsubsec:vertical_var_production}}


For $\zeta>0$, the fundamental modulations to $\langle v'^2 \rangle^+_{x,z,t}$ production occur outside the viscous region of the roughness sublayer as seen from the inner-scaled, averaged vertical variance production,  $\langle P_{22} \rangle^+_{x}$ in figure~\ref{fig:tke_P22_new}.  \hlll{The extent of this production deviates farther from zero with increase in wave steepness, $\zeta$ as both the streamwise ($\langle d\langle v\rangle_{z,t}/dx  \rangle^+_{x}$) and vertical ($\langle d\langle v\rangle_{z,t}/dy  \rangle^+_{x}$) gradients of the mean vertical velocity} (see figures~\ref{fig:prof_dvdx_avg2_new} and \ref{fig:prof_dvdy_avg2_new}) \hlll{increasingly depart from zero due to local heterogeneity. Consequently, turbulence production, $\langle P^{v'u'}_{22} \rangle^+_{x}=-\langle \langle v'u'\rangle_{z,t} d\langle v\rangle_{z,t}/dx  \rangle^+_{x}$ generates vertical variance in the buffer layer  and destroys some of it in the viscous layer below} (figure~\ref{fig:tke_P22_dvdx_new}). \hlll{Similarly, $\langle P^{v'v'}_{22} \rangle^+_{x}=-\langle \langle v'v'\rangle_{z,t} d\langle v\rangle_{z,t}/dy  \rangle^+_{x}$ destroys variance in the viscous layer and generates above it} (figure~\ref{fig:tke_P22_dvdy_new}). 
\hlll{The $\langle v'^2 \rangle^+_{x,z,t}$ production, $\langle {\mathcal{P}_{22}} \rangle_x^+$ has a dominant contribution from $\langle {\mathcal{P}^{u'v'}_{22}} \rangle_x^+$ compared to $\langle {\mathcal{P}^{u'v'}_{22}} \rangle_x^+$}  (figures~\ref{fig:tke_P22_new} and \ref{fig:tke_P22_dvdx_new}). 
\hlll{This is consistent with trends in magnitudes of $\langle d\langle v\rangle_{z,t}/dx  \rangle^+_{x}$ and $\langle d\langle v\rangle_{z,t}/dy  \rangle^+_{x}$ (see figures~\ref{fig:prof_dvdx_avg2_new} and \ref{fig:prof_dvdy_avg2_new}), i.e. $\langle d\langle v\rangle_{z,t}/dy  \rangle^+_{x} > \langle d\langle v\rangle_{z,t}/dx  \rangle^+_{x}$ and $\langle {\mathcal{P}^{u'v'}_{22}} \rangle_x^+ > \langle {\mathcal{P}^{v'v'}_{22}} \rangle_x^+$. This suggests that the production process depends strongly on gradients of vertical velocity although the precise nature of this relationship needs to be explored.} 
	
%
In addition to the surface-induced production, the dominant contribution to the local generation of $\langle v'^2\rangle^+_{x,z,t}$, $\langle {\Lambda}_{22} \rangle_x^+$ comes from  $\langle \mathcal{R}_{22} \rangle_x^+$.
\hlll{For these small values of $\zeta$, $\langle \mathcal{R}_{22} \rangle_x^+$ and $\langle \lambda_{22} \rangle_x^+$ are similar in structure with that for homogeneous flat channel turbulence.}
\hlll{We observe that closer to the surface in the viscous layer, $\langle \mathcal{R}_{22} \rangle_x^+ < 0$ due to splat events from wall blockage, i.e. $\langle v'^2\rangle^+_{x,z,t}$ is converted to $\langle u'^2\rangle^+_{x,z,t}$ and $\langle w'^2\rangle^+_{x,z,t}$. 
 }
Away from the surface in the buffer layer, $\langle \mathcal{R}_{22} \rangle_x^+ > 0$ enables return to isotropy. Increase in $\zeta$ enhances this pressure-rate-of-strain mechanism resulting in faster growth of the vertical variance ${\langle {v^{\prime}}^2 \rangle^+_{x,z,t}}$ through the buffer layer (figure~\ref{fig:prof_vvar_avg2_new}) and return to isotropy as evidenced by the location of peak $\langle \mathcal{R}_{22} \rangle_x^+$ moving closer to the surface. 
\begin{figure}[ht!]
\centering
\mbox{
\subfigure[\label{fig:SurfaceDispersion_in_ABL-turbulence_P22_a}]{\includegraphics[width=0.21\textwidth]{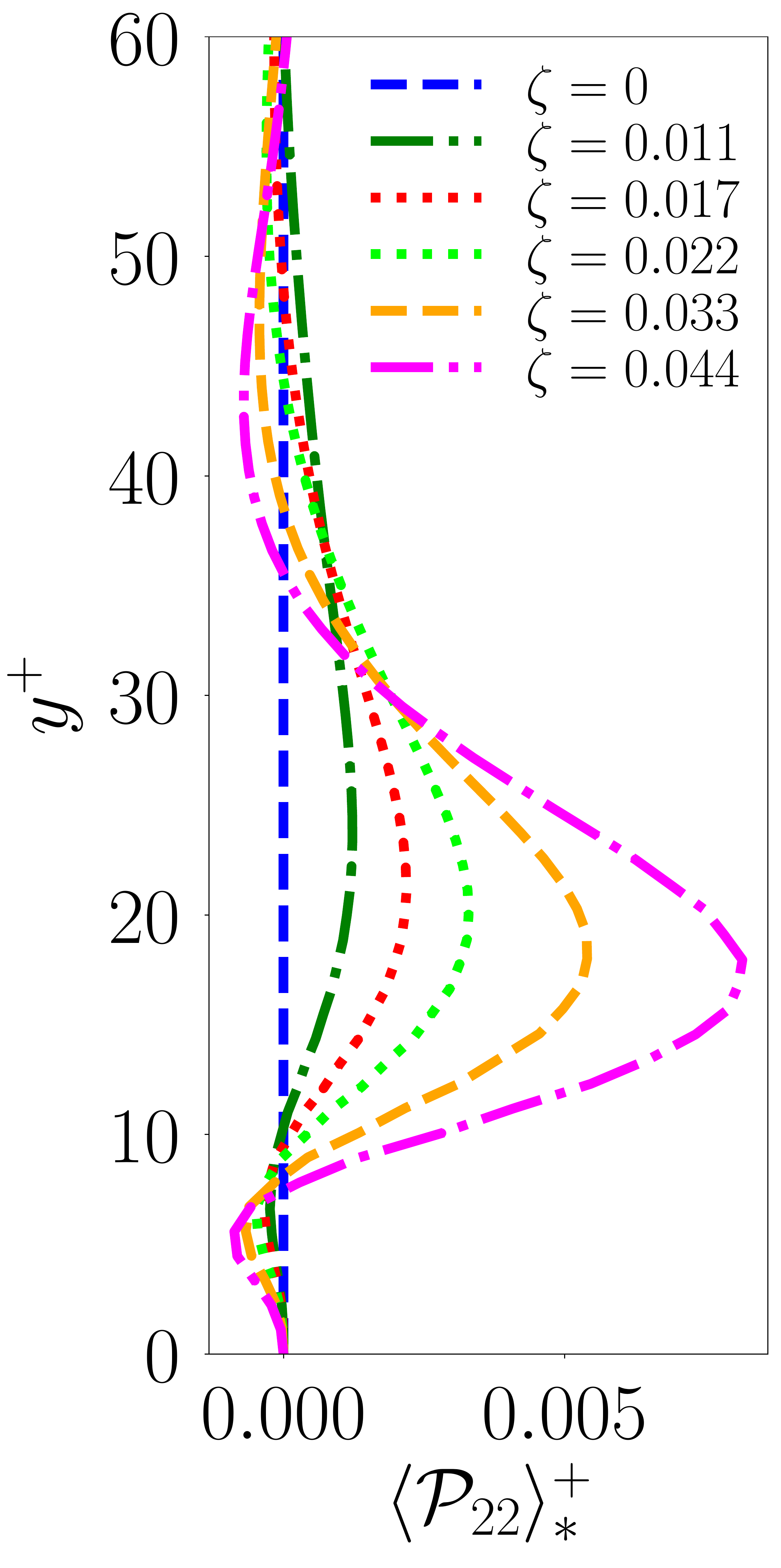}}
\subfigure[\label{fig:SurfaceDispersion_in_ABL-turbulence_P22_b}]{\includegraphics[width=0.21\textwidth]{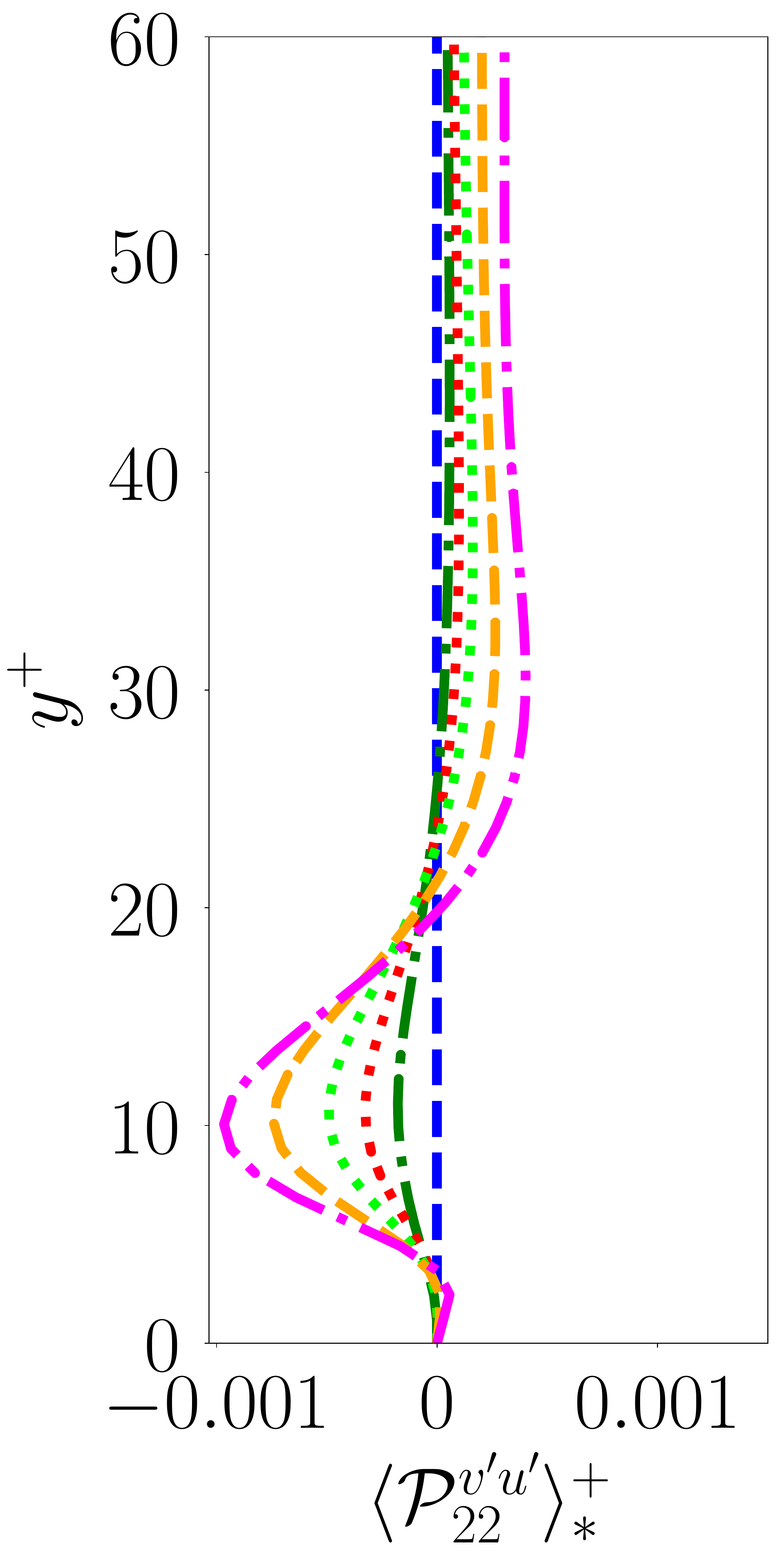}}
\subfigure[\label{fig:SurfaceDispersion_in_ABL-turbulence_P22_c}]{\includegraphics[width=0.21\textwidth]{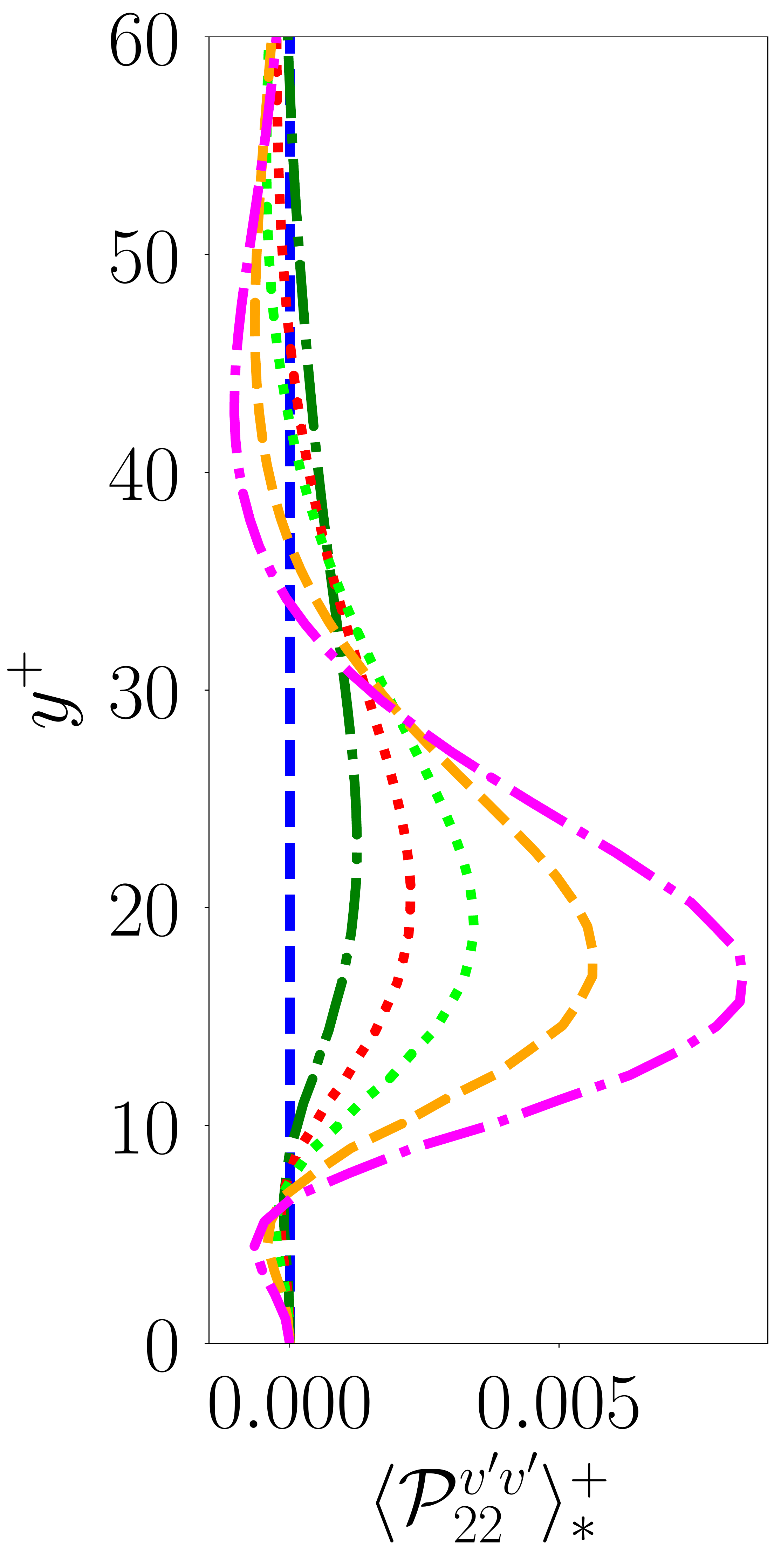}}
}
\mbox{
\subfigure[\label{fig:SurfaceDispersion_in_ABL-turbulence_P22_d}]{\includegraphics[width=0.21\textwidth]{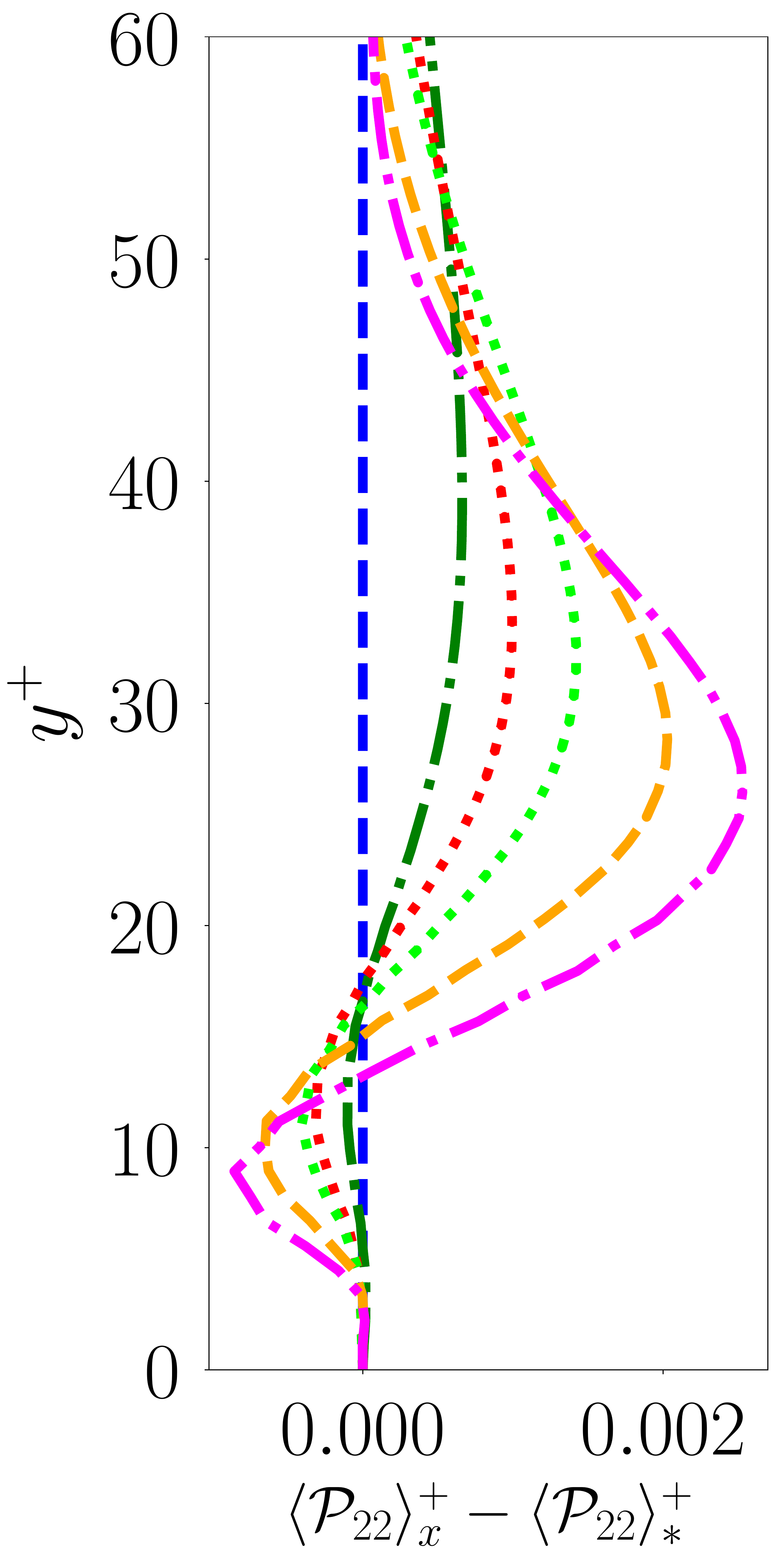}}
\subfigure[\label{fig:SurfaceDispersion_in_ABL-turbulence_P22_e}]{\includegraphics[width=0.21\textwidth]{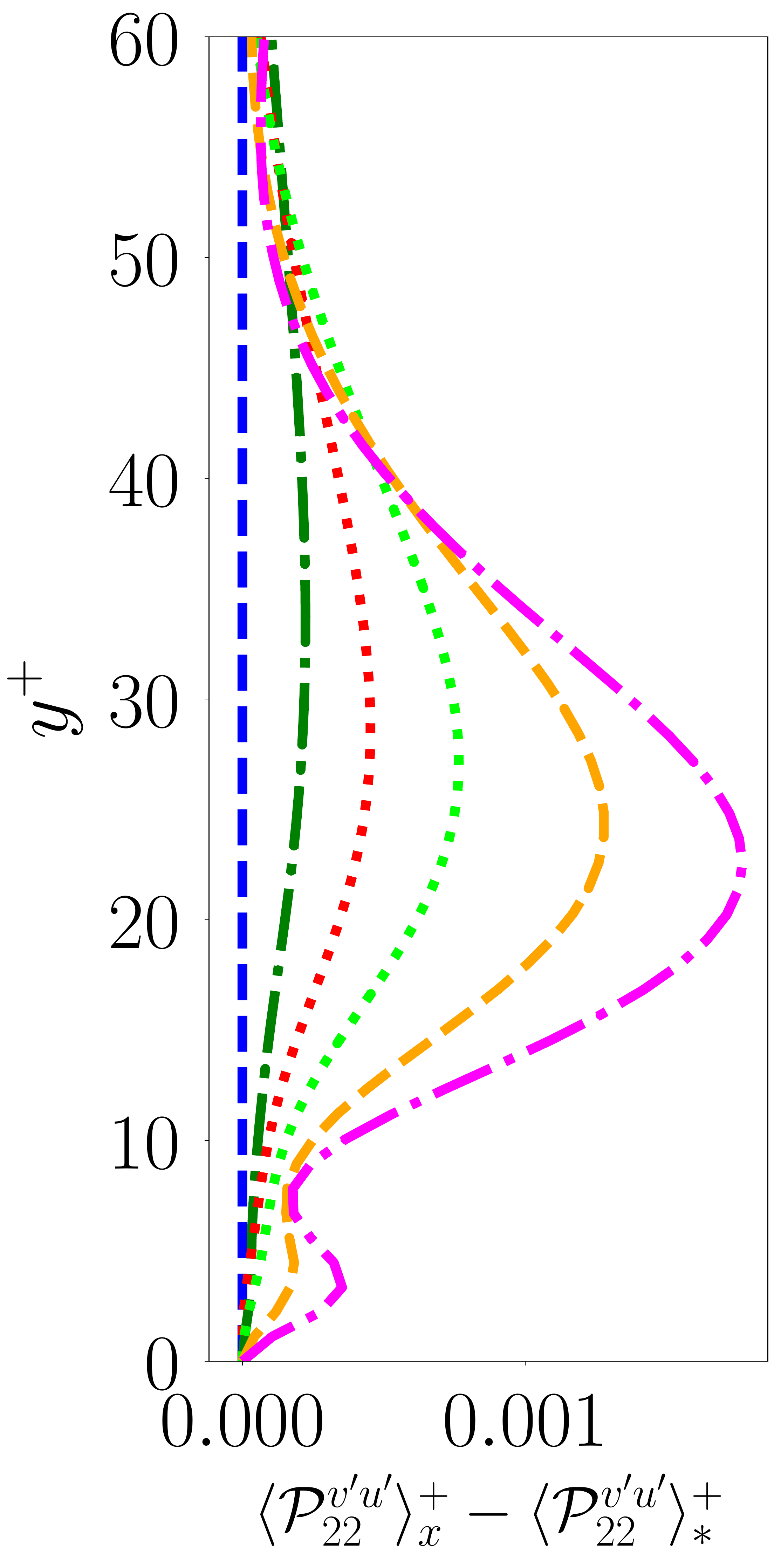}}
\subfigure[\label{fig:SurfaceDispersion_in_ABL-turbulence_P22_f}]{\includegraphics[width=0.21\textwidth]{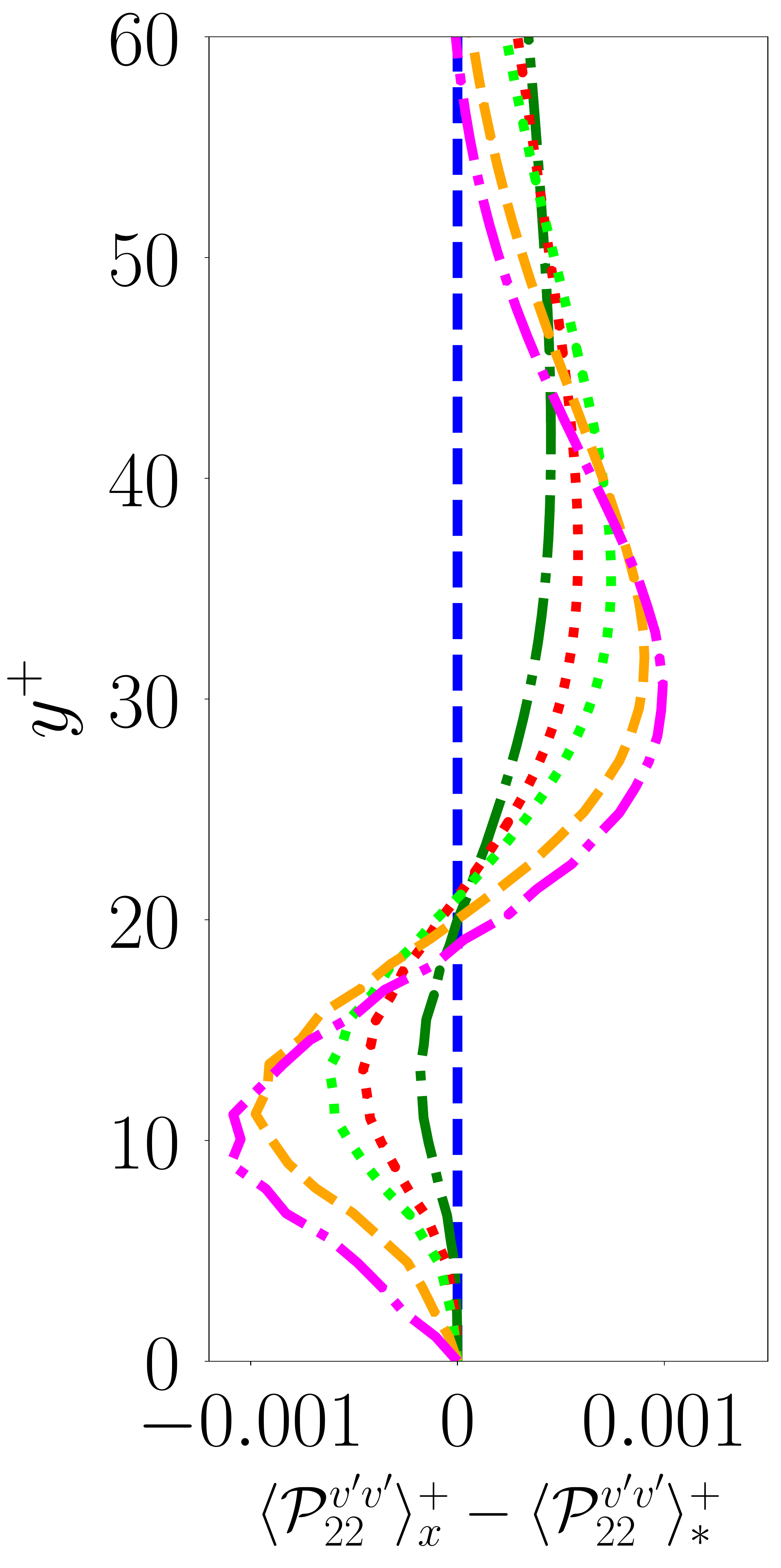}}
}
\caption{{Schematic showing pseudo-production estimates of vertical variance (top row) using product of double-averaged Reynolds stress and mean strain rate (denoted by a `$*$' subscript) and their deviations from true double-averaged production (bottom row). Top row: (a) total pseudo-production, $\langle \mathcal{P}_{22} \rangle_*^+$  (b) component, $\langle \mathcal{P}^{v'u'}_{11} \rangle_*^+$ and (c) component, $\langle \mathcal{P}^{v'v'}_{11} \rangle_*^+$. Bottom row: deviations (d) $\langle \mathcal{P}_{11} \rangle_x^+$-$\langle \mathcal{P}_{11} \rangle_*^+$, (e) $\langle \mathcal{P}^{v'u'}_{11} \rangle_x^+$-$\langle \mathcal{P}^{v'u'}_{11} \rangle_*^+$ and (f) $\langle \mathcal{P}^{v'v'}_{11} \rangle_x^+$-$\langle \mathcal{P}^{v'v'}_{11} \rangle_*^+$ .}\label{fig:SurfaceDispersion_in_ABL-turbulence_P22}}
\end{figure}

\paragraph{Averaged Production of Surface Undulations:}
$\langle \mathcal{P}_{22} \rangle^+_{x}$ originates solely from heterogeneity effects through $\langle P^{v'u'}_{22} \rangle^+_{x}=-\langle \langle v'u'\rangle_{z,t} d\langle v\rangle_{z,t}/dx  \rangle^+_{x}$ and $\langle P^{v'v'}_{22} \rangle^+_{x}=-\langle \langle v'v'\rangle_{z,t} d\langle v\rangle_{z,t}/dy  \rangle^+_{x}$ as shown in figures~\ref{fig:tke_P22_dvdx_new} and \ref{fig:tke_P22_dvdy_new} respectively.
Unlike $\langle \mathcal{P}_{11} \rangle^+_{x}$, both these components for $\langle \mathcal{P}_{22} \rangle^+_{x}$ produce vertical variance in the buffer layer (i.e. reinforce each other) due to non-zero streamwise and vertical gradients of $\langle v \rangle^+_{z,t}$ (see figures~\ref{fig:prof_dvdx_avg2_new} and \ref{fig:prof_dvdy_avg2_new})\cmnt{ arising from surface undulations}. In fact, $\zeta$ directly enters, $\langle \mathcal{P}_{22} \rangle^+_{x}$ as it relates to the ratio of $\langle d\langle v\rangle_{z,t}/dx  \rangle^+_{x}$ and $\langle d\langle v\rangle_{z,t}/dy  \rangle^+_{x}$. 
Given that (i) $\langle v'u'\rangle_{x,z,t} <0$ (everywhere except for a small region near the surface at higher $\zeta$ as shown in figure~\ref{fig:prof_upvpzoom_avg2_new}) and $\langle v'v'\rangle_{x,z,t} >0$ across the TBL and (ii) $\langle d\langle v\rangle_{z,t}/dx  \rangle^+_{x} \gtrsim 0$ and $\langle d\langle v\rangle_{z,t}/dy  \rangle^+_{x} < 0$ in the buffer layer, one can approximately estimate the production $\langle \mathcal{P}^{u'v'}_{22} \rangle^+_{*}=-\langle u'v'\rangle^+_{x,z,t}d\langle v\rangle^+_{x,z,t}/dx^+$ and $\langle \mathcal{P}^{v'v'}_{22} \rangle^+_{*}=-\langle v'v'\rangle^+_{x,z,t}d\langle v\rangle^+_{x,z,t}/dy^+$ (see figures~\ref{fig:SurfaceDispersion_in_ABL-turbulence_P22_b} and \ref{fig:SurfaceDispersion_in_ABL-turbulence_P22_c} respectively). This yields $\langle \mathcal{P}^{u'v'}_{22} \rangle^+_{*} > 0$ and $\langle \mathcal{P}^{v'v'}_{22} \rangle^+_{*} > 0$ away from the surface and destroys variance closer to the wall which is qualitatively consistent with the trends for true estimates, $\langle \mathcal{P}^{u'v'}_{22} \rangle^+_{x}$ and $\langle \mathcal{P}^{v'v'}_{22} \rangle^+_{x}$ (see figures~\ref{fig:tke_P22_dvdx_new} and \ref{fig:tke_P22_dvdy_new}). 
However, there exists noticeable quantitative discrepancy, $\sim 50\%$ error between $\langle \mathcal{P}^{u'v'}_{22} \rangle^+_{x}$ and $\langle \mathcal{P}^{u'v'}_{22} \rangle^+_{*}$ and $\sim 20\%$ error between $\langle \mathcal{P}^{v'v'}_{22} \rangle^+_{x}$ and $\langle \mathcal{P}^{v'v'}_{22} \rangle^+_{*}$. These deviations represents dispersion effects on production given by $\langle \mathcal{P}^{u'v'}_{22} \rangle^+_{x}-\langle \mathcal{P}^{u'v'}_{22} \rangle^+_{*}=-\langle \langle u'v'\rangle^+_{z,t} d\langle v\rangle^+_{z,t}/dx^+  \rangle_{x}+\langle u'v'\rangle^+_{x,z,t}d\langle v\rangle^+_{x,z,t}/dx^+$  and $\langle \mathcal{P}^{u'v'}_{22} \rangle^+_{x}-\langle \mathcal{P}^{v'v'}_{22} \rangle^+_{*}=-\langle \langle v'v'\rangle^+_{z,t} d\langle v\rangle^+_{z,t}/dy^+  \rangle_{x}+\langle v'v'\rangle^+_{x,z,t}d\langle v\rangle^+_{x,z,t}/dy^+$ (quantified in figures~\ref{fig:SurfaceDispersion_in_ABL-turbulence_P22_d}-\ref{fig:SurfaceDispersion_in_ABL-turbulence_P22_f}) which are both zero in the absence of surface heterogeneity. Figure~\ref{fig:SurfaceDispersion_in_ABL-turbulence_P22} shows the pseudo-production estimates in the top row and the corresponding dispersion contribution in the bottom row.  
\hlll{The extent of deviations only increase with $\zeta$.} 
Naturally both the surface-induced production (figures~\ref{fig:tke_P22_new}-\ref{fig:tke_P22_dvdy_new}) as well as the dispersion profiles (figures~\ref{fig:SurfaceDispersion_in_ABL-turbulence_P22_e}-\ref{fig:SurfaceDispersion_in_ABL-turbulence_P22_f}) can be used to characterize the roughness sublayer height. In this case the average estimate is $\sim 4a^+ \approx 50$ for the former and $\sim 5a^+ \approx 60$ in the latter with no visible dependence on $\zeta$. 


\begin{figure}[ht!]
	\centering
	\mbox{
		\subfigure[$\mathcal{P}_{22}^+$\label{fig:P22-contours-a}]{\includegraphics[width=0.27\textwidth]{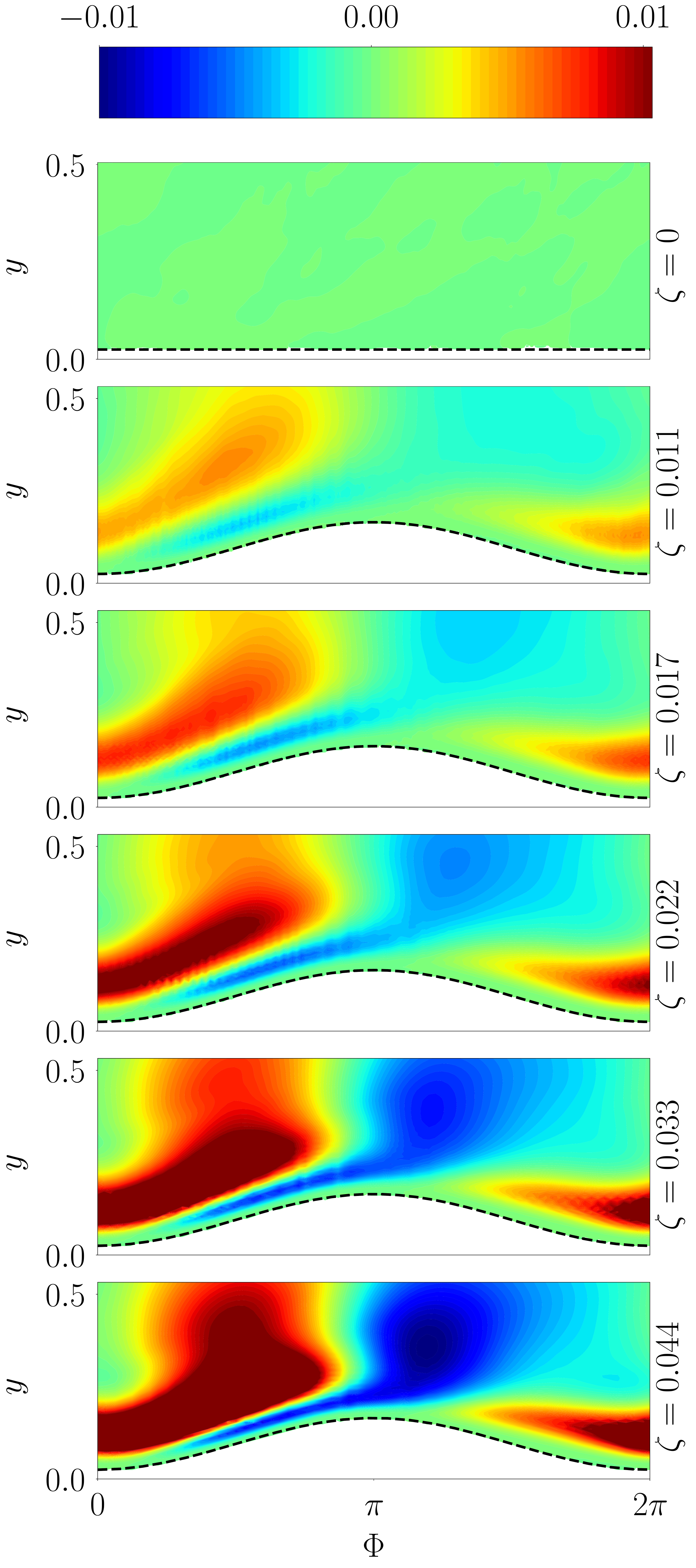}}
		\subfigure[${\mathcal{P}_{22}^{v'u'}}^+$\label{fig:P22-contours-b}]{\includegraphics[width=0.27\textwidth]{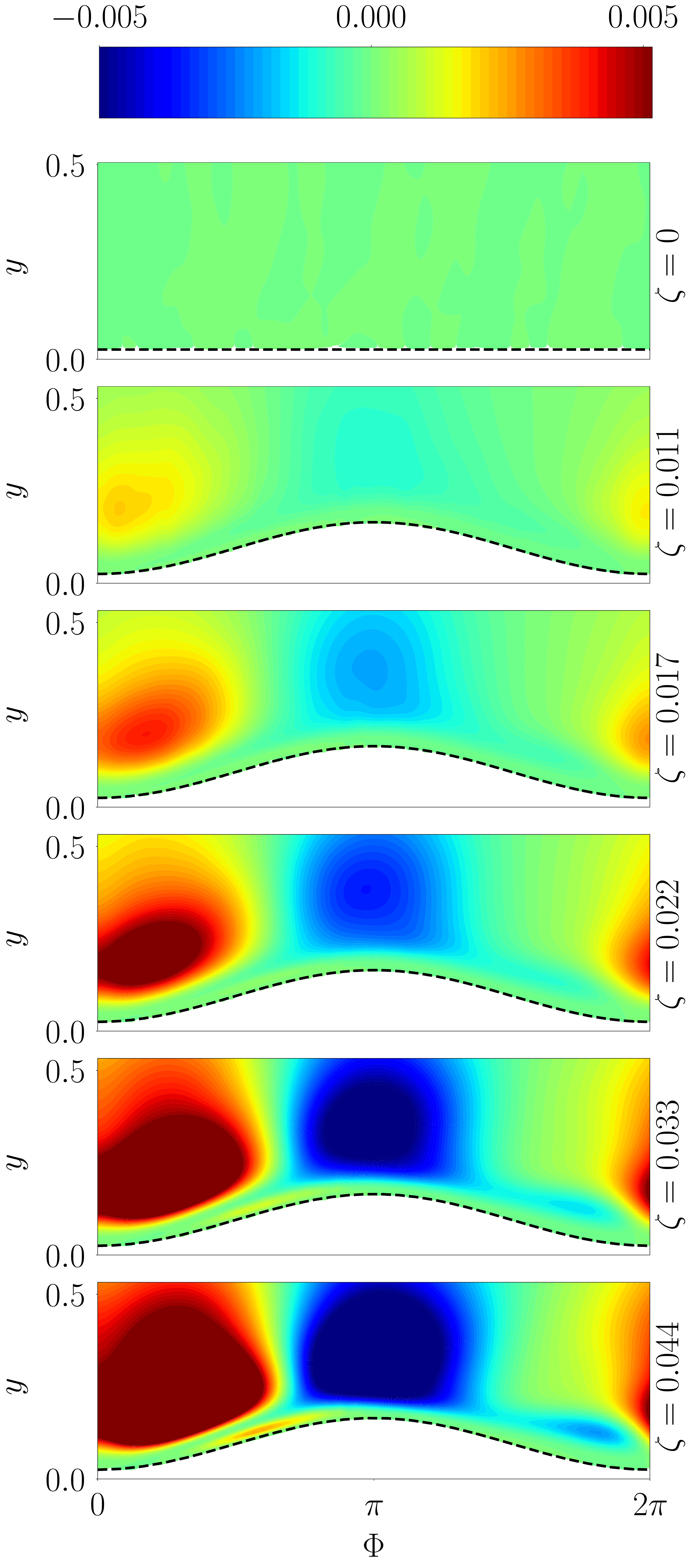}}
		\subfigure[${\mathcal{P}_{22}^{v'v'}}^+$\label{fig:P22-contours-c}]{\includegraphics[width=0.27\textwidth]{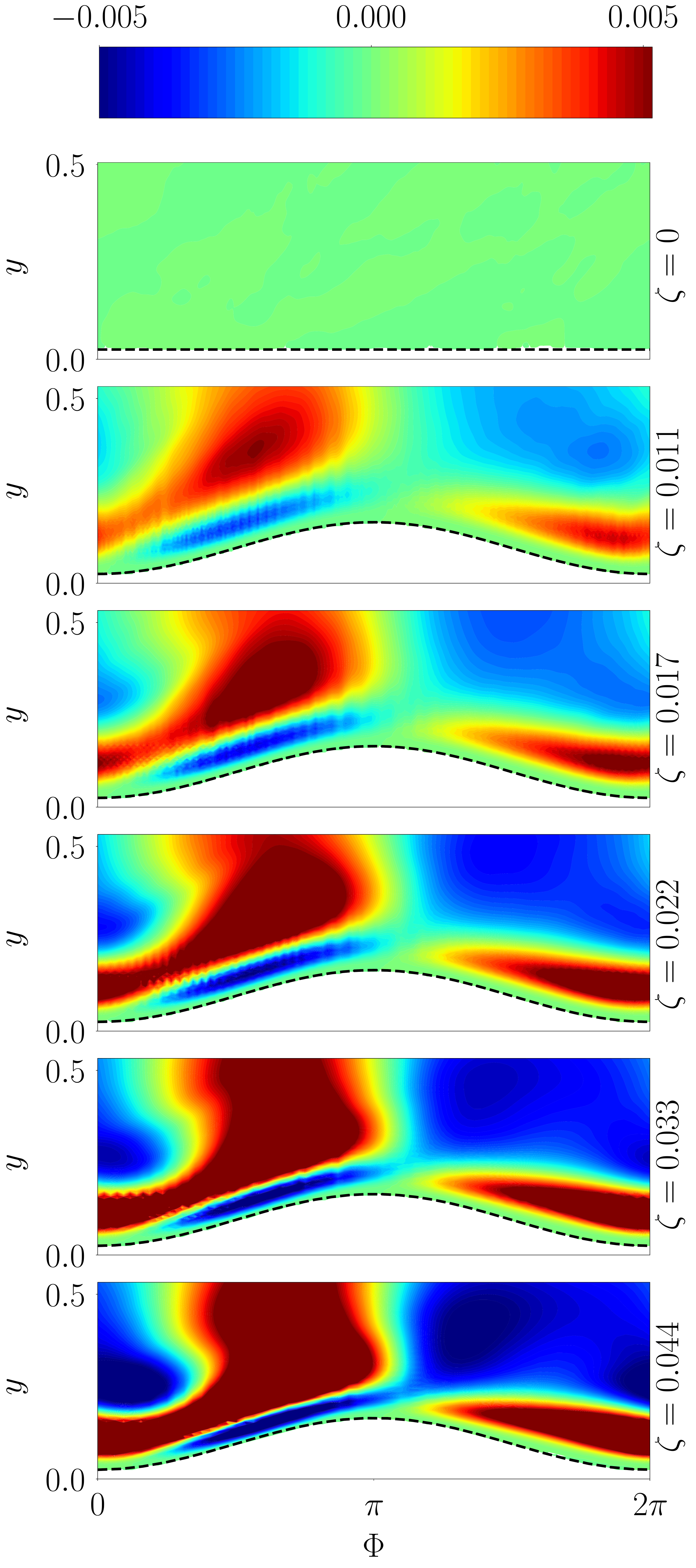}}
	}
	\caption{Contours of inner--scaled (single) averaged vertical variance production, $\mathcal{P}_{22}^+$ (a) and its components, ${\mathcal{P}_{22}^{v'u'}}^+$ (b) and ${\mathcal{P}_{22}^{v'v'}}^+$ (c) as a function of $\Phi$ and $y$, where, $\Phi=2\pi x/\lambda$. All the plots show the near surface region, i.e. $y/\delta \leq 0.5$ with $\delta=1$ in our computations.\label{fig:P22-contours}
	}
\end{figure}

\paragraph{Two-dimensional Structure of $\langle v'^2\rangle^+_{z,t}$ Production:}
%
%
For a deeper understanding of the production mechanisms we look at the inner-scaled two-dimensional production contours of $\mathcal{P}^+_{22} (x,y)$, ${\mathcal{P}^{u'v'}_{22}}^+ (x,y)$ 
and ${\mathcal{P}^{v'v'}_{22}}^+ (x,y)$ based on averaging along homogeneous direction ($z$) and over a stationary window ($t$) of the turbulent flow in figure~\ref{fig:P22-contours}.  
We clearly see from the isocontours  (figures~\ref{fig:P22-contours-a} and \ref{fig:P22-contours-b}) that qualitatively ${\mathcal{P}_{22}}^+\approx {{\mathcal{P}^{v'v'}_{22}}^+}$ and this primary contribution originates comes from interaction of vertical variance, $\langle {v'v'}\rangle^+_{z,t}$ with the vertical gradient of $\langle v ' \rangle^+_{z,t}$. The secondary contribution, ${{\mathcal{P}^{v'u'}_{22}}^+}$ arises from  the interaction of vertical turbulent momentum flux, $\langle {v'u'}\rangle^+_{z,t}$ with the streamwise gradient of $\langle v ' \rangle^+_{z,t}$. 
%
Figure~\ref{fig:P22-contours} shows inner-scaled production over the $y-\phi$ space where $\phi$ is a non-dimensional streamwise phase coordinate ($\phi = 2\pi x/\lambda$) and $y=y/\delta$ ($\delta=1$). We clearly observe that the  production contour shape is mostly qualitatively invariant with respect to the wave shape for different $\zeta=2a/\lambda > 0$ indicating that the size of the structures scale with $\lambda$. However, the inner-scaled magnitude of production (i.e. the colors in figure~\ref{fig:P22-contours}) depends on the wave slope with higher $\zeta$ producing stronger peaks and troughs. 
As before, we characterize the regions near and far from the surface using the wall-normal distance relative to the local surface height, $d$. 

\begin{figure}[ht!]
\centering
\hspace{-2em}
\mbox{
\subfigure[$\langle u'v'\rangle_{z,t}^+$\label{fig:cont_uvcovar_2}]{\includegraphics[width=0.25\textwidth]{plots//mean_contour/uvcovar_contour_full_1wave.pdf}}
\subfigure[$\langle dv/dx \rangle_{z,t}^+$\label{fig:cont_dvdx}]{\includegraphics[width=0.25\textwidth]{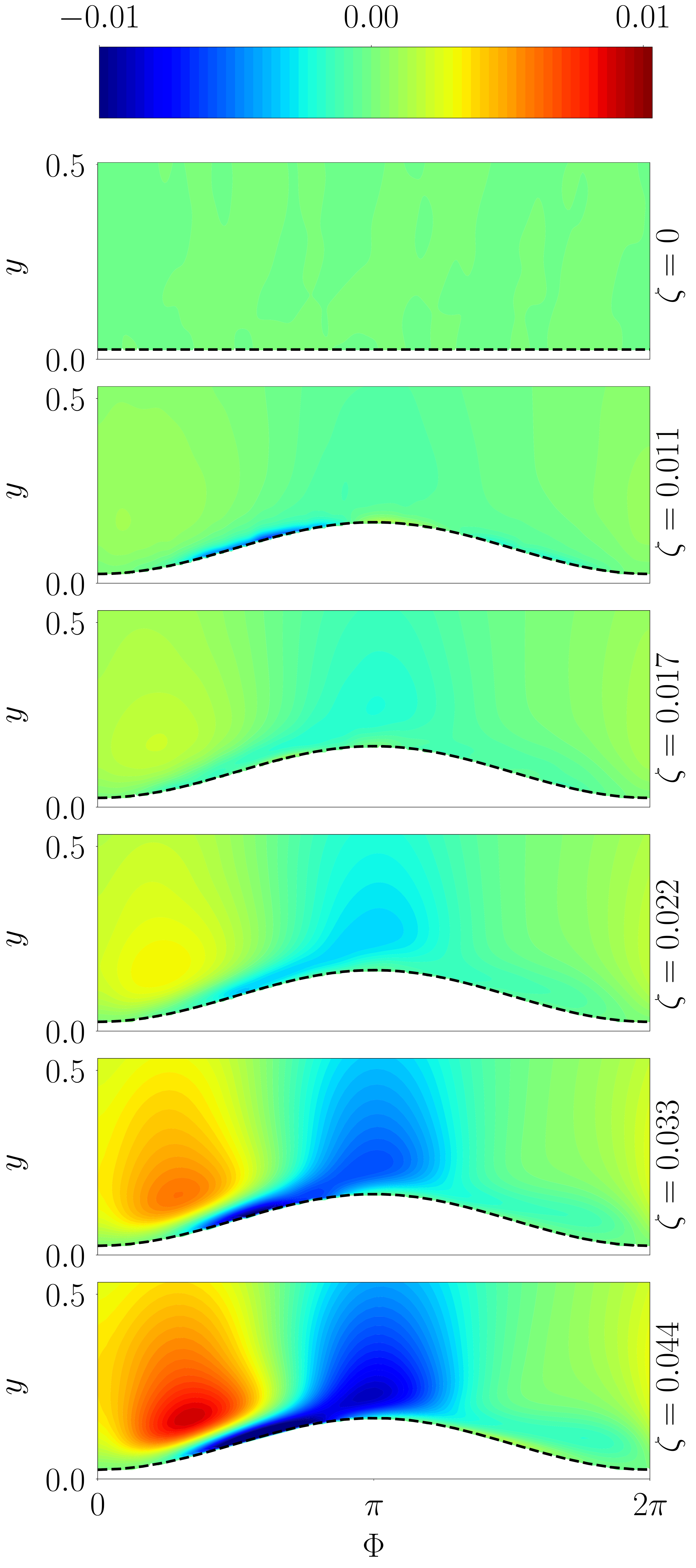}}
\subfigure[$\langle v^{\prime 2}\rangle_{z,t}^+$\label{fig:cont_vvar}]{\includegraphics[width=0.25\textwidth]{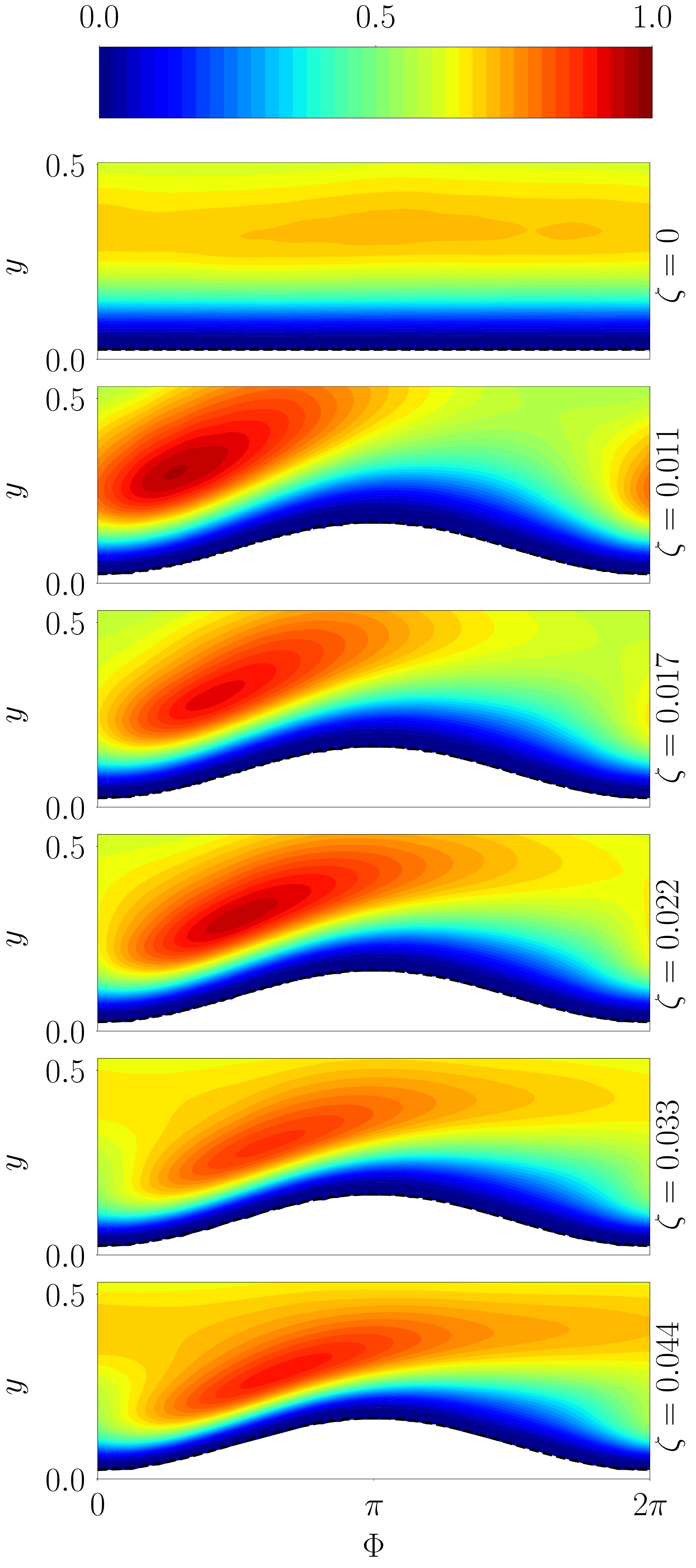}}
\subfigure[$\langle dv/dy \rangle_{z,t}^+$\label{fig:cont_dvdy}]{\includegraphics[width=0.25\textwidth]{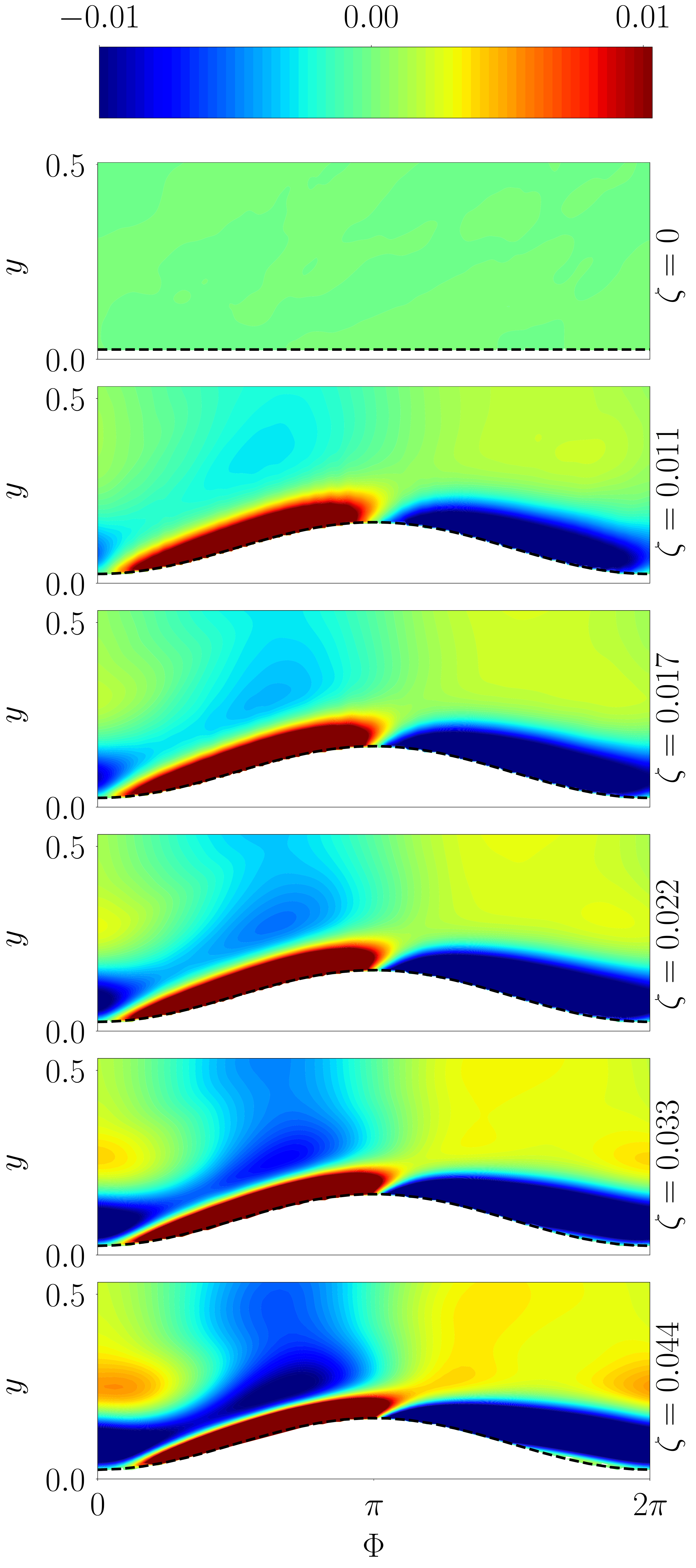}}
}
\caption{\label{fig:cont_ReynoldsStressStrainRate_vvariance} Contours of inner-scaled spanwise and temporally averaged (a) covariance, $\langle u'v' \rangle^+_{z,t}$ (b) streamwise gradient of $\langle v \rangle^+_{z,t}$, (c) vertical variance, $\langle {v'}^2 \rangle^+_{z,t}$ and (d) vertical gradient of $\langle v \rangle^+_{z,t}$. The cyan region closer to the wall in (a) identifies regions of $\langle u'v' \rangle_{z,t} > 0$.
}
\end{figure}

Looking at the structure of $\mathcal{P}^+_{22} (x,y)$\cmnt{ and ${\mathcal{P}^{v'v'}_{22}}^+ (x,y)$}, we see that the bulk of $\langle {v'}^2\rangle^+_{z,t}$ production from surface undulations occur along the slopes of the wave with slight departure from the wall ($d/\delta \gtrsim 0.05$) due to wall blockage. 
Visually, this structure of $\mathcal{P}^+_{22}$ is derived primarily from ${\mathcal{P}^{v'v'}_{22}}^+$ with a secondary contribution from ${\mathcal{P}^{v'u'}_{22}}^+$.
Correlating figure~\ref{fig:P22-contours-b} with figures~\ref{fig:cont_uvcovar_2}-\ref{fig:cont_dvdx} we see that ${\mathcal{P}^{v'u'}_{22}}^+$ is qualitatively characterized by $d\langle v\rangle^+_{z,t}/dx^+$. Similarly, correlating figure~\ref{fig:P22-contours-c} with figures~\ref{fig:cont_vvar}-\ref{fig:cont_dvdy} we see that ${\mathcal{P}^{v'v'}_{22}}^+$ is characterized by $d\langle v\rangle^+_{z,t}/dy^+$. Given this dependence on the strain rate structure, it is worth looking at the structure of mean vertical velocity, $\langle v \rangle_{z,t}$ generated by the streamlines curving over the wavy undulations. This creates a region of upward flow ($\langle v \rangle_{z,t} > 0$) over the windward slope and downward flow ($\langle v \rangle_{z,t} < 0$) along the leeward slope. These vertical flow structures mostly represent surface form/shape influences and therefore, extend through most the roughness layer\cmnt{ THIS MIGHT REQUIRE PROOF OF MEAN VELOCITY CONTOURS}. This is true for $d\langle v\rangle^+_{z,t}/dx^+$ (see figure~\ref{fig:cont_dvdx}) as well which varies gradually along the vertical direction. The larger gradient, $d\langle v\rangle^+_{z,t}/dy^+$ (due to small $\zeta$) mostly represents near-wall shear stress and therefore, decreases rapidly away from the surface (see figure~\ref{fig:cont_dvdy}). Therefore, $d\langle v\rangle^+_{z,t}/dx^+$ in spite of being smaller in magnitude, persists farther away from the surface. Although, it is in the buffer and inertial regions that $\langle u'v' \rangle^+_{z,t}$ and $\langle v'v' \rangle^+_{z,t}$ achieve their maximum magnitudes (figures~\ref{fig:cont_uvcovar_2}-\ref{fig:cont_vvar}) respectively, the strain rate structure causes the bulk of production from ${\mathcal{P}^{u'v'}_{22}}^+$ to occur further away from the surface than ${\mathcal{P}^{v'v'}_{22}}^+$ (see figures~\ref{fig:P22-contours-b}-\ref{fig:P22-contours-c}). The larger production of $\langle {v'}^2\rangle^+_{z,t}$ by ${\mathcal{P}^{v'v'}_{22}}^+$ occurs close t the surface along the leeward slope and away from the surface in the windward side. The smaller production from ${\mathcal{P}^{u'v'}_{22}}^+$ occurs away from the surface along the windward slope of the wave to supplement that from ${\mathcal{P}^{v'v'}_{22}}^+$.  In summary, the production of $\langle {v'}^2\rangle^+_{z,t}$ dominates in the buffer layer over the windward slope and closer to the surface along the leeward slope. Given the preponderance on strain rates, the higher slopes naturally generate stronger variance production.

\subsection{Spanwise Variance, $\langle w'^2 \rangle^+_{x,z,t}$ \label{subsec:spanwise_var}}
\begin{figure}[ht!]
	\centering
	\mbox{
		\subfigure[\label{fig:prof_wvar_avg2_new60}]{\includegraphics[width=0.21\textwidth]{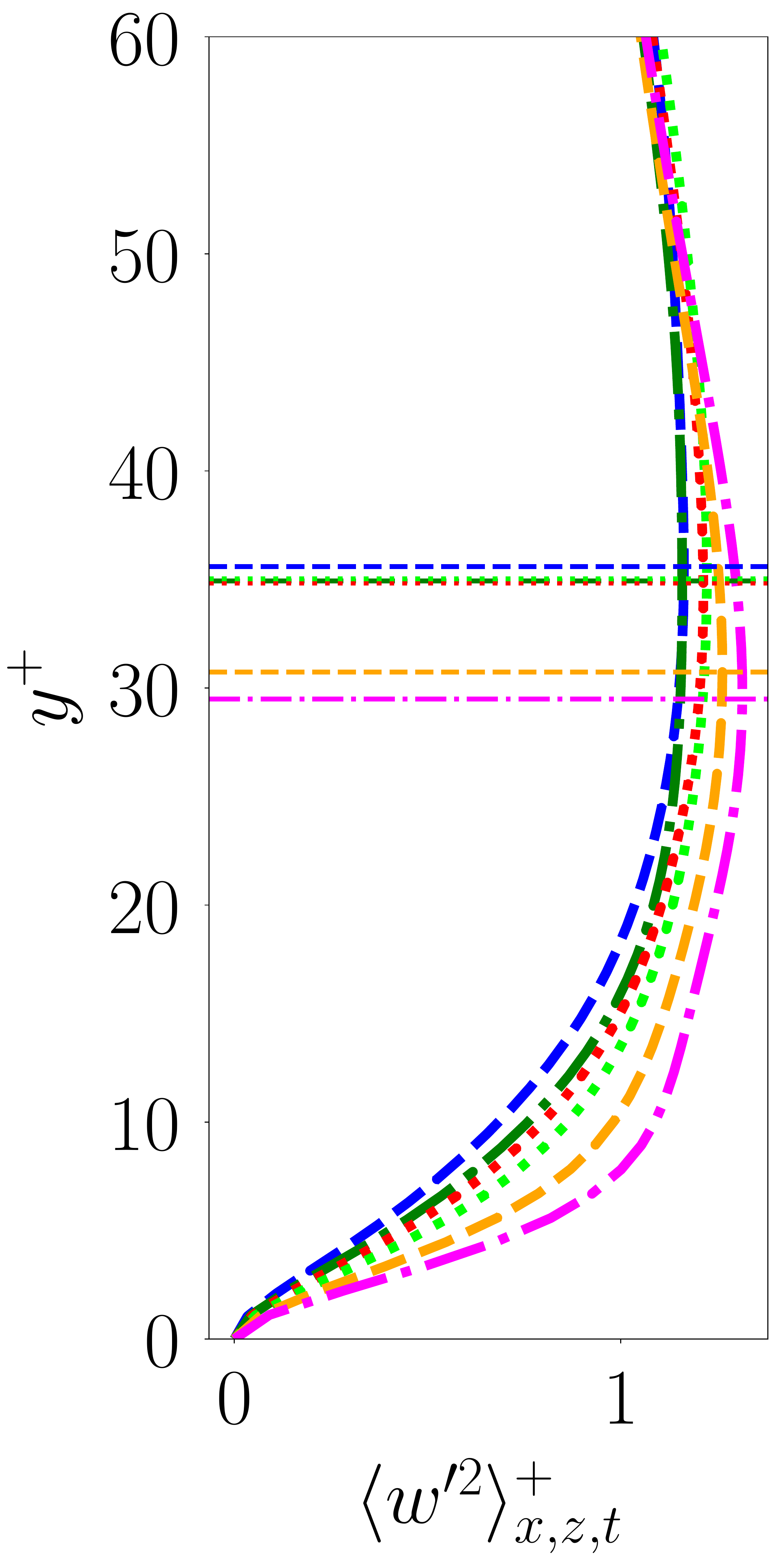}}\hspace{0em}
		\subfigure[\label{fig:tke_P33_new}]{\includegraphics[width=0.21\textwidth]{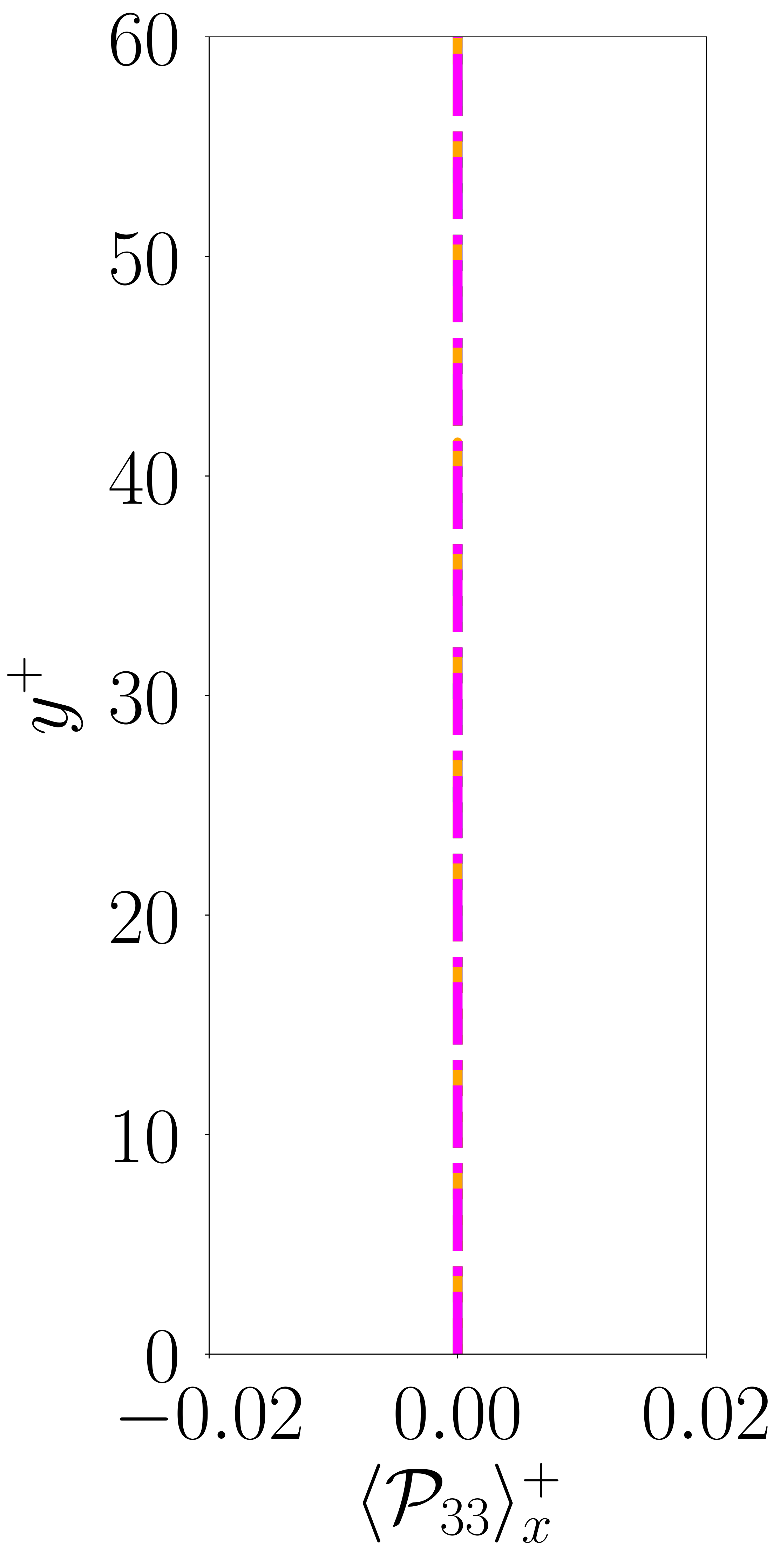}}\hspace{0em}
		\subfigure[\label{fig:tke_E33_new}]{\includegraphics[width=0.21\textwidth]{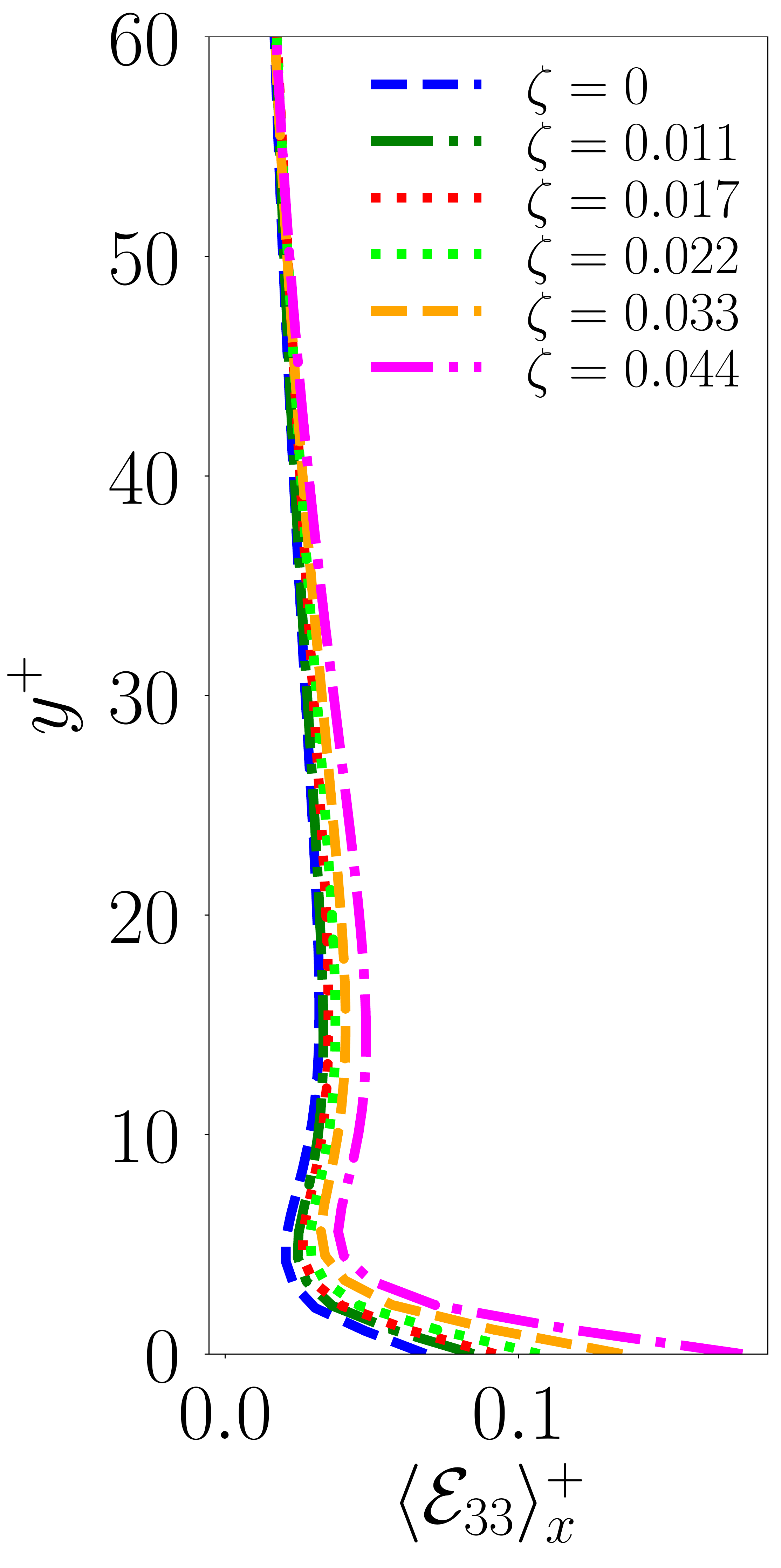}}
	}
	\mbox{
		\subfigure[\label{fig:tke_R33_new}]{\includegraphics[width=0.21\textwidth]{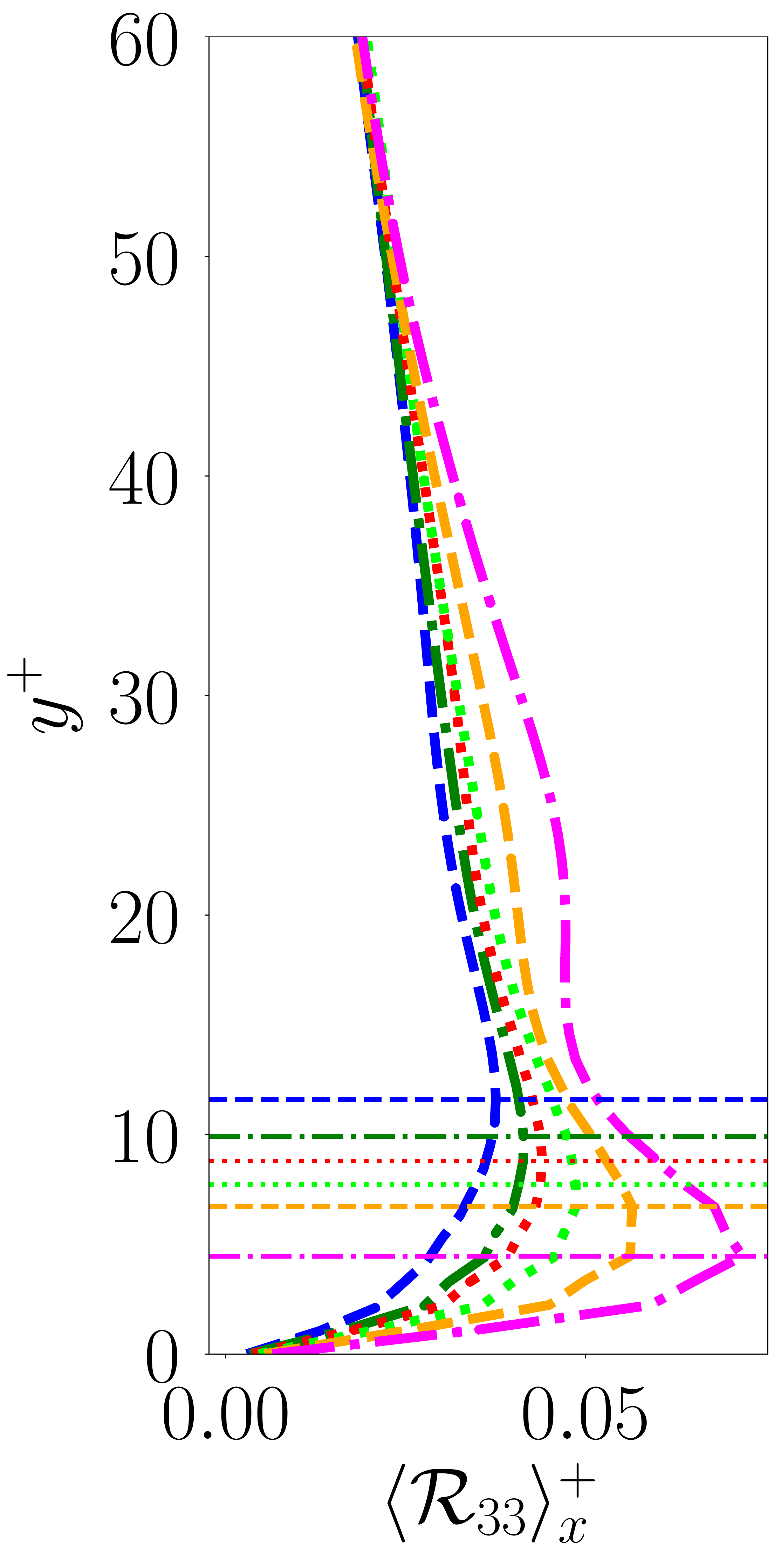}}\hspace{0em}
		\subfigure[\label{fig:tke_L33_new}]{\includegraphics[width=0.21\textwidth]{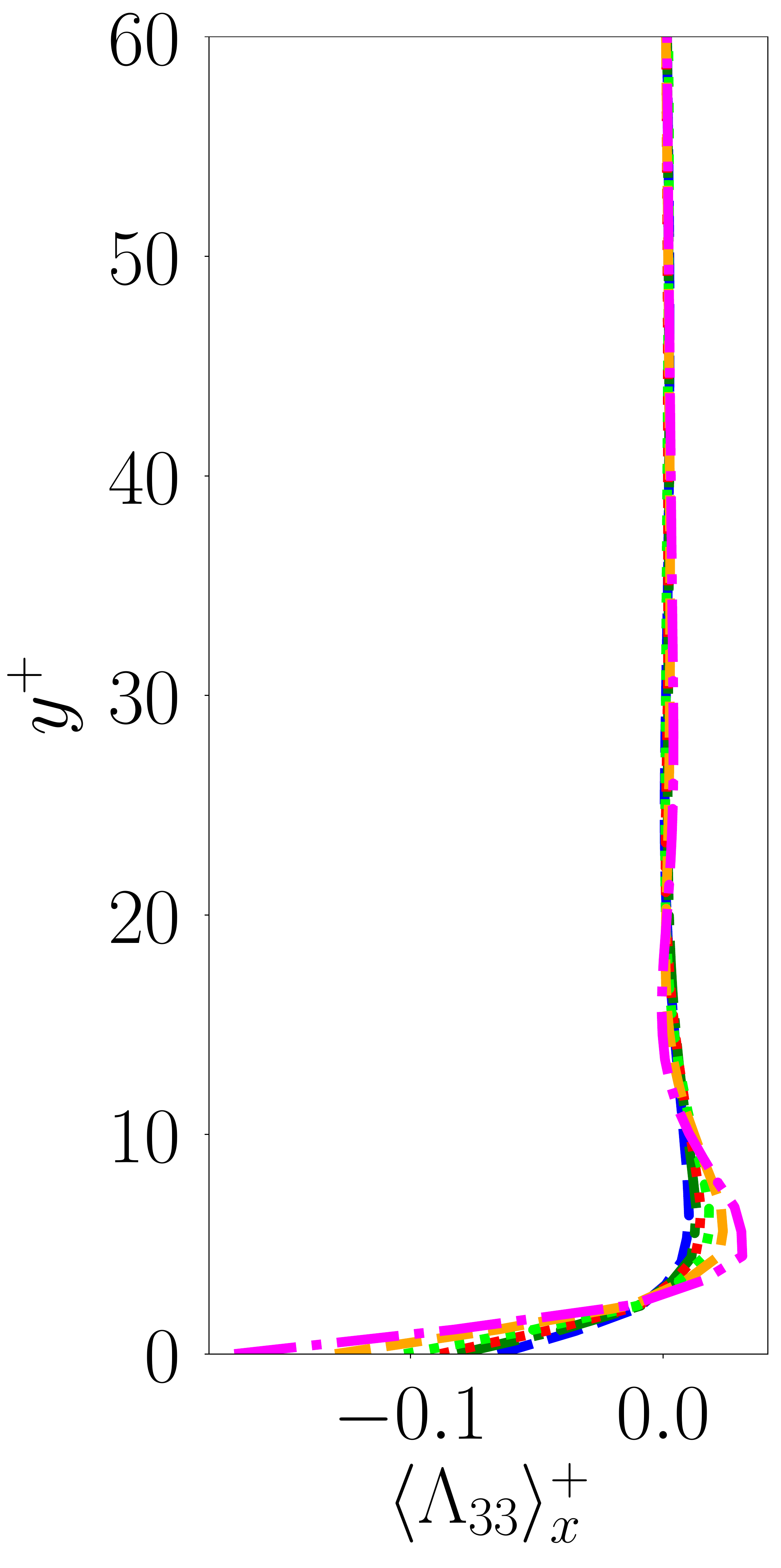}}\hspace{0em}
		\subfigure[\label{fig:tke_D33_new}]{\includegraphics[width=0.21\textwidth]{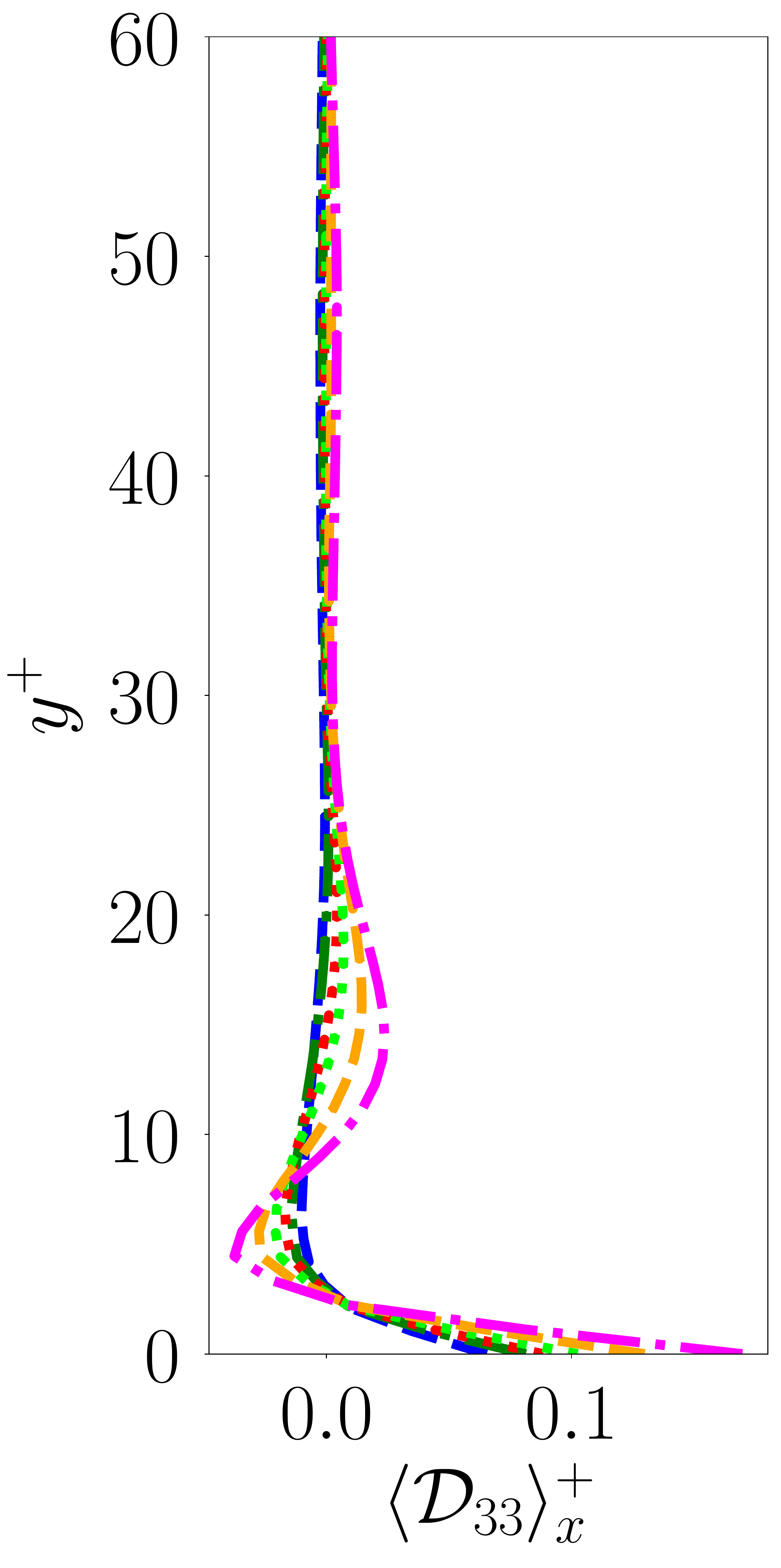}}
	}
	\caption{\label{fig:wvar_production_new} 
		Schematic illustration of the  wall-normal variation of inner-scaled spanwise variance (a) along with the corresponding double averaged production (b), dissipation (c), pressure-rate-of-strain (d), cumulative effect $\langle \Lambda_{33} \rangle^+_{x}=\langle\mathcal{P}_{33}\rangle^+_x-\langle\mathcal{E}_{33}\rangle^+_x+\langle\mathcal{R}_{33}\rangle^+_x$ in (e), and diffusion (f). The horizontal lines correspond to the vertical location of maximum/minimum value for a chosen statistic. 
	}
\end{figure}

Similar to ${\langle {v^{\prime}}^2 \rangle^+_{x,z,t}}$, the inner-scaled spanwise variance, ${\langle w^{\prime 2} \rangle^+_{x,z,t}}$ also shows stronger growth (see figures~\ref{fig:prof_wvar_avg2_new} and \ref{fig:prof_wvar_avg2_new60}) through the viscous and lower regions of the buffer layer with steeper surface undulations. \hlll{ Additionally, ${\langle w^{\prime 2} \rangle^+_{x,z,t}}$ peaks in the buffer layer ($y^+ \approx 35$) and this peak location is lower than that observed for ${\langle v^{\prime 2} \rangle^+_{x,z,t}}$ due to faster growth near the surface} (see figure~\ref{fig:prof_variances_avg2_new}) in the absence of wall blockage. \hlll{Further, the peak magnitudes of ${\langle w^{\prime 2} \rangle^+_{x,z,t}}$ increase with $\zeta$ while the locations of the peak decrease with $\zeta$ suggesting surface-enhanced return to isotopy. Unlike the trends for ${\langle {v^{\prime}}^2 \rangle^+_{x,z,t}}$, the peak magnitudes and locations for ${\langle w^{\prime 2} \rangle^+_{x,z,t}}$ show increased sensitivity to $\zeta$ as they occur well within the roughness sublayer. Particularly, we see two clusters of peak locations corresponding to  $\zeta=0.0-0.022$ and $\zeta=0.033-0.044$.  This suggests possible dependence of ${\langle w^{\prime 2} \rangle^+_{x,z,t}}$ on case specific flow features such as flow separation at higher $\zeta$ which in turn depends on the Reynolds number. However, the variation in peak magnitudes of ${\langle w^{\prime 2} \rangle^+_{x,z,t}}$ with $\zeta$ is systematic}. 

\subsubsection{Dynamics of Spanwise Variance Transport\label{subsubsec:spanwise_var_transport}}
Unlike both the streamwise (${\langle {u^{\prime}}^2 \rangle^+_{x,z,t}}$) and vertical (${\langle {v^{\prime}}^2 \rangle^+_{x,z,t}}$) variances, the spanwise variance (${\langle {w^{\prime}}^2 \rangle^+_{x,z,t}}$) is not generated through Reynolds stress-strain rate interactions ($\langle \mathcal{P}_{33} \rangle_x^+=0$ in figure~\ref{fig:tke_P33_new}) due to spanwise homogeneity. Nevertheless, the transport process is still sensitive to $\zeta$ due to coupling across the different velocity components. 
For example, wall-blockage converts vertical motions into other components \cmnt{(note $\langle \mathcal{R}_{22} \rangle_x^+ \lesssim 0$  in figure~\ref{fig:tke_R22_new})} for all $\zeta$.
In fact, we clearly see $\langle \mathcal{R}_{11} \rangle_x^+ \lesssim 0$, $\langle \mathcal{R}_{22} \rangle_x^+ \lesssim 0$ and $\langle \mathcal{R}_{33} \rangle_x^+ > 0$ for $\zeta \geq 0$ near the surface in figures~\ref{fig:tke_R11_new}, \ref{fig:tke_R22_new} and \ref{fig:tke_R33_new} respectively. Away from the surface, we have $\langle \mathcal{R}_{11} \rangle_x^+ < 0$, $\langle \mathcal{R}_{22} \rangle_x^+ > 0$ and $\langle \mathcal{R}_{33} \rangle_x^+ > 0$. 
%
\hlll{
This dynamics is governed by $\langle \mathcal{R}_{11} \rangle_x^+ + \langle \mathcal{R}_{22} \rangle_x^+ + \langle \mathcal{R}_{33} \rangle_x^+ = 0$ for incompressible flow which leaves $\langle \mathcal{R}_{33} \rangle_x^+ >0$ (and $\langle \mathcal{R}_{11} \rangle_x^+ \leq 0$) through the TBL. 
In the current study ${\langle {w^{\prime}}^2 \rangle^+_{x,z,t}}$  is generated closer to the surface using two different  pressure-rate-of-strain mechanisms: (i) conversion of diffused ${\langle {v^{\prime}}^2 \rangle^+_{x,z,t}}$ and (ii) conversion of ${\langle {u^{\prime}}^2 \rangle^+_{x,z,t}}$ produced by interaction of mean shear with Reynolds stress.  With increase in effective wave-slope, $\zeta$, $\langle \mathcal{R}_{11} \rangle_x^+$, $\langle \mathcal{R}_{22} \rangle_x^+$ and $\langle \mathcal{R}_{33} \rangle_x^+$ not only increase in magnitude through the roughness layer but also assume peak values closer to the surface indicating return to isotropy nearer to the wall. The effect of surface undulations on the different $\langle \mathcal{R} \rangle_x^+$ profiles is felt until $y^+=3a^+$. } 

\begin{figure}[ht!]
	\centering
	\mbox{
		\subfigure[$ \mathcal{R}_{11}^+$\label{fig:R11-contours}]{\includegraphics[width=0.27\textwidth]{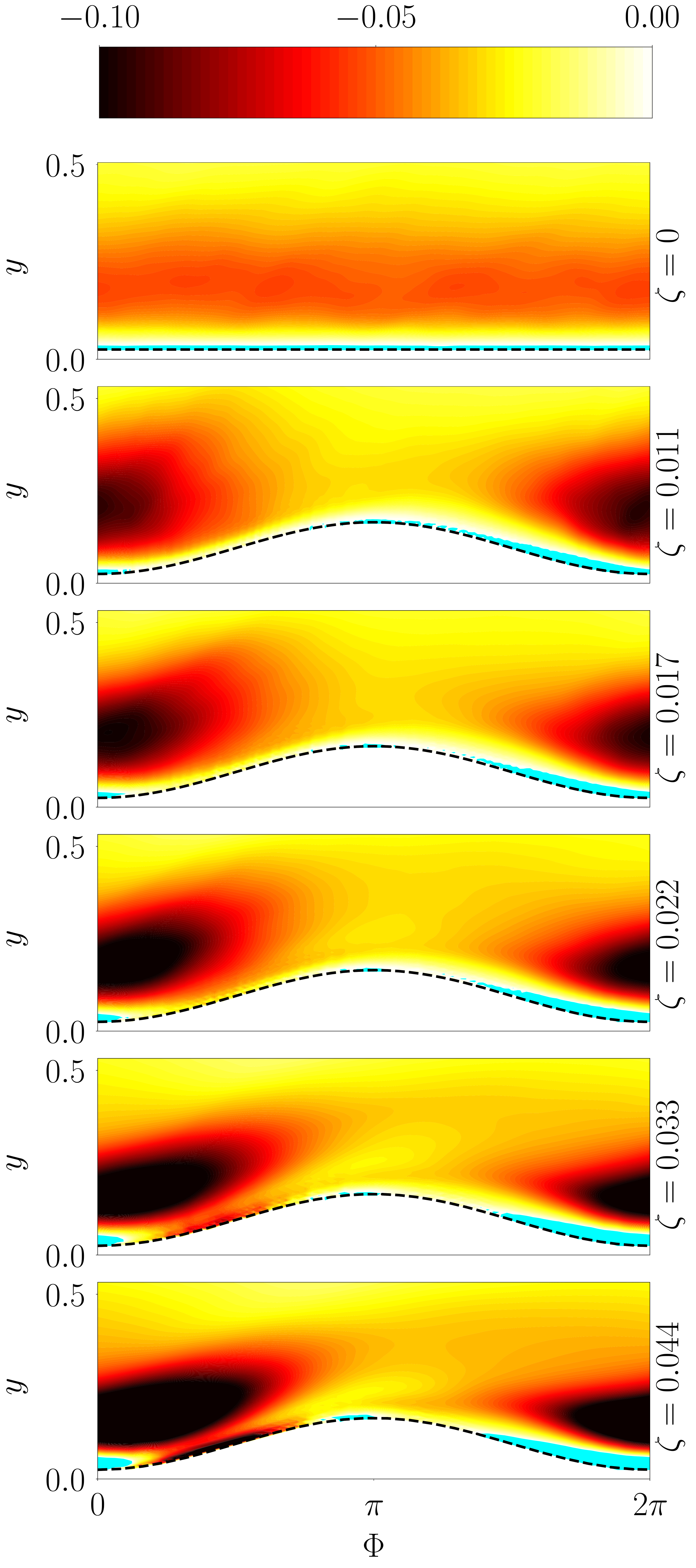}}
		\subfigure[$ \mathcal{R}_{22}^+$\label{fig:R22-contours}]{\includegraphics[width=0.27\textwidth]{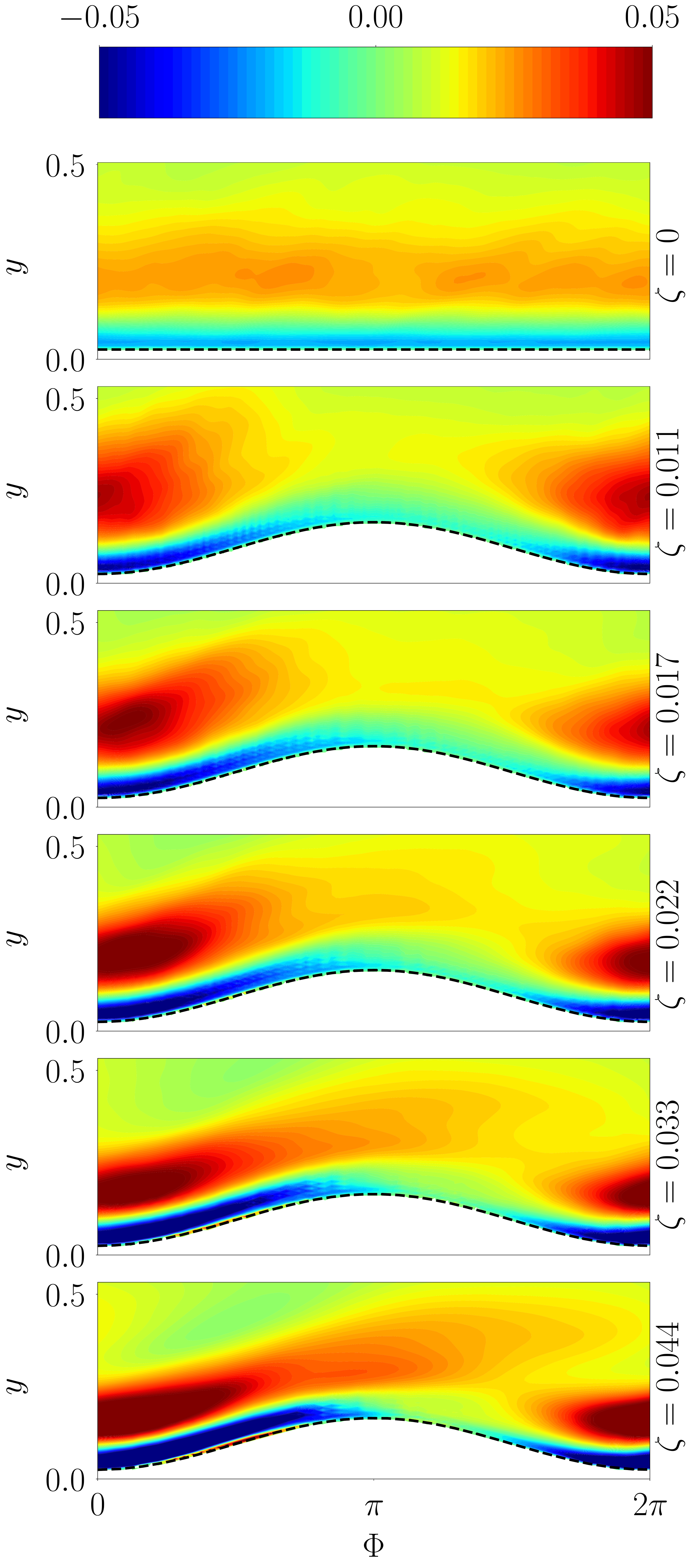}}
		\subfigure[$ \mathcal{R}_{33}^+$\label{fig:R33-contours}]{\includegraphics[width=0.27\textwidth]{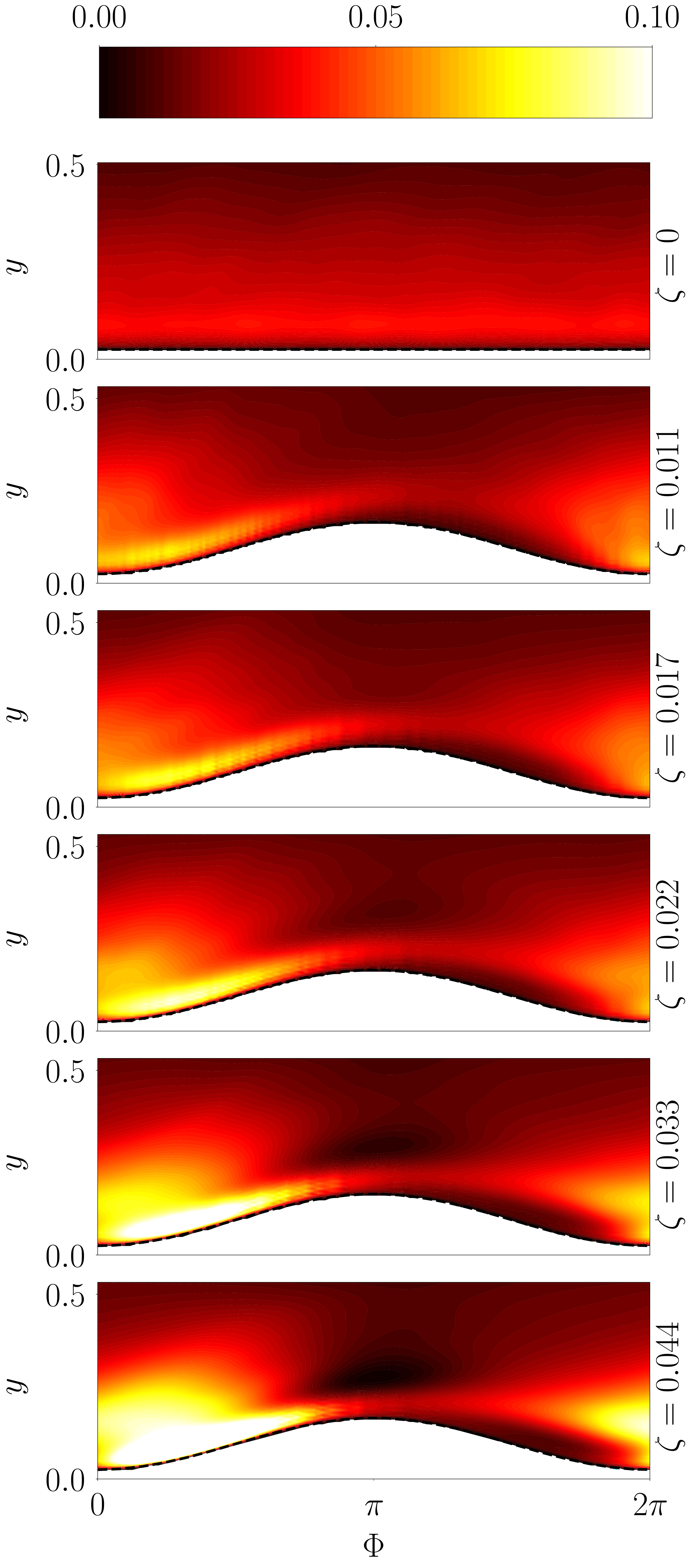}}
	}
	\caption{\label{fig:Rii-contours}Inner scaled contours of pressure-rate-of-strain terms from the different variance transport equations, namely, ${\langle {u^{\prime}}^2 \rangle^+_{z,t}}$ (a), ${\langle {v^{\prime}}^2 \rangle^+_{z,t}}$ (b) and ${\langle {w^{\prime}}^2 \rangle^+_{x,z,t}}$ (c). We note that (a) and (c) have the same color scheme but with opposite signs. In (b) blue to red  represents negative to positive. In (a), cyan region represent positive, $\mathcal{R}^+_{11}$.  
}
\end{figure}
\vspace{-2em}
\subsubsection{Two-dimensional Variation of  Pressure-rate-of-strain Term $\mathcal{R}_{ii}^+ (x,y)$, $ii=11,22,33$\label{subsubsec:spanwise_var_2DStructure_R}}
\vspace{-1em}
Given that the averaged spanwise variance, ${\langle {w^{\prime}}^2 \rangle^+_{x,z,t}}$ is generated purely by conversion of ${\langle {u^{\prime}}^2 \rangle^+_{x,z,t}}$ and ${\langle {v^{\prime}}^2 \rangle^+_{x,z,t}}$, it is important to understand the key mechanisms underlying two-dimensional structure of the pressure-rate-of-strain terms $\langle \mathcal{R}_{ii} \rangle_x^+ (y)$, i.e. $\mathcal{R}_{ii}^+ (x,y)$ for $ii=11,22,33$\cmnt{ (note that subscript $ii$ here does not imply summation)}. 
Figure~\ref{fig:Rii-contours} shows the different inner-scaled pressure-rate-of-strain terms over the $y-\phi$ space where $\phi = 2\pi x/\lambda$ and $y=y/\delta$ with $\delta=1$. Once again, the  qualitative structure of $\mathcal{R}_{ii}^+ (x,y)$ for $ii=11,22,33$ remains grossly invariant in this $y-\phi$ space to different $\zeta=2a/\lambda >0$ while the magnitudes depend on wave slope. 
%
In figure~\ref{fig:R11-contours}, we see that $\mathcal{R}_{11}^+ < 0$ over the entire $y-\phi$ space except near the surface in the wave trough ($\phi \approx 3\pi/2 -9\pi/4$ for $\zeta=0.044$) where 
$\mathcal{R}_{11}^+ > 0$ (cyan region near the surface in figure~\ref{fig:R11-contours}) representing net conversion of energy from vertical motions ($v'$) to streamwise motions ($u'$) through splat events. 
Such splat events are well known even in TBL over flat surfaces (see figure~\ref{fig:R11-contours}, $\zeta=0$) with  $\mathcal{R}_{11}^+ > 0$ (cyan region) and $\mathcal{R}_{22}^+ < 0$ (see figure~\ref{fig:R22-contours}, $\zeta=0$) close to the wall. These effects are highly pronounced in buoyant atmospheric boundary layer flows~\cite{jayaraman2018transition,jayaraman2014transition} with significant updrafts and downdrafts. For the wavy surfaces, we observe $\mathcal{R}_{11}^+ > 0$ over flat (at the crest and trough of the wave for $\zeta>0$ and everywhere for $\zeta=0$) and concave ($\phi \approx 3\pi/2 -5\pi/2$) regions (for $\zeta>0$).
At smaller values of $\zeta$, the gradual surface slope allows for widespread splat events along the leeward surface (cyan region in figure~\ref{fig:R11-contours})\cmnt{ over $\phi \approx \pi -2\pi$}. The streamwise extent of this positive $\mathcal{R}_{11}^+$ (cyan layer) region decreases with increase in $\zeta$ from $\phi \approx \pi-9\pi/4$ for $\zeta=0.011$ to $\phi \approx 3\pi/2 -9\pi/4$ for $\zeta=0.044$.  

On the windward side before the crest ($\phi \approx \pi/4 -\pi$ for $\zeta>0$), $\mathcal{R}_{11}^+ < 0$ (and this magnitude increases with $\zeta$) indicating that splat-type events are less likely in this region. This region corresponds to a favorable pressure gradient zone with $\langle u'v'\rangle^+_{z,t} >0$ (figure~\ref{fig:cont_uvcovar_2}) associated with surface-induced $\langle u'^2\rangle^+_{z,t}$ generation. The inner-scaled pressure-rate-of-strain terms here are such that $\mathcal{R}_{11}^+ < 0$, $\mathcal{R}_{22}^+ < 0$ and $\mathcal{R}_{33}^+ > 0$ indicating conversion of $\langle u'^2\rangle^+_{z,t}$ and $\langle v'^2\rangle^+_{z,t}$ (from diffusion) to $\langle w'^2\rangle^+_{z,t}$. Naturally, this is a region of strong spanwise variance generation (see figures~\ref{fig:R11-contours}-\ref{fig:R33-contours}) which is enhanced further by $\zeta$.  
At sufficiently large $\zeta$, the flow behind the crest ($\pi < \phi \lesssim 3\pi/2$) is impacted by the surface curvature through the pressure field in a way which pushes the splat dynamics further downstream (closer to the trough) causing $\mathcal{R}_{11}^+$ to be weakly negative. Instead, in this region the pressure-rate-of-strain terms tend to be small with $\mathcal{R}_{11}^+ \lesssim 0$, $\mathcal{R}_{33}^+ \gtrsim 0$ and $\mathcal{R}_{22}^+ \lesssim 0 \textrm{ or} \gtrsim 0$ depending on $\zeta$. This is not surprising given the very little $\langle v'^2 \rangle^+_{z,t}$ generation (see $\mathcal{P}_{22}^+ \approx 0$ in figure~\ref{fig:P22-contours}) as compared to significant $\langle u'^2 \rangle^+_{z,t}$ generation ($\mathcal{P}_{11}^+ > 0$ in figure~\ref{fig:P11-contours}) over this region. 
%
Thus, we end up with three distinct regions along the wavy surface with different pressure-rate-of-strain dynamics.

\begin{figure}[ht!]
	\centering
	\mbox{
		\subfigure[\label{fig:PressureStrainRateRatios-3-2}]{\includegraphics[width=0.2\textwidth]{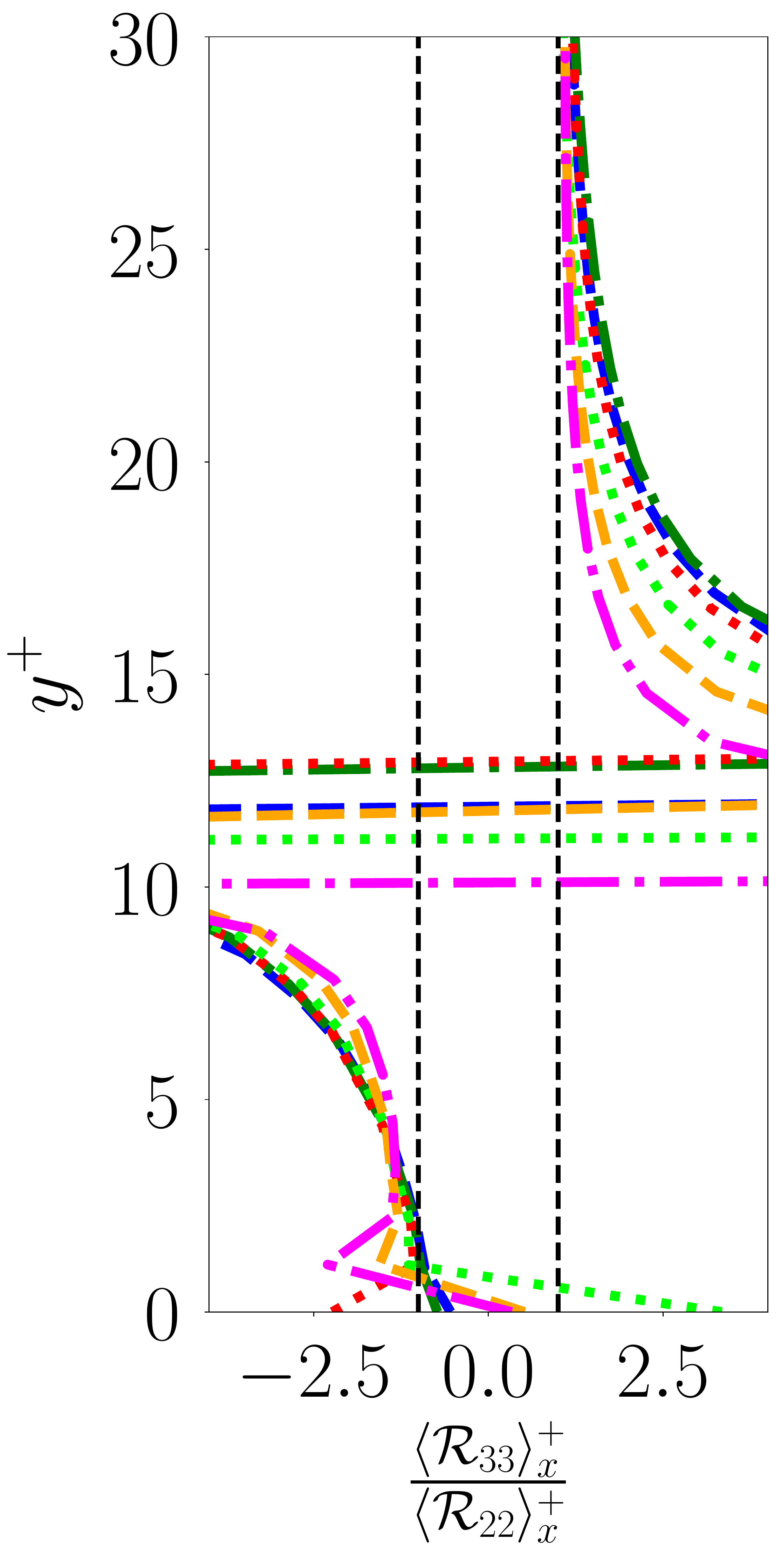}}\hspace{1em}
		\subfigure[\label{fig:PressureStrainRateRatios-3-1}]{\includegraphics[width=0.2\textwidth]{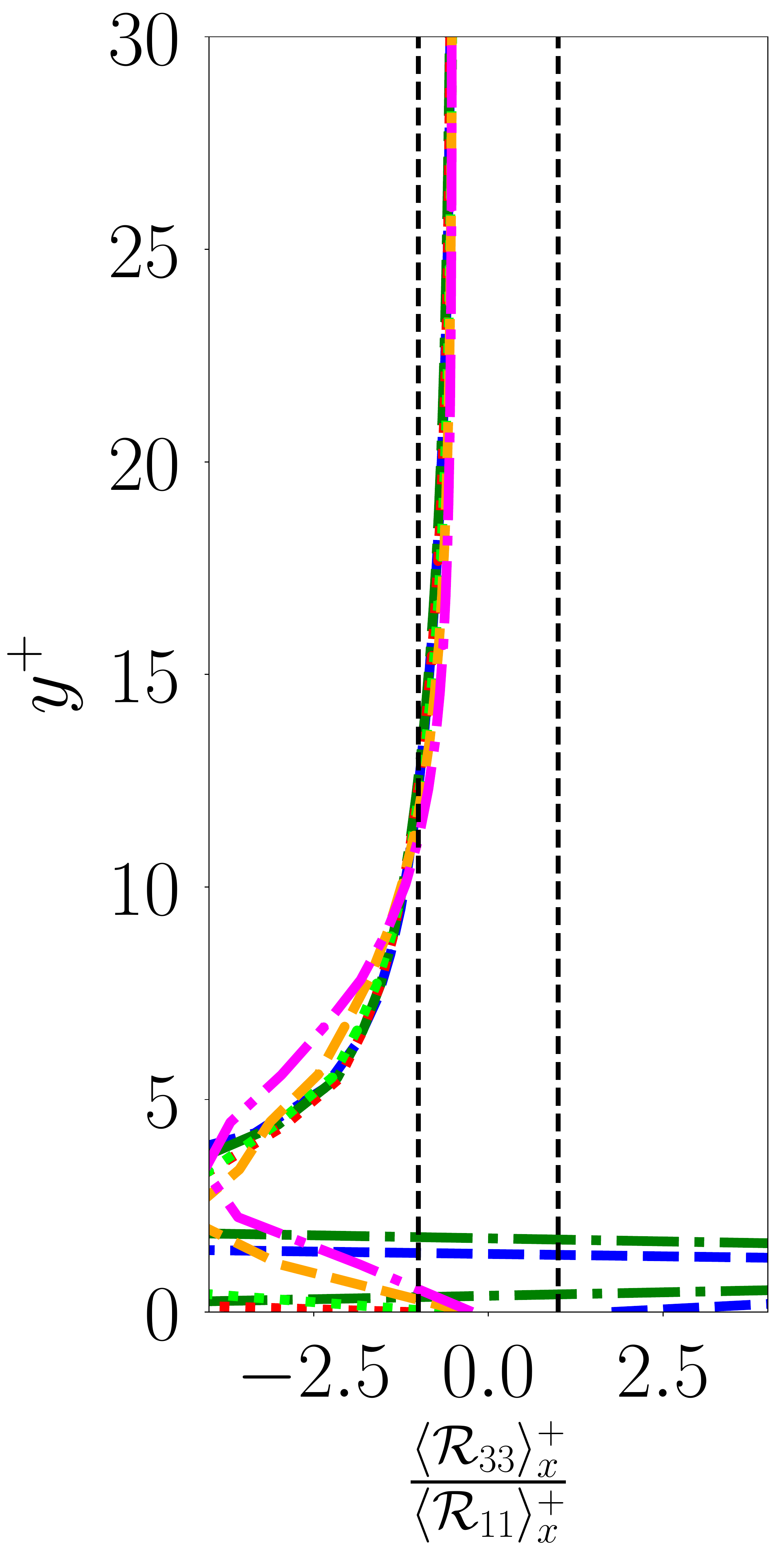}}\hspace{1em}
		\subfigure[\label{fig:PressureStrainRateRatios-2-1}]{\includegraphics[width=0.2\textwidth]{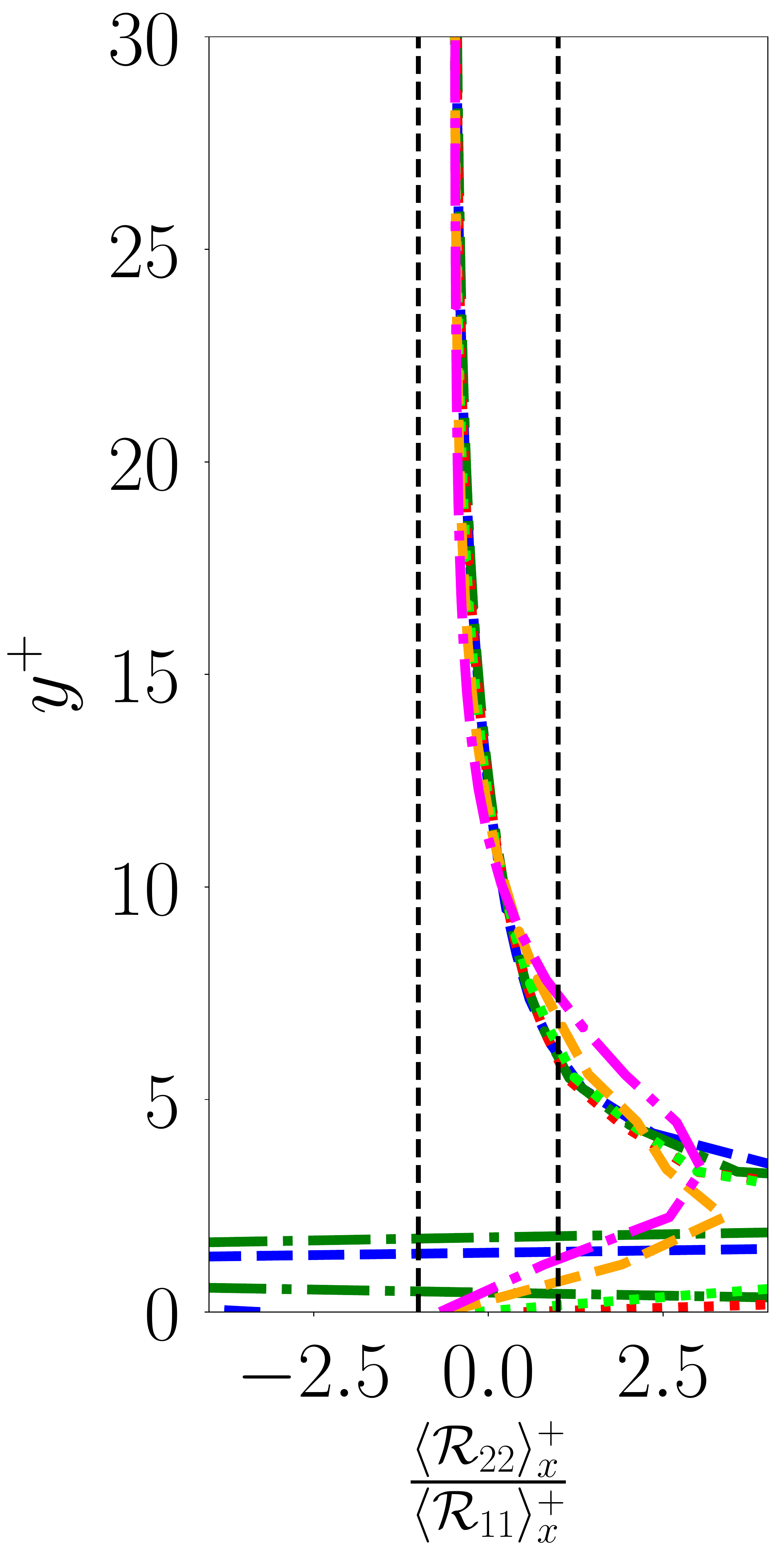}}
	}
	\caption{{Vertical variation of the ratio of different diagonal elements of averaged pressure-rate-of-strain tensor corresponding to streamwise (a) vertical (b) and spanwise (c) variance.}\label{fig:PressureStrainRateRatios}}
\end{figure}

\hlll{Away from the surface, the pressure-rate-of-strain term pushes the flow towards isotropy with the dominant $\langle u'^2 \rangle^+_{z,t}$ increasingly converted to $\langle v'^2 \rangle^+_{z,t}$ and $\langle w'^2 \rangle^+_{z,t}$ such that $\mathcal{R}_{11}^+<0$, $\mathcal{R}_{22}^+>0$ and $\mathcal{R}_{33}^+>0$.
In fact, the regions with large magnitudes of $\mathcal{R}_{11}^+$ and $\mathcal{R}_{22}^+$ (and consequently, $\mathcal{R}_{33}^+$) mostly coincide with those of significant variances, $\langle u'^2 \rangle^+_{z,t}$ (figure~\ref{fig:cont_uvar}) and $\langle v'^2 \rangle^+_{z,t}$ (figure~\ref{fig:cont_vvar}) respectively. Further, as $\zeta$ increases, both the inner-scaled variances, $\langle u'^2 \rangle^+_{z,t}$ (figure~\ref{fig:cont_uvar}) and $\langle u'^2 \rangle^+_{z,t}$ (figure~\ref{fig:cont_vvar}) show smaller magnitudes locally while $\mathcal{R}_{11}^+$, $\mathcal{R}_{22}^+$ and $\mathcal{R}_{33}^+$ increase. This suggests more rapid pressure-rate-of-strain dynamics in spite of the smaller normalized variances at higher $\zeta$.    
} 
\hlll{In fact, this return to isotropy occurs much closer to the mean surface height at higher $\zeta$ as evidenced by the faster approach of the ratio 
$\mathcal{R}_{33}^+/\mathcal{R}_{22}^+$ to unity} (in figure~\ref{fig:PressureStrainRateRatios-3-2}) at higher $\zeta$.
\hlll{These ratios of double-averaged pressure-rate-of-strain terms show that sufficiently far away from the surface, rates of conversion of energy of $u'$ fluctuations to that of $v'$ and $w'$ are equal, i.e. $\mathcal{R}_{33}^+/\mathcal{R}_{22}^+\sim 1$} (figure~\ref{fig:PressureStrainRateRatios-3-2}) and therefore, $\mathcal{R}_{22}^+/\mathcal{R}_{11}^+=\mathcal{R}_{33}^+/\mathcal{R}_{11}^+=0.5$ (figures~\ref{fig:PressureStrainRateRatios-2-1} and \ref{fig:PressureStrainRateRatios-3-1}). 

\vspace{-2em}
\section{Conclusion \label{sec:conclusion}}
We present a DNS-based study of turbulence structure over non-flat surfaces, with emphasis on diagonal components of the Reynolds stress and terms that govern their evolution, especially in the region of the TBL affected by surface heterogeneity. For this reason, the high-fidelity DNS is carried out at smaller turbulent Reynolds numbers ($Re_{\tau}=180$) between two infinitely parallel plates with two-dimensional, streamwise oriented wavy walls that are phase locked between the lower and upper surfaces. We characterize the shape of different wavy surfaces using an effective slope measure denoted by $\zeta=2a/\lambda$. Fixing the roughness Reynolds number and wave amplitude $a$, we vary $\zeta$ over $0-0.044$ corresponding to mildly separated flows.
%
	%
Consistent with literature on rough wall turbulence, the streamwise mean velocity structure indicates a characteristic downward shift of the logarithmic region indicating increased flow drag with wave slope, $\zeta$. This is accompanied by sustained upward vertical flow in the lower roughness sublayer and downward flow in the buffer layer whose magnitude increases with $\zeta$.  All this impacts the near surface turbulence production processes as evidenced from the inner-scaled turbulence production, $\langle \mathcal{P}_{ii} \rangle^+$ which shows buffer layer modulation with increasing wave slope, $\zeta$. In fact, we observe that $\zeta$ changes the proportion of form drag relative to viscous drag for a fixed roughness Reynolds number, $a^+$. 
%

The effect of surface undulations on the TBL is to enhance near-surface mixing and reduce anisotropy of the buffer region at higher $\zeta$\cite{khan2019statistical}. Specifically, the surface undulations reduce the inner-scaled streamwise variance, $\langle u'^2 \rangle^+_{x,z,t} $ while enhancing inner-scaled vertical ($\langle v'^2 \rangle^+_{x,z,t} $ ) and spanwise ($\langle w'^2 \rangle^+_{x,z,t} $) variances in the surface layer of the TBL. Thus, the flow becomes increasingly isotropic closer to the wall at higher steepness. 
For these shallow wavy surfaces, the streamwise variance is predominantly generated from shear-induced mechanisms, i.e. ${\mathcal{P}^{u'v'}_{11}}^+$ driven by the strain rate term, $d\langle u \rangle^+_{z,t}/dy^+$. Nevertheless, the surface-induced contribution, ${\mathcal{P}^{u'u'}_{11}}^+$ driven by $d\langle u \rangle^+_{z,t}/dx^+$ impacts key aspects of the two-dimensional production structure, especially near the windward wavy surface. This secondary term can be related to production from surface-induced dispersive stresses and determines key trends observed in the double-averaged (one-dimensional) statistics, $\langle {\mathcal{P}^{u'v'}_{11}} \rangle_x^+$ and $\langle u'^2 \rangle^+_{x,z,t}$. 
%
Spatially, the streamwise variance ($\langle u'^2 \rangle^+_{z,t}$) generation occurs in the leeward side the wave in the buffer layer in response to the large inner-scaled gradients, $d\langle u \rangle^+_{z,t}/dy^+$ in the wake of wave crest with strong turbulent mixing. Along the windward region of the wavy surface, $\langle u'^2 \rangle^+_{z,t}$ is produced close to the wall due to surface-slope-induced positive covariance, i.e. $\langle u'v' \rangle^+_{z,t} >0$. As the wave slope ($\zeta$) increases, it enhances the strain rate terms $d\langle u \rangle^+_{z,t}/dy^+$ and $d\langle u \rangle^+_{z,t}/dx^+$ and thereby, variance production.      

Unlike the TBL over a flat surface, small amounts of vertical variance is produced from Reynolds stress-mean strain rate interactions that arise purely from surface heterogeneity effects. In this case, the mean strain rate terms responsible for production, $d\langle v \rangle^+_{z,t}/dy^+$ and $d\langle v \rangle^+_{z,t}/dx^+$ are both small with $d\langle v \rangle^+_{z,t}/dy^+ \gg d\langle v \rangle^+_{z,t}/dx^+$ (since $\zeta \ll 1$). Therefore, as observed for $\langle u'^2 \rangle^+_{z,t}$, the vertical mean velocity gradient, $d\langle v \rangle^+_{z,t}/dy^+$ determines the qualitative structure of the production term, ${\mathcal{P}_{22}}^+$. Spatially, $\langle v'^2 \rangle^+_{z,t}$ is produced along both the leeward (near the surface) and windward (near and away from the surface) sides of the wave with the latter being more dominant. Despite this production, the primary source of $\langle v'^2 \rangle^+_{z,t}$ is through the pressure-rate-of-strain mechanism which generates vertical fluctuations in the upper buffer layer. Near the surface, the same mechanism converts vertical fluctuations into streamwise and spanwise fluctuations on account of wall blockage. As the wave slope increases, both the double-averaged production, $\langle {\mathcal{P}_{22}} \rangle_x^+$  and pressure-rate-of-strain term, $\langle {\mathcal{R}_{22}} \rangle_x^+$ increase in magnitudes closer to the surface resulting in faster growth of $\langle v'^2 \rangle^+_{x,z,t}$ (through the lower buffer layer) resulting in a downward shift in peak $\langle v'^2 \rangle^+_{x,z,t}$\cmnt{ which represents the cumulative effect of the wave slope-dependent buffer layer dynamics including enhanced production and conversion from streamwise velocity fluctuations}.
As spanwise fluctuations are not as severely blocked by the wall as vertical fluctuations, $\langle w'^2 \rangle^+_{x,z,t}$ grows faster (relative to vertical variance) with $y^+$ and thereby causing it to peak in the lower buffer layer. Due to spanwise homogeneity, the production of $\langle w'^2 \rangle^+_{z,t}$, $ {\mathcal{P}_{33}}^+$ from Reynolds stress-strain rate interactions is zero. Therefore, generation of $\langle w'^2 \rangle^+_{z,t}$ occurs through the pressure-rate-of-strain term, ${\mathcal{R}_{33}}^+$ that converts streamwise and vertical fluctuations into spanwise turbulence, especially along the windward side of the wavy surface. With increase in wave slope, this conversion process is enhanced, especially within the viscous and buffer layers. The 2D structure of the pressure-rate-of-strain terms near the surface show three very distinct phase ($\phi$)-dependent conversion mechanisms along the wave. These include splat-type phenomena along the leeward side and surface-induced generation of spanwise fluctuations over the windward side.  Away from the surface, the well known return-to-isotropy mechanism is observed in regions of strong vertical and streamwise variance. 
We also explore different metrics for quantifying the vertical extent of the surface (or roughness) layer, i.e. height beyond which surface effects are not observed. We look at surface induced variance double-averaged production ($\langle \mathcal{P}^{u'u'}_{11} \rangle^+_{x}$, $\langle \mathcal{P}_{22} \rangle^+_{x}$) as well as dispersion in this production structure (such as $\langle \mathcal{P}^{u'u'}_{11} \rangle^+_{x}-\langle \mathcal{P}^{u'u'}_{11} \rangle^+_{*}$) to make these characterizations. The roughness layer height varies between $\sim 3a^+-5a^+$ which is larger than earlier reported~\cite{florens2013defining} values of $\sim 2a^+$ using other metrics. More research is needed to assess such variability in estimates and Reynolds number sensitivity of these conclusions.    
\vspace{-1em}
\section{References}
\vspace{-1em}
\bibliography{pof_production}

\begin{thebibliography}{55}%
\makeatletter
\providecommand \@ifxundefined [1]{%
 \@ifx{#1\undefined}
}%
\providecommand \@ifnum [1]{%
 \ifnum #1\expandafter \@firstoftwo
 \else \expandafter \@secondoftwo
 \fi
}%
\providecommand \@ifx [1]{%
 \ifx #1\expandafter \@firstoftwo
 \else \expandafter \@secondoftwo
 \fi
}%
\providecommand \natexlab [1]{#1}%
\providecommand \enquote  [1]{``#1''}%
\providecommand \bibnamefont  [1]{#1}%
\providecommand \bibfnamefont [1]{#1}%
\providecommand \citenamefont [1]{#1}%
\providecommand \href@noop [0]{\@secondoftwo}%
\providecommand \href [0]{\begingroup \@sanitize@url \@href}%
\providecommand \@href[1]{\@@startlink{#1}\@@href}%
\providecommand \@@href[1]{\endgroup#1\@@endlink}%
\providecommand \@sanitize@url [0]{\catcode `\\12\catcode `\$12\catcode
  `\&12\catcode `\#12\catcode `\^12\catcode `\_12\catcode `\%12\relax}%
\providecommand \@@startlink[1]{}%
\providecommand \@@endlink[0]{}%
\providecommand \url  [0]{\begingroup\@sanitize@url \@url }%
\providecommand \@url [1]{\endgroup\@href {#1}{\urlprefix }}%
\providecommand \urlprefix  [0]{URL }%
\providecommand \Eprint [0]{\href }%
\providecommand \doibase [0]{https://doi.org/}%
\providecommand \selectlanguage [0]{\@gobble}%
\providecommand \bibinfo  [0]{\@secondoftwo}%
\providecommand \bibfield  [0]{\@secondoftwo}%
\providecommand \translation [1]{[#1]}%
\providecommand \BibitemOpen [0]{}%
\providecommand \bibitemStop [0]{}%
\providecommand \bibitemNoStop [0]{.\EOS\space}%
\providecommand \EOS [0]{\spacefactor3000\relax}%
\providecommand \BibitemShut  [1]{\csname bibitem#1\endcsname}%
\let\auto@bib@innerbib\@empty
\bibitem [{\citenamefont {Schultz}(2007)}]{schultz2007effects}%
  \BibitemOpen
  \bibfield  {author} {\bibinfo {author} {\bibfnamefont {M.~P.}\ \bibnamefont
  {Schultz}},\ }\bibfield  {title} {\enquote {\bibinfo {title} {Effects of
  coating roughness and biofouling on ship resistance and powering},}\
  }\href@noop {} {\bibfield  {journal} {\bibinfo  {journal} {Biofouling}\
  }\textbf {\bibinfo {volume} {23}},\ \bibinfo {pages} {331--341} (\bibinfo
  {year} {2007})}\BibitemShut {NoStop}%
\bibitem [{\citenamefont {Darcy}(1857)}]{darcy1857recherches}%
  \BibitemOpen
  \bibfield  {author} {\bibinfo {author} {\bibfnamefont {H.}~\bibnamefont
  {Darcy}},\ }\href@noop {} {\emph {\bibinfo {title} {Recherches
  exp{\'e}rimentales relatives au mouvement de l'eau dans les tuyaux}}}\
  (\bibinfo  {publisher} {Mallet-Bachelier},\ \bibinfo {year}
  {1857})\BibitemShut {NoStop}%
\bibitem [{\citenamefont {Nikuradse}(1950)}]{nikuradse1950laws}%
  \BibitemOpen
  \bibfield  {author} {\bibinfo {author} {\bibfnamefont {J.}~\bibnamefont
  {Nikuradse}},\ }\href@noop {} {\emph {\bibinfo {title} {Laws of flow in rough
  pipes}}}\ (\bibinfo  {publisher} {National Advisory Committee for Aeronautics
  Washington, DC},\ \bibinfo {year} {1950})\BibitemShut {NoStop}%
\bibitem [{\citenamefont {Colebrook}\ \emph {et~al.}(1939)\citenamefont
  {Colebrook}, \citenamefont {Blench}, \citenamefont {Chatley}, \citenamefont
  {Essex}, \citenamefont {Finniecome}, \citenamefont {Lacey}, \citenamefont
  {Williamson},\ and\ \citenamefont {Macdonald}}]{colebrook1939correspondence}%
  \BibitemOpen
  \bibfield  {author} {\bibinfo {author} {\bibfnamefont {C.~F.}\ \bibnamefont
  {Colebrook}}, \bibinfo {author} {\bibfnamefont {T.}~\bibnamefont {Blench}},
  \bibinfo {author} {\bibfnamefont {H.}~\bibnamefont {Chatley}}, \bibinfo
  {author} {\bibfnamefont {E.}~\bibnamefont {Essex}}, \bibinfo {author}
  {\bibfnamefont {J.}~\bibnamefont {Finniecome}}, \bibinfo {author}
  {\bibfnamefont {G.}~\bibnamefont {Lacey}}, \bibinfo {author} {\bibfnamefont
  {J.}~\bibnamefont {Williamson}},\ and\ \bibinfo {author} {\bibfnamefont
  {G.}~\bibnamefont {Macdonald}},\ }\bibfield  {title} {\enquote {\bibinfo
  {title} {Correspondence. turbulent flow in pipes, with particular reference
  to the transition region between the smooth and rough pipe laws.(includes
  plates).}}\ }\href@noop {} {\bibfield  {journal} {\bibinfo  {journal}
  {Journal of the Institution of Civil engineers}\ }\textbf {\bibinfo {volume}
  {12}},\ \bibinfo {pages} {393--422} (\bibinfo {year} {1939})}\BibitemShut
  {NoStop}%
\bibitem [{\citenamefont {Moody}(1944)}]{moody1944friction}%
  \BibitemOpen
  \bibfield  {author} {\bibinfo {author} {\bibfnamefont {L.~F.}\ \bibnamefont
  {Moody}},\ }\bibfield  {title} {\enquote {\bibinfo {title} {Friction factors
  for pipe flow},}\ }\href@noop {} {\bibfield  {journal} {\bibinfo  {journal}
  {Trans. ASME}\ }\textbf {\bibinfo {volume} {66}},\ \bibinfo {pages}
  {671--684} (\bibinfo {year} {1944})}\BibitemShut {NoStop}%
\bibitem [{\citenamefont {Shockling}, \citenamefont {Allen},\ and\
  \citenamefont {Smits}(2006)}]{shockling2006roughness}%
  \BibitemOpen
  \bibfield  {author} {\bibinfo {author} {\bibfnamefont {M.}~\bibnamefont
  {Shockling}}, \bibinfo {author} {\bibfnamefont {J.}~\bibnamefont {Allen}},\
  and\ \bibinfo {author} {\bibfnamefont {A.}~\bibnamefont {Smits}},\ }\bibfield
   {title} {\enquote {\bibinfo {title} {Roughness effects in turbulent pipe
  flow},}\ }\href@noop {} {\bibfield  {journal} {\bibinfo  {journal} {Journal
  of Fluid Mechanics}\ }\textbf {\bibinfo {volume} {564}},\ \bibinfo {pages}
  {267--285} (\bibinfo {year} {2006})}\BibitemShut {NoStop}%
\bibitem [{\citenamefont {Hultmark}\ \emph {et~al.}(2013)\citenamefont
  {Hultmark}, \citenamefont {Vallikivi}, \citenamefont {Bailey},\ and\
  \citenamefont {Smits}}]{hultmark2013logarithmic}%
  \BibitemOpen
  \bibfield  {author} {\bibinfo {author} {\bibfnamefont {M.}~\bibnamefont
  {Hultmark}}, \bibinfo {author} {\bibfnamefont {M.}~\bibnamefont {Vallikivi}},
  \bibinfo {author} {\bibfnamefont {S.}~\bibnamefont {Bailey}},\ and\ \bibinfo
  {author} {\bibfnamefont {A.}~\bibnamefont {Smits}},\ }\bibfield  {title}
  {\enquote {\bibinfo {title} {Logarithmic scaling of turbulence in smooth-and
  rough-wall pipe flow},}\ }\href@noop {} {\bibfield  {journal} {\bibinfo
  {journal} {Journal of Fluid Mechanics}\ }\textbf {\bibinfo {volume} {728}},\
  \bibinfo {pages} {376--395} (\bibinfo {year} {2013})}\BibitemShut {NoStop}%
\bibitem [{\citenamefont {Chan}\ \emph {et~al.}(2015)\citenamefont {Chan},
  \citenamefont {MacDonald}, \citenamefont {Chung}, \citenamefont {Hutchins},\
  and\ \citenamefont {Ooi}}]{chan2015systematic}%
  \BibitemOpen
  \bibfield  {author} {\bibinfo {author} {\bibfnamefont {L.}~\bibnamefont
  {Chan}}, \bibinfo {author} {\bibfnamefont {M.}~\bibnamefont {MacDonald}},
  \bibinfo {author} {\bibfnamefont {D.}~\bibnamefont {Chung}}, \bibinfo
  {author} {\bibfnamefont {N.}~\bibnamefont {Hutchins}},\ and\ \bibinfo
  {author} {\bibfnamefont {A.}~\bibnamefont {Ooi}},\ }\bibfield  {title}
  {\enquote {\bibinfo {title} {A systematic investigation of roughness height
  and wavelength in turbulent pipe flow in the transitionally rough regime},}\
  }\href@noop {} {\bibfield  {journal} {\bibinfo  {journal} {Journal of Fluid
  Mechanics}\ }\textbf {\bibinfo {volume} {771}},\ \bibinfo {pages} {743--777}
  (\bibinfo {year} {2015})}\BibitemShut {NoStop}%
\bibitem [{\citenamefont {Coleman}\ \emph {et~al.}(2007)\citenamefont
  {Coleman}, \citenamefont {Nikora}, \citenamefont {McLean},\ and\
  \citenamefont {Schlicke}}]{coleman2007spatially}%
  \BibitemOpen
  \bibfield  {author} {\bibinfo {author} {\bibfnamefont {S.}~\bibnamefont
  {Coleman}}, \bibinfo {author} {\bibfnamefont {V.~I.}\ \bibnamefont {Nikora}},
  \bibinfo {author} {\bibfnamefont {S.}~\bibnamefont {McLean}},\ and\ \bibinfo
  {author} {\bibfnamefont {E.}~\bibnamefont {Schlicke}},\ }\bibfield  {title}
  {\enquote {\bibinfo {title} {Spatially averaged turbulent flow over square
  ribs},}\ }\href@noop {} {\bibfield  {journal} {\bibinfo  {journal} {Journal
  of Engineering Mechanics}\ }\textbf {\bibinfo {volume} {133}},\ \bibinfo
  {pages} {194--204} (\bibinfo {year} {2007})}\BibitemShut {NoStop}%
\bibitem [{\citenamefont {Flack}, \citenamefont {Schultz},\ and\ \citenamefont
  {Shapiro}(2005)}]{flack2005experimental}%
  \BibitemOpen
  \bibfield  {author} {\bibinfo {author} {\bibfnamefont {K.~A.}\ \bibnamefont
  {Flack}}, \bibinfo {author} {\bibfnamefont {M.~P.}\ \bibnamefont {Schultz}},\
  and\ \bibinfo {author} {\bibfnamefont {T.~A.}\ \bibnamefont {Shapiro}},\
  }\bibfield  {title} {\enquote {\bibinfo {title} {Experimental support for
  townsend’s reynolds number similarity hypothesis on rough walls},}\
  }\href@noop {} {\bibfield  {journal} {\bibinfo  {journal} {Physics of
  Fluids}\ }\textbf {\bibinfo {volume} {17}},\ \bibinfo {pages} {035102}
  (\bibinfo {year} {2005})}\BibitemShut {NoStop}%
\bibitem [{\citenamefont {Schultz}\ and\ \citenamefont
  {Flack}(2005)}]{schultz2005outer}%
  \BibitemOpen
  \bibfield  {author} {\bibinfo {author} {\bibfnamefont {M.}~\bibnamefont
  {Schultz}}\ and\ \bibinfo {author} {\bibfnamefont {K.}~\bibnamefont
  {Flack}},\ }\bibfield  {title} {\enquote {\bibinfo {title} {Outer layer
  similarity in fully rough turbulent boundary layers},}\ }\href@noop {}
  {\bibfield  {journal} {\bibinfo  {journal} {Experiments in Fluids}\ }\textbf
  {\bibinfo {volume} {38}},\ \bibinfo {pages} {328--340} (\bibinfo {year}
  {2005})}\BibitemShut {NoStop}%
\bibitem [{\citenamefont {Schultz}\ and\ \citenamefont
  {Flack}(2007)}]{schultz2007rough}%
  \BibitemOpen
  \bibfield  {author} {\bibinfo {author} {\bibfnamefont {M.}~\bibnamefont
  {Schultz}}\ and\ \bibinfo {author} {\bibfnamefont {K.}~\bibnamefont
  {Flack}},\ }\bibfield  {title} {\enquote {\bibinfo {title} {The rough-wall
  turbulent boundary layer from the hydraulically smooth to the fully rough
  regime},}\ }\href@noop {} {\bibfield  {journal} {\bibinfo  {journal} {Journal
  of Fluid Mechanics}\ }\textbf {\bibinfo {volume} {580}},\ \bibinfo {pages}
  {381--405} (\bibinfo {year} {2007})}\BibitemShut {NoStop}%
\bibitem [{\citenamefont {Schultz}\ and\ \citenamefont
  {Flack}(2009)}]{schultz2009turbulent}%
  \BibitemOpen
  \bibfield  {author} {\bibinfo {author} {\bibfnamefont {M.~P.}\ \bibnamefont
  {Schultz}}\ and\ \bibinfo {author} {\bibfnamefont {K.~A.}\ \bibnamefont
  {Flack}},\ }\bibfield  {title} {\enquote {\bibinfo {title} {Turbulent
  boundary layers on a systematically varied rough wall},}\ }\href@noop {}
  {\bibfield  {journal} {\bibinfo  {journal} {Physics of Fluids}\ }\textbf
  {\bibinfo {volume} {21}},\ \bibinfo {pages} {015104} (\bibinfo {year}
  {2009})}\BibitemShut {NoStop}%
\bibitem [{\citenamefont {Flack}\ and\ \citenamefont
  {Schultz}(2014)}]{flack2014roughness}%
  \BibitemOpen
  \bibfield  {author} {\bibinfo {author} {\bibfnamefont {K.~A.}\ \bibnamefont
  {Flack}}\ and\ \bibinfo {author} {\bibfnamefont {M.~P.}\ \bibnamefont
  {Schultz}},\ }\bibfield  {title} {\enquote {\bibinfo {title} {Roughness
  effects on wall-bounded turbulent flows},}\ }\href@noop {} {\bibfield
  {journal} {\bibinfo  {journal} {Physics of Fluids}\ }\textbf {\bibinfo
  {volume} {26}},\ \bibinfo {pages} {101305} (\bibinfo {year}
  {2014})}\BibitemShut {NoStop}%
\bibitem [{\citenamefont {Flack}, \citenamefont {Schultz},\ and\ \citenamefont
  {Connelly}(2007)}]{flack2007examination}%
  \BibitemOpen
  \bibfield  {author} {\bibinfo {author} {\bibfnamefont {K.}~\bibnamefont
  {Flack}}, \bibinfo {author} {\bibfnamefont {M.}~\bibnamefont {Schultz}},\
  and\ \bibinfo {author} {\bibfnamefont {J.}~\bibnamefont {Connelly}},\
  }\bibfield  {title} {\enquote {\bibinfo {title} {Examination of a critical
  roughness height for outer layer similarity},}\ }\href@noop {} {\bibfield
  {journal} {\bibinfo  {journal} {Physics of Fluids}\ }\textbf {\bibinfo
  {volume} {19}},\ \bibinfo {pages} {095104} (\bibinfo {year}
  {2007})}\BibitemShut {NoStop}%
\bibitem [{\citenamefont {Flack}\ and\ \citenamefont
  {Schultz}(2010)}]{flack2010review}%
  \BibitemOpen
  \bibfield  {author} {\bibinfo {author} {\bibfnamefont {K.~A.}\ \bibnamefont
  {Flack}}\ and\ \bibinfo {author} {\bibfnamefont {M.~P.}\ \bibnamefont
  {Schultz}},\ }\bibfield  {title} {\enquote {\bibinfo {title} {Review of
  hydraulic roughness scales in the fully rough regime},}\ }\href@noop {}
  {\bibfield  {journal} {\bibinfo  {journal} {Journal of Fluids Engineering}\
  }\textbf {\bibinfo {volume} {132}},\ \bibinfo {pages} {041203} (\bibinfo
  {year} {2010})}\BibitemShut {NoStop}%
\bibitem [{\citenamefont {Jim{\'e}nez}(2004)}]{jimenez2004turbulent}%
  \BibitemOpen
  \bibfield  {author} {\bibinfo {author} {\bibfnamefont {J.}~\bibnamefont
  {Jim{\'e}nez}},\ }\bibfield  {title} {\enquote {\bibinfo {title} {Turbulent
  flows over rough walls},}\ }\href@noop {} {\bibfield  {journal} {\bibinfo
  {journal} {Annu. Rev. Fluid Mech.}\ }\textbf {\bibinfo {volume} {36}},\
  \bibinfo {pages} {173--196} (\bibinfo {year} {2004})}\BibitemShut {NoStop}%
\bibitem [{\citenamefont {De~Angelis}, \citenamefont {Lombardi},\ and\
  \citenamefont {Banerjee}(1997)}]{de1997direct}%
  \BibitemOpen
  \bibfield  {author} {\bibinfo {author} {\bibfnamefont {V.}~\bibnamefont
  {De~Angelis}}, \bibinfo {author} {\bibfnamefont {P.}~\bibnamefont
  {Lombardi}},\ and\ \bibinfo {author} {\bibfnamefont {S.}~\bibnamefont
  {Banerjee}},\ }\bibfield  {title} {\enquote {\bibinfo {title} {Direct
  numerical simulation of turbulent flow over a wavy wall},}\ }\href@noop {}
  {\bibfield  {journal} {\bibinfo  {journal} {Physics of Fluids}\ }\textbf
  {\bibinfo {volume} {9}},\ \bibinfo {pages} {2429--2442} (\bibinfo {year}
  {1997})}\BibitemShut {NoStop}%
\bibitem [{\citenamefont {Cherukat}\ \emph {et~al.}(1998)\citenamefont
  {Cherukat}, \citenamefont {Na}, \citenamefont {Hanratty},\ and\ \citenamefont
  {McLaughlin}}]{cherukat1998direct}%
  \BibitemOpen
  \bibfield  {author} {\bibinfo {author} {\bibfnamefont {P.}~\bibnamefont
  {Cherukat}}, \bibinfo {author} {\bibfnamefont {Y.}~\bibnamefont {Na}},
  \bibinfo {author} {\bibfnamefont {T.}~\bibnamefont {Hanratty}},\ and\
  \bibinfo {author} {\bibfnamefont {J.}~\bibnamefont {McLaughlin}},\ }\bibfield
   {title} {\enquote {\bibinfo {title} {Direct numerical simulation of a fully
  developed turbulent flow over a wavy wall},}\ }\href@noop {} {\bibfield
  {journal} {\bibinfo  {journal} {Theoretical and computational fluid
  dynamics}\ }\textbf {\bibinfo {volume} {11}},\ \bibinfo {pages} {109--134}
  (\bibinfo {year} {1998})}\BibitemShut {NoStop}%
\bibitem [{\citenamefont {Bhaganagar}, \citenamefont {Kim},\ and\ \citenamefont
  {Coleman}(2004)}]{bhaganagar2004effect}%
  \BibitemOpen
  \bibfield  {author} {\bibinfo {author} {\bibfnamefont {K.}~\bibnamefont
  {Bhaganagar}}, \bibinfo {author} {\bibfnamefont {J.}~\bibnamefont {Kim}},\
  and\ \bibinfo {author} {\bibfnamefont {G.}~\bibnamefont {Coleman}},\
  }\bibfield  {title} {\enquote {\bibinfo {title} {Effect of roughness on
  wall-bounded turbulence},}\ }\href@noop {} {\bibfield  {journal} {\bibinfo
  {journal} {Flow, turbulence and combustion}\ }\textbf {\bibinfo {volume}
  {72}},\ \bibinfo {pages} {463--492} (\bibinfo {year} {2004})}\BibitemShut
  {NoStop}%
\bibitem [{\citenamefont {Chau}\ and\ \citenamefont
  {Bhaganagar}(2012)}]{chau2012understanding}%
  \BibitemOpen
  \bibfield  {author} {\bibinfo {author} {\bibfnamefont {L.}~\bibnamefont
  {Chau}}\ and\ \bibinfo {author} {\bibfnamefont {K.}~\bibnamefont
  {Bhaganagar}},\ }\bibfield  {title} {\enquote {\bibinfo {title}
  {Understanding turbulent flow over ripple-shaped random roughness in a
  channel},}\ }\href@noop {} {\bibfield  {journal} {\bibinfo  {journal}
  {Physics of Fluids}\ }\textbf {\bibinfo {volume} {24}},\ \bibinfo {pages}
  {115102} (\bibinfo {year} {2012})}\BibitemShut {NoStop}%
\bibitem [{\citenamefont {Napoli}, \citenamefont {Armenio},\ and\ \citenamefont
  {De~Marchis}(2008)}]{napoli2008effect}%
  \BibitemOpen
  \bibfield  {author} {\bibinfo {author} {\bibfnamefont {E.}~\bibnamefont
  {Napoli}}, \bibinfo {author} {\bibfnamefont {V.}~\bibnamefont {Armenio}},\
  and\ \bibinfo {author} {\bibfnamefont {M.}~\bibnamefont {De~Marchis}},\
  }\bibfield  {title} {\enquote {\bibinfo {title} {The effect of the slope of
  irregularly distributed roughness elements on turbulent wall-bounded
  flows},}\ }\href@noop {} {\bibfield  {journal} {\bibinfo  {journal} {Journal
  of Fluid Mechanics}\ }\textbf {\bibinfo {volume} {613}},\ \bibinfo {pages}
  {385--394} (\bibinfo {year} {2008})}\BibitemShut {NoStop}%
\bibitem [{\citenamefont {Leonardi}, \citenamefont {Orlandi},\ and\
  \citenamefont {Antonia}(2007)}]{leonardi2007properties}%
  \BibitemOpen
  \bibfield  {author} {\bibinfo {author} {\bibfnamefont {S.}~\bibnamefont
  {Leonardi}}, \bibinfo {author} {\bibfnamefont {P.}~\bibnamefont {Orlandi}},\
  and\ \bibinfo {author} {\bibfnamefont {R.~A.}\ \bibnamefont {Antonia}},\
  }\bibfield  {title} {\enquote {\bibinfo {title} {Properties of d-and k-type
  roughness in a turbulent channel flow},}\ }\href@noop {} {\bibfield
  {journal} {\bibinfo  {journal} {Physics of fluids}\ }\textbf {\bibinfo
  {volume} {19}},\ \bibinfo {pages} {125101} (\bibinfo {year}
  {2007})}\BibitemShut {NoStop}%
\bibitem [{\citenamefont {Leonardi}\ and\ \citenamefont
  {Castro}(2010)}]{leonardi2010channel}%
  \BibitemOpen
  \bibfield  {author} {\bibinfo {author} {\bibfnamefont {S.}~\bibnamefont
  {Leonardi}}\ and\ \bibinfo {author} {\bibfnamefont {I.~P.}\ \bibnamefont
  {Castro}},\ }\bibfield  {title} {\enquote {\bibinfo {title} {Channel flow
  over large cube roughness: a direct numerical simulation study},}\
  }\href@noop {} {\bibfield  {journal} {\bibinfo  {journal} {Journal of Fluid
  Mechanics}\ }\textbf {\bibinfo {volume} {651}},\ \bibinfo {pages} {519--539}
  (\bibinfo {year} {2010})}\BibitemShut {NoStop}%
\bibitem [{\citenamefont {De~Marchis}\ and\ \citenamefont
  {Napoli}(2012)}]{de2012effects}%
  \BibitemOpen
  \bibfield  {author} {\bibinfo {author} {\bibfnamefont {M.}~\bibnamefont
  {De~Marchis}}\ and\ \bibinfo {author} {\bibfnamefont {E.}~\bibnamefont
  {Napoli}},\ }\bibfield  {title} {\enquote {\bibinfo {title} {Effects of
  irregular two-dimensional and three-dimensional surface roughness in
  turbulent channel flows},}\ }\href@noop {} {\bibfield  {journal} {\bibinfo
  {journal} {International Journal of Heat and Fluid Flow}\ }\textbf {\bibinfo
  {volume} {36}},\ \bibinfo {pages} {7--17} (\bibinfo {year}
  {2012})}\BibitemShut {NoStop}%
\bibitem [{\citenamefont {Thakkar}, \citenamefont {Busse},\ and\ \citenamefont
  {Sandham}(2018)}]{thakkar2018direct}%
  \BibitemOpen
  \bibfield  {author} {\bibinfo {author} {\bibfnamefont {M.}~\bibnamefont
  {Thakkar}}, \bibinfo {author} {\bibfnamefont {A.}~\bibnamefont {Busse}},\
  and\ \bibinfo {author} {\bibfnamefont {N.}~\bibnamefont {Sandham}},\
  }\bibfield  {title} {\enquote {\bibinfo {title} {Direct numerical simulation
  of turbulent channel flow over a surrogate for nikuradse-type roughness},}\
  }\href@noop {} {\bibfield  {journal} {\bibinfo  {journal} {Journal of Fluid
  Mechanics}\ }\textbf {\bibinfo {volume} {837}} (\bibinfo {year}
  {2018})}\BibitemShut {NoStop}%
\bibitem [{\citenamefont {Busse}, \citenamefont {Thakkar},\ and\ \citenamefont
  {Sandham}(2017)}]{busse2017reynolds}%
  \BibitemOpen
  \bibfield  {author} {\bibinfo {author} {\bibfnamefont {A.}~\bibnamefont
  {Busse}}, \bibinfo {author} {\bibfnamefont {M.}~\bibnamefont {Thakkar}},\
  and\ \bibinfo {author} {\bibfnamefont {N.}~\bibnamefont {Sandham}},\
  }\bibfield  {title} {\enquote {\bibinfo {title} {Reynolds-number dependence
  of the near-wall flow over irregular rough surfaces},}\ }\href@noop {}
  {\bibfield  {journal} {\bibinfo  {journal} {Journal of Fluid Mechanics}\
  }\textbf {\bibinfo {volume} {810}},\ \bibinfo {pages} {196--224} (\bibinfo
  {year} {2017})}\BibitemShut {NoStop}%
\bibitem [{\citenamefont {Jayaraman}\ and\ \citenamefont
  {Brasseur}(2014)}]{jayaraman2014transition}%
  \BibitemOpen
  \bibfield  {author} {\bibinfo {author} {\bibfnamefont {B.}~\bibnamefont
  {Jayaraman}}\ and\ \bibinfo {author} {\bibfnamefont {J.}~\bibnamefont
  {Brasseur}},\ }\bibfield  {title} {\enquote {\bibinfo {title} {Transition in
  atmospheric turbulence structure from neutral to convective stability
  states},}\ }in\ \href@noop {} {\emph {\bibinfo {booktitle} {32nd ASME Wind
  Energy Symposium}}}\ (\bibinfo {year} {2014})\ p.\ \bibinfo {pages}
  {0868}\BibitemShut {NoStop}%
\bibitem [{\citenamefont {Jayaraman}\ and\ \citenamefont
  {Brasseur}(2018)}]{jayaraman2018transition}%
  \BibitemOpen
  \bibfield  {author} {\bibinfo {author} {\bibfnamefont {B.}~\bibnamefont
  {Jayaraman}}\ and\ \bibinfo {author} {\bibfnamefont {J.~G.}\ \bibnamefont
  {Brasseur}},\ }\bibfield  {title} {\enquote {\bibinfo {title} {Transition in
  atmospheric boundary layer turbulence structure from neutral to moderately
  convective stability states and implications to large-scale rolls},}\
  }\href@noop {} {\bibfield  {journal} {\bibinfo  {journal} {arXiv preprint
  arXiv:1807.03336}\ } (\bibinfo {year} {2018})}\BibitemShut {NoStop}%
\bibitem [{\citenamefont {Khan}\ and\ \citenamefont
  {Jayaraman}(2019)}]{khan2019statistical}%
  \BibitemOpen
  \bibfield  {author} {\bibinfo {author} {\bibfnamefont {S.}~\bibnamefont
  {Khan}}\ and\ \bibinfo {author} {\bibfnamefont {B.}~\bibnamefont
  {Jayaraman}},\ }\bibfield  {title} {\enquote {\bibinfo {title} {Statistical
  structure and deviations from equilibrium in wavy channel turbulence},}\
  }\href@noop {} {\  (\bibinfo {year} {2019})}\BibitemShut {NoStop}%
\bibitem [{\citenamefont {Coceal}\ \emph {et~al.}(2006)\citenamefont {Coceal},
  \citenamefont {Thomas}, \citenamefont {Castro},\ and\ \citenamefont
  {Belcher}}]{coceal2006mean}%
  \BibitemOpen
  \bibfield  {author} {\bibinfo {author} {\bibfnamefont {O.}~\bibnamefont
  {Coceal}}, \bibinfo {author} {\bibfnamefont {T.}~\bibnamefont {Thomas}},
  \bibinfo {author} {\bibfnamefont {I.}~\bibnamefont {Castro}},\ and\ \bibinfo
  {author} {\bibfnamefont {S.}~\bibnamefont {Belcher}},\ }\bibfield  {title}
  {\enquote {\bibinfo {title} {Mean flow and turbulence statistics over groups
  of urban-like cubical obstacles},}\ }\href@noop {} {\bibfield  {journal}
  {\bibinfo  {journal} {Boundary-Layer Meteorology}\ }\textbf {\bibinfo
  {volume} {121}},\ \bibinfo {pages} {491--519} (\bibinfo {year}
  {2006})}\BibitemShut {NoStop}%
\bibitem [{\citenamefont {Hama}(1954)}]{hama1954boundary}%
  \BibitemOpen
  \bibfield  {author} {\bibinfo {author} {\bibfnamefont {F.~R.}\ \bibnamefont
  {Hama}},\ }\bibfield  {title} {\enquote {\bibinfo {title} {Boundary layer
  characteristics for smooth and rough surfaces},}\ }\href@noop {} {\bibfield
  {journal} {\bibinfo  {journal} {Trans. Soc. Nav. Arch. Marine Engrs.}\
  }\textbf {\bibinfo {volume} {62}},\ \bibinfo {pages} {333--358} (\bibinfo
  {year} {1954})}\BibitemShut {NoStop}%
\bibitem [{\citenamefont {Townsend}(1980)}]{townsend1980structure}%
  \BibitemOpen
  \bibfield  {author} {\bibinfo {author} {\bibfnamefont {A.}~\bibnamefont
  {Townsend}},\ }\href@noop {} {\emph {\bibinfo {title} {The structure of
  turbulent shear flow}}}\ (\bibinfo  {publisher} {Cambridge university
  press},\ \bibinfo {year} {1980})\BibitemShut {NoStop}%
\bibitem [{\citenamefont {Raupach}, \citenamefont {Antonia},\ and\
  \citenamefont {Rajagopalan}(1991)}]{raupach1991rough}%
  \BibitemOpen
  \bibfield  {author} {\bibinfo {author} {\bibfnamefont {M.}~\bibnamefont
  {Raupach}}, \bibinfo {author} {\bibfnamefont {R.}~\bibnamefont {Antonia}},\
  and\ \bibinfo {author} {\bibfnamefont {S.}~\bibnamefont {Rajagopalan}},\
  }\bibfield  {title} {\enquote {\bibinfo {title} {Rough-wall turbulent
  boundary layers},}\ }\href@noop {} {\bibfield  {journal} {\bibinfo  {journal}
  {Applied mechanics reviews}\ }\textbf {\bibinfo {volume} {44}},\ \bibinfo
  {pages} {1--25} (\bibinfo {year} {1991})}\BibitemShut {NoStop}%
\bibitem [{\citenamefont {Volino}, \citenamefont {Schultz},\ and\ \citenamefont
  {Flack}(2011)}]{volino2011turbulence}%
  \BibitemOpen
  \bibfield  {author} {\bibinfo {author} {\bibfnamefont {R.~J.}\ \bibnamefont
  {Volino}}, \bibinfo {author} {\bibfnamefont {M.~P.}\ \bibnamefont
  {Schultz}},\ and\ \bibinfo {author} {\bibfnamefont {K.~A.}\ \bibnamefont
  {Flack}},\ }\bibfield  {title} {\enquote {\bibinfo {title} {Turbulence
  structure in boundary layers over periodic two-and three-dimensional
  roughness},}\ }\href@noop {} {\bibfield  {journal} {\bibinfo  {journal}
  {Journal of Fluid Mechanics}\ }\textbf {\bibinfo {volume} {676}},\ \bibinfo
  {pages} {172--190} (\bibinfo {year} {2011})}\BibitemShut {NoStop}%
\bibitem [{\citenamefont {Volino}, \citenamefont {Schultz},\ and\ \citenamefont
  {Flack}(2009)}]{volino2009turbulence}%
  \BibitemOpen
  \bibfield  {author} {\bibinfo {author} {\bibfnamefont {R.}~\bibnamefont
  {Volino}}, \bibinfo {author} {\bibfnamefont {M.}~\bibnamefont {Schultz}},\
  and\ \bibinfo {author} {\bibfnamefont {K.}~\bibnamefont {Flack}},\ }\bibfield
   {title} {\enquote {\bibinfo {title} {Turbulence structure in a boundary
  layer with two-dimensional roughness},}\ }\href@noop {} {\bibfield  {journal}
  {\bibinfo  {journal} {Journal of Fluid Mechanics}\ }\textbf {\bibinfo
  {volume} {635}},\ \bibinfo {pages} {75--101} (\bibinfo {year}
  {2009})}\BibitemShut {NoStop}%
\bibitem [{\citenamefont {Krogstad}\ and\ \citenamefont
  {Efros}(2012)}]{krogstad2012turbulence}%
  \BibitemOpen
  \bibfield  {author} {\bibinfo {author} {\bibfnamefont {P.-{\AA}.}\
  \bibnamefont {Krogstad}}\ and\ \bibinfo {author} {\bibfnamefont
  {V.}~\bibnamefont {Efros}},\ }\bibfield  {title} {\enquote {\bibinfo {title}
  {About turbulence statistics in the outer part of a boundary layer developing
  over two-dimensional surface roughness},}\ }\href@noop {} {\bibfield
  {journal} {\bibinfo  {journal} {Physics of Fluids}\ }\textbf {\bibinfo
  {volume} {24}},\ \bibinfo {pages} {075112} (\bibinfo {year}
  {2012})}\BibitemShut {NoStop}%
\bibitem [{\citenamefont {Thorsness}\ and\ \citenamefont
  {Hanratty}(1977)}]{thorsness1977turbulent}%
  \BibitemOpen
  \bibfield  {author} {\bibinfo {author} {\bibfnamefont {C.}~\bibnamefont
  {Thorsness}}\ and\ \bibinfo {author} {\bibfnamefont {T.}~\bibnamefont
  {Hanratty}},\ }\bibfield  {title} {\enquote {\bibinfo {title} {Turbulent flow
  over wavy surfaces},}\ }in\ \href@noop {} {\emph {\bibinfo {booktitle} {Proc.
  Symp. Turbulent Flows, Pennsylvania State Univ}}}\ (\bibinfo {year}
  {1977})\BibitemShut {NoStop}%
\bibitem [{\citenamefont {Zilker}, \citenamefont {Cook},\ and\ \citenamefont
  {Hanratty}(1977)}]{zilker1977influence}%
  \BibitemOpen
  \bibfield  {author} {\bibinfo {author} {\bibfnamefont {D.~P.}\ \bibnamefont
  {Zilker}}, \bibinfo {author} {\bibfnamefont {G.~W.}\ \bibnamefont {Cook}},\
  and\ \bibinfo {author} {\bibfnamefont {T.~J.}\ \bibnamefont {Hanratty}},\
  }\bibfield  {title} {\enquote {\bibinfo {title} {Influence of the amplitude
  of a solid wavy wall on a turbulent flow. part 1. non-separated flows},}\
  }\href@noop {} {\bibfield  {journal} {\bibinfo  {journal} {Journal of Fluid
  Mechanics}\ }\textbf {\bibinfo {volume} {82}},\ \bibinfo {pages} {29--51}
  (\bibinfo {year} {1977})}\BibitemShut {NoStop}%
\bibitem [{\citenamefont {Zilker}\ and\ \citenamefont
  {Hanratty}(1979)}]{zilker1979influence}%
  \BibitemOpen
  \bibfield  {author} {\bibinfo {author} {\bibfnamefont {D.~P.}\ \bibnamefont
  {Zilker}}\ and\ \bibinfo {author} {\bibfnamefont {T.~J.}\ \bibnamefont
  {Hanratty}},\ }\bibfield  {title} {\enquote {\bibinfo {title} {Influence of
  the amplitude of a solid wavy wall on a turbulent flow. part 2. separated
  flows},}\ }\href@noop {} {\bibfield  {journal} {\bibinfo  {journal} {Journal
  of Fluid Mechanics}\ }\textbf {\bibinfo {volume} {90}},\ \bibinfo {pages}
  {257--271} (\bibinfo {year} {1979})}\BibitemShut {NoStop}%
\bibitem [{\citenamefont {Hudson}, \citenamefont {Dykhno},\ and\ \citenamefont
  {Hanratty}(1996)}]{hudson1996turbulence}%
  \BibitemOpen
  \bibfield  {author} {\bibinfo {author} {\bibfnamefont {J.~D.}\ \bibnamefont
  {Hudson}}, \bibinfo {author} {\bibfnamefont {L.}~\bibnamefont {Dykhno}},\
  and\ \bibinfo {author} {\bibfnamefont {T.}~\bibnamefont {Hanratty}},\
  }\bibfield  {title} {\enquote {\bibinfo {title} {Turbulence production in
  flow over a wavy wall},}\ }\href@noop {} {\bibfield  {journal} {\bibinfo
  {journal} {Experiments in Fluids}\ }\textbf {\bibinfo {volume} {20}},\
  \bibinfo {pages} {257--265} (\bibinfo {year} {1996})}\BibitemShut {NoStop}%
\bibitem [{\citenamefont {Buckles}, \citenamefont {Hanratty},\ and\
  \citenamefont {Adrian}(1984)}]{buckles1984turbulent}%
  \BibitemOpen
  \bibfield  {author} {\bibinfo {author} {\bibfnamefont {J.}~\bibnamefont
  {Buckles}}, \bibinfo {author} {\bibfnamefont {T.~J.}\ \bibnamefont
  {Hanratty}},\ and\ \bibinfo {author} {\bibfnamefont {R.~J.}\ \bibnamefont
  {Adrian}},\ }\bibfield  {title} {\enquote {\bibinfo {title} {Turbulent flow
  over large-amplitude wavy surfaces},}\ }\href@noop {} {\bibfield  {journal}
  {\bibinfo  {journal} {Journal of Fluid Mechanics}\ }\textbf {\bibinfo
  {volume} {140}},\ \bibinfo {pages} {27--44} (\bibinfo {year}
  {1984})}\BibitemShut {NoStop}%
\bibitem [{\citenamefont {Perry}, \citenamefont {Lim},\ and\ \citenamefont
  {Henbest}(1987)}]{perry1987experimental}%
  \BibitemOpen
  \bibfield  {author} {\bibinfo {author} {\bibfnamefont {A.}~\bibnamefont
  {Perry}}, \bibinfo {author} {\bibfnamefont {K.}~\bibnamefont {Lim}},\ and\
  \bibinfo {author} {\bibfnamefont {S.}~\bibnamefont {Henbest}},\ }\bibfield
  {title} {\enquote {\bibinfo {title} {An experimental study of the turbulence
  structure in smooth-and rough-wall boundary layers},}\ }\href@noop {}
  {\bibfield  {journal} {\bibinfo  {journal} {Journal of Fluid Mechanics}\
  }\textbf {\bibinfo {volume} {177}},\ \bibinfo {pages} {437--466} (\bibinfo
  {year} {1987})}\BibitemShut {NoStop}%
\bibitem [{\citenamefont {Leonardi}\ \emph {et~al.}(2003)\citenamefont
  {Leonardi}, \citenamefont {Orlandi}, \citenamefont {Smalley}, \citenamefont
  {Djenidi},\ and\ \citenamefont {Antonia}}]{leonardi2003direct}%
  \BibitemOpen
  \bibfield  {author} {\bibinfo {author} {\bibfnamefont {S.}~\bibnamefont
  {Leonardi}}, \bibinfo {author} {\bibfnamefont {P.}~\bibnamefont {Orlandi}},
  \bibinfo {author} {\bibfnamefont {R.}~\bibnamefont {Smalley}}, \bibinfo
  {author} {\bibfnamefont {L.}~\bibnamefont {Djenidi}},\ and\ \bibinfo {author}
  {\bibfnamefont {R.}~\bibnamefont {Antonia}},\ }\bibfield  {title} {\enquote
  {\bibinfo {title} {Direct numerical simulations of turbulent channel flow
  with transverse square bars on one wall},}\ }\href@noop {} {\bibfield
  {journal} {\bibinfo  {journal} {Journal of Fluid Mechanics}\ }\textbf
  {\bibinfo {volume} {491}},\ \bibinfo {pages} {229--238} (\bibinfo {year}
  {2003})}\BibitemShut {NoStop}%
\bibitem [{\citenamefont {Nakato}\ \emph {et~al.}(1985)\citenamefont {Nakato},
  \citenamefont {Onogi}, \citenamefont {Himeno}, \citenamefont {Tanaka},\ and\
  \citenamefont {Suzuki}}]{nakato1985resistance}%
  \BibitemOpen
  \bibfield  {author} {\bibinfo {author} {\bibfnamefont {M.}~\bibnamefont
  {Nakato}}, \bibinfo {author} {\bibfnamefont {H.}~\bibnamefont {Onogi}},
  \bibinfo {author} {\bibfnamefont {Y.}~\bibnamefont {Himeno}}, \bibinfo
  {author} {\bibfnamefont {I.}~\bibnamefont {Tanaka}},\ and\ \bibinfo {author}
  {\bibfnamefont {T.}~\bibnamefont {Suzuki}},\ }\bibfield  {title} {\enquote
  {\bibinfo {title} {Resistance due to surface roughness},}\ }in\ \href@noop {}
  {\emph {\bibinfo {booktitle} {Proceedings of the 15th Symposium on Naval
  Hydrodynamics}}}\ (\bibinfo {year} {1985})\ pp.\ \bibinfo {pages}
  {553--568}\BibitemShut {NoStop}%
\bibitem [{\citenamefont {Laizet}\ and\ \citenamefont
  {Lamballais}(2009)}]{laizet2009high}%
  \BibitemOpen
  \bibfield  {author} {\bibinfo {author} {\bibfnamefont {S.}~\bibnamefont
  {Laizet}}\ and\ \bibinfo {author} {\bibfnamefont {E.}~\bibnamefont
  {Lamballais}},\ }\bibfield  {title} {\enquote {\bibinfo {title} {High-order
  compact schemes for incompressible flows: A simple and efficient method with
  quasi-spectral accuracy},}\ }\href@noop {} {\bibfield  {journal} {\bibinfo
  {journal} {Journal of Computational Physics}\ }\textbf {\bibinfo {volume}
  {228}},\ \bibinfo {pages} {5989--6015} (\bibinfo {year} {2009})}\BibitemShut
  {NoStop}%
\bibitem [{\citenamefont {Peskin}(1972)}]{peskin1972flow}%
  \BibitemOpen
  \bibfield  {author} {\bibinfo {author} {\bibfnamefont {C.~S.}\ \bibnamefont
  {Peskin}},\ }\bibfield  {title} {\enquote {\bibinfo {title} {Flow patterns
  around heart valves: a numerical method},}\ }\href@noop {} {\bibfield
  {journal} {\bibinfo  {journal} {Journal of computational physics}\ }\textbf
  {\bibinfo {volume} {10}},\ \bibinfo {pages} {252--271} (\bibinfo {year}
  {1972})}\BibitemShut {NoStop}%
\bibitem [{\citenamefont {Parnaudeau}\ \emph {et~al.}(2004)\citenamefont
  {Parnaudeau}, \citenamefont {Lamballais}, \citenamefont {Heitz},\ and\
  \citenamefont {Silvestrini}}]{parnaudeau2004combination}%
  \BibitemOpen
  \bibfield  {author} {\bibinfo {author} {\bibfnamefont {P.}~\bibnamefont
  {Parnaudeau}}, \bibinfo {author} {\bibfnamefont {E.}~\bibnamefont
  {Lamballais}}, \bibinfo {author} {\bibfnamefont {D.}~\bibnamefont {Heitz}},\
  and\ \bibinfo {author} {\bibfnamefont {J.~H.}\ \bibnamefont {Silvestrini}},\
  }\bibfield  {title} {\enquote {\bibinfo {title} {Combination of the immersed
  boundary method with compact schemes for dns of flows in complex geometry},}\
  }in\ \href@noop {} {\emph {\bibinfo {booktitle} {Direct and Large-Eddy
  Simulation V}}}\ (\bibinfo  {publisher} {Springer},\ \bibinfo {year} {2004})\
  pp.\ \bibinfo {pages} {581--590}\BibitemShut {NoStop}%
\bibitem [{\citenamefont {Kim}\ and\ \citenamefont
  {Moin}(1985)}]{kim1985application}%
  \BibitemOpen
  \bibfield  {author} {\bibinfo {author} {\bibfnamefont {J.}~\bibnamefont
  {Kim}}\ and\ \bibinfo {author} {\bibfnamefont {P.}~\bibnamefont {Moin}},\
  }\bibfield  {title} {\enquote {\bibinfo {title} {Application of a
  fractional-step method to incompressible navier-stokes equations},}\
  }\href@noop {} {\bibfield  {journal} {\bibinfo  {journal} {Journal of
  computational physics}\ }\textbf {\bibinfo {volume} {59}},\ \bibinfo {pages}
  {308--323} (\bibinfo {year} {1985})}\BibitemShut {NoStop}%
\bibitem [{\citenamefont {Gautier}, \citenamefont {Laizet},\ and\ \citenamefont
  {Lamballais}(2014)}]{gautier2014dns}%
  \BibitemOpen
  \bibfield  {author} {\bibinfo {author} {\bibfnamefont {R.}~\bibnamefont
  {Gautier}}, \bibinfo {author} {\bibfnamefont {S.}~\bibnamefont {Laizet}},\
  and\ \bibinfo {author} {\bibfnamefont {E.}~\bibnamefont {Lamballais}},\
  }\bibfield  {title} {\enquote {\bibinfo {title} {A dns study of jet control
  with microjets using an immersed boundary method},}\ }\href@noop {}
  {\bibfield  {journal} {\bibinfo  {journal} {International Journal of
  Computational Fluid Dynamics}\ }\textbf {\bibinfo {volume} {28}},\ \bibinfo
  {pages} {393--410} (\bibinfo {year} {2014})}\BibitemShut {NoStop}%
\bibitem [{\citenamefont {Kim}, \citenamefont {Moin},\ and\ \citenamefont
  {Moser}(1987)}]{kim1987turbulence}%
  \BibitemOpen
  \bibfield  {author} {\bibinfo {author} {\bibfnamefont {J.}~\bibnamefont
  {Kim}}, \bibinfo {author} {\bibfnamefont {P.}~\bibnamefont {Moin}},\ and\
  \bibinfo {author} {\bibfnamefont {R.}~\bibnamefont {Moser}},\ }\bibfield
  {title} {\enquote {\bibinfo {title} {Turbulence statistics in fully developed
  channel flow at low reynolds number},}\ }\href@noop {} {\bibfield  {journal}
  {\bibinfo  {journal} {Journal of fluid mechanics}\ }\textbf {\bibinfo
  {volume} {177}},\ \bibinfo {pages} {133--166} (\bibinfo {year}
  {1987})}\BibitemShut {NoStop}%
\bibitem [{\citenamefont {Nagib}\ and\ \citenamefont
  {Chauhan}(2008)}]{nagib2008variations}%
  \BibitemOpen
  \bibfield  {author} {\bibinfo {author} {\bibfnamefont {H.~M.}\ \bibnamefont
  {Nagib}}\ and\ \bibinfo {author} {\bibfnamefont {K.~A.}\ \bibnamefont
  {Chauhan}},\ }\bibfield  {title} {\enquote {\bibinfo {title} {Variations of
  von k{\'a}rm{\'a}n coefficient in canonical flows},}\ }\href@noop {}
  {\bibfield  {journal} {\bibinfo  {journal} {Physics of Fluids}\ }\textbf
  {\bibinfo {volume} {20}},\ \bibinfo {pages} {101518} (\bibinfo {year}
  {2008})}\BibitemShut {NoStop}%
\bibitem [{\citenamefont {Pope}(2001)}]{pope2001turbulent}%
  \BibitemOpen
  \bibfield  {author} {\bibinfo {author} {\bibfnamefont {S.~B.}\ \bibnamefont
  {Pope}},\ }\href@noop {} {\enquote {\bibinfo {title} {Turbulent flows},}\ }
  (\bibinfo {year} {2001})\BibitemShut {NoStop}%
\bibitem [{\citenamefont {Ganju}\ \emph {et~al.}(2019)\citenamefont {Ganju},
  \citenamefont {Davis}, \citenamefont {Bailey},\ and\ \citenamefont
  {Brehm}}]{ganju2019direct}%
  \BibitemOpen
  \bibfield  {author} {\bibinfo {author} {\bibfnamefont {S.}~\bibnamefont
  {Ganju}}, \bibinfo {author} {\bibfnamefont {J.}~\bibnamefont {Davis}},
  \bibinfo {author} {\bibfnamefont {S.~C.}\ \bibnamefont {Bailey}},\ and\
  \bibinfo {author} {\bibfnamefont {C.}~\bibnamefont {Brehm}},\ }\bibfield
  {title} {\enquote {\bibinfo {title} {Direct numerical simulations of
  turbulent channel flows with sinusoidal walls},}\ }in\ \href@noop {} {\emph
  {\bibinfo {booktitle} {AIAA Scitech 2019 Forum}}}\ (\bibinfo {year} {2019})\
  p.\ \bibinfo {pages} {2141}\BibitemShut {NoStop}%
\bibitem [{\citenamefont {Florens}, \citenamefont {Eiff},\ and\ \citenamefont
  {Moulin}(2013)}]{florens2013defining}%
  \BibitemOpen
  \bibfield  {author} {\bibinfo {author} {\bibfnamefont {E.}~\bibnamefont
  {Florens}}, \bibinfo {author} {\bibfnamefont {O.}~\bibnamefont {Eiff}},\ and\
  \bibinfo {author} {\bibfnamefont {F.}~\bibnamefont {Moulin}},\ }\bibfield
  {title} {\enquote {\bibinfo {title} {Defining the roughness sublayer and its
  turbulence statistics},}\ }\href@noop {} {\bibfield  {journal} {\bibinfo
  {journal} {Experiments in fluids}\ }\textbf {\bibinfo {volume} {54}},\
  \bibinfo {pages} {1500} (\bibinfo {year} {2013})}\BibitemShut {NoStop}%
\end{thebibliography}%

\section*{Appendix}
\vspace{-2em}
\begin{figure}[ht!]
	\centering
	\mbox{
		\subfigure[\label{fig:prof_ConvectiveTransport_C11}]{\includegraphics[width=0.21\textwidth]{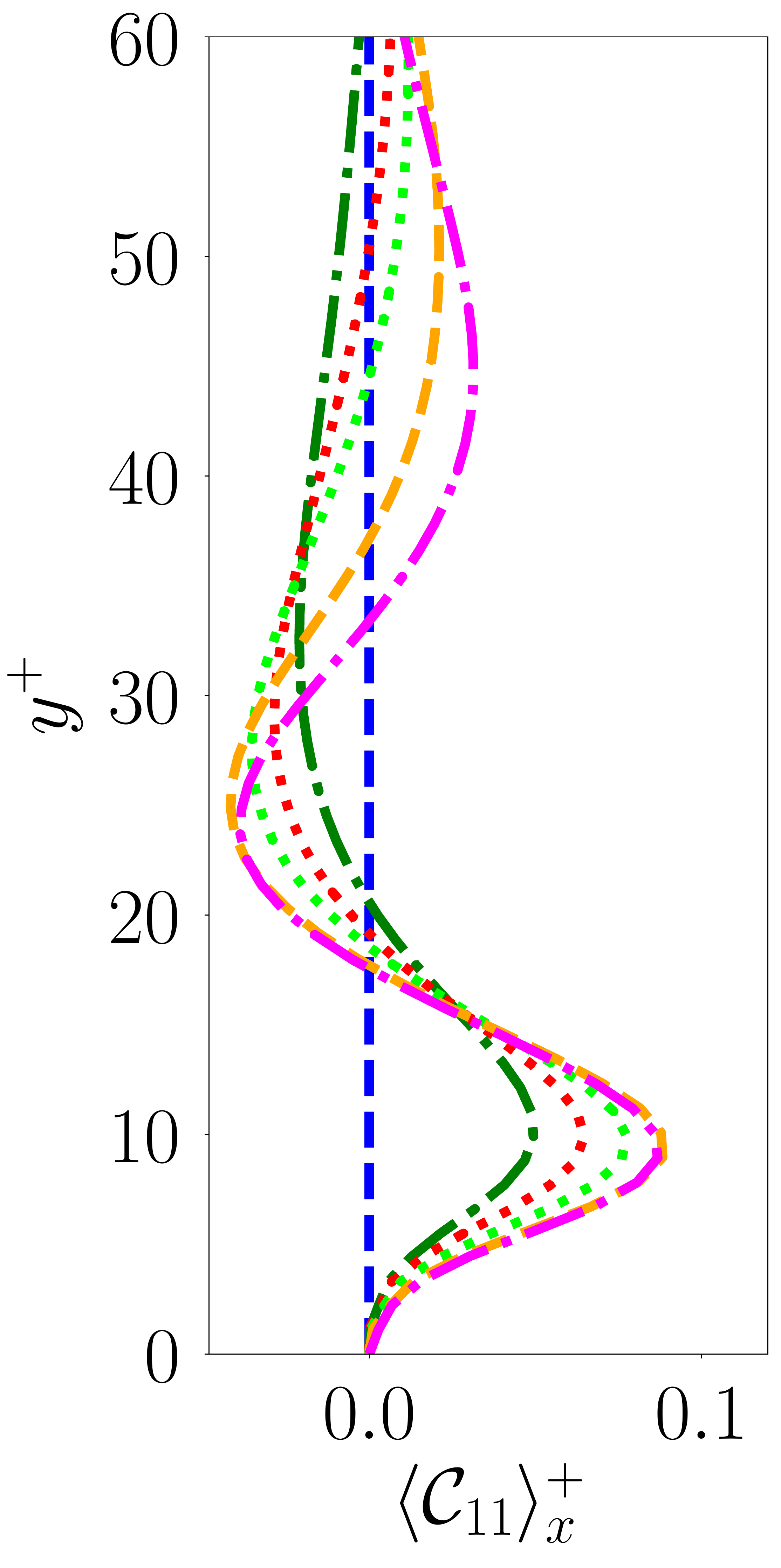}}
		\subfigure[\label{fig:prof_ConvectiveTransport_C22}]{\includegraphics[width=0.21\textwidth]{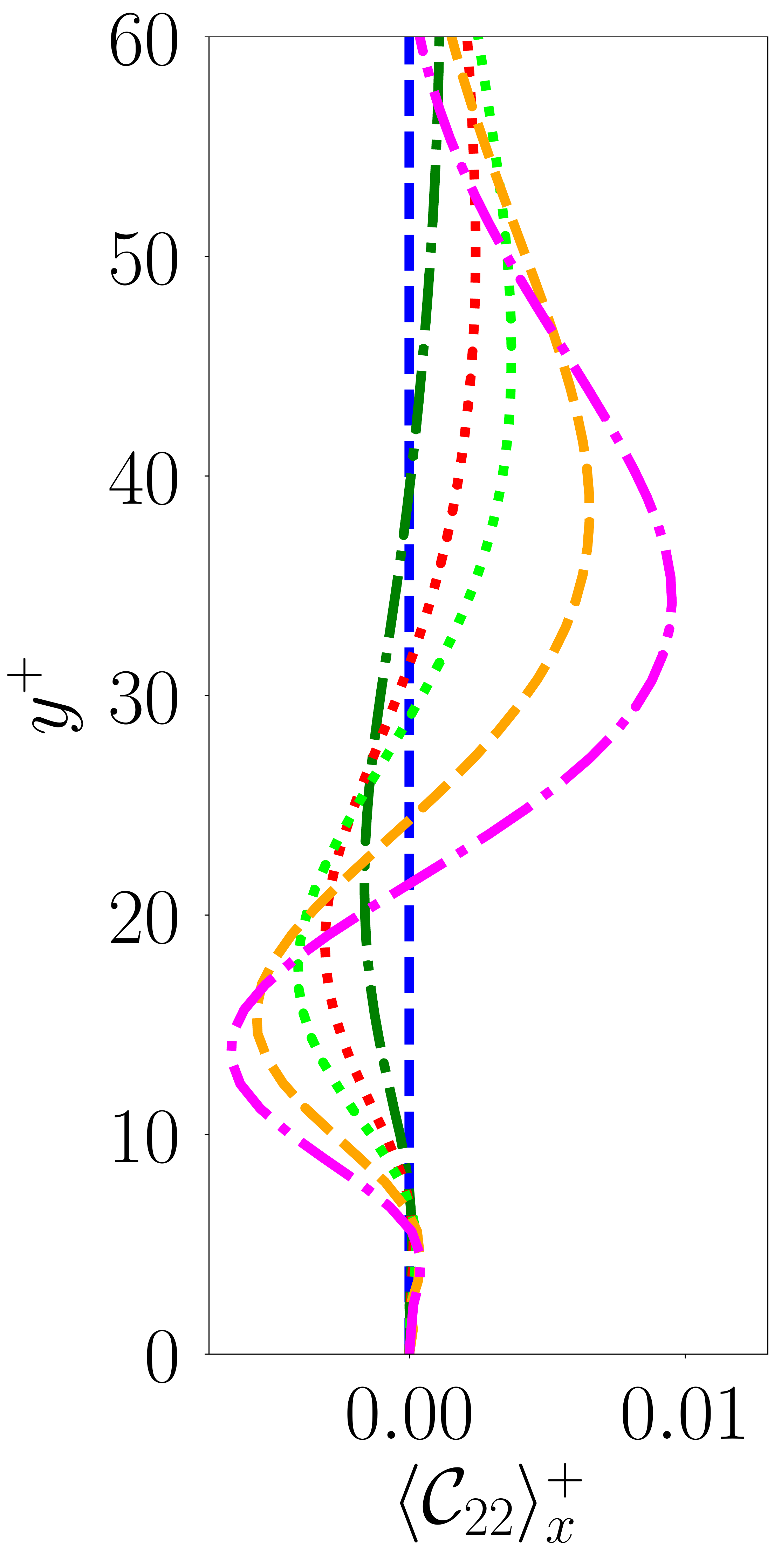}}
		\subfigure[\label{fig:prof_ConvectiveTransport_C33}]{\includegraphics[width=0.21\textwidth]{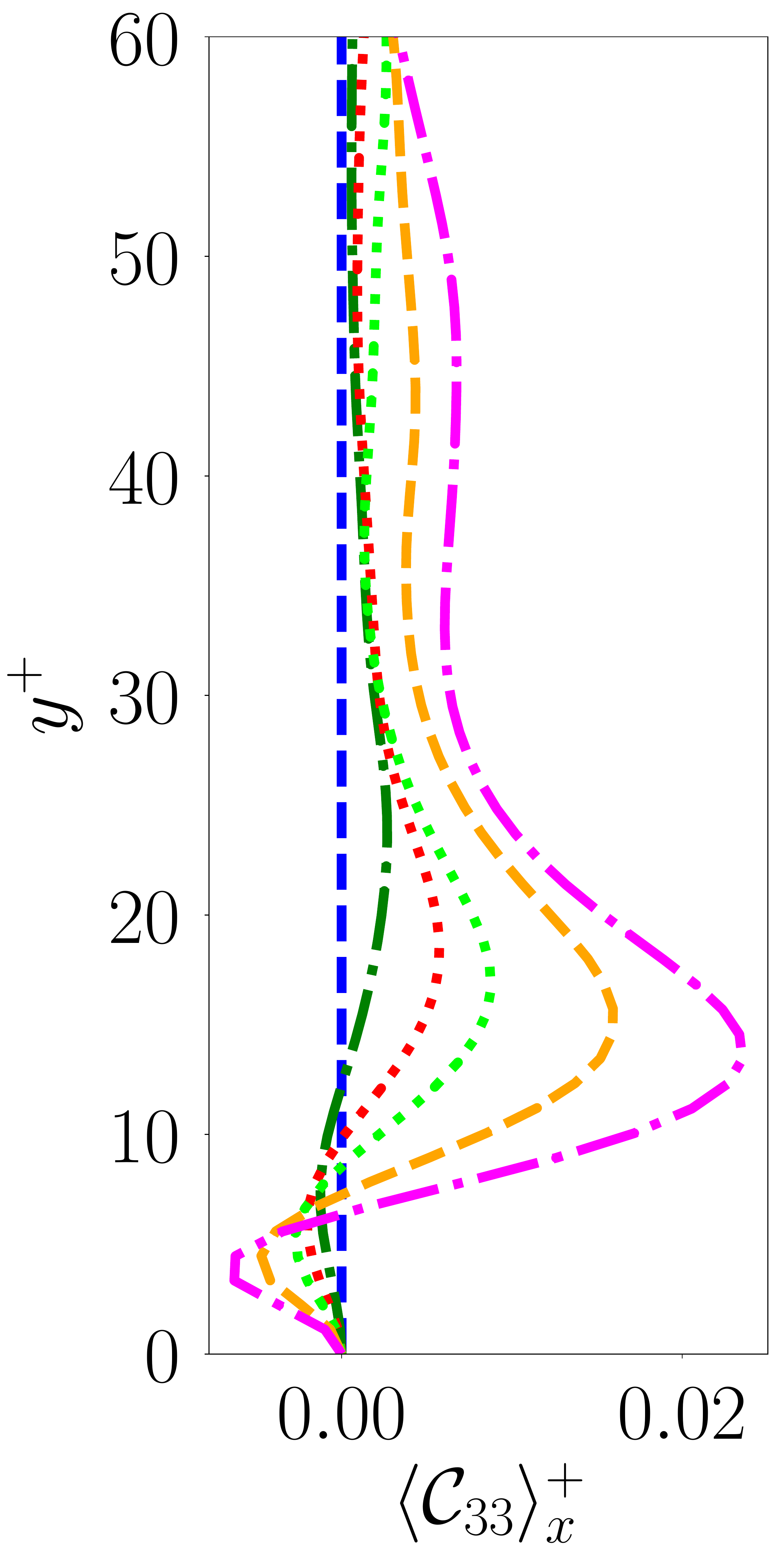}}
	}
	\caption{Vertical variation of the inner scaled averaged convective transport of streamwise (a), vertical (b) and spanwise (c) variance. Mathematical description of the convective terms can be found in section~\ref{sec:results-secondorder}. \label{fig:prof_ConvectiveTransport}}.
\end{figure}
\end{document}